\documentclass[a4paper, 12pt]{article}
\usepackage{braket,comment,float}

\usepackage[utf8]{inputenc}
\usepackage{a4wide}
\usepackage{amsmath}
\usepackage{cite}
\usepackage{units}
\usepackage[usenames,dvipsnames]{xcolor} 
\usepackage{amssymb}
\usepackage{amsfonts}
\usepackage{longtable}
\usepackage{booktabs}
\usepackage{makecell}
\usepackage{multirow}
\usepackage{pdflscape}
\usepackage{textcomp}
\usepackage{gensymb}
\usepackage{blkarray,bigstrut}
\usepackage{array}
\usepackage{enumerate}
\usepackage[small,bf]{caption}
\usepackage{graphicx}
\usepackage{hyperref}
\usepackage{mathbbol}
\usepackage{appendix}
\usepackage{verbatim}
\usepackage{geometry}
\usepackage{makecell}
\usepackage{listings}
\usepackage{mathtools}
\usepackage{dsfont}
\usepackage{arydshln}
\usepackage{tabularx}
\usepackage{mathrsfs} 
\usepackage{slashed}
\usepackage[abbreviations,symbols,nogroupskip,nonumberlist,automake]{glossaries-extra}
\usepackage{tikz}
\usepackage[compat=1.1.0]{tikz-feynman}
\tikzfeynmanset{warn luatex=false}
\usetikzlibrary{fit,matrix}
\usepackage{fancyhdr}
\usepackage{afterpage}
\usepackage{cleveref}
\usepackage[normalem]{ulem}

\makeglossaries

\glsxtrnewsymbol[description={number of isosinglet $q$-type vector-like quarks ($q=u,d$)}]{nq}{\ensuremath{n_q}}
\glsxtrnewsymbol[description={quark mixing matrix, of dimension $(3+n_u)\times (3+n_d)$}]{V}{\ensuremath{V}}
\glsxtrnewsymbol[description={$3\times 3$ CKM matrix (upper-left block of $V$, not necessarily unitary)}]{VCKM}{\ensuremath{V_\textrm{\normalfont CKM}}}
\glsxtrnewsymbol[description={exact matrix parameterizing deviations from unitarity}]{X}{\ensuremath{X}}
\glsxtrnewsymbol[description={approximate matrix parameterizing deviations from unitarity}]{Theta}{\ensuremath{\Theta}}
\glsxtrnewsymbol[description={matrices controlling FCNC ($q=u,d$)}]{Fq}{\ensuremath{F^q}}
\glsxtrnewsymbol[description={$q$-type quark mass matrix ($q=u,d$)}]{Mq}{\ensuremath{\mathcal{M}_q}}
\glsxtrnewsymbol[description={first 3 rows of $\mathcal{V}^q_\chi$ ($q=u,d$; $\chi=L,R$)}]{A}{\ensuremath{A^q_\chi}}
\glsxtrnewsymbol[description={last $n_q$ rows of $\mathcal{V}^q_\chi$ ($q=u,d$; $\chi=L,R$)}]{B}{\ensuremath{B^q_\chi}}
\glsxtrnewsymbol[description={$(3+n_q)$-dimensional unitary diagonalization matrix ($q=u,d$; $\chi=L,R$)}]{mV}{\ensuremath{\mathcal{V}^q_\chi}}

\newabbreviation{LHC}{LHC}{Large Hadron Collider}
\newabbreviation{VLQ}{VLQ}{vector-like quark}
\newabbreviation{NP}{NP}{new physics}
\newabbreviation{SM}{SM}{Standard Model}
\newabbreviation{CKM}{CKM}{Cabibbo-Kobayashi-Maskawa}
\newabbreviation{PMNS}{PMNS}{Pontecorvo-Maki-Nakagawa-Sakata}
\newabbreviation{FCNC}{FCNC}{flavour-changing neutral currents}
\newabbreviation{VEV}{VEV}{vacuum expectation value}
\newabbreviation{EWSB}{EWSB}{electroweak symmetry breaking}
\newabbreviation{BR}{BR}{branching ratio} 
\newabbreviation{NB}{NB}{Nelson-Barr} 
\newabbreviation{ED}{ED}{extra dimensions}
\newabbreviation{SMEFT}{SMEFT}{SM effective field theory}
\newabbreviation{CPV}{CPV}{CP violation}
\newabbreviation[description={weak basis}]{WB}{WB}{weak-basis}
\newabbreviation{BBP}{BBP}{Bento-Branco-Parada}

\newabbreviation{KK}{KK}{Kaluza-Klein}
\newabbreviation[
  longplural={Grand Unified Theories}
]{GUT}{GUT}{Grand Unified Theory}
\newabbreviation{UV}{UV}{ultraviolet}
\newabbreviation{RS}{RS}{Randall-Sundrum}
\newabbreviation{HNL}{HNL}{heavy neutral leptons}

\setabbreviationstyle[acronym]{long-short}
\glslocalunseteach{LHC}

\crefformat{pluralequation}{eqs.~(#2#1#3)}
\Crefformat{pluralequation}{Eqs.~(#2#1#3)}

\crefmultiformat{pluralequation}{eqs.~(#2#1#3)}{ and~(#2#1#3)}{, (#2#1#3)}{ and~(#2#1#3)}

\providecommand{\eps}{\epsilon}
\providecommand{\cM}{\mathcal{M}}
\providecommand{\cV}{\mathcal{V}}
\providecommand{\om}{\overline{m\hspace{-.2ex}}\hspace{.2ex}}
\providecommand{\oM}{\overline{\!M}\mkern-2mu}
\providecommand{\oYd}{\overline{Y}\!_d}
\providecommand{\oYu}{\overline{Y}\!_u}
\newcommand\oY{\overline{Y}\!}
\providecommand{\oR}{\overline{\!R}\,}
\providecommand{\oK}{\overline{\!K}\,}
\providecommand{\eq}[1]{\begin{equation} #1 \end{equation}}
\providecommand{\eqali}[1]{\begin{equation}\begin{aligned} #1
    \end{aligned}\end{equation}}
\providecommand{\subeqali}[2][]{\begin{subequations}\begin{align}
#2    \end{align}\end{subequations}}
\providecommand{\mtrx}[1]{\begin{pmatrix} #1 \end{pmatrix}}
\providecommand{\ml}[1]{\mbox{\large $#1$}}
\providecommand{\ums}[2][1]{\ml{\tfrac{#1}{#2}}}

\providecommand{\thetabar}{\bar{\theta}}

\providecommand{\cO}{\mathcal{O}}

\providecommand{\bs}[1]{\boldsymbol{#1}}
\providecommand{\aver}[1]{\langle #1 \rangle}

\providecommand{\ZZ}{\mathbb{Z}}

\usepackage{easybmat}
\usepackage{bigdelim}

\makeatletter
\newcommand{\overleftrightsmallarrow}{\mathpalette{\overarrowsmall@\leftrightarrowfill@}}
\newcommand{\overrightsmallarrow}{\mathpalette{\overarrowsmall@\rightarrowfill@}}
\newcommand{\overleftsmallarrow}{\mathpalette{\overarrowsmall@\leftarrowfill@}}
\newcommand{\overarrowsmall@}[3]{%
  \vbox{%
    \ialign{%
      ##\crcr
      #1{\smaller@style{#2}}\crcr
      \noalign{\nointerlineskip}%
      $\m@th\hfil#2#3\hfil$\crcr
    }%
  }%
}
\def\smaller@style#1{%
  \ifx#1\displaystyle\scriptstyle\else
    \ifx#1\textstyle\scriptstyle\else
      \scriptscriptstyle
    \fi
  \fi
}
\makeatother
\newcommand{\te}[1]{\overleftrightsmallarrow{#1}}

\providecommand{\btheta}{\bar{\theta}}
\providecommand{\lag}{\mathscr{L}}

\newcommand\topstrutN[1][1.2ex]{\setlength\bigstrutjot{#1}{\bigstrut[t]}}
\newcommand\botstrutN[1][0.9ex]{\setlength\bigstrutjot{#1}{\bigstrut[b]}}

\newcommand{\id}{\mathds{1}}
\DeclareMathOperator{\diag}{diag}
\DeclareMathOperator{\tr}{Tr}
\DeclareMathOperator{\im}{Im}
\DeclareMathOperator{\re}{Re}

\newcommand{\pH}{\vartheta}
\newcommand{\br}{\text{BR}}
\newcommand{\effHam}{\mathcal{H}_\text{eff}}

\newcommand\myunder[2]{\mathrlap{\smash{\underbrace{\phantom{%
    \begin{matrix} #2 \end{matrix}}}_{\mbox{$#1$}}}}#2}
\newcommand\myrightbrace[2]{%
\left.\vphantom{\begin{matrix} #1 \end{matrix}}\right\}#2}

\allowdisplaybreaks
\numberwithin{equation}{section}

\begin{document}
\begin{titlepage}
\begin{center}

${ }$\vspace{0.2cm}

{\large \bf {Vector-like Singlet Quarks: a Roadmap}}

\vspace{1.0cm}

\renewcommand*{\thefootnote}{\fnsymbol{footnote}}
Jo\~ ao M. Alves$^{~a,}$\footnote[1]{\href{mailto:j.magalhaes.alves@tecnico.ulisboa.pt}{\tt j.magalhaes.alves@tecnico.ulisboa.pt}}
G. C. Branco$^{~a,}$\footnote[2]{\href{mailto:gbranco@tecnico.ulisboa.pt}{\tt gbranco@tecnico.ulisboa.pt}}
A. L. Cherchiglia$^{~b,c}$\footnote[3]{\href{mailto:alche@unicamp.br}{\tt alche@unicamp.br}}
C. C. Nishi$^{~d,}$\footnote[4]{\href{mailto:celso.nishi@ufabc.edu.br}{\tt celso.nishi@ufabc.edu.br}}\\[2mm]
J. T. Penedo$^{~a,}$\footnote[5]{\href{mailto:joao.t.n.penedo@tecnico.ulisboa.pt}{\tt joao.t.n.penedo@tecnico.ulisboa.pt}} 
Pedro~M.~F.~Pereira$^{~a,}$\footnote[6]{\href{mailto:pedromanuelpereira@tecnico.ulisboa.pt}{\tt pedromanuelpereira@tecnico.ulisboa.pt}}
M. N. Rebelo$^{~a,}$\footnote[7]{\href{mailto:rebelo@tecnico.ulisboa.pt}{\tt rebelo@tecnico.ulisboa.pt}} and
J.~I.~Silva-Marcos$^{~a,}$\footnote[8]{\href{mailto:juca@cftp.tecnico.ulisboa.pt}{\tt juca@cftp.tecnico.ulisboa.pt}}
\renewcommand*{\thefootnote}{\arabic{footnote}}
\setcounter{footnote}{0}

\vspace{0.5cm}

$^{a}$~Centro de F\'{\i}sica Te\'orica de Part\'{\i}culas, CFTP, Departamento de F\'{\i}sica,\\
Instituto Superior T\'ecnico, Universidade de Lisboa,\\
Av.~Rovisco Pais 1, 1049-001 Lisboa, Portugal
\\[2mm]
$^{b}$~Instituto de Física Gleb Wataghin, Universidade Estadual de Campinas, \\
Rua Sérgio Buarque de Holanda, 777, Campinas, SP, Brasil\\[2mm]
$^{c}$~Departamento de F\'isica Te\'orica y del Cosmos, Universidad de Granada, \\
Campus de Fuentenueva, E–18071 Granada, Spain\\[2mm]
$^{d}$~Centro de Matemática, Computação e Cognição,\\
Universidade Federal do ABC -- UFABC, 09.210-170,
Santo André, SP, Brazil
\end{center}

\vskip 0.7cm

\begin{abstract}
We review the theory and phenomenology of isosinglet vector-like quarks (VLQs). In recent years, interest in VLQs has been increasing, due to their contributions to new physics effects that can be tested in experiments at LHC and High-Luminosity LHC. The similarities of models with isosinglet VLQs and the seesaw framework in the leptonic sector are pointed out. The existence of VLQs leads to flavour-changing neutral currents at tree level and deviations from unitarity of the CKM matrix, introducing rich phenomenological implications. These new effects are naturally suppressed by the masses of the new quarks, that are constrained to be above the electroweak scale. In addition, striking new effects can be achieved with the inclusion of an extra complex scalar singlet. Such a minimal extension of the SM can give rise to new sources of CP violation with profound theoretical implications, allowing for a solution to the strong CP problem and a possible explanation for the baryon asymmetry of the Universe. We list and explain strong motivations to consider this class of models. We also briefly review how models with VLQs can be matched to the SM effective field theory (SMEFT). A detailed analysis of flavour observables that can be affected by the presence of VLQs is presented. Current bounds from collider searches of VLQs are summarized. We point out that the discovery of VLQs can be within the reach of present or future colliders being planned.
\end{abstract}

\end{titlepage}

\tableofcontents

\clearpage
\glstocfalse
\printglossaries

\vfill
\pagebreak
\section{Introduction}

The \gls*{SM} is thought to be the low-energy approximation of a more fundamental theory yet to be clarified. As the search for new physics beyond the SM progresses, the validity of the model is put under stringent scrutiny, while hints of deviations are expected to help guide this quest.

From the bottom-up, one of the simplest extensions of the SM consists in the introduction of \glspl*{VLQ}. VLQs are quarks with left- and right-handed components transforming in the same way under the SM gauge group $SU(3)_c\times SU(2)_L\times U(1)_Y$. While these extensions have to agree with the present experimental data, they also give rise to a rich variety of \gls*{NP} phenomena, which may be at the reach of the \gls*{LHC} and the next generation of experiments. There are seven different types of VLQ multiplets which can mix with the SM quarks through Yukawa couplings with the Higgs doublet~\cite{delAguila:2000aa}. These involve quark isosinglets, doublets or triplets of $SU(2)_L$. In this review, we focus exclusively on models with isosinglet quarks, which have the same quantum numbers as the SM right-handed quarks so that mixing follows naturally.
They mostly give corrections to the left-handed currents%
\footnote{Doublets induce right-handed currents.}
coupling to the $Z$ and $W$ bosons, similarly to the triplets but with a different sign.

From the top-down, vector-like fermions are also an important outcome of some of the most prominent \gls*{ED} models, such as the Randall and Sundrum warped geometry models~\cite{Randall:1999ee}. Indeed, vector-like fermions form an integral part of these ED models. For instance, when integrating out higher dimensions to obtain an effective four-dimensional action, several Kaluza-Klein modes of the quark fields become massive and appear as VLQs, with large masses of at least $\mathcal{O}(\unit{TeV})$.%
\footnote{
See e.g.~\cite{Carena:2007tn,Gopalakrishna:2011ef,Barcelo:2011wu,Gopalakrishna:2013hua} for the collider phenomenology of VLQs in the context of ED models.
}
Likewise, in some Grand Unified models, VLQs are often part of the spectrum. This is the case of $E_6$, whose fundamental representation includes an isosinglet down-type VLQ (see~\cite{Hewett:1988xc} and references within; see also~\cite{Sultansoy:2006cw,Kang:2007ib} for examples).
Notice that the introduction of VLQs does not worsen the gauge hierarchy problem~\cite{Ramond:1981jx} and is anomaly-free.
Their presence may aid in stabilizing the Higgs electroweak vacuum (see e.g.~\cite{Blum:2015rpa,Gopalakrishna:2018uxn,Hiller:2022rla}).
Also, in composite Higgs scenarios, not-too-heavy VLQs may be needed to allow for a realistic Higgs mass~\cite{Matsedonskyi:2012ym}.
More details on some ultraviolet motivations for VLQs are provided in~\cref{app:UV}.

\vskip 2mm

Bare mass terms for VLQs are gauge invariant and therefore are not protected by the gauge symmetry. Thus, their scale, $M$, can be significantly larger than the electroweak scale, $m$. The fact that VLQ masses are not fully generated by the couplings to scalar doublets contrasts with the situation of a hypothetical fourth (chiral) quark generation, which has already been ruled out~\cite{Erler:2004nh,Chen:2012wz,Djouadi:2012ae,Kuflik:2012ai,Eberhardt:2012gv}. Upon gauge symmetry breaking, mass terms of scale $m$ are generated. The presence of these terms together with the bare mass terms involving VLQs and some of the standard-like quarks leads to mixing among both types of quarks. As a result the charged-current couplings are no longer given by the $3\times 3$ unitary \gls*{CKM} matrix. Ordering the quarks by their masses and allowing the scale $M$ to be much larger than $m$ leads to naturally small deviations from unitarity of the $3\times 3$ block of the mixing matrix that plays the role of the SM CKM matrix. The crucial point is that this $3\times 3$ matrix is no longer unitary
and this violation of unitarity has a definite sign: there is a deficit.%
\footnote{Triplet VLQs induce violations of opposite sign.}

The small deviations from unitarity are proportional to the ratio $m/M$ of the two different scales. Another effect of the extended mixing involving both types of quarks is that the $Z$-mediated neutral currents may no longer be diagonal in flavour. Therefore, models with VLQs generically predict $Z$-mediated \gls*{FCNC}. These are also suppressed by the ratio $m/M$. In simple scenarios, the strength of FCNC can be expressed in terms of the deviations from unitarity of the $3\times 3$ CKM matrix. The Higgs couplings to SM quarks are also modified and Higgs-mediated FCNC may be generated. Their pattern is similar to that of $Z$-mediated FCNC, but involves additional factors of ratios of quark masses and the electroweak \gls*{VEV} $v$. For transitions involving only light quarks, these ratios act as further suppression factors. Clearly, it is the possibility of having two different scales for the mass terms that allows for a natural suppression of effects beyond the SM. Basic notation, together with some detailed derivations for models with isosinglet VLQs can be found in~\cref{sec:notation}.

The fact that, in the presence of VLQs, the $3\times 3$ CKM matrix is no longer unitary and there are FCNC at tree level may seem somewhat surprising, especially to those unfamiliar with VLQs. At this stage, it is useful to compare the situation in the lepton and quark sectors. The SM predicts massless neutrinos and no leptonic mixing. However, experiment has shown that neutrinos do have non-vanishing masses and do mix. The simplest extension of the SM which allows for neutrino masses and leptonic mixing consists of introducing right-handed neutrinos. Like VLQs, the latter also allow for gauge-invariant bare mass terms, although of Majorana nature. The presence of both Majorana and Dirac mass terms in the neutrino sector automatically leads, through the seesaw mechanism, to naturally small light neutrino masses and small deviations of $3\times 3$ unitarity of the \gls*{PMNS} matrix as well as naturally suppressed but non-vanishing tree-level FCNC in the neutrino sector. The situation in the quark and lepton sectors with VLQs and the seesaw mechanism presents many similarities. Further discussion can be found in~\cref{sec:neutrinos}.

\vskip 2mm

A remarkable aspect of \gls*{CPV} in the quark sector is the fact that in the SM there is no CP violation in the limit of degenerate masses of at least two same-charge quarks. At very high energies, for example at the TeV scale, there is no possibility of distinguishing between a $d$ quark jet and an $s$ quark jet. Therefore CP violation in the SM effectively disappears at this scale. A distinctive feature of extensions with VLQs is the fact that there is CP violation even in the limit $m_d=m_s$. Weak-basis and Higgs-basis invariants play an important role in the study of flavour and Higgs physics beyond the SM. In the presence of VLQs, it is possible to construct CP-odd weak-basis invariants which do not vanish in the above limit~\cite{delAguila:1997vn,Branco:1999fs,Albergaria:2022zaq}. Some of these invariants are introduced in~\cref{sec:WBphy}, where effective (squared) mass matrices are also discussed. 

\vskip 2mm

One of the strongest motivations for the introduction of singlet VLQs is the fact that they may provide one of the simplest solutions to the so-called CKM unitarity problem~\cite{Seng:2018yzq,Seng:2018qru,Belfatto:2019swo,Czarnecki:2019mwq,Cheung:2020vqm,Seng:2020wjq,Hayen:2020cxh,Shiells:2020fqp,FlavourLatticeAveragingGroupFLAG:2021npn,Crivellin:2022rhw}. Recently, using improved values for the form factors and radiative corrections associated to the relevant meson and neutron decay processes, the unitarity condition of the first row of the CKM matrix was analysed and a clear deviation from unitarity was found. In~\cref{sec:devunit}, we review the CKM unitarity problem and present some of the simplest solutions based on VLQs~\cite{Belfatto:2019swo,Belfatto:2021jhf,Branco:2021vhs,Botella:2021uxz}. This is done with the help of an especially useful exact parameterization of deviations from unitarity introduced in the literature by some of the authors of this review~\cite{Agostinho:2017wfs,Branco:2019avf}. Additionally, we discuss the perturbativity constraints on the deviations from unitarity and the intimate connection between the latter and FCNC. We should note that singlet VLQs by themselves may be responsible for solving other recent problems, see, for instance,~\cite{Cao:2022mif} related to the $W$-mass measurement by the CDF collaboration~\cite{CDF:2022hxs}.%
\footnote{
A VLQ doublet may help solve the CKM unitarity problem and, at the same time, account for the new $W$-mass measurement, while a simultaneous explanation of both anomalies using VLQ singlets is disfavoured~\cite{Belfatto:2023tbv}.
}

\vskip 2mm

Models with VLQs have a very rich phenomenology. Their presence affects a plethora of flavour observables such as meson mixing, oblique and electroweak parameters, among others. This is expected since, as already explained, in models with VLQs the presence of FCNC mediated by the $Z$ or Higgs bosons is unavoidable. Moreover, since the CKM matrix is no longer unitary, weak processes are also modified. In~\cref{sec:pheno}, we collect general formulas for a set of flavour observables affected by VLQs, which may help the reader perform their own phenomenological studies. We also briefly review how models with VLQs can be matched to the \gls*{SMEFT} in~\cref{sec:SMEFT}. 

\vskip 2mm

In the SM there is only one Higgs doublet and CP is violated explicitly through the introduction of complex coefficients in the Yukawa sector. Adding a second scalar doublet or a complex singlet are the two simplest extensions of the SM allowing for spontaneous CP violation putting CP breaking and electroweak symmetry breaking on the same footing. For CP to be spontaneously violated it must be a good symmetry of the Lagrangian, broken only by the vacuum (which must be complex). An important challenge for any realistic model of spontaneous CP violation is to generate both a CP-violating vacuum phase and a complex CKM matrix. Scalar singlets can couple directly to isosinglet VLQs, which in turn can mix with the SM-like quarks. Therefore, introducing a scalar singlet that acquires a complex VEV together with at least one VLQ is enough to generate a realistic CKM matrix, as will be shown in~\cref{sec:strongCP}. Since the singlet VEV introduces a new mass scale, in general the isosinglet VLQ may acquire mass terms of scale $M^\prime$ through the coupling to the scalar singlet, in addition to the bare mass terms of scale $M$. 
Interestingly, the limit in which only the bare mass of the isosinglet quark becomes very large leads to an unrealistic CKM matrix, showing that an interplay between the CP breaking scale and the VLQ quark masses must be present.

Another motivation for the introduction of VLQs is the fact that they provide one of the simplest solutions to the strong CP problem without axions, as suggested by Nelson and Barr~\cite{Nelson:1983zb,Barr:1984qx}. The strong CP problem arose from the 't Hooft solution to the U(1) problem~\cite{tHooft:1976rip,tHooft:1976snw}. An essential point of the solution is the inclusion in the Lagrangian of a CP-violating term with coefficient $\theta_{\text{QCD}}$ originating from the QCD vacuum. However, CP violation has not been observed in the strong interactions. Furthermore, $\theta_{\text{QCD}}$ is a free parameter. Experimentally, what is measurable is the combination $\overline{\theta} = \theta_\text{QCD} - \theta_{\text{weak}} $ which is an angle ranging from 0 to 2$\pi$. In fact, within the SM the electric dipole moment of the neutron, which is CP, P and T violating, is proportional to $\overline{\theta}$. The parameter $\theta_{\text{weak}}$ is given by $\theta_{\text{weak}} = \arg(\det \mathcal{M}_u \times \det \mathcal{M}_d) $, where $\mathcal{M}_{u,d}$ are the quark mass matrices. The experimental bound on the electric dipole moment of the neutron requires $\overline{\theta} < 5 \times 10^{-11}$~\cite{Abel:2020pzs}. This implies a strong cancellation between the two terms $\theta_\text{QCD}$ and $\theta_{\text{weak}}$ or the extreme smallness of both. Either way, this puzzling situation constitutes the strong CP problem. An elegant solution was proposed by Peccei and Quinn~\cite{Peccei:1977hh,Peccei:1977ur} which leads to the existence of axions. Another solution~\cite{Georgi:1978xz,Beg:1978mt,Mohapatra:1978fy} consists of assuming that CP is a good symmetry of the Lagrangian, which implies that $\theta_{\text{QCD}}$ vanishes in a natural way, and that ${\cal M}_u$ and ${\cal M}_d$ are such that $\theta_{\text{weak}}$ vanishes at least at tree level. A particular class of models that satisfy these conditions are the ones suggested by Nelson and Barr and a minimal realization of \gls*{NB} models was put forward in~\cite{Bento:1991ez}. \Cref{sec:strongCP} also discusses these aspects.

\vskip 2mm

Finally, a strong motivation to consider models with VLQs is the fact that there is the possibility of discovering them at present and future colliders, namely at the LHC and its high-luminosity upgrade. Experimental searches often rely on assumptions on VLQ \glspl*{BR}, and it is often assumed that the new heavy quarks preferentially couple to third-generation quarks. The analysis performed in this review is not restricted by this hypothesis. In VLQ models explaining deviations from unitarity of the CKM matrix (see~\cref{sec:devunit}), the decay of the heavy quark into the first generation may actually be favoured, going against the above assumption. Moreover, most searches focus on pair-produced VLQs. This mode of production only depends on QCD and is therefore model-independent. However, in the general case, the more model-dependent single production of VLQs may also need to be taken into account. In~\cref{sec:searches}, a detailed description of past and present experimental searches is presented, highlighting the main hypotheses of each search. An outlook for future searches is also given.

\vskip 2mm

This review addresses several of the fundamental physical implications of models with VLQs, which have attracted a lot of attention in the literature. For an earlier review, see, for instance, Ref.~\cite{Barger:1995dd} for the case of isosinglets only, and Ref.~\cite{Aguilar-Saavedra:2013qpa} for a comprehensive analysis of all seven possible representations and references therein. We hope to convince the reader that VLQs remain at present one of the most plausible extensions of the SM.

\vfill
\clearpage

\section{Setup and Notation}
\label{sec:notation}

There are only seven different types of \gls*{VLQ} multiplets which may mix with the \gls*{SM} quarks through Yukawa couplings with the Higgs doublet~\cite{delAguila:2000aa,delAguila:2000rc}.
They consist of isosinglets, doublets or triplets, all coloured in the fundamental representation.%
\footnote{For leptons, i.e.~colour singlets, there are only six possibilities~\cite{delAguila:2008pw}.}
The complete quantum numbers are listed in~\cref{tab:VLQ:irreps}.
In this review we shall be mainly concerned with extensions of the quark sector with vector-like singlet quarks $U$ and $D$, 
also sometimes denoted, here and in the literature, as $T$ and $B$, respectively. 
%
%
\begin{table}[t!]
  \centering
  \begin{tabular}{cccccccc}
    \toprule    
    Multiplet
 & $U$ & $D$ & $\mtrx{U\cr D}$ &  $\mtrx{X\cr U}$ &  $\mtrx{D\cr Y}$ &  $\mtrx{X\cr U\cr D}$ &  $\mtrx{U \cr D \cr Y}$
\\ \midrule
$SU(2)_L$ & $\mathbf{1}$ & $\mathbf{1}$ & $\mathbf{2}$ & $\mathbf{2}$ & $\mathbf{2}$ & $\mathbf{3}$ & $\mathbf{3}$
\\
$U(1)_Y$ & $2/3$ & $-1/3$ & $1/6$ & $7/6$ & $-5/6$ & $2/3$ & $-1/3$
\\ \bottomrule
  \end{tabular}
  \caption{
  The only VLQ multiplets that couple to the SM quarks through Yukawa interactions ($Q=I_3+Y$).}
  \label{tab:VLQ:irreps}
\end{table}
%

\subsection{Mass matrices}
\label{sec:massmatrices}
We start by settling our notation for the case of $n_u$ up ($U^0_{Lr}$, $U^0_{Rr}$) and $n_d$ down ($D^0_{Ls}$, $D^0_{Rs}$) vector-like isosinglet quarks. 
The flavour indices $r,s$ run from 1 to $n_u$ or $ n_d$, respectively, while the `0' superscript refers to the flavour basis.
With the introduction of these additional quarks, the Yukawa interactions in this basis read
\begin{equation}
\label{eq:lagrangian}
  \begin{split}
  \lag_{\text{Y}} \,=\, 
  -\,\Big[
  &\,\,\overline Q_{Li}^{0} \, \tilde\Phi \, \left(Y_u\right)_{ij}  \, u^0_{Rj}
  \,+\, \overline Q_{Li}^{0} \, \tilde\Phi \, \left(\oYu\right)_{is} \, U^0_{Rs} 
  \\ +
  &\,\,\overline Q_{Li}^{0} \, \Phi       \, \left(Y_d\right)_{ij}  \, d^0_{Rj} 
  \,+\,\overline Q_{Li}^{0} \, \Phi       \, \left(\oYd\right)_{ir} \, D^0_{Rr}
  \,\Big] + \text{h.c.}\,,
  \end{split}
\end{equation}
where the indices $i$ and $j$ run from 1 to 3 as in the SM and $Y_d$, $Y_u$ are the SM Yukawa couplings. Here, $Q_L = ( u_L  \,\,\, d_L)^T$ are the SM quark doublets, $u_R$ and $d_R$ are the SM quark singlets, $\Phi$ is the Higgs doublet and $\tilde\Phi = i \tau_2 \Phi^*$ is its $SU(2)_L$ conjugate. The matrices ${\oYd}$ and ${\oYu}$ denote the new Yukawa couplings to the extra right-handed fields.

In general, the following mass terms are also allowed and must therefore be introduced,
\begin{equation}
\label{eq:baremasses}
\begin{split}
  \lag_\text{b.m.} \,=\, 
  -\,\Big[
  &\,\, \overline U^0_{Lr} \, \left(\,\oM_u\right)_{r i} \, u^0_{Ri}
  \,+\, \overline U^0_{Lr} \, \left(M_u\right)_{r r'}   \, U^0_{Rr'}
  \\+
  &\,\, \overline D^0_{Ls} \, \left(\,\oM_d\right)_{s i} \, d^0_{Ri}
  \,+\, \overline D^0_{Ls} \, \left(M_d\right)_{s s'}   \, D^0_{Rs'}
  \,\Big] + \text{h.c.}\,,
\end{split}
\end{equation}
where $\,\oM_q$ and $M_q$ ($q=u,d$) are, respectively, $\gls*{nq}\times 3$ and $n_q \times n_q$ general complex matrices. At this level, they are understood as bare mass (b.m.) terms. Within specific models they may originate e.g.~from the VEV of a complex scalar singlet.

After spontaneous \gls*{EWSB}, one has
$\Phi = \big( 0  \,\,\,  \frac{v+h}{\sqrt{2}}\big)^T$
in the unitary gauge, where $h$ is the Higgs field and $v \simeq 246\;\unit{GeV}$.
Additional mass terms are generated from $\lag_\text{Y}$ of~\cref{eq:lagrangian}. All mass terms can be collected in a compact form,
\begin{equation}
\lag_\text{mass} \,=\, 
- \begin{pmatrix}
    \overline{u}^0_L &  \overline{U}^0_L
  \end{pmatrix}
  \,\mathcal{M}_u\,
  \begin{pmatrix}
    u^0_{R} \\[1mm] U^0_{R} 
  \end{pmatrix}
- \begin{pmatrix}
    \overline{d}^0_L &  \overline{D}^0_L
  \end{pmatrix}
  \,\mathcal{M}_d\,
  \begin{pmatrix}
    d^0_{R} \\[1mm] D^0_{R} 
  \end{pmatrix}
+ \text{h.c.}\,,
\end{equation}
and the matrices $\gls*{Mq}$ ($q=u,d$) explicitly read
\begin{align}
\renewcommand{\arraystretch}{1}
\setlength{\extrarowheight}{6pt}
\mathcal{M}_q \,=\,
\left(\begin{array}{c;{2pt/2pt}c}
\,m_q\,\, & \,\om_q\,
\\[1.5mm] \hdashline[2pt/2pt]
\!\myunder{\scriptstyle 3}{\,\,\,\oM_q\,\,\,}\! &
\!\myunder{\scriptstyle n_q}{\,\,\,M_q\,\,\,}\!\\[1mm]
\end{array}\right) \!\!\!
\begin{array}{l}
    \myrightbrace{\overline{M}_q}{\scriptstyle \,3}\\[0mm]
    \myrightbrace{M_q}{\scriptstyle \, n_q}
\end{array}
\,,
\label{eq:genmass}
\\[-3mm] \nonumber
\end{align}
where $m_q = \frac{v}{\sqrt{2}} Y_q$ and $\om_q = \frac{v}{\sqrt{2}} \overline{Y}\!_q$.
It is important to emphasize that, in general, the matrices $\mathcal{M}_q$ are not symmetric nor Hermitian. A priori, a hierarchy $\oM_q \sim M_q \gg \om_q \sim m_q$ may be expected.
In what follows we omit the indices $q=u,d$ whenever there is no risk of confusion.

The mass matrices $\mathcal{M}_q$ can be diagonalized by biunitary transformations --- their singular value decompositions --- of the form
\begin{align}
\renewcommand{\arraystretch}{1}
\setlength{\extrarowheight}{6pt}
\allowdisplaybreaks[0]
    {\mathcal{V}^q_L}^\dagger
    \,\,\mathcal{M}_q\,\,
    \mathcal{V}_R^q
    \,=\, \mathcal{D}_q\,,
    \quad \text{with } \mathcal{D}_q\,=\,
    \left(\begin{array}{c;{2pt/2pt}c}
      \,\,\,d_q\,\,\,\,\, & \,\,\,0\,
      \\[1.5mm] \hdashline[2pt/2pt]
      \!\myunder{\scriptstyle 3}{\,\,\,\,0\,\,\,\,\, }\! &
      \myunder{\scriptstyle \,\,\,n_q}{\,\,\,\,D_q\,\,\,\,}\!\\[1mm]
    \end{array}\right) \!\!\!
    \begin{array}{l}
      \myrightbrace{0}{\scriptstyle \,3}\\[0mm]
      \myrightbrace{D}{\scriptstyle \, n_q}
    \end{array}
\,,
\label{eq:diagonalization}
\\[-1mm] \nonumber
\end{align}
where $\mathcal{V}_{L,R}^q$ correspond to the four unitary rotations (two for each sector) which connect the flavour and physical (mass) bases. The diagonal matrices $d_q$ and $D_q$ contain the light $(d_q)_i \geq 0$ ($i = 1,2,3$) and heavy $(D_q)_r > 0$ ($r = 1,\ldots,n_q$) quark masses. It should be clear from the context that $D_q$ represents a matrix and not a down-type quark field. Where there is no risk of confusion, we will further employ the notations $d_u = \diag(m_u,m_c,m_t)$, $d_d = \diag(m_d,m_s,m_b)$, $D_u = \diag(m_{U1},\ldots,m_{Un_u})$, and $D_d = \diag(m_{D1},\ldots,m_{Dn_d})$.

\vskip 2mm

Independently of how one parameterizes the four diagonalization matrices $\gls*{mV}$ ($q=u,d$; $\chi=L,R$), it proves convenient to split them into two blocks,
\begin{equation}
\cV^q_\chi \,=\,
\setlength{\extrarowheight}{1.2pt}
    \left(\begin{array}{c}
     { }\\[-4mm]
      \qquad \gls*{A} \qquad 
      \\[2mm] \hdashline[2pt/2pt]
       { }\\[-4mm]
      \qquad \gls*{B} \qquad 
      \\[2mm]
    \end{array}\right) 
    \setlength{\extrarowheight}{6pt}
        \begin{array}{l}
      \myrightbrace{A}{\scriptstyle \,3}\\[0mm]
      \myrightbrace{B}{\scriptstyle \, n_q}
    \end{array}
\,,
\label{def:V:AB}
\end{equation}
where $A^q_\chi$ is a $3 \times (3 + n_q)$ matrix and $B^q_\chi$ is an $n_q \times (3+ n_q)$ matrix.
The flavour basis and the physical basis (no `0' superscript) are thus connected via
\begin{equation}
\begin{pmatrix}
\,u^0_\chi\, \\[1mm] 
U^0_\chi
\end{pmatrix} 
\,=\,
\begin{pmatrix}
\,A^u_\chi\, \\[1mm] 
B^u_\chi
\end{pmatrix} 
\begin{pmatrix}
\,u_\chi\, \\[1mm] 
U_\chi
\end{pmatrix} \,,
\qquad
\begin{pmatrix}
\,d^0_\chi\, \\[1mm] 
D^0_\chi
\end{pmatrix} 
\,=\,
\begin{pmatrix}
\,A^d_\chi\, \\[1mm] 
B^d_\chi
\end{pmatrix} 
\begin{pmatrix}
\,d_\chi\, \\[1mm] 
D_\chi
\end{pmatrix} \,,
\label{eq:connect}
\end{equation}
with $\chi = L,R$.
Within each sector, the diagonalization relation~\eqref{eq:diagonalization} then allows us to
write the mass matrix blocks of~\cref{eq:genmass} in terms of the physical masses as
\begin{equation}
\label[pluralequation]{eq:mdefs}
\begin{aligned}
  m_q   &= A^q_{L} \,\mathcal{D}_q\, {A_{R}^q}^\dagger\,, \quad 
  \om_q = A^q_{L} \,\mathcal{D}_q\, {B_{R}^q}^\dagger\,, \\[2mm]
  \oM_q &= B^q_{L} \,\mathcal{D}_q\, {A_{R}^q}^\dagger\,, \quad
  M_q   = B^q_{L} \,\mathcal{D}_q\, {B_{R}^q}^\dagger\,.
\end{aligned}
\end{equation}
Finally, unitarity of the $\cV^q_\chi$ implies
\begin{subequations}\begin{align}
\begin{pmatrix}
\,A^q_\chi\, \\[2mm] 
B^q_\chi
\end{pmatrix} \begin{pmatrix}
\,{A^q_\chi}^\dagger & {B^q_\chi}^\dagger\,
\end{pmatrix} 
\,&=\,
\begin{pmatrix}
\,A^q_\chi\,{A^q_\chi}^\dagger & A^q_\chi\,{B^q_\chi}^\dagger\,\\[2mm] 
\,B^q_\chi\,{A^q_\chi}^\dagger & B^q_\chi\,{B^q_\chi}^\dagger\,
\end{pmatrix} 
\,=\,
\begin{pmatrix}
\,\id_{3}\, & \,0\, \\[2mm] 
\,0\, & \,\id_{n_q}\,
\end{pmatrix}\,, 
 \label{eq:unitAB02}
\\[2mm]
\begin{pmatrix}
\,{A^q_\chi}^\dagger & {B^q_\chi}^\dagger\,
\end{pmatrix} \begin{pmatrix}
\,A^q_\chi\, \\[2mm] 
B^q_\chi
\end{pmatrix}
\,&=\,
\,
{A^q_\chi}^\dagger\,A^q_\chi+ 
{B^q_\chi}^\dagger\,B^q_\chi
\,
\,=\,
\,
\id_{3+n_q}
\,,
 \label{eq:unitAB2}
\end{align}\end{subequations}
for each $q=u,d$ and $\chi = L,R$.

\subsection{Gauge and Higgs interactions}
\label{sec:gaugeinteractions}

Electromagnetic interactions of quarks, including the VLQ isosinglets, are described by the Lagrangian 
\begin{equation}
    \lag_A = - e\, J^\mu_\text{em} A_\mu\,,
\end{equation}
with
\begin{equation}
\begin{aligned}
J^\mu_\text{em} &\,\equiv\,
\frac{2}{3}  
\Big( \overline{u_{i}^0} \gamma^\mu  {u_{i}^0}
+ \overline{U_{r}^0} \gamma^\mu {U_{r}^0}  \Big)  
- \frac{1}{3} 
\Big(\overline{d_{i}^0} \gamma^\mu  {d_{i}^0}
+ \overline{D_{s}^0} \gamma^\mu {D_{s}^0} \Big)\\[2mm]
 &\,=\, 
\frac{2}{3}  
\Big( \overline{u_{i}} \gamma^\mu  {u_{i}}
+ \overline{U_{r}} \gamma^\mu {U_{r}}  \Big)  
- \frac{1}{3} 
\Big(\overline{d_{i}} \gamma^\mu  {d_{i}}
+ \overline{D_{s}} \gamma^\mu {D_{s}} \Big)
\,,
\end{aligned}
\end{equation}
where we have defined $\psi \equiv \psi_L + \psi_R$ for $\psi\in\{u_i^{(0)},d_i^{(0)},U_r^{(0)},D_s^{(0)}\}$. As before, the index $i$ runs from 1 to 3 as in the SM, while the indices $r,s$ run from 1 to $n_{u,d}$. Note that the structure of the electromagnetic current $J^\mu_\text{em}$ is unchanged in going from the flavour to the mass basis.

\vskip 2mm

The charged current Lagrangian in the flavour basis takes the form
\begin{equation}
   \lag_W\,=\,
   -\frac{g}{\sqrt{2}}\overline{u_{Li}^0} \gamma^\mu 
{d_{Li}^0} W_\mu^+ + \text{h.c.}\,,
 \end{equation}
where $W_\mu^+$ denotes the $W$ boson field. 
In the mass eigenstate basis, it becomes
\begin{equation}
\label{eq:W-coupling}
   \lag_W \,=\, -\frac{g}{\sqrt{2}}
 \begin{pmatrix}
\overline{u}_L & \overline{U}_L 
\end{pmatrix}\,
V\,    \gamma^\mu
\begin{pmatrix}
d_L \\[2mm] D_L 
\end{pmatrix} 
    W_\mu^+
    \,+\,\text{h.c.}\,,
\end{equation}
with $V$ the $(3+n_u)\times (3+n_d)$ non-unitary mixing matrix, defined as
\begin{equation}\label{eq:VCKM}
    V \equiv {A_L^u}^\dagger A^d_L\,,
\end{equation}
see~\cref{eq:connect}.
Note that the mixing matrix $\gls*{V}$ --- rectangular whenever $n_u\neq n_d$ --- is not unitary in general.%
\footnote{It can however be written as the upper-left block of an auxiliary $(3+n_u+n_d) \times (3+n_u+n_d)$ unitary matrix, see~\cref{eq:uparam-param} and Ref.~\cite{Branco:1992wr}.
\label{foot:part_of_unitary}}
The role of the \gls*{CKM} quark mixing matrix, denoted $\gls*{VCKM}$, is played by the upper-left $3 \times 3$ block of $V$. Thus, $V_\text{CKM}$ is also not unitary in general. 

\vskip 2mm

The weak neutral current Lagrangian in the flavour basis is given by
\begin{equation}
\lag_Z \,=\, -\frac{g}{2 \cos{\theta_W}}   \left[
  \overline{u_{Li}^0} \gamma^\mu {u_{Li}^0}
- \overline{d_{Li}^0} \gamma^\mu {d_{Li}^0}
- 2 \sin^2{\theta_W} J^\mu_\text{em} \right]
Z_\mu\,,
\end{equation}
where $Z_\mu$ denotes the $Z$ boson field. 
Unlike right-handed neutrinos, the VLQs couple to the photon and to the $Z$ boson already in the flavour basis --- even though they are singlets of $SU(2)_L$ --- since they carry non-zero weak hypercharge. The hypercharges are assigned in such a way that the VLQs acquire the same electrical charges as their SM counterparts, allowing for mixing.
While there are no $Z$-mediated FCNC in the flavour basis, FCNC do arise in the mass eigenstate basis, where $\lag_Z$ reads
\begin{equation}
\label{eq:zmass}
\lag_Z = -\frac{g}{2 \cos{\theta_W}}   \left[
 \begin{pmatrix}\overline{u}_L & \overline{U}_L \end{pmatrix}
\,F^u\, \gamma^\mu
\begin{pmatrix} u_L \\[2mm] U_L \end{pmatrix}
-
 \begin{pmatrix}\overline{d}_L & \overline{D}_L \end{pmatrix}
\,F^d\, \gamma^\mu
\begin{pmatrix} d_L \\[2mm] D_L \end{pmatrix}
- 2 \sin^2{\theta_W} J^\mu_\text{em} \right]
Z_\mu\,,
\end{equation}
with%
\footnote{
The matrices $F$ are often denoted in the literature as 
$z$~\cite{Branco:1992wr,Frampton:1999xi}, $Z$~\cite{Branco:1995us,Branco:1999fs}, or $X$~\cite{delAguila:2000aa,Aguilar-Saavedra:2013qpa}.}
\begin{equation}
\begin{aligned}
F^u \,&\equiv\, {A_L^u}^\dagger A_L^u \,=\,
\id - {B_L^u}^\dagger B_L^u
\,=\, VV^\dagger \,,\\
F^d \,&\equiv\, {A_L^d}^\dagger A_L^d  \,=\,
\id - {B_L^d}^\dagger B_L^d
\,=\, V^\dagger V\,.
\end{aligned}
\label[pluralequation]{eq:Fud}
\end{equation}
$F^{u,d}$ are $(3+ n_{u,d})$-dimensional Hermitian matrices controlling FCNC.
The latter are sensitive to the off-diagonal elements of $F^{u,d}$ (see also~\cref{sec:du-FCNC}). \Cref{eq:Fud} thus show that there is a strong connection between the strength of FCNC and deviations of $V$ from unitarity.
 
\vskip 2mm

The couplings to the Higgs field, $h$, in the flavour basis are obtained from~\cref{eq:lagrangian}, following EWSB:
\begin{equation}
\label{eq:LH0}
  \begin{aligned}
  \lag_h \,=\, 
  -\,\Big[
  &\,\,\overline u_{Li}^{0} \,
  \left(m_u\right)_{ij}  \, u^0_{Rj}
  \,+\, \overline u_{Li}^{0} \,
  \left(\om_u\right)_{ir} \, U^0_{Rr} 
  \\+
  &\,\,\overline d_{Li}^{0} \,
  \left(m_d\right)_{ij}  \, d^0_{Rj} 
  \,+\,\overline d_{Li}^{0} \, 
  \left(\om_d\right)_{is} \, D^0_{Rs}
  \,\Big]\, \frac{h}{v}
  + \text{h.c.}\,.
  \end{aligned}
\end{equation}
To switch to the mass eigenstate basis one uses~\cref{eq:connect,eq:mdefs}, which yield 
\begin{equation}
  \overline u_{L}^{0} \, m_u  \, u^0_{R}
  \,+\,
  \overline u_{L}^{0} \, \om_u \, U^0_{R} 
\quad \rightarrow\quad
\begin{pmatrix}\overline{u}_L & \overline{U}_L \end{pmatrix}
\,{A_{L}^{u}}^{\dag} A_{L}^{u}\, \mathcal{D}_u\,
\big(
{A^u_R}^\dagger A^u_R+ 
{B^u_R}^\dagger B^u_R
\big)
\begin{pmatrix} u_R \\[2mm] U_R \end{pmatrix}\,
\end{equation}
and an analogous transformation for the down-type quark terms. Via the unitarity relation of~\cref{eq:unitAB2}, $\lag_h$ then simplifies to
\begin{equation}
\label{eq:LH}
\lag_h = - \left[
 \begin{pmatrix}\overline{u}_L & \overline{U}_L \end{pmatrix}
\,F^u\, \mathcal{D}_u
\begin{pmatrix} u_R \\[2mm] U_R \end{pmatrix}
\,+\,
 \begin{pmatrix}\overline{d}_L & \overline{D}_L \end{pmatrix}
\,F^d\, \mathcal{D}_d
\begin{pmatrix} d_R \\[2mm] D_R \end{pmatrix}
\right] \frac{h}{v}
  + \text{h.c.}
\end{equation}
in the mass basis.
Note that in general $F^{u,d} \neq \id$ and there are $Z$- and Higgs-mediated FCNC. The strength of the latter is controlled by the off-diagonal entries of the matrices $F^{u,d}$ and, additionally, by the ratios in the diagonals of $\mathcal{D}_{u,d}/v$, in contrast with the case of $Z$-mediated FCNC. Thus, for transitions involving {\it only} the lighter quarks $u$ and $c$, an additional suppression --– by a factor of $m_u/v \sim 10^{-5}$ or $m_c/v\sim 10^{-2}$ --– is present.
We will not consider extended Higgs sectors unless explicitly stated.%
\footnote{
For works considering VLQs in the context of extended Higgs sectors or composite Higgs scenarios see, for instance, Refs.~\cite{Arhrib:2016rlj,Benbrik:2019zdp,Banerjee:2022izw,Banerjee:2022xmu,Bhardwaj:2022nko}.
}

We close this section with a discussion of the similarities between VLQs and neutrinos in the seesaw context.

\subsection{Comparison between VLQ and Majorana neutrino effects}
\label{sec:neutrinos}
VLQs arise in very plausible extensions of the Standard Model, appearing in a variety of frameworks. One of the salient features of VLQs is the fact that they lead to the violation of the dogma that no $Z$-mediated flavour-changing neutral currents should exist at tree level. Furthermore, they lead to the violation of unitarity of the matrix equivalent to the SM CKM matrix. They also lead to Higgs-mediated FCNC at tree level even without extending the SM scalar sector. This was mentioned in the Introduction and shown in~\cref{sec:gaugeinteractions}. The strong suppression of these beyond-the-SM (BSM) effects, as required by experiment, is obtained as a result of the presence of two mass scales: the scale $v$ of electroweak symmetry breaking and the scale $M$ of bare mass terms for VLQs, which are invariant under the gauge group. Since the latter mass terms are not protected by the gauge symmetry, $M$ can be considerably larger than $v$. At this stage, it is worth making an analogy to what one encounters in the neutrino sector, in extensions of the SM where non-vanishing but naturally small neutrino masses are generated by the seesaw mechanism. 

In the leptonic sector there are the charged-lepton sector and the neutrino sector. For the charged-lepton sector we can also introduce new charged vector-like leptons, in perfect analogy to what can be done for quarks. Some recent examples of studies with isosinglet vector-like charged leptons can be found in Refs.~\cite{Chala:2020odv,Crivellin:2020ebi,Athron:2021iuf,Cherchiglia:2021syq} with implications for charged lepton flavour violating processes, for the muon anomalous magnetic moment or for leptonic CP violation.

In the neutrino sector the situation is significantly different. In the SM, neutrinos are strictly massless due to the fact that no right-handed neutrinos are present. In the seesaw framework, right-handed neutrino singlets $\nu_R$ are introduced, often in the number of three but, at present, two would suffice to generate the observed pattern of masses and mixing. These right-handed neutrinos are sterile in the flavour basis meaning that they are singlets of all the gauge group symmetries of the SM, to wit $SU(3)_c$, $SU(2)_L$ and ${U(1)}_{Y}$. In the seesaw framework their mixing with the SM-like neutrinos is very suppressed and as a result the heavy mass eigenstates are almost sterile. This is the reason why the designation of sterile neutrinos for the heavy neutrino mass eigenstates is still often kept. 

There is a completely new feature in the neutrino sector with the addition of sterile right-handed neutrinos, which is the possibility of introducing Majorana mass terms. Majorana mass terms for right-handed neutrinos are bilinear products involving $\nu_R$. In this extended context, the only fermions for which such terms are allowed are the sterile neutrinos. Majorana mass terms violate any $U(1)$ symmetry such as electric charge as well as a leptonic number and, therefore, bare Majorana mass terms are only allowed for electrically neutral fermions.

Majorana spinors are a combination of a single Weyl spinor such that $\Psi^c = \Psi$ with 
\begin{equation}
\Psi^c \equiv C \overline{\Psi^T}\,, \qquad C= i \gamma_2 \gamma_0 \,,
\end{equation}
defining particle--antiparticle conjugation. The explicit form of $C$ above is valid in the Dirac and Weyl representations (see e.g.~\cite{Pal:2010ih}).
The right-handed and left-handed components of a Majorana spinor are related by:
\begin{equation} \label{eq:maj}
\Psi = \Psi_L + \Psi_R = \Psi_L + (\Psi_L)^c = (\Psi_R)^c + \Psi_R\,, \qquad \Psi_R= (\Psi_L)^c =  (\Psi^c)_R \,.
\end{equation}
The most general expressions allow for the possibility of introducing a phase factor in the relation between $\Psi^c$ and $\Psi$~\cite{Akhmedov:1999uz}. The particle--antiparticle conjugation of a right-handed field is a left-handed field and vice-versa.

In the seesaw framework, the Dirac and the Majorana mass terms of the neutrinos may be combined into a compact expression in terms of a larger mass matrix $\cal{M}$ in the fermion basis $n_L = ( \nu_L^0, (\nu_R^0)^c)$: 
\begin{equation}
 {\cal L}_{m_\nu} =-\overline{{\nu}_{L}^0} m \nu_{R}^0 +
\frac{1}{2} \nu_{R}^{0T} C^\dagger M \nu_{R}^0 + {\rm h.c.}\,
= \frac{1}{2} n_{L}^{T} C^\dagger {\cal M}^* n_L +  {\rm h.c.}\,.
\end{equation}
When three right-handed neutrinos are introduced,
the mass matrix $\mathcal{M}$ is of dimension $6 \times 6$ and is given by
\begin{equation}
 {\cal M} = \begin{pmatrix}
     0 & m \\[2mm]
m^T & M  \end{pmatrix}
\,.
\end{equation}
Comparing this matrix $\cal{M}$ with the matrices ${\cal M}_d$, ${\cal M}_u$ of the quark sector one should immediately notice that by definition $\cal{M}$ is a symmetric matrix, which does not need to be the case for ${\cal M}_d$, ${\cal M}_u$. Also, the same fields appear as factors on the left- and right-handed side of $\cal{M}$. In the case of ${\cal M}_d$, ${\cal M}_u$ these are independent spinors. This fact has implications for the possible choices of weak bases. Furthermore, in the minimal seesaw framework $\cal{M}$ has a zero block in the upper left-hand corner leading to physical implications.

The introduction of a vector-like isosinglet neutrino with independent left- and right-handed components in analogy to what is done in the other fermionic sectors is equivalent to the introduction of two right-handed sterile neutrino states or else to the introduction of two left-handed sterile neutrino states. It is just a question of naming. It is the Majorana character of the neutrinos in the seesaw framework that allows to introduce an odd number of such singlet states.

The minimal number of right-handed neutrino singlets required to account for the observed pattern of light neutrino masses is two. In this case one of the three light neutrinos has zero mass at tree level and there are two heavy neutrinos which may be accessible at colliders. In the literature such heavy neutrinos are often referred to as \gls*{HNL}. From what has just been presented one might expect these two heavy neutrinos to be Majorana particles. However in the case of exact degeneracy of the heavy neutrinos there are two different ways of pairing the right- and left-handed components of the two Weyl spinors. One possibility is given by~\cref{eq:maj}:
\begin{equation}
\Psi^i_M = \Psi^i_L + (\Psi^i_L)^c = (\Psi^i_R)^c + \Psi^i_R\,,  \qquad i = 1,2\,,
\end{equation}
while the other possibility is:
\begin{equation}
\begin{aligned}
\Psi_D &= \frac{1}{\sqrt{2}} ( \Psi^1_M + i \Psi^2_M ) = \frac{1}{\sqrt{2}} \left[ (\Psi^1_L+ i \Psi^2_L )
+ (\Psi^1_R+ i \Psi^2_R ) \right]\,, \\ 
\Psi^c_D &= \frac{1}{\sqrt{2}} ( \Psi^1_M - i \Psi^2_M ) = \frac{1}{\sqrt{2}} \left[ (\Psi^1_L- i \Psi^2_L )
+ (\Psi^1_R - i \Psi^2_R ) \right]\,.
\end{aligned}
\end{equation}
The latter corresponds to having a Dirac spinor and its anti-particle.  In the case of almost degeneracy of the masses of two HNL these may form quasi-Dirac fermions which approximately conserve the total lepton number and therefore could in principle be distinguished from the case of two Majorana fermions via their decays involving charged leptons~\cite{deGouvea:2021rpa}. The electroweak production of HNL can occur via $W$, $Z$ or $h$ vertices associated to the production of a charged lepton ($W$) or light neutrinos ($Z$, $h$). Before their decay there is the possibility of coherent heavy neutrino--anti-neutrino oscillations in the pseudo-Dirac case. These oscillations are responsible for deviations from the exact lepton number conserving limit. This has been a topic widely discussed in the literature both for physics at the LHC as well as for physics at future colliders, see e.g.~\cite{Anamiati:2016uxp,Tastet:2019nqj,Hernandez:2018cgc,Drewes:2022rsk,Antusch:2023jsa,Kwok:2023dck}.

Despite all these differences there are fundamental similarities between models with VLQs and the seesaw mechanism which are related to the fact that there are two different mass scales that play a similar role in the theory in both cases.

\vfill
\clearpage

\section{Weak Bases and Physical Parameters}
\label{sec:WBphy}

\subsection{Weak-basis transformations in the SM}
\label{sec:WBSM}

In the Standard Model, the flavour structure of Yukawa couplings is not constrained by gauge invariance. The Yukawa couplings $Y_u$ and $Y_d$ are $3 \times 3$ arbitrary complex matrices and therefore contain 36 independent parameters. It is clear that there is a large redundancy which results from the fact that one can perform \gls*{WB} transformations of the type 
\begin{equation}
\label{eq:SMWB}
\begin{array}{l@{\qquad}l}
  u^0_L \,\to\, W_L  \, u^0_L
  \,,
  &d^0_L \,\to\, W_L  \, d^0_L
  \,,\\[2mm]
  u^0_R \,\to\, W^u_R  \, u^0_R
  \,,
  &d^0_R \,\to\, W^d_R  \, d^0_R
\,,
\end{array}
\end{equation}
while keeping the charged and neutral gauge currents invariant.
Here, $W_L$ and $W_R^{u,d}$ are three $3\times 3$ unitary matrices in flavour space.
The important point is that although WB transformations may help simplify the Lagrangian in the flavour basis by reducing the number of independent parameters, these transformations do not change the physics, i.e.~two sets of quark mass matrices related by a WB transformation describe the same quark masses, mixing and physical couplings.
It follows that physical processes can only depend on WB-invariant quantities. The latter are particularly useful in the analysis of CP violation (see~\cref{sec:CPWB}), as they distinguish between spurious and physical phases~\cite{Branco:1999fs}.

Under the WB change of~\cref{eq:SMWB}, the Yukawa matrices are transformed as
\begin{equation}
\begin{aligned}
Y_u  \quad &\to \quad Y'_u \,=\, W_L^\dagger\, Y_u\, W^u_R \,,  \\
Y_d  \quad &\to \quad Y'_d \,=\, W_L^\dagger\, Y_d\, W^d_R \,.
\end{aligned}
\end{equation}
By performing WB transformations it is possible to make both $Y_u$ and $Y_d$ Hermitian without loss of generality ({\it SM Hermitian WB}). In this WB, the Yukawa matrices still contain 18 independent parameters.

A more convenient basis choice, where the physical parameter count becomes apparent, corresponds to having one of the Yukawa matrices diagonal and real, while the other is Hermitian, say
\begin{equation}
\label{eq:YuDiagSMWB}
Y_u \sim \mbox{diagonal}\,, \qquad    Y_d \sim  \mbox{Hermitian}\,.                            
\end{equation}
In this WB, $Y_u$ contains 3 real parameters while $Y_d$ contains 6 real parameters and (still) 3 phases. Note, however, that the combination
\begin{equation}
\varphi_d \,=\, \arg \left[({Y_d})_{12} ({Y_d})_{23} ({Y_d})_{31} \right]\,
\end{equation}
is invariant under the simultaneous rephasing of up and down quarks $W_L = W_R^u = W_R^d = \diag(e^{i\alpha_1},e^{i\alpha_2},e^{i\alpha_3})$ which does not spoil~\cref{eq:YuDiagSMWB}. Such a rephasing can be used to remove 2 of the 3 phases in $Y_d$.%
\footnote{Naturally, a chain of WB transformations defines a single WB change.}
One is then left with 9 real parameters and 1 phase, $\varphi_d$. Therefore, in this {\it SM minimal WB}, the matrices $Y_u$ and $Y_d$ contain a total of 10 parameters, the same number appearing in the physical basis with 6 quark masses, 3 mixing angles and 1 CP-violation phase parameterizing the standard CKM matrix.

As we have seen, WB transformations can be used to make some elements of the Yukawa and mass matrices vanish (texture zeros), thereby reducing the number of parameters. It is important to distinguish between zeros that are simply the result of a WB choice and zeros that carry physical content, leading to special predictions. As an example, consider the following ansatz by Fritzsch~\cite{Fritzsch:1977vd}
\begin{equation}
\label{eq:Fritzsch}
Y_u \sim \begin{pmatrix} 
0 & \times & 0 \\
\times & 0 & \times \\
0 & \times & \times
\end{pmatrix} \text{ Hermitian}\,, \qquad
Y_d \sim \begin{pmatrix} 
0 & \times & 0 \\
\times & 0 & \times \\
0 & \times & \times
\end{pmatrix} \text{ Hermitian}\,,
\end{equation}
in which the lightest quarks acquire their masses via so-called nearest neighbour interactions (NNI).
One can check that the matrices $Y_u$ and $Y_d$ only contain 8 physical parameters since, as before, 2 phases can be removed by rephasing. Therefore, the ansatz in~\cref{eq:Fritzsch} is not a simple WB choice. It implies correlations between quark masses and mixing, that have by now been ruled out experimentally. However, it has been shown that having $Y_u$ and $Y_d$ in the NNI form while lifting the requirements of Hermiticity does indeed amount to a particular WB choice within the SM~\cite{Branco:1988iq}. In this {\it NNI weak basis}, the Yukawa matrices contain a total of 12 independent parameters, after exhausting the rephasing freedom.

In theories beyond the SM, the set of allowed WB transformations,~\cref{eq:SMWB}, may need to be modified. For instance, preserving gauge currents in left-right-symmetric models forces $W_R^u = W_R^d$. In this case, the NNI ansatz no longer corresponds to a WB choice even when the requirement of Hermiticity is lifted.
The interesting case of WB transformations in the context of the SM augmented by VLQs is discussed in what follows.

\subsection{Weak-basis transformations in the presence of VLQs}
\label{sec:basis_VLQ}
Consider, as before, the SM extended by $n_u$ up- and $n_d$ down-type vector-like isosinglet quarks $U^0_{L,R}$ and $D^0_{L,R}$ (see~\cref{sec:notation}).
As established in~\cref{eq:lagrangian,eq:baremasses}, the corresponding flavour sector looks like
\begin{equation}
\label{eq:LF}
  \begin{split}
  \lag_{F} \,=\, 
  -\,\Big[
  &\,\, \overline Q_{L}^{0} \, \tilde\Phi \, Y_u  \, u^0_{R}
  \,+\, \overline Q_{L}^{0} \, \tilde\Phi \, \oYu \, U^0_{R} 
  \,+\, \overline U^0_{L}   \, \,\oM_u    \, u^0_{R}
  \,+\, \overline U^0_{L}   \, M_u        \, U^0_{R}
  \\ +
  &\,\, \overline Q_{L}^{0} \, \Phi       \, Y_d  \, d^0_{R} 
  \,+\, \overline Q_{L}^{0} \, \Phi       \, \oYd \, D^0_{R}
  \,+\, \overline D^0_{L}   \, \,\oM_d    \, d^0_{R}
  \,+\, \overline D^0_{L}   \, M_d        \, D^0_{R}
  \,\Big] + \text{h.c.}\,,
  \end{split}
\end{equation}
where all matrices are complex.
The most general WB transformation in this context takes the form
\begin{equation}
\label{eq:VLQWB}
\begin{array}{l@{\qquad}l}
  u^0_L \,\to\, W_L  \, u^0_L
  \,,
  &d^0_L \,\to\, W_L  \, d^0_L
  \,,\\[2mm]
  U^0_L \,\to\, W_L^U  \, U^0_L
  \,,
  &D^0_L \,\to\, W_L^D  \, D^0_L
  \,,\\[3mm]
  \begin{pmatrix}
     u^0_R \\[1mm] U^0_R 
  \end{pmatrix}
  \,\to\, \mathcal{W}_R^u 
  \begin{pmatrix}
     u^0_R \\[1mm] U^0_R 
  \end{pmatrix}
  \,,
  &\begin{pmatrix}
     d^0_R \\[1mm] D^0_R 
  \end{pmatrix}
  \,\to\, \mathcal{W}_R^d 
  \begin{pmatrix}
     d^0_R \\[1mm] D^0_R 
  \end{pmatrix}
\,,
\end{array}
\end{equation}
where $W_L \sim 3\times 3$, $W_L^{U,D} \sim n_{u,d}\times n_{u,d}$ and $\mathcal{W}^{u,d}_R \sim (3+n_{u,d})\times (3+n_{u,d})$ are unitary matrices in flavour space.
Note that, as before, the transformations of $u_L^0$ and $d_L^0$ are correlated in a WB change. This is not required for the transformations of the $U_L^0$ and $D_L^0$ isosinglet fields. Moreover, the transformation matrices for right-handed isosinglet fields $\mathcal{W}^{u,d}_R$ are enlarged with respect to the SM case. They allow for the mixing of $u_R^0$ and $U_R^0$ (and similarly of $d_R^0$ and $U_R^0$),
since the $3+n_q$ right-handed quarks in each sector ($q=u,d$) carry the same quantum numbers.

Although these and the following statements are presented for the case $n_f=3$, where $n_f$ is the number of SM quark families, they can be directly generalized and applied to the case $n_f \neq 3$.

\subsubsection{Useful weak bases}
\label{sec:wbzero}
Different WB choices allow for particular forms of the extended quark mass matrices $\mathcal{M}_{q}$, defined in~\cref{eq:genmass}. 
It is more convenient to deal directly with the mass matrices in this context. These contain contributions from both Yukawa and bare mass terms, see~\mbox{\cref{eq:lagrangian,eq:baremasses}}, unlike in the SM case where mass matrices $m_q = \frac{v}{\sqrt{2}} Y_q$ share the properties of the Yukawa coupling matrices.
For generic $n_u$ and $n_d$, under the WB change of~\cref{eq:VLQWB}, the mass matrices are transformed as
\begin{equation}
\begin{aligned}
\mathcal{M}_u \,=\,
\begin{pmatrix}
m_u & \om_u  \\[2mm]
\oM_u & M_u
\end{pmatrix}  \quad &\to \quad \mathcal{M}'_u \,=\, 
\begin{pmatrix}
W_L^\dagger \, m_u & W_L^\dagger \, \om_u  \\[2mm]
{W_L^U}^\dagger \, \oM_u & {W_L^U}^\dagger \, M_u
\end{pmatrix}
\mathcal{W}_R^u
\,,  \\[2mm]
\mathcal{M}_d \,=\,
\begin{pmatrix}
m_d & \om_d  \\[2mm]
\oM_d & M_d
\end{pmatrix}  \quad &\to \quad \mathcal{M}'_d \,=\, 
\begin{pmatrix}
W_L^\dagger \, m_d & W_L^\dagger \, \om_d  \\[2mm]
{W_L^D}^\dagger \, \oM_d & {W_L^D}^\dagger \, M_d
\end{pmatrix}
\mathcal{W}_R^d
\,.
\end{aligned}
\end{equation}

The large unitary rotations $\mathcal{W}^q_R$ appearing on the right-hand side of these transformations are quite helpful in shaping the mass matrices $\mathcal{M}_q$ and can be used to make some part of it vanish, under certain conditions.
In particular, note that any complex square matrix $\mathcal{M}$ can be written as the product of a lower-triangular matrix $L$ and a unitary matrix $W$ to its right, $\mathcal{M} = LW$, or as the product of an upper-triangular matrix $R$ and a (in general different) unitary matrix $W'$ to its right, i.e.~as $\mathcal{M} = RW'$.%
\footnote{This follows from the well-known QR decomposition, according to which any complex square matrix $A$ can be written in the form $A=QR$, with $Q$ an unitary matrix and $R$ an upper (or right) triangular matrix. By decomposing $\mathcal{M}^T$ in this way, one finds the decomposition $\mathcal{M} = LQ'$, where $Q'$ is unitary and $L$ is lower (or left) triangular. Similarly, by QR-decomposing $\mathcal{M}^T \mathcal{P}$, where $\mathcal{P}$ is the anti-diagonal permutation matrix, one can show that it is possible to write $\mathcal{M} = R'Q''$, where $Q''$ is unitary and $R'$ is upper triangular.}
The matrices $\mathcal{W}^u_R$ and $\mathcal{W}^d_R$ can then be used to cancel the $W^{(\prime)}$ unitary factors and render the mass matrices $\mathcal{M}_u$ and $\mathcal{M}_d$ lower or upper triangular, as desired. They can further be used to permute the columns of these matrices.
As a result:
\begin{itemize}
\item A WB where $\om_q=0$ exists for all $n_q$ (\it vanishing $\,\om$ WB).
\item A WB where $\oM_q=0$ exists for all $n_q$ (\it vanishing $\,\oM$ WB).
\item A WB where $m_q=0$ is guaranteed to exist only if $n_q \geq n_f = 3$.
\item A WB where $M_q=0$ is guaranteed to exist only if $n_q \leq n_f = 3$.
\end{itemize}
In general, one can impose at most one of these conditions at a time via a WB choice,
i.e.~generically no more than one of the sub-matrices $m$, $\om$, $\oM$, or $M$ can be made to vanish simultaneously. The vanishing conditions may however be enforced independently in the up and down sectors.
Thus, in each sector, one is always free to consider a weak basis where the mass matrix takes the form%
\footnote{Here and in what follows, it should be clear that sub-matrices are unrelated across WB definitions, in spite of a common notation.}
\begin{equation}
\label{eq:WBzeros}
\mathcal{M} \,=\,
\begin{pmatrix}
m & \om  \\[2mm]
0 & M
\end{pmatrix} \!\!\qquad\text{or}\qquad
\mathcal{M} \,=\,
\begin{pmatrix}
m & 0 \\[2mm]
\oM & M
\end{pmatrix} \,.
\end{equation}
Moreover, within each sector, the lower-right matrix blocks $M$ can always be brought to a diagonal form $D'$ with real and non-negative entries, through a second WB transformation. This is accomplished by rotating away the left and right unitary matrices in the singular value decomposition of $M$, via appropriate choices of $W^{U,D}_L$ and $\mathcal{W}^{u,d}_R$ (now taken to be block diagonal). Such a procedure may affect $\oM$, if present, which at this point is a general complex matrix.
Instead, the upper-left matrix blocks $m$ in each sector cannot be simultaneously brought to a diagonal form via the same procedure, as one is limited by a single $W_L$ matrix, common to both the up and down sectors.
As a result, starting e.g.~from upper-triangular mass matrices $\mathcal{M}_u$ and $\mathcal{M}_d$, one can then at most arrive at a WB for which a unitary rotation remains in one of the sectors (say the down sector), namely
\begin{equation}
\label{eq:WBMorozumi}
\mathcal{M}_u \,=\,
\begin{pmatrix}
d_u' & \om_u  \\[2mm]
0 & D_u'
\end{pmatrix} \,, \qquad
\mathcal{M}_d \,=\,
\begin{pmatrix}
V' d_d' & \om_d  \\[2mm]
0 & D_d'
\end{pmatrix} \,.
\end{equation}
Here ({\it minimal vanishing $\,\oM$ WB}), $d_q'$ and $D_q'$ are diagonal matrices with real non-negative entries, $V'$ is the anticipated residual unitary rotation, and $\om_q$ are general complex matrices.
A closely-related WB choice has been made in Ref.~\cite{Branco:1992wr},
where $m_d$ has been kept general complex.
Note that the diagonal matrices are primed since, in general, they differ from those containing the physical masses, defined in~\cref{eq:diagonalization}.
Nevertheless, one can check that $|\det(d_q')\det(D_q')|=|\det(d_q)\det(D_q)|$, which equals the product of all $q$-type quark masses.
Similarly, one is free to choose a WB where
\begin{equation}
\label{eq:WBlower}
\mathcal{M}_u \,=\,
\begin{pmatrix}
d_u' & 0  \\[2mm]
\oM_u & D_u'
\end{pmatrix} \,, \qquad
\mathcal{M}_d \,=\,
\begin{pmatrix}
V' d_d' & 0  \\[2mm]
\oM_d & D_d'
\end{pmatrix}
\end{equation}
({\it minimal vanishing $\,\om$ WB}),
with $d_q'$ and $D_q'$ diagonal, $V'$ unitary and now $\oM_q$ general complex.
Finally, by the above procedure, it further follows that one may easily arrange for weak bases like
\begin{equation}
\label{eq:WBLavoura}
\mathcal{M}_u \,=\,
\begin{pmatrix}
d_u' & 0  \\[2mm]
\oM_u & D_u'
\end{pmatrix}  \,, \qquad
\mathcal{M}_d \,=\,
\begin{pmatrix}
V' d_d' & \om_d  \\[2mm]
0 & D_d'
\end{pmatrix}
\end{equation}
({\it mixed minimal WB}), where both $\om_u$ and $\oM_d$ are made to vanish.
A closely related WB choice has been made in Ref.~\cite{Branco:1986my},
where again $m_d$ has been kept general complex.

By considering e.g.~\cref{eq:WBMorozumi}, one sees that for arbitrary $n_u$ and $n_d$ it is not in general possible to bring any of the matrices $\mathcal{M}_u$ or $\mathcal{M}_d$ to a diagonal form by virtue of a WB transformation. 
In other words, a diagonal mass matrix in the flavour basis does not generically correspond to a simple WB choice: it is a physical statement, like the Fritzsch ansatz.
In the presence of VLQs, the exception arises in the limit cases with either $n_u = 0$ or $n_d = 0$, for which one can clearly diagonalize one of the mass matrices, $\mathcal{M}_u$ or $\mathcal{M}_d$, respectively. As an example, Ref.~\cite{Bento:1991ez} considered the simple case with $n_u = 0$ and $n_d = 1$ and a WB was chosen where $\mathcal{M}_u = d_u$ diagonal, while $\mathcal{M}_d$ was given as in~\cref{eq:WBlower}.

\subsubsection{Parameter counting}
\label{sec:par_counting}
In the WB choices of~\cref{eq:WBMorozumi,eq:WBlower,eq:WBLavoura}, the physical parameter count becomes apparent. Looking at e.g.~\cref{eq:WBMorozumi}, one finds
\begin{itemize}
\item $n_f + n_q = 3+n_q$ real parameters (moduli) in the matrices $d_q'$ and $D_q'$ of each sector,
\item $n_fn_q = 3 n_q$ complex parameters in each $\om_q$ matrix, i.e.~$3n_q$ moduli and $3n_q$ phases in each sector, and
\item $n_f^2 = 9$ parameters in the unitary matrix $V'$, corresponding to $\frac{1}{2}n_f(n_f-1) = 3$ moduli and $\frac{1}{2}n_f(n_f+1) = 6$ phases.
\end{itemize}
Note however that, as in the SM case, some WB rephasing freedom remains and not all of the above phases are physical.
To arrive at a WB with the minimal number of parameters, consider WB rephasings where the unitary matrices $W_L$, $W_L^{U,D}$ and $\mathcal{W}_R^{u,d}$ are diagonal matrices of phases. One may then
\begin{itemize}
\item remove $2n_f-1 = 5$ phases from $V'$ by use of the first $n_f = 3$ phases of $\mathcal{W}_R^d$ (commuting with $d_d'$) and of $W_L$, whose effect on $d_u'$ may be cancelled by $\mathcal{W}_R^u$, 
\item remove $n_q$ phases from $\om_q$ in each sector via the last $n_q$ phases of $\mathcal{W}_R^q$ (see also~\cite{Branco:1992wr}), whose effects on $D_q'$ may be cancelled by the corresponding $W_L^{U,D}$.
\end{itemize}
Therefore, in this {\it minimal WB} the mass matrices display a total of $N_\text{phys} = 10 + 6(n_u+n_d)$ parameters, while keeping the form of~\cref{eq:WBMorozumi}. Their breakdown into moduli and phases is summarized in~\cref{tab:params}.
%
\begin{table}[t]
  \centering
  \begin{tabular}{lccc}
    \toprule
    & Moduli & Phases & Total \\
    \midrule
$\mathcal{M}_u$ (minimal WB) & $3+4 n_u $ & $2n_u$ & $3+6n_u$ \\
$\mathcal{M}_d$ (minimal WB) & $6+4 n_d$ & $1+2n_d$ & $7+6n_d$ \\
    \midrule
Total & $9+4(n_u+n_d)$ & $1+2(n_u+n_d)$ & $10+6(n_u+n_d)$ \\
    \bottomrule
  \end{tabular}
  \caption{Physical parameter count in the presence of isosinglet VLQs ($n_f = 3$).}
  \label{tab:params}
\end{table}
%
Not only does one recover the SM result when $n_d=n_u=0$, but one also expects each singlet vector-like quark to bring about 6 new physical parameters: 4 moduli (1 mass and 3 mixing angles) and 2 physical phases.

\vskip 2mm

It is instructive to rederive these results making use of the spurion formalism~\cite{Berger:2008zq}.
We start by noting that the $N_\text{general} = N^\text{moduli}_\text{general} + N^\text{phases}_\text{general}$ parameters in the most general $\mathscr{L}_{F}$ of~\cref{eq:LF}, with
\begin{equation}
  N^\text{moduli}_\text{general}=N^\text{phases}_\text{general}=\big(n_f+n_u\big)^2+\big(n_f+n_d\big)^2\,,
\end{equation}
 break the flavour symmetry of these models,
\begin{equation}
  G_F=U\big(n_f\big)_{q_L}\times U\big(n_u\big)_{U_L}\times U\big(n_d\big)_{D_L}\times U\big(n_f+n_u\big)_{u_R,U_R} \times U\big(n_f+n_d\big)_{d_R,D_R}\,,
\end{equation}
down to the $U(1)_B$ of baryon number.
Since an element of $U(n)$ is identified by $\frac{1}{2}n(n-1)$ moduli and $\frac{1}{2}n(n+1)$ phases, for a total of $n^2$ parameters, the spurions $Y_q$, $\oY_q$, $\,\oM_q$, and $M_q$ break $N_\text{broken} = N^\text{moduli}_\text{broken}+N^\text{phases}_\text{broken}$ of its generators, where
\begin{equation}
\begin{split}
  N^\text{moduli}_\text{broken}&=\frac{3}{2}n_f\big(n_f-1\big)+n_u\big(n_u-1\big)+n_d\big(n_d-1\big)+n_f\big(n_u+n_d\big)\,,\\
  N^\text{phases}_\text{broken}&=\frac{3}{2}n_f\big(n_f+1\big)+n_u\big(n_u+1\big)+n_d\big(n_d+1\big)+n_f\big(n_u+n_d\big)-1\,.
\end{split}
\end{equation}
Given the well-established result~\cite{Santamaria:1993ah}
\begin{equation}
  N_\text{phys}=N_\text{general}-N_\text{broken}\,,
\end{equation}
we conclude that the flavour sector of these models is parameterized by a total of $N_\text{phys} = N^\text{moduli}_\text{phys} + N^\text{moduli}_\text{phys}$ physical parameters, divided as
\begin{equation}
\begin{split}
\label{eq:fsp-counting}
  N^\text{moduli}_\text{phys}&=\frac{1}{2}n_f\big(n_f+3\big)+\big(n_f+1\big)\big(n_u+n_d\big)=9+4\big(n_u+n_d\big)\,,\\
  N^\text{phases}_\text{phys}&=\frac{1}{2}\big(n_f-1\big)\big(n_f-2\big)+\big(n_f-1\big)\big(n_u+n_d\big)=1+2\big(n_u+n_d\big)\,,
\end{split}
\end{equation}
for $n_f = 3$, as found above (cf.~\cref{tab:params}).

\subsection{CP violation}
\label{sec:CPWB}

\subsubsection{Recalling the analysis in the SM}
CP violation arises as a conflict between the CP transformations required by the gauge interactions and those required by the Yukawa interactions. Indeed, one defines the most general CP transformation%
\footnote{The need to use the most general CP transformation reflects the fact that gauge interactions do not distinguish the $u$, $c$, $t$ quarks as well as the $d$, $s$, $b$ quarks and this has to be taken into account in the definition of CP.}
as that which leaves the gauge interactions invariant~\cite{Grimus:1995zi}. One then investigates the restrictions imposed by CP invariance on the Yukawa couplings. In the context of the SM, it can be readily shown that CP invariance implies~\cite{Bernabeu:1986fc}
\begin{equation}
I_\text{SM} \,\equiv\, \tr\left[
Y_u Y_u^\dagger ,\, Y_d Y_d^\dagger
\right]^3 \,=\, 0\,,
\label{eq:ISM}
\end{equation}
where $Y_{u,d}$ are the Yukawa couplings, i.e.~$I_\text{SM} = 0$ is a necessary condition for CP invariance. The quantity $I_\text{SM}$ is a weak-basis invariant, so it can be evaluated in any WB.
While it trivially vanishes for one generation, for two generations it can be shown that $I_\text{SM}  = 0$ follows from the Hermiticity of $Y_q Y_q^\dagger$ ($q=u,d$).%
\footnote{Indeed, note that $[H_1,H_2]^2 \propto \mathbb{1}$ for $H_{1,2}$ arbitrary $2\times 2$ Hermitian matrices.}
We postpone the discussion on the possibility of strong CP violation to~\cref{sec:strongCP}.

For three generations,~\cref{eq:ISM} implies non-trivial restrictions on the Yukawa couplings. As one can evaluate $I_\text{SM}$ in any WB, consider the basis where $Y_u$ is diagonal and real. One finds
\begin{equation}
\begin{aligned}
I_\text{SM} &\,=\, 6i\,(y_1^2-y_2^2)(y_2^2-y_3^2)(y_3^2-y_1^2)\,\im\left(H_{12}H_{23}H_{31}\right)
\\[2mm]
&\,\xrightarrow{\text{\tiny EWSB}}\, \frac{6i}{v^{12}} \,(m_c^2 - m_u^2)(m_t^2 - m_c^2)(m_t^2- m_u^2) \\
& \qquad\qquad\quad
(m_b^2 - m_s^2)(m_b^2 - m_d^2)(m_s^2- m_d^2) \,\im\left( V_{us} V_{cb} V_{ub}^* V_{cs}^*\right)
\,,
\end{aligned}
\label{eq:ISM2}
\end{equation}
where $Y_u = \diag(y_1,y_2,y_3)$ and $H \equiv Y_d Y_d^\dagger$. Moreover, in this case~\cref{eq:ISM} is both a necessary and sufficient condition to have CP invariance in the SM. 
Note also that in the three-generation case
this condition is equivalent to $\det [M_uM_u^\dagger, M_dM_d^\dagger] = 0$~\cite{Jarlskog:1985ht,Jarlskog:1985cw}, since this determinant is proportional to $I_\text{SM}$.
For more than three generations,~\cref{eq:ISM} continues to be a necessary condition for having CP invariance but it is no longer a sufficient condition (see also~\cite{Branco:1999fs}).

In the first line of~\cref{eq:ISM2}, the invariant $I_\text{SM}$ is written in terms of Yukawa couplings alone and the analysis is done at the Lagrangian level, prior to EWSB. Obviously, it can also be done after the breaking, as evidenced by the second line of~\cref{eq:ISM2}, where the invariant is expressed in terms of quark masses and mixing. What is worth emphasizing is that the CP conditions can be studied at the Lagrangian level, without having to refer to quark masses and mixing. 
The important result of Kobayashi and Maskawa~\cite{Kobayashi:1973fv} could have been derived without much effort by considering a general CP transformation without having to invoke the matrix $V_\text{CKM}$.

Note that CP violation is present in the SM quark sector if and only if any of the rephasing-invariant functions of the CKM matrix is not real~\cite{Branco:1999fs}. Examples of such functions are \emph{quartets} like $Q_{uscb} \equiv V_{us} V_{cb} V_{ub}^* V_{cs}^*$. The fact that there is a single physical phase parameter in the SM quark sector, as shown in~\cref{sec:WBSM}, implies the equality of the imaginary parts of all quartets up to a sign (see also~\cite{Jarlskog:1985ht,Jarlskog:1985cw,Dunietz:1985uy}). Defining $I \equiv \im Q_{uscb}$, it is clear that the condition $I=0$ is not only consistent with CP invariance, but also implies it, as can be seen from~\cref{eq:ISM2}. 
Furthermore, the quantity $|I|$ has a geometrical interpretation, as it corresponds to twice the area of any unitarity triangle. Taking into account the magnitudes of the moduli of $V_\text{CKM}$ elements, the standard and most useful unitarity triangle is defined by the unitarity relation
\begin{equation}
V_{ud}V_{ub}^* + V_{cd} V_{cb}^* + V_{td} V_{tb}^* = 0\,,
\label{eq:triangle}
\end{equation}
with each term in the sum corresponding to a side of the triangle in the complex plane.

The experimental determination of the physical CP-violating phases entering the quark mixing matrix provides stringent tests of the SM. Without imposing unitarity, the $3\times 3$ matrix $V_\text{CKM}$ contains 4 independent rephasing-invariant phases, since 5 out of the 9 initial phases can be removed by rephasing the standard quark fields as discussed above. We stress that this result is completely general: it does not depend on the presence of extra generations and holds true \emph{even in the presence of VLQs}.
These 4 phases can taken to be~\cite{Aleksan:1994if,Botella:2002fr}:
\begin{equation}
\begin{aligned}
\beta &\,\equiv\, \arg\left(-V_{cd}V_{tb}V_{cb}^*V_{td}^*\right)\,,\\
\gamma &\,\equiv\, \arg\left(-V_{ud}V_{cb}V_{ub}^*V_{cd}^*\right)\,,\\
\chi \text{ or } \beta_s &\,\equiv\, \arg\left(-V_{cb}V_{ts}V_{cs}^*V_{tb}^*\right)\,,\\
\chi' \text{ or } \beta_K &\,\equiv\, \arg\left(-V_{us}V_{cd}V_{ud}^*V_{cs}^*\right)\,.
\end{aligned}
\label{eq:CKMphases}
\end{equation}
One also typically defines
$\alpha \equiv \arg\left(-V_{td}V_{ub}V_{tb}^*V_{ud}^*\right)$, which however is not independent from the phases above since $\alpha + \beta + \gamma = \pi$ (mod $2\pi$) by definition.%
\footnote{In the notations of Ref.~\cite{Aleksan:1994if}, one has $\omega_1 = \pi - \alpha$, $\omega_2 = \pi - \beta$, $\omega_3 = \pi -\chi$, $\omega_4 = \pi -\chi'$ (mod $2\pi$). In Ref.~\cite{Branco:1999fs}, $\chi$ and $\chi'$ are denoted $\epsilon$ and $\epsilon'$, not to be confused with the CP violation parameters in neutral kaon decays. It is common to find the alternative notation $(\phi_1,\phi_2,\phi_3) = (\beta,\alpha,\gamma)$ in the literature (see e.g.~\cite{Bigi:2000yz}).}

The matrix $V_\text{CKM}$ describing the mixing among the known quarks thus contains in general 13 measurable quantities: 9 moduli and 4 rephasing-invariant phases. In the SM limit where it is unitary, there are $n_f^2 = 9$ exact and independent relations among these quantities, such as that of~\cref{eq:triangle}, leaving a total of 4 independent parameters. While typically chosen as the standard 3 mixing angles and 1 CP-violation phase, these 4 parameters may alternatively be taken to be e.g.~i) the phases in~\cref{eq:CKMphases}, or ii) four independent moduli. 
Indeed, it has been shown in Ref.~\cite{Aleksan:1994if} that, if one assumes $3\times 3$ unitarity of the $V_\text{CKM}$, the rephasing-invariant phases can be used to reconstruct the full CKM matrix. In the SM limit, the phases $\alpha$, $\beta$ and $\gamma$ also correspond to the inner angles of the aforementioned standard unitarity triangle.
Instead, one may reconstruct the $3\times 3$ unitary $V_\text{CKM}$, up to a two-fold ambiguity and including the strength of CP violation measured by $|I|$, by taking as input 4 independent moduli $|V_{ij}|$~\cite{Botella:1985gb,Branco:1987mj}. An additional sign is needed for the full reconstruction, since the moduli do not fix the sign of CP violation.
Finally, recall that physical processes can only depend on WB-invariant quantities. It has been shown~\cite{Branco:1987mj} that in the SM any four independent moduli of $V_\text{CKM}$ can be computed from the four WB invariants 
\begin{equation}
\tr (H_u H_d)\,, \quad  \tr (H_u H^2_d)\,, \quad \tr (H^2_u H_d)\,, \quad \tr (H^2_u H^2_d)\,,
\label{eq:SMWBinv}
\end{equation}
and the six known quark masses, with $H_q \equiv Y_q Y_q^\dagger$ ($q=u,d$).

\subsubsection{CP violation in the presence of VLQs}

Consider once more the addition of VLQs to the SM.
To determine the restrictions on quark mass matrices implied by CP invariance, one has to consider the most general CP transformation which leaves the gauge currents invariant~\cite{Branco:1986my}:
\begin{equation}
\label{eq:VLQCP}
\begin{array}{l@{\qquad}l}
  u^0_L \,\to\, \mathcal{U}_L \,\gamma_0 \,C \, \overline{u^0_L}^T 
  \,,
  &d^0_L \,\to\, \mathcal{U}_L \,\gamma_0 \,C \, \overline{d^0_L}^T 
  \,,\\[2mm]
  U^0_L \,\to\, \mathcal{U}_L^U \,\gamma_0 \,C \, \overline{U^0_L}^T 
  \,,
  &D^0_L \,\to\, \mathcal{U}_L^D \,\gamma_0 \,C \, \overline{D^0_L}^T 
  \,,\\[3mm]
  \begin{pmatrix}
     u^0_R \\[1mm] U^0_R 
  \end{pmatrix}
  \,\to\, \mathcal{U}_R^u  \,\gamma_0 \,C 
  \begin{pmatrix}
     \,\overline{u^0_R}^T \\ \,\overline{U^0_R}^T
  \end{pmatrix}
  \,,
  &\begin{pmatrix}
     d^0_R \\[1mm] D^0_R 
  \end{pmatrix}
  \,\to\, \mathcal{U}_R^d  \,\gamma_0 \,C 
  \begin{pmatrix}
     \,\overline{d^0_R}^T \\ \,\overline{D^0_R}^T
  \end{pmatrix}
\,,
\end{array}
\end{equation}
where $\mathcal{U}_L \sim 3\times 3$, $\mathcal{U}_L^{U,D} \sim n_{u,d}\times n_{u,d}$, and $\mathcal{U}^{u,d}_R \sim (3+n_{u,d})\times (3+n_{u,d})$ are unitary matrices acting in flavour space and $C$ is the charge conjugation matrix.
The need to include these matrices reflects the fact that gauge interactions do not distinguish the various families.

In order for the quark mass matrices to be CP invariant, the following conditions have to be satisfied, independently of the mechanism responsible for generating quark masses:
\begin{equation}
\label[pluralequation]{eq:CPconditions}
\begin{array}{l@{\qquad}l}
   \mathcal{U}_L^\dagger \,\begin{pmatrix} m_u & \om_u \end{pmatrix}\, \mathcal{U}_R^u \,= \begin{pmatrix} m_u^* & \om_u^* \end{pmatrix} \,, &
   \mathcal{U}_L^\dagger \,\begin{pmatrix} m_d & \om_d \end{pmatrix}\, \mathcal{U}_R^d \,= \begin{pmatrix} m_d^* & \om_d^* \end{pmatrix} \,, 
   \\[3mm]
   \left(\mathcal{U}_L^U\right)^\dagger \,\begin{pmatrix} \,\oM_u & M_u \end{pmatrix}\, \mathcal{U}_R^u \,= \begin{pmatrix} \,\oM_u^{\,*} & M_u^* \end{pmatrix} \,, &
   \left(\mathcal{U}_L^D\right)^\dagger \,\begin{pmatrix} \,\oM_d & M_d \end{pmatrix}\, \mathcal{U}_R^d \,= \begin{pmatrix} \,\oM_d^{\,*} & M_d^* \end{pmatrix} \,.
\end{array}
\end{equation}
This set of relations represents the necessary and sufficient conditions for having CP invariance in the charged and neutral gauge interactions. These conditions are WB independent in the following sense: if one finds unitary matrices $\mathcal{U}$ satisfying~\cref{eq:CPconditions} in a given WB, then one can find unitary matrices such that these relations hold in any other basis, cf.~\cref{eq:VLQWB}.
Despite being WB independent,~\cref{eq:CPconditions} are not very practical. From them one can derive necessary conditions for CP invariance in gauge interactions expressed in terms of WB invariants, namely~\cite{Branco:1986my}
\begin{equation}
    \tr\left[\left(h_d\right)^a,\,\left(h_u\right)^b\right]^r = 0\,,\quad
    \tr\left[\left(h_d'\right)^a,\,\left(\mathcal{H}_d\right)^b\right]^r = 0\,,\quad
    \tr\left[\left(h_u'\right)^a,\,\left(\mathcal{H}_u\right)^b\right]^r = 0\,,
\label{eq:CPinvariants}
\end{equation}
with $r$ odd, $r \ge 3$, and $a,b$ arbitrary positive integers and where we have introduced the Hermitian matrices
\begin{equation}
    h_q  \equiv m_q m_q^\dagger + \om_q \om_q^\dagger\,,\quad
    h_q' \equiv \begin{pmatrix}
     m_q^\dagger m_q & m_q^\dagger \om_q  \\[1mm]
     \om_q^\dagger m_q & \om_q^\dagger \om_q \end{pmatrix}\,,\quad
    \mathcal{H}_q  \equiv \begin{pmatrix}
     \,\oM_q^{\,\dagger} \,\oM_q & \,\oM_q^{\,\dagger} M_q  \\[1mm]
     M_q^\dagger \,\oM_q & M_q^\dagger M_q \end{pmatrix}\,.
\end{equation}
In the SM limit, the first family of invariants in~\cref{eq:CPinvariants} simply yields the CP-odd $I_\text{SM}$ of~\cref{eq:ISM} at the lowest non-trivial order (up to a factor of $v^{12}$). The remaining families of CP-odd invariants arise only in the presence of down- and up-type VLQs, respectively. 

Other useful and independent CP-odd WB invariants can be constructed in this context (see e.g.~\cite{Albergaria:2022zaq}), such as 
\begin{equation}
\label{eq:newinv}
\im \tr \left[\mathcal{H}_u 
\begin{pmatrix}     m_u^\dagger \\[1mm]     \om_u^\dagger \end{pmatrix}
\, h_d \,\begin{pmatrix}     m_u & \om_u\end{pmatrix}\,h_u'\right]  \,,
\quad
\im \tr \left[\mathcal{H}_d 
\begin{pmatrix}     m_d^\dagger \\[1mm]     \om_d^\dagger \end{pmatrix}
\, h_u \,\begin{pmatrix}     m_d & \om_d\end{pmatrix}\,h_d'\right]  \,,
\end{equation}
which are a straightforward generalization of the invariant $I_1$ of Ref.~\cite{delAguila:1997vn} for an arbitrary number of isosinglet VLQs.
Finding a set of necessary and sufficient conditions for CP invariance in gauge interactions in terms of WB invariants is complicated in general.%
\footnote{The case $n_d = 1$ and $n_u=0$ has been analysed in Ref.~\cite{delAguila:1997vn}, where
a complete set of necessary and sufficient conditions for CP conservation has been given in terms of 7 independent CP-odd invariants.}
One can also construct CP-even WB invariants analogous to those presented in~\cref{eq:SMWBinv} for the SM case, e.g.
\begin{equation}
\tr \left(\left(h_d\right)^a \left(h_u\right)^b\right) \,, \quad
\tr \left(\left(h'_d\right)^a \left(\mathcal{H}_d\right)^b\right) \,, \quad
\tr \left(\left(h'_u\right)^a \left(\mathcal{H}_u\right)^b\right)\,, 
\end{equation}
with $a$ and $b$ positive integers. A sufficiently large number of independent CP-even and CP-odd WB invariants may be used to reconstruct the quark mixing matrix $V$~\cite{Albergaria:2022zaq}.

\vskip 2mm

The existence of VLQs naturally brings about new sources of CP violation. To illustrate this point, consider as an example the case where a single up-type VLQ is added to the SM ($n_u = 1$, $n_d = 0$) and one takes the extreme chiral limit $m_u = m_c = m_d = m_s = 0$. 
In such a limit $I_\text{SM} = 0$, i.e.~there is no CP violation in the SM since the masses of same-charge quarks are degenerate.
However, the interactions with VLQs allow for CP-violating phases which are physical even in this chiral limit (see also~\cite{delAguila:1997vn}). Indeed, for $n_u = 1$ and $n_d = 0$ one can choose a WB where $\mathcal{M}_d$ is diagonal via an appropriate choice of $W_L$ and $\mathcal{W}^d_R$, and where $\mathcal{M}_u = \mathcal{V}^u_L\, \mathcal{D}_u$ given that $\mathcal{V}_R^u$ can be cancelled by a suitable $\mathcal{W}^u_R$. The up-quark mass matrix then has two vanishing columns in the chiral limit. Using the remaining rotation and rephasing freedoms, one can arrive at a WB where
\begin{equation}
\mathcal{M}_u \sim \begin{pmatrix} 
0 & 0 & \times & 0 \\
0 & 0 & 0 & \times \\
0 & 0 & \times & \ast\\
0 & 0 & \times & \times
\end{pmatrix} \,, \qquad
\mathcal{M}_d \sim \begin{pmatrix} 
0 & 0 & 0 \\
0 & 0 & 0 \\
0 & 0 & m_b
\end{pmatrix} \,,
\end{equation}
and all entries are real except ``$\ast$'', which is complex and whose phase cannot be removed in general. Its presence implies the non-vanishing of the first invariant in~\cref{eq:newinv}, which turns out to be proportional to $\im(\ast)$ and $m_b^2$, signalling a violation of the CP symmetry beyond the SM.

The chiral limit is not only physically natural for studying CP violation beyond the SM but phenomenologically relevant in practice. At very high energy collisions, for example at the TeV scale, light fermion masses are negligible and there is no possibility of distinguishing light quark jets.
Sizeable CP violation at high energy is expected to have its origin beyond the SM.

\vskip 2mm

Finally, the extraction of physical phases from data needs to be reexamined in the presence of VLQs.
Recall that the mixing matrix $V$ is $(3+n_u) \times (3+n_d)$ and non-unitary in general. As explained in the previous subsection, its upper-left $3\times 3$ block $V_\text{CKM}$ contains 4 independent physical phases, see~\cref{eq:CKMphases}.
Without loss of generality, one may adopt the following phase convention~\cite{Branco:1999fs}
\vskip -3mm
\begin{equation}
V = 
  \tikz[baseline=(M.west)]{%
    \node[matrix of math nodes,matrix anchor=west,left delimiter=(,right delimiter=),ampersand replacement=\&] (M) {%
{}|V_{ud}{}| \& {}|V_{us}{}|\,e^{i\chi'} \& {}|V_{ub}{}|\,e^{- i \gamma} \& \,\cdots\, \\
-{}|V_{cd}{}| \& {}|V_{cs}{}| \& {}|V_{cb}{}| \& 	\cdots \\
{}|V_{td}{}|\,e^{-i \beta} \& -{}|V_{ts}{}|\,e^{i \chi} \& {}|V_{tb}{}| \& \cdots \\
\vdots \& \vdots \& \vdots \& \ddots\\
    };
    \node[draw,fit=(M-3-1)(M-1-3),inner sep = -1pt,label={\scriptsize $V_\text{CKM}$}] {};
  }\,.
\end{equation}

NP effects may interfere in how these phases are obtained from experiment. If NP contributions are negligible in weak processes where the SM contributes at tree level, as is the case in models of VLQs, then the extractions of $\gamma$ and of moduli from the first two rows of $V_\text{CKM}$ are not disturbed. However, the extraction of $\beta$ and $\chi$ from $B$ decays may be contaminated by NP effects, with experiments being instead sensitive to the combinations $\bar\beta \equiv \beta - \phi_d$ and $\bar\chi \equiv \chi + \phi_s$. Here, $\phi_q$ ($q=d,s$) are phases parameterizing NP-induced deviations from the SM in the neutral meson systems $B^0_q$--$\overline{B^0_q}$. For instance, the decay $B_d^0 \to J/\psi \, K$ turns out to be sensitive to $\sin 2(\bar\beta - \chi') \simeq \sin 2\bar\beta$, while the decay $B_s^0 \to J/\psi \, \phi$ measures $\sin 2\bar\chi$.
We should note that the measurement of these phases involves the interplay of CPV in mixing and decays. In the presence of FCNC there will be new contributions to the
decay amplitudes which makes the extraction of these phases even more complicated.

In the context of the SM or NP scenarios where $V_\text{CKM}$ is unitary, the remaining phase $\chi'$ is constrained by unitarity to be extremely small, namely $\chi' \sim 6 \times 10^{-4}$~\cite{Aguilar-Saavedra:2004roc}. If $V_\text{CKM}$ is not unitary but can nevertheless be written as part of a unitary matrix --- as is the case for models of VLQs (cf.~\cref{foot:part_of_unitary}) --- a bound on $\chi'$ can still be placed~\cite{Kurimoto:1997ex, Branco:1999fs}.
In particular, taking the CKM moduli to vary in their $3\sigma$ ranges~\cite{ParticleDataGroup:2020ssz}, one finds $|\chi'| \lesssim 0.06 \simeq 3^\circ$. This phase can be neglected in most cases of interest.
Similarly, the phase $\chi$ can be made significantly larger than the SM expectation $\chi \sim 0.02$ in models with up-type VLQs, but not in models with down-type VLQs alone~\cite{Aguilar-Saavedra:2004roc}. Up-type VLQs have also been used to alleviate tensions in the data via a non-zero $\phi_d$~\cite{Botella:2012ju}. As we will see in what follows, both up- and down-type VLQs may address apparent deviations from unitarity in the first row of $V_\text{CKM}$~\cite{Belfatto:2019swo,Belfatto:2021jhf,Branco:2021vhs}.

Note that possible deviations from unitarity modify the familiar SM unitary triangle and, e.g.~in the case of extensions with only one VLQ, give rise to various quadrangles~\cite{delAguila:1997vn} instead. Subsequently, instead of finding only one CP-odd invariant, proportional to $\im\left( V_{us} V_{cb} V_{ub}^* V_{cs}^*\right)$, one obtains several rephasing-invariants from other quartets, i.e.~proportional to $\im\left( V_{ij} V_{km} V_{im}^* V_{kj}^*\right)$, which in general are not equal but may be partly related.

\subsection{The effective light-quark squared mass matrices}
\label{sec:light-eff}
We close this section by pointing out how VLQs can leave an imprint in the structure of light quark masses and mixing, even in the limit of very heavy VLQ masses.
To illustrate this, consider a region of the physical parameter space
where the VLQ-related mass terms are much larger than the electroweak scale.
Under this assumption, the effective mass matrices for the three SM generations can be reliably obtained from the approximate block diagonalization of the full mass matrix including VLQs.

As rotations of the right-handed fields are not physically observable, we can concentrate on the squared Hermitian mass matrices $\mathcal{H}\equiv \mathcal{M}\mathcal{M}^\dagger$, with $\mathcal{M}_{(q)} \sim (3+n_q)\times(3+n_q)$ given in the form of~\cref{eq:genmass}. For each sector, we then define the shorthands
\begin{equation}
\mathcal{H}=
\begin{pmatrix}
h & p \\[1mm]
p^{\dagger } & H
\end{pmatrix}
\equiv 
\begin{pmatrix}
mm^{\dagger }+\overline{m}\,\overline{m}^{\dagger } & m\overline{M}^{\dagger }+%
\overline{m}M^{\dagger } \\ 
\overline{M} m^{\dagger }+M \overline{m}^{\dagger}
& \overline{M}\,\overline{M}^{\dagger }+MM^{\dagger }
\end{pmatrix} \,,
\label{h1}
\end{equation}
where $h\sim 3\times 3$ is the light block.
One expects $|h|\ll|p|\ll |H|$ in the sense of singular values.
Therefore, at leading order we can consider the block diagonalization
\begin{equation}
\cV^\dagger\, \mathcal{H}\, \cV
\,\simeq\,
\mtrx{h_{\rm eff} & 0\cr 0& H_{\rm eff}}
\,,
\end{equation}
with the matrix $\cV$ approximately given by
\begin{equation}\label{eq:light-V}
\cV\simeq\left( 
\begin{array}{cc}
\id_3 & \pH \\ 
-\pH^\dag & \id
\end{array}
\right)\,,
\end{equation}
having defined $\pH \equiv pH^{-1}$ to ensure the zero blocks. The non-zero effective blocks at this order are
\begin{equation}
\label{eq:heff}
h_{\rm eff}= h-pH^{-1}p^\dag\,,
\qquad
H_{\rm eff}= H\,.
\end{equation}
The former is the effective $3\times 3$ squared mass matrix for the SM-like quarks.
More explicitly, one has
\begin{equation}
\label{eq:explicitheff}
h_\text{eff} =  mm^{\dagger }+\overline{m%
}\,\overline{m}^{\dagger }-\left( m\overline{M}^{\dagger }+\overline{m}%
M^{\dagger }\right) \left( \overline{M}\,\overline{M}^{\dagger }+MM^{\dagger
}\right) ^{-1}\left( \overline{M} m^{\dagger }+M \overline{m}^{\dagger}\right)  \,.
\end{equation}
 From this expression, it is clear that in general there is no suppression of the third term, i.e.~all terms may contribute equally to the effective squared Hermitian mass matrices of the light quarks, since in the limit of large $M$ and $\oM$ the ``numerator'' compensates the suppression of the ``denominator''.

From~\cref{eq:explicitheff} one can also see that, in the WB in which $\oM=0$ one simply has 
\begin{equation}
    h_{\rm eff} = mm^\dagger \qquad\text{(vanishing $\overline{M}$ WB)}
\end{equation}
and the contributions to light masses from the light-heavy mixing have been incorporated in the new matrix $m$.
Next-to-leading-order corrections to $h_{\rm eff}$ can be derived. See~\cref{app:improved_seesaw} for a systematic expansion in this WB or Ref.~\cite{Grimus:2000vj} for the diagonalization starting from exact relations.

Consider finally the following toy scenario that illustrates how the inclusion of a VLQ may generate a mass for a massless SM quark and also generate (previously absent) mixing among SM quarks. 
Suppose one has three generations of standard quarks such that the $3\times 3$ mass matrix in both sectors has the form $m=\diag(0,0,m_{3})$. Obviously, there is only one massive generation and there is no quark mixing.
Now suppose an extra VLQ is added to one of the sectors such that, in that sector, the full mass matrix has the form 
\begin{equation}
\mathcal{M=}\left( 
\begin{array}{llll}
0 & 0 & 0 & 0 \\ 
0 & 0 & 0 & \overline{m}_{2} \\ 
0 & 0 & m_{3} & \overline{m}_{3} \\ 
0 & \oM_2 & 0 & M
\end{array}
\right)
\label{M1}
\end{equation}
in a certain basis, which may e.g.~be a consequence of flavour symmetry.
For simplicity, all parameters are taken to be real and $\overline{m}_{2}, \overline{m}_{3} < m_{3}$. 
Computing the resulting effective squared mass matrix via~\cref{eq:heff}, we obtain
\begin{equation}
h_{\rm eff}= \diag (0,0,m_3^2) +
\zeta^2
\begin{pmatrix}
0 & 0 & 0 \\ 
0 & \overline{m}_{2}^{2} & \overline{m}_{2}\,\overline{m}_{3} \\ 
0 & \overline{m}_{2}\,\overline{m}_{3} & 
\overline{m}_{3}^{2}
\end{pmatrix}\,,\quad \text{with } 
\zeta^{2}\equiv\frac{{\oM_2}^2}{{\oM_2}^2+M^2}\,.
\label{eq:heff1}
\end{equation}
One sees that, in the approximation $\zeta\ll 1$, one of the massless quarks has become massive, with mass $\zeta\, \om_2$, and an effective contribution to the mixing matrix also arises in that sector, namely
\begin{equation}
    \tilde{V}^\dag \,h_{\rm eff}\,\tilde{V}\, \simeq\, \diag (0,\,\zeta^2 \,\overline{m}_2^2,\, m_3^2)\,,
\end{equation}
with
\begin{equation}
\tilde{V}= \begin{pmatrix}
1 & 0 & 0 \\ 
0 & 1 & \frac{\overline{m}_{2}\,\overline{m}_{3}}{m_{3}^{2}} \zeta^2\\ 
0 & -\frac{\overline{m}_{2}\,\overline{m}_{3}}{m_{3}^{2}} \zeta^2 & 1
\end{pmatrix}\,.
\label{eq:mv}
\end{equation}
The existence of VLQs can thus be responsible for lighter-generation quark masses and mixing angles, whose smallness is connected and controlled by VLQ-related parameters --- in this case $\zeta$. While the above is a toy example, it hints at how the observed flavour structures (including the observed quark CP violation) may have their origin in simple vector-like extensions of the SM fermion sector. See also~\cite{Bjorken:2002vt} for a model built adopting a similar approach, where all masses of the standard down-quarks are generated by mixing with down-type VLQs.

\vfill
\clearpage

\section{Deviations from Unitarity}
\label{sec:devunit}
 
A central feature in VLQ extensions of the SM is the possible deviation from unitarity of the $3\times 3$ mixing matrix of SM quarks.
In the SM, the $3\times 3$ CKM matrix controlling the charged currents is unitary. Therefore, if one detects any deviation from the presumed unitarity, this will clearly constitute evidence for the existence of new physics beyond the Standard Model~\cite{Czarnecki:2004cw,Belfatto:2019swo,Belfatto:2021jhf,Crivellin:2021bkd,FlavourLatticeAveragingGroupFLAG:2021npn}. 
Specifically for singlet VLQs, which are the focus of our review, this deviation is a deficit in unitarity, contrasting to triplet VLQs which lead to an excess.
This feature is crucial in view of the CKM unitarity problem that will be discussed next.

\subsection{The CKM unitarity problem}\label{sec:CKMUP}
Imposing unitarity of the CKM matrix translates, for the first row, into
\begin{equation}
    |V_{ud}|^{2}+|V_{us}|^{2}+|V_{ub}|^{2}=1\,.  
\end{equation}
However, considering the fact that $|V_{ub}|^2$ is much smaller than the other two entries, with $|V_{ub}|^2 \simeq 1.6\times 10^{-5}$, the following approximate relation
\begin{equation}\label{eq:unitary}
    |V_{ud}|^{2}+|V_{us}|^{2} \simeq 1\,,
\end{equation}
is valid up to $\mathcal{O}\left(|V_{ub}|^2\right)$.
Current experimental data and the control of theoretical uncertainties have allowed to determine $|V_{ud}|$, $|V_{us}|$ and $|V_{ud}|/|V_{us}|$ with considerable precision. These results, together with the approximate relation in~\cref{eq:unitary}, provide stringent tests of CKM unitarity. 

At energies much smaller than $M_{W}$, the CKM matrix entries $|V_{ud}|$ and $|V_{us}|$ determine the strength of the effective charged-current interaction which describes leptonic decays of hadrons involving the valence quarks $u,d$ and $s$. 
While neutron decays~\cite{UCNt:2021pcg} are mostly relevant for $|V_{ud}|$, the entry $|V_{us}|$ is essentially calculated from kaon decay data.
The ratio $\left| V_{us}/V_{ud}\right| $ can be independently determined by comparing rates of certain radiative pion and kaon decays. 

Recently, the use of improved values for the form factors and radiative corrections associated to the relevant meson and neutron decay processes has resulted in a downward shift of the central value for $|V_{ud}|$, leading to tension with the first row CKM unitarity constraint~\cite{Seng:2018yzq,Seng:2018qru,Czarnecki:2019mwq,Seng:2020wjq,Hayen:2020cxh,Shiells:2020fqp,Cirigliano:2022yyo}.
Taking into account these new theory calculations with reduced hadronic uncertainties, the authors of Ref.~\cite{Belfatto:2019swo} find:
\begin{equation}\label{eq:cabibbo2}
    \begin{array}{ccc}
         |V_{us}|=0.22333(60)\,,\quad  & |V_{ud}|=0.97370(14)\,,\quad  & \left|V_{us}/V_{ud}\right| =0.23130(50)\,.
    \end{array}
\end{equation}
These results deviate from the unitarity condition in~\cref{eq:unitary} by
more than $4\sigma $, disfavouring the SM CKM unitarity at this level.
In fact, the values in~\cref{eq:cabibbo2} are compatible with
\begin{equation}\label{eq:unitary2}
    |V_{ud}|^{2}+|V_{us}|^{2}=1-\delta\,,
\end{equation}
where 
\begin{equation}\label{eq:delta}
    \sqrt{\delta} =0.04\pm 0.01\, \quad(95\%\text{ C.L.})\,.
\end{equation}
Thus, one finds that $\delta$ is of $\mathcal{O}(|V_{cb}|^2)$. 
Recent lattice results from the FLAG collaboration~\cite{FlavourLatticeAveragingGroupFLAG:2021npn} give support to these findings. 
This discrepancy between the SM and experimental data is referred to as the CKM unitarity problem or Cabibbo (angle) anomaly.%
\footnote{Note however that a test of Cabibbo angle universality, i.e.~the equality of the Cabibbo angle as extracted from independent processes, goes beyond a test of CKM unitarity~\cite{Grossman:2019bzp}.}

New physics beyond the SM is required to account for this discrepancy, should it hold. A possible solution is the introduction of an extra mixing channel.
One way to achieve this effect is the addition of a singlet vector-like quark to the SM.
An extra VLQ would lead to a larger mixing matrix, with the effective CKM mixing expressed as a $3\times 3$ non-unitary sub-matrix involving the standard quarks.
In the literature, both models with one extra down-type VLQ or an extra up-type VLQ were considered in this context~\cite{Belfatto:2019swo,Belfatto:2021jhf,Branco:2021vhs,Botella:2021uxz}.

\subsection{From an exact parameterization to approximate ones}
\label{sec:exact_param_X}
In this subsection, we will first present an exact parameterization for the quark mixing matrix $V$ in the presence of VLQs. Then, we will introduce an approximation which is only valid for small deviations from unitarity. Parameterizations have no physical content. Nevertheless, different parameterizations may be particularly suited to discuss different aspects of a given problem. Hence, their usefulness depends on the context. 
Approximations may be very useful to simplify expressions and extract physical insights. However, one must take care with their domain of validity. In particular, some approximations that are widely used in a context where the additional particles are very heavy fail when the scale of their masses approaches the electroweak scale. 

Consider the $(3+n_q)\times(3+n_q)$ unitary matrices $\cV^q_\chi$ introduced in~\cref{sec:massmatrices} to diagonalize the mass matrices $\cM_q$ (in the rest of this subsection, we will often drop the indices $q=u,d$ and $\chi=L,R$ for simplicity).
It was shown, initially for Majorana neutrinos in~\cite{Agostinho:2017wfs,Branco:2019avf}, that one may decompose any unitary matrix in the following manner,
\begin{equation}\label{eq:uparam-exact}
    \cV =
    \begin{pmatrix}\,\,\, A\,\,\, \\B\end{pmatrix}=
    \begin{pmatrix}K&0\\0&\oK\end{pmatrix}\begin{pmatrix}\id&X\\-X^\dagger&\id\end{pmatrix}=
    \renewcommand{\arraystretch}{1}\setlength{\extrarowheight}{6pt}
    \left(\begin{array}{c;{2pt/2pt}c}
      \,\,\,K\,\,\,\,\, & \,\,\,K\,X\,\\[1.5mm] \hdashline[2pt/2pt]
      \!\myunder{\scriptstyle 3}{-\oK{X}^\dag\,\, }\! &\myunder{\scriptstyle\,\,\,n_q }{\quad\,\,\oK\quad\,\,}\!\\[1mm]
    \end{array}\right)\!\!\!
    \begin{array}{l}%
      \myrightbrace{-\oK\,X}{\scriptstyle\,3}\\[0mm]
      \myrightbrace{\quad\oK\quad}{\scriptstyle\,n_q}
    \end{array}\,,\\[4mm]
\end{equation}
where $\gls*{X}$ is a $3\times n_q$ matrix.
This parameterization is completely general and valid whenever the $3\times3$ matrix $K$ and the $n_q\times n_q$ matrix $\,\oK$ are non-singular.
With this notation, the $(3+n_u) \times (3+n_d)$ quark mixing matrix $V$ that controls the charged currents is of the form
\begin{equation}\label{eq:uparam-exactV}
    V=A^{u\dagger}_LA^d_L=\renewcommand{\arraystretch}{1}\setlength{\extrarowheight}{6pt}
    \left(\begin{array}{c;{2pt/2pt}c}
      \,\,\,V_\text{CKM}\,\,\,\,\, & \,\,\, V_\text{CKM}\,{X_L^d}
      \\[1.5mm] \hdashline[2pt/2pt]
      \!\!\myunder{\scriptstyle 3}{\,\,\,\,{X_L^u}^\dag \,V_\text{CKM}\,\,\,\,\, }\!\!\! &
      \!\myunder{\scriptstyle \,\,\,n_d}{\,\,\,{X_L^u}^\dag \,V_\text{CKM}\,{X_L^d}\,\,\,\,}\!\!\!\\[1mm]
    \end{array}\right)\!\!\!
    \begin{array}{l}%
      \myrightbrace{V_\text{CKM}\,{X_L^d}^\dagger}{\scriptstyle \,3}\\[0mm]
      \myrightbrace{V_\text{CKM}\,{X_L^d}^\dagger}{\scriptstyle \, n_u}
    \end{array},\\[4mm]
\end{equation}
where the $3\times3$ CKM matrix is given by
\begin{equation}\label{eq:uparam-exactCKM}
    V_\text{CKM}=K^{u\dagger}_LK^d_L\,.
\end{equation}
From the decomposition in~\cref{eq:uparam-exact}, we check that each large unitary matrix $\cV$ has three ingredients: $K$, $\oK$ and $X$.
Unitarity of $\cV$ further relates $K$, $\oK$ and $X$ as
\begin{equation}\label{eq:uparam-exactK}
    K=V_K\big(\id_3+XX^\dagger\big)^{-1/2}\,,\quad\oK=V_{\oK}\big(\id_{n_q}+X^\dagger X\big)^{-1/2}\,,
\end{equation}
with $V_K$ a $3\times3$ unitary matrix and $V_{\oK}$ a $n_q\times n_q$ unitary matrix.
Hence, $X,V_K,V_{\oK}$ fully determine $\cV$.
The combinations $(\id_3+XX^\dagger)^{-1/2}$ and $(\id_{n_q}+X^\dagger X)^{-1/2}$ are manifestly Hermitian.
When there are no VLQs in a sector of the model, 
the respective $K^q_L$ is unitary ($X^{q}_L =0$).
As a result, deviations from unitarity in $V_\text{CKM}$ become directly related to those of the other $K^{q'}_L$, thus giving additional importance to~\cref{eq:uparam-exactK}.
Since this last $K^{q'}_L$ becomes unitary when the corresponding $X=0$, one may then see $X$ as a simple way of parameterizing all physical deviations from $3\times 3$ unitarity.

\vskip 2mm
To derive useful approximations, it is convenient to simplify~\cref{eq:uparam-exact,eq:uparam-exactK} and consider a specific WB. We start by noting that the WB transformation
\begin{equation}\label{eq:uparam-wbt}
\begin{array}{l@{\qquad}l}
  u^0_L \,\to\, V_{K^u_L}  \, u^0_L
  \,,
  &d^0_L \,\to\, V_{K^u_L}  \, d^0_L
  \,,\\[2mm]
  U^0_L \,\to\, V_{\oK\!^u_L}  \, U^0_L
  \,,
  &D^0_L \,\to\, V_{\oK\!^d_L}  \, D^0_L
  \,,
\end{array}
\end{equation}
removes the matrices $V_{K^u_L}$, $V_{\oK\!^d_L}$ and $V_{\oK\!^u_L}$ from~\cref{eq:uparam-exactK} without modifying the charged currents.
Then, we expand
\begin{equation}\label{eq:uparam-approx}
    \big(\id_3+XX^\dagger\big)^{-1/2}\simeq\id_3-\frac{1}{2}XX^\dagger\,,\quad            \big(\id_n+X^\dagger X\big)^{-1/2}\simeq\id_n-\frac{1}{2}X^\dagger X\,,
\end{equation}
for small $X$ (i.e.~small deviations from unitarity) 
to write~\cref{eq:uparam-exact} in the following way:
\begin{equation}\label{eq:uparam-approxVX}
\begin{split}
    \cV^d_L&\simeq
    \begin{pmatrix}V'&0\\0&\id_n\end{pmatrix}
    \begin{pmatrix}\id_3-\frac{1}{2}\Theta_d\Theta^\dagger_d&\Theta_d\\ -\Theta^\dagger_d&\id_{n_d}-\frac{1}{2}\Theta^\dagger_d\Theta_d\end{pmatrix}\,,\\[1mm]
    \cV^u_L&\simeq
    \begin{pmatrix}\id_3-\frac{1}{2}\Theta_u\Theta^\dagger_u&\Theta_u\\                      -\Theta^\dagger_u&\id_{n_u}-\frac{1}{2}\Theta^\dagger_u\Theta_u\end{pmatrix}\,,
\end{split}\end{equation}
where we have defined $V' = V_{K^u_L}^\dagger V_{K^d_L}$. Furthermore, we have replaced the notation $X^q_L$ by $\gls*{Theta}_q$, to highlight the fact that an approximation has been made. This convention will be used in what follows.%
\footnote{While the approximation of~\eqref{eq:light-V} in~\cref{sec:light-eff} matches the lower-order version of~\cref{eq:uparam-approxVX} with $\Theta\to\pH$ and $V' = \id$, the focus there is only on the block diagonalization of $\mathcal{M}\mathcal{M}^\dagger$ (see also~\cref{app:improved_seesaw}).}
With this notation, the quark mixing matrix is given by
\begin{equation}\label{eq:uparam-approxV}
    V\simeq\begin{pmatrix}                                                                     V'-\frac{1}{2}\Theta_u\Theta^\dagger_uV'-\frac{1}{2}V'\Theta_d\Theta^\dagger_d&        V'\Theta_d\\\Theta^\dagger_uV'&\Theta^\dagger_uV'\Theta_d\end{pmatrix}\,,
\end{equation}
where the $3\times 3$ unitary matrix $V'$ is related to the CKM matrix $V_\text{CKM}$ through
\begin{equation}
\label{eq:approxCKM}
        V_\text{CKM}\simeq                                       \bigg(\id_3-\frac{1}{2}\Theta_u\Theta^\dagger_u\bigg)\,V' \,           \bigg(\id_3-\frac{1}{2}\Theta_d\Theta^\dagger_d\bigg)\,,
\end{equation}
an expression valid up to terms of order $\Theta^3_q$.
The deviation from unitarity is evident for nonzero $\Theta_u$ or $\Theta_d$.

\vskip 2mm
In the literature, a polar decomposition is often employed to split the CKM matrix in an Hermitian part $H_{L,R}$ and a unitary part $U_\text{CKM}$:%
\footnote{See e.g.~\cite{delAguila:2000aa}, and also~\cite{Fernandez-Martinez:2007iaa} in the context of neutrinos.}
\begin{equation}
    V_\text{CKM}\,=\,U_\text{CKM}\,H_R\,=\,H_L\,U_\text{CKM}\,,
\end{equation}
with the Hermitian matrices being trivially related through $H_R=U^\dagger_\text{CKM}H_LU_\text{CKM}$.
It is then common to make an analogy with the deviations from unitarity in the neutrino sector by defining
\begin{equation}
    H_L=\id_3+\eta_L\,,\quad H_R=\id_3+\eta_R\,,
\end{equation}
such that $\eta_{R,L}$ parameterize all deviations from unitarity in the upper $3\times3$ block of the quark mixing matrix.
In general, one cannot relate these couplings to the previous notation in a trivial manner.
Nevertheless, when no down-type VLQs are present in the theory, we find
\begin{equation}
    \eta_L=\big(\id_3+X^u_LX^{u\dagger}_L\big)^{-1/2}-\id_3                                   \simeq-\frac{1}{2}\Theta_u\Theta^\dagger_u\,.
\end{equation}
Likewise, the expression
\begin{equation}
    \eta_R=\big(\id_3+X^d_L X^{d\dagger}_L\big)^{-1/2}-\id_3                                     \simeq-\frac{1}{2}\Theta_d\Theta^\dagger_d
\end{equation}
applies in models without up-type VLQs.

\vskip 2mm
While all of the previous treatments can be useful in their own right, one often favours the use of angles and phases as in the traditional PDG parameterization~\cite{Chau:1984fp,ParticleDataGroup:2020ssz}.
With that in mind, it is useful to embed $V$ in a larger unitary matrix $\mathcal{U}$ for which such parameterizations are known.
In Ref.~\cite{Branco:1992wr}, this was achieved with the auxiliary
\begin{equation}\label{eq:uparam-param}
    \mathcal{U}=\begin{pmatrix}V&B^{u\dagger}_L\\B^d_L&0\end{pmatrix}\,,
\end{equation}
where the $B^q_L$ were defined in~\cref{sec:massmatrices}
(see also~\cref{eq:uparam-exact}).
This embedding will be the starting point for the parameterizations of the quark mixing matrix of~\cref{sec:ph-parameterizations}.

\subsection{Flavour-changing neutral currents}
\label{sec:du-FCNC}

Before proceeding, let us comment on standard quark FCNC and their relation to deviations from unitarity of $V_\text{CKM}$, exploiting also the above parameterization.
As shown throughout~\cref{sec:gaugeinteractions}, in models with singlet VLQs the FCNC for the standard quarks in each sector are entirely determined by $F^d=V^\dagger V$ and $F^u=VV^\dagger$.
Clearly, both are absent, i.e.~one has the upper-left $3\times 3$ block of each $\gls*{Fq}$ equal to $\id_3$, when $V_\text{CKM}$, the upper-left $3\times 3$ block of the full quark mixing matrix $V$, is unitary.
Therefore, FCNC are intimately connected to 
the previously mentioned deviations from unitarity.
With the exact parameterization~\eqref{eq:uparam-exact}, such behaviour is made explicit.
In fact, one can show that both $F^q$ ($q=u,d$) can be written in the compact form:
\begin{equation}\begin{aligned}\label{eq:duFCNC-exact}
    F^q&=\renewcommand{\arraystretch}{1}\setlength{\extrarowheight}{6pt}
    \left(\begin{array}{c;{2pt/2pt}c}
      \,\,\,\left(\id+X^q_{L}{X^q_{L}}^{\dagger }\right) ^{-1}\,\,\,\,\, & \,\,\,\left(\id+X^q_{L}{X^q_{L}}^{\dagger }\right) ^{-1}{X_L^q}
      \\[1.5mm] \hdashline[2pt/2pt]
      \!\!\myunder{\scriptstyle 3}{\,\,\,\,{X_L^q}^\dag \left(\id+X^q_L {X^q_L}^{\dagger }\right) ^{-1}\,\,\,\,\, }\!\!\! &
      \!\myunder{\scriptstyle \,\,\,n_q}{\,\,\,{X_L^q}^\dag \left(\id+X^q_L{X^q_L}^{\dagger }\right) ^{-1}{X_L^q}\,\,\,\,}\!\!\!\\[1mm]
    \end{array}\right) \!\!\!
    \begin{array}{l}%
      \myrightbrace{\big({K_L^q}^\dagger K_L^q\big){X_L^q}^\dagger}{\scriptstyle \,3}\\[0mm]
      \myrightbrace{X_L^q \big({K_L^q}^\dagger K_L^q\big){X_L^q}^\dagger}{\scriptstyle \, n_q}
    \end{array}\\[-1mm]
\,.
\end{aligned}\end{equation}
\vskip 2mm
\noindent
When the deviations from unitarity are small, this relation may be approximated by
\begin{equation}\label{eq:duFCNC-approx}
    F^q\simeq                                                                              \begin{pmatrix}\id_3-\Theta_q\Theta^\dagger_q&\Theta_q\\                                  \Theta^\dagger_q&\Theta^\dagger_q\Theta_q\end{pmatrix}\,,
\end{equation}
with the replacement $X_L^q \to \Theta_q$, an expression valid up to terms of order $\Theta^3_q$.
Note that in the limit $\Theta_q \to 0$, $F^q$ approaches the diagonal matrix $\diag(1,1,1,0,\ldots,0)$.

\vskip 2mm

In order to better illustrate the relation between FCNC and deviations from unitarity, we introduce the parameters
\begin{equation}\label{eq:duFCNC-Delta}
    \Delta^d_n\equiv 1-\sum^{3+n_u}_{j=1}\big|V_{jn}\big|^2\,,
    \qquad                             
    \Delta^u_n\equiv 1-\sum^{3+n_d}_{j=1}\big|V_{nj}\big|^2\,.
\end{equation}
Here, $\Delta^d_n$ and $\Delta^u_n$ test the unitarity of, respectively, the $n$-th column and $n$-th row of $V$, with $n=1,2,3$.
Since $F^d=V^\dagger V$ and $F^u=VV^\dagger$, they are also given by
\begin{equation}
    \Delta^q_n=1-F^q_{nn}\,.
\end{equation}
\Cref{eq:duFCNC-exact,eq:duFCNC-approx} allow us to write
\begin{equation}
    \Delta^q_n=1-\Big(\id_3+X^q_LX^{q\dagger}_L\Big)^{-1}_{nn}\simeq                    \big(\Theta_q\Theta^\dagger_q\big)_{nn}=\sum^{n_q}_{j=1}
    \big|(\Theta_q)_{nj}\big|^2\,.
\end{equation}
At this stage, we should point out that these parameters are not easily measurable.
Instead, 
experimentally one tends to test the unitarity of $V_\text{CKM}$ directly (as presented in~\cref{sec:CKMUP}).
With that in mind, we probe the unitarity of its columns with
\begin{equation}
    \delta^d_n\equiv 1-\sum^3_{j=1}\big|V_{jn}\big|^2=1-|V_{un}|^2-|V_{cn}|^2-|V_{tn}|^2
\end{equation}
and that of its rows through
\begin{equation}
    \delta^u_n\equiv 1-\sum^3_{j=1}\big|V_{nj}\big|^2=1-|V_{nd}|^2-|V_{ns}|^2-|V_{nb}|^2\,.
\end{equation}
By comparing these expressions with~\cref{eq:duFCNC-Delta}, we can relate the 
parameters $\Delta^q_n$ to the more experimental ones $\delta^q_n$,
\begin{equation}
    \delta^d_n-\Delta^d_n=\sum^{n_u}_{j=1}\big|V_{j+3,n}\big|^2\,,\qquad                       \delta^u_n-\Delta^u_n=\sum^{n_d}_{j=1}\big|V_{n,j+3}\big|^2\,.
\end{equation}
By using~\cref{eq:uparam-exactV,eq:uparam-approxV}, we can further develop these relations into
\begin{equation}\begin{split}
    \delta^d_n-\Delta^d_n&=                                                          \sum^{n_u}_{j=1}\big|(X^{u\dagger}_LV_\text{CKM})_{jn}\big|^2\simeq                \sum^{n_u}_{j=1}\big|(\Theta^\dagger_uV')_{jn}\big|^2\,,\\
    \delta^u_n-\Delta^u_n&=                                                          \sum^{n_d}_{j=1}\big|(V_\text{CKM}X^d_L)_{nj}\big|^2\simeq                         \sum^{n_d}_{j=1}\big|(V'\Theta_d)_{nj}\big|^2\,.
\end{split}\end{equation}
Finally, 
it is instructive to consider 
the following limit scenarios:
\begin{itemize}
    \item 
    When there are only down-type VLQs in the theory ($n_u=0$, $n_d \neq 0$), the rows of $V$ are orthonormal, which gives $F^u = \id_3$ and $\Delta^u_n=0$.
    However, since these rows have more than three entries, the rows of the $3 \times 3$ $V_\text{CKM}$ generically deviate from unitarity, i.e.~they are not orthonormal, $\delta^u_n=\sum^{n_d}_{j=1}|V_{n,j+3}|^2\neq0$. 
    Meanwhile, since in these models $V$ is a $3\times (3+n_d)$ matrix,
    $\delta^d_n=\Delta^d_n$, and the columns of $V_\text{CKM}$ are also not expected to be orthonormal.
    \item
    When there are only up-type VLQs in the theory ($n_u \neq 0$, $n_d=0$), the columns of $V$ are orthonormal, i.e.~$F^d = \id_3$ and $\Delta^d_n=0$.
    Nevertheless, $\delta^d_n=\sum^{n_u}_{j=1}|V_{j+3,n}|^2\neq0$ generically.
    Meanwhile, since in these models $V$ is a $(3+n_u)\times3$ matrix, $\delta^u_n=\Delta^u_n$. As in the previous case, neither the rows nor the columns of the $3\times 3$ $V_\text{CKM}$ are expected to be orthonormal in general.
\end{itemize}
In the following section, we focus on useful parameterizations for $V$ and $F^{u,d}$ in such limit cases, where only one VLQ isosinglet is added to the SM.

\subsection{Parameterizations for models with only one VLQ}
\label{sec:ph-parameterizations}

In models where a single down-type VLQ $B$ is introduced ($n_d=1$, $n_u=0$), $B^u_L$ has no physical meaning and the corresponding columns are absent from the embedding of~\cref{eq:uparam-param}, which reads 
\begin{equation}
\mathcal{U} = \begin{pmatrix} V \\ B_L^d \end{pmatrix}\,,
\end{equation}
thus establishing
the quark mixing matrix $V$ as \emph{the first three rows} of a $4\times4$ unitary matrix $\mathcal{U}$.
For this scenario, we consider the following parameterization,
\begin{equation}\label{eq:SVLQ-Udown}
    \mathcal{U}=\mkern-4mu
    \begingroup
    \setlength\arraycolsep{2pt}
    \begin{pmatrix}
    &&&0\\&V_\text{PDG}&&0\\&&&0\\0&0&0&1
    \end{pmatrix}
    \mkern-8mu
    \begin{pmatrix}
    c_{14}&&&-s_{14}e^{-i\delta_{14}}\\&1\\&&1\\s_{14}e^{i\delta_{14}}&&&c_{14}
    \end{pmatrix}
    \mkern-8mu
    \begin{pmatrix}
    1\\&c_{24}&&-s_{24}e^{-i\delta_{24}}\\&&1\\&s_{24}e^{i\delta_{24}}&&c_{24}
    \end{pmatrix}
    \mkern-8mu
    \begin{pmatrix}
    1\\&1\\&&c_{34}&-s_{34}\\&&s_{34}&c_{34}
    \end{pmatrix}
    \,,
    \endgroup
\end{equation}
where $V_\text{PDG}$ is the standard PDG parameterization of a $3\times3$ unitary matrix~\cite{Chau:1984fp,ParticleDataGroup:2020ssz},
\begin{equation}\label{eq:SVLQ-PDG}
    V_\text{PDG}=\mkern-4mu
    \begin{pmatrix}
    c_{12}c_{13}&s_{12}c_{13}&s_{13}e^{-i\delta}\\
    -s_{12}c_{23}-c_{12}s_{23}s_{13}e^{i\delta}&c_{12}c_{23}-s_{12}s_{23}s_{13}e^{i\delta}&s_{23}c_{13}\\
    s_{12}s_{23}-c_{12}c_{23}s_{13}e^{i\delta}&-c_{12}s_{23}-s_{12}c_{23}s_{13}e^{i\delta}&c_{23}c_{13}
    \end{pmatrix}\,,
\end{equation}
and $(c_{ij},s_{ij})=(\cos\theta_{ij},\sin\theta_{ij})$.
With this choice, the traditional picture of $V_\text{CKM}$ is recovered when all $\theta_{i4}\to0$.
Meanwhile, in these models the matrices which control all tree-level FCNC are given by
\begin{equation}\label{eq:SVLQ-FD}
    F^d=\mkern-4mu
    \begin{pmatrix}
    1-s^2_{14}&-s_{14}c_{14}s_{24}e^{-i(\delta_{14}-\delta_{24})}&-s_{14}c_{14}c_{24}s_{34}e^{-i\delta_{14}}&-s_{14}c_{14}c_{24}c_{34}e^{-i\delta_{14}}\\
    -s_{14}c_{14}s_{24}e^{i(\delta_{14}-\delta_{24})}&1-c^2_{14}s^2_{24}&-c^2_{14}s_{24}c_{24}s_{34}e^{-i\delta_{24}}&-c^2_{14}s_{24}c_{24}c_{34}e^{-i\delta_{24}}\\
    -s_{14}c_{14}c_{24}s_{34}e^{i\delta_{14}}&-c^2_{14}s_{24}c_{24}s_{34}e^{i\delta_{24}}&1-c^2_{14}c^2_{24}s^2_{34}&-c^2_{14}c^2_{24}s_{34}c_{34}\\
    -s_{14}c_{14}c_{24}c_{34}e^{i\delta_{14}}&-c^2_{14}s_{24}c_{24}c_{34}e^{i\delta_{24}}&-c^2_{14}c^2_{24}s_{34}c_{34}&1-c^2_{14}c^2_{24}c^2_{34}
    \end{pmatrix}
\end{equation}
and $F^u=\id_3$.
 For small angles, $F^d$ becomes
 \begin{equation}\label{eq:SVLQ-FD:exp}
    F^d\simeq\mkern-4mu
    \begin{pmatrix}
    1-\theta^2_{14}&-\theta_{14}\theta_{24}e^{-i(\delta_{14}-\delta_{24})}&-\theta_{14}\theta_{34}e^{-i\delta_{14}}&-\theta_{14}e^{-i\delta_{14}}\\
    -\theta_{14}\theta_{24}e^{i(\delta_{14}-\delta_{24})}&1-\theta^2_{24}&-\theta_{24}\theta_{34}e^{-i\delta_{24}}&-\theta_{24}e^{-i\delta_{24}}\\
    -\theta_{14}\theta_{34}e^{i\delta_{14}}&-\theta_{24}\theta_{34}e^{i\delta_{24}}&1-\theta^2_{34}&-\theta_{34}\\
    -\theta_{14}e^{i\delta_{14}}&-\theta_{24}e^{i\delta_{24}}&-\theta_{34}&\theta^2_{14}+\theta^2_{24}+\theta^2_{34}
    \end{pmatrix}\mkern-4mu
\end{equation}
plus terms of order $\mathcal{O}(\theta^3_{i4})$.

\vskip 2mm

In theories in which a single up-type VLQ $T$ is introduced, it is $B^d_L$ and the corresponding rows that are absent.
Then,~\cref{eq:uparam-param} 
reads 
\begin{equation}
\mathcal{U} = \begin{pmatrix} V & B_L^{u \dagger} \end{pmatrix}\,,
\end{equation}
and
relates the quark mixing matrix $V$ with \emph{the first three columns} of the $4\times4$ unitary matrix $\mathcal{U}$.
For this scenario, we parameterize $\mathcal{U}$ with the six mixing angles and three complex phases in
\begin{equation}\label{eq:SVLQ-Uup}
    \mathcal{U}=\mkern-4mu
    \begingroup
    \setlength\arraycolsep{2pt}
    \begin{pmatrix}
    1\\&1\\&&c_{34}&s_{34}\\&&-s_{34}&c_{34}
    \end{pmatrix}
    \mkern-8mu
    \begin{pmatrix}
    1\\&c_{24}&&s_{24}e^{-i\delta_{24}}\\&&1\\&-s_{24}e^{i\delta_{24}}&&c_{24}
    \end{pmatrix}
    \mkern-8mu
    \begin{pmatrix}
    c_{14}&&&s_{14}e^{-i\delta_{14}}\\&1\\&&1\\-s_{14}e^{i\delta_{14}}&&&c_{14}
    \end{pmatrix}
    \mkern-8mu
    \begin{pmatrix}
    &&&0\\&V_\text{PDG}&&0\\&&&0\\0&0&0&1
    \end{pmatrix}
    \,.
    \endgroup
\end{equation}
Once again, $V_\text{CKM}$ is described by the traditional PDG picture when all $\theta_{i4}\rightarrow 0$ (see also~\cite{Botella:1985gb}).
With this parameterization, the matrices which control the tree-level FCNC of these models are $F^d=\id_3$ and
\begin{equation}\label{eq:SVLQ-FU}
    F^u=\mkern-4mu
    \begin{pmatrix}
    1-s^2_{14}&-s_{14}c_{14}s_{24}e^{-i(\delta_{14}-\delta_{24})}&-s_{14}c_{14}c_{24}s_{34}e^{-i\delta_{14}}&-s_{14}c_{14}c_{24}c_{34}e^{-i\delta_{14}}\\
    -s_{14}c_{14}s_{24}e^{i(\delta_{14}-\delta_{24})}&1-c^2_{14}s^2_{24}&-c^2_{14}s_{24}c_{24}s_{34}e^{-i\delta_{24}}&-c^2_{14}s_{24}c_{24}c_{34}e^{-i\delta_{24}}\\
    -s_{14}c_{14}c_{24}s_{34}e^{i\delta_{14}}&-c^2_{14}s_{24}c_{24}s_{34}e^{i\delta_{24}}&1-c^2_{14}c^2_{24}s^2_{34}&-c^2_{14}c^2_{24}s_{34}c_{34}\\
    -s_{14}c_{14}c_{24}c_{34}e^{i\delta_{14}}&-c^2_{14}s_{24}c_{24}c_{34}e^{i\delta_{24}}&-c^2_{14}c^2_{24}s_{34}c_{34}&1-c^2_{14}c^2_{24}c^2_{34}
    \end{pmatrix}
    \,.
\end{equation}
For small angles, we neglect $\mathcal{O}(\theta^3_{i4})$ terms  to express the latter as
\begin{equation}\label{eq:SVLQ-FU:exp}
    F^u\simeq
    \begin{pmatrix}
    1-\theta^2_{14}&-\theta_{14}\theta_{24}e^{-i(\delta_{14}-\delta_{24})}&-\theta_{14}\theta_{34}e^{-i\delta_{14}}&-\theta_{14}e^{-i\delta_{14}}\\
    -\theta_{14}\theta_{24}e^{i(\delta_{14}-\delta_{24})}&1-\theta^2_{24}&-\theta_{24}\theta_{34}e^{-i\delta_{24}}&-\theta_{24}e^{-i\delta_{24}}\\
    -\theta_{14}\theta_{34}e^{i\delta_{14}}&-\theta_{24}\theta_{34}e^{i\delta_{24}}&1-\theta^2_{34}&-\theta_{34}\\
    -\theta_{14}e^{i\delta_{14}}&-\theta_{24}e^{i\delta_{24}}&-\theta_{34}&\theta^2_{14}+\theta^2_{24}+\theta^2_{34}
    \end{pmatrix}
    \,.
\end{equation}
Notice that, due to the parameterizations chosen here,~\cref{eq:SVLQ-FD,eq:SVLQ-FU} are identical.
This is, however, a simple textural coincidence, as the corresponding models are unrelated. In practice, the experimental fits should prefer different values of the $\theta_{i4}$ in each case.

\vskip 2mm

With~\cref{eq:duFCNC-approx},~\cref{eq:SVLQ-FD:exp} is translated into the language of~\cref{sec:exact_param_X} through
\begin{equation}\label{eq:SVLQ-Tdown}
    \Theta_d=-\theta_{14}e^{-i\delta_{14}}\,,\quad\Theta_s=-\theta_{24}e^{-i\delta_{24}}\,,\quad\Theta_b=-\theta_{34}\,,
\end{equation}
where we have expanded $\Theta_d$ as $(\Theta_d,\Theta_s,\Theta_b)^T$.
Note that $\Theta_d$ may denote both a $3\times 1$
complex vector and its first entry.
We adhere to this notation unless there is risk of confusion.
Meanwhile, after comparing~\cref{eq:SVLQ-FU:exp} with~\eqref{eq:duFCNC-approx}, we find
\begin{equation}\label{eq:SVLQ-Tup}
    \Theta_u=-\theta_{14}e^{-i\delta_{14}}\,,\quad\Theta_c=-\theta_{24}e^{-i\delta_{24}}\,,\quad\Theta_t=-\theta_{34}\,,
\end{equation}
where the angles and phases now refer to the parameters introduced in~\cref{eq:SVLQ-Uup}.
Like before, we have expanded the vector $\Theta_u$ as $(\Theta_u,\Theta_c,\Theta_t)^T$.

\subsection{Mass matrices and perturbativity}
\label{sec:perturb}

Up to this point, within the present section, we have not committed to a particular form of the quark mass matrices or WB (see~\cref{sec:wbzero}).
Using the exact parameterization introduced in~\cref{sec:exact_param_X}, one can relate the mass matrices $\mathcal{M}$, the matrices $K$, $\oK$ and $X$ parameterizing the mixing, and the physical quark masses. For a given sector $q$ and starting from the bidiagonalization relation ${\mathcal{V}_L}^\dagger\mathcal{M}\mathcal{V}_R = \mathcal{D}$ of~\cref{eq:diagonalization}, one finds
\begin{equation}
\label{eq:perturb-diagonalization}
\begin{split}
    m=K_L\big(d+X_LDX^\dagger_R\big)K^\dagger_R\,,&\quad\om=K_L\big(X_LD-dX_R\big)\,\oK\!^\dagger_R\,,\\
    \oM=\oK\!_L\big(DX^\dagger_R-X^\dagger_Ld\big)K^\dagger_R\,,&\quad M=\oK\!_L\big(D+X^\dagger_LdX_R\big)\,\oK\!^\dagger_R\,.
\end{split}
\end{equation}
These are the same relations given in~\cref{eq:mdefs}, but written in terms of a specific parameterization for $\mathcal{V}_L$ and $\mathcal{V}_R$.
Depending on the WB, these rotations may be further connected.
Namely, it follows that 
\begin{align}
    d X_L &= X_R D \quad\text{in the vanishing } \oM \text{ WB}\,,  \\[1mm]
  \text{and}\quad   d X_R &= X_L D \quad\text{in the vanishing } \om \text{ WB}\,.
\end{align}    

Concentrating on the more physical left-handed rotation, one may relate $X_L$ to the four matrix blocks in $\mathcal{M}$, starting from the relation $\mathcal{V}_L^\dagger \mathcal{M} \mathcal{M}^\dagger \mathcal{V}_L= \mathcal{D}^2$ . This is done in~\cref{app:improved_seesaw}, where an improved seesaw expansion is developed. Employing the definitions of~\cref{eq:uparam-exactK}, 
one finds, to first order,
\begin{equation}
X_L \simeq \Theta = V_{K_L}^\dagger \left(\om M^{-1}\right) V_{\oK\!_L}\,,   
\end{equation}
in the vanishing $\oM$ WB, leading to
\begin{equation}
 V_\text{CKM}\simeq   
 V_{K^u_L}^\dagger
 \bigg(\id_3-\frac{1}{2}
 \left(\om_uM^{-1}_u\right)\left(\om_uM^{-1}_u\right)^\dagger\bigg)\bigg(\id_3-\frac{1}{2}\left(\om_dM^{-1}_d\right)\left(\om_dM^{-1}_d\right)^\dagger\bigg) 
 V_{K^d_L}
\end{equation}
if that basis is considered for both sectors, cf.~\cref{eq:approxCKM}. 
Simpler expressions are obtained in a WB closely related to that of~\cref{eq:WBMorozumi}, namely the WB defined by
\begin{equation}
\mathcal{M}_u \,=\,
\begin{pmatrix}
d_u' & \om_u  \\[2mm]
0 & D_u'
\end{pmatrix} \,, \qquad
\mathcal{M}_d \,=\,
\begin{pmatrix}
V' d_d' & V'\,\om_d  \\[2mm]
0 & D_d'
\end{pmatrix} \,,
\end{equation}
or in the WB defined by
\begin{equation}\label{eq:WB2}
\mathcal{M}_u \,=\,
\begin{pmatrix}
V'^\dagger d_u' & V'^\dagger\,\om_u  \\[2mm]
0 & D_u'
\end{pmatrix} \,, \qquad
\mathcal{M}_d \,=\,
\begin{pmatrix}
 d_d' & \om_d  \\[2mm]
0 & D_d'
\end{pmatrix} \,.
\end{equation}
In either of these bases, one finds the simpler expressions
\begin{equation}\label{eq:uparam-approxT}
\Theta_u=
   \om_u D'^{-1}_u\,,
    \qquad
        \Theta_d=
    \om_d D'^{-1}_d\,,
\end{equation}
to first order in the expansion of~\cref{app:improved_seesaw}, which can be inserted into~\cref{eq:approxCKM}.
These results clearly illustrate the fact that, within VLQ models, small deviations from unitarity are proportional to the ratio between the scales entering the upper and lower rows of the mass matrices. A priori, this corresponds to the ratio between the electroweak scale $v$ and the new VLQ mass scale.
In practice, the requirement of perturbative couplings 
ensures that the deviations from unitarity are bound by this ratio.  

\vskip 2mm

To close, we discuss the role that perturbativity of the Yukawa couplings of SM quarks to the Higgs boson plays in the mixing matrices we have defined.
To make the interplay between deviations from unitarity, mass scales and perturbativity explicit, we focus on the upper-left block on both sides of the relation $\cM_q\cM^\dagger_q=\cV^q_L\mathcal{D}^2_q\cV^{q\dagger}_L$, finding
\begin{equation}\label{eq:perturb-bidiagonalization}
    m_qm^\dagger_q+\om_q\om^\dagger_q=K^q_L\big(d^2_q+X^q_LD^2_qX^{q\dagger}_L\big)K^{q\dagger}_L\,.
\end{equation}
When combined with~\cref{eq:uparam-exactK}, this relation implies
\begin{equation}\label{eq:perturb-trace}
\begin{aligned}    
    \tr\big(m_qm^\dagger_q+\om_q\om^\dagger_q\big)&=
    \tr \Big[\big(d^2_q+X^q_LD^2_qX^{q\dagger}_L\big)\big(1+X^q_LX^{q\dagger}_L\big)^{-1}\Big]\\
   &\simeq    
   \sum^3_{i=1}m^2_{q_i}+\sum^3_{i=1}\sum^{n_q}_{j=1}\big|(\Theta_q)_{ij}\big|^2M^2_{q_j}
    \,,
\end{aligned}
\end{equation}
where in the second line we have considered an approximation valid for small deviations from unitarity. Here, $m_{q_i}$ and $M_{q_j}$ refer, respectively, to the the light and heavy quark masses in a given sector.
At this stage, perturbativity comes into play through the definitions $m_q=\frac{v}{\sqrt{2}}Y_q$ and $\om_q=\frac{v}{\sqrt{2}}\overline{Y}\!_q$ 
and via the requirement that
\begin{equation}
    \tr\big(m_qm^\dagger_q+\om_q\om^\dagger_q\big)<\rho\, m^2_t\,,
\end{equation}
for some $\mathcal{O}(1-10)$ numerical factor $\rho$.
Combining this bound with~\cref{eq:perturb-trace}, one finds
\begin{equation}\label{eq:perturb-bound}
    \sum_{i,j}\big|(\Theta_q)_{ij}\big|^2 M^2_{q_j}<\rho\,m^2_t
    -\sum_i
    m^2_{q_i}\,,
\end{equation}
showing how the deviations from unitarity are limited by the ratio between the electroweak and VLQ scales in perturbative models.

For illustrative purposes, consider a model
with a single up-type VLQ $T$, for which
\begin{equation}\label{eq:perturb-traceU}
\sum_{i,j}\big|(\Theta_u)_{ij}\big|^2 M^2_{u_j}
= \left(|\Theta_u|^2+|\Theta_c|^2+|\Theta_t|^2\right)m^2_T
\end{equation}
holds. Here, we again have expanded the $3\times1$ matrix $\Theta_u$ as $(\Theta_u,\Theta_c,\Theta_t)^T$.
In such a model,~\cref{eq:perturb-bound} looks like
\begin{equation}
    \sqrt{|\Theta_u|^2+|\Theta_c|^2+|\Theta_t|^2}
    \lesssim\frac{m_t}{m_T}\,,
\end{equation}
which shows that all deviations from unitarity must be smaller than $\mathcal{O}(m_t/m_T)$, as expected.

\vfill
\clearpage

\section{Phenomenology}
\label{sec:pheno}

In this section we present a comprehensive description of observables affected by the presence of VLQs.%
\footnote{
For constraints on simplified VLQ benchmark scenarios, where some mixing matrix elements are assumed to vanish, see e.g.~\cite{Vatsyayan:2020jan}.
}
We will mainly focus on electroweak precision quantities and low-energy effects such as neutral meson mixing and meson decays. Rare decays of the top quark, oblique parameters as well as the $R_{b}$ ratio (related to the decay of the $Z$ boson) will also be described. 

It is important to notice that some of the quantities to be described in this section may depend on tree-level contributions, while
others involve loop calculations. Each model has its particular contributions.
In~\cref{sec:ph-fits} global fits performed in the presence of one VLQ of up or down type will be presented, which clearly show the importance of the distinct observables. Finally, we review in~\cref{sec:SMEFT} the effective field theory description, with emphasis on the matching of models with VLQs to the SMEFT.

\subsection{Observables}\label{sec:ph-observables}
In this section, we gather the observables which are typically used to constrain the parameter space of an extension of the SM with an arbitrary number of singlet VLQs.
While not all are equally important, we leave that discussion for~\cref{sec:ph-fits}.

Throughout this section, we will employ the well-known Inami-Lim functions~\cite{Inami:1980fz,Buchalla:1995vs}
\begin{equation}\begin{split}
    S_0(x_i,x_j)&=\frac{x_ix_j}{4}\bigg[-\frac{3}{(1-x_i)(1-x_j)}\\
    &+\frac{4-8x_i+x^2_i}{(1-x_i)^2(x_i-x_j)}\log x_i +\frac{4-8x_j+x^2_j}{(1-x_j)^2(x_j-x_i)}\log x_j\bigg]\,,\\
    S_0(x)&=\frac{x}{4}\bigg[\frac{4-11x+x^2}{(1-x)^2}-\frac{6x^2}{(1-x)^3}\log x\bigg]\,,\\
    X_0(x)&=\frac{x}{8}\bigg[\frac{x+2}{x-1}+\frac{3x-6}{(1-x)^2}\log x\bigg]\,,\\
    Y_0(x)&=\frac{x}{8}\bigg[\frac{x-4}{x-1}+\frac{3x}{(1-x)^2}\log x\bigg]\,,\\
    Z_0(x)&=\frac{x(108-259x+163x^2-18x^3)}{144(1-x)^3}+ \frac{24x^4-6x^3-63x^2+50x-8}{72(1-x)^4}\log x\,,\\
    E_0(x)&=\frac{x(18-11x-x^2)}{12(1-x)^3}-\frac{4-16x+9x^2}{6(1-x)^4}\log x\,,\\
	\end{split}\end{equation}
together with the usual shorthands (save minor exceptions, e.g.~\cref{sec:D0sys})
\begin{equation}
    \lambda^\alpha_{ij}\equiv V^*_{\alpha i}V_{\alpha j}\,,\quad x_i\equiv m^2_i/m^2_W\,,
\end{equation}
and the loop functions
\begin{equation}\label{eq:functionN}
    N\big(x_i,x_j\big)=\frac{x_ix_j}{8}\bigg(\frac{\log x_i-\log x_j}{x_i-x_j}\bigg)\,.
\end{equation}
For $x_i=x_j$, \cref{eq:functionN} has the following limit,
\begin{equation}
    N\big(x_i\big)\equiv N\big(x_i,x_i\big)=\frac{x_i}{8}\,.
\end{equation}

\subsubsection{Rare top decays}
In the SM, the decays $t\rightarrow Z(h)q$ are heavily inhibited as they are generated at one-loop through a version of the GIM mechanism~\cite{Eilam:1990zc}.
However, in models with VLQs both of those suppression factors are circumvented via their tree-level FCNC which, at leading order, produce
\begin{equation}\label{eq:rtd-width}\begin{split}
    \Gamma\big(t\rightarrow Zq\big)&=\frac{G_Fm^3_t}{16\pi\sqrt{2}}\lambda\big[r_Z,r_q\big] \Big[\big(1-r^2_q\big)^2+r^2_Z\big(1+r^2_q\big)-2r^4_Z\Big]\big|F^u_{qt}\big|^2\,,\\
    \Gamma\big(t\rightarrow hq\big)&=\frac{G_Fm^3_t}{16\pi\sqrt{2}}\lambda\big[r_h,r_q\big] \Big[\big(1+r^2_q\big)^2+4r^2_q-r^2_h\big(1+r^2_q\big)\Big]\big|F^u_{qt}\big|^2\,.
\end{split}\end{equation}
In the expression above, we have introduced the ratios $r_i=m_i/m_t$ alongside the function
\begin{equation}\label{eq:lambda-function}
    \lambda^2\big[a,b\big]=1-2\big(a^2+b^2\big)+\big(a^2-b^2\big)^2\,.
\end{equation}
Since $\br(t\rightarrow W^+b)\simeq1$ and $r_q\ll1$, we can employ~\cref{eq:rtd-width} to obtain the approximations
\begin{equation}\label{eq:rtd-branchingratio}\begin{split}
    \br\big(t\rightarrow Zq\big)&\simeq\frac{1-3r^4_Z+2r^6_Z}{1-3r^4_W+2r^6_W} \frac{|F^u_{qt}|^2}{2|V_{tb}|^2}\,,\\
    \br\big(t\rightarrow hq\big)&\simeq\frac{(1-r^2_h)^2}{1-3r^4_W+2r^6_W} \frac{|F^u_{qt}|^2}{2|V_{tb}|^2}\,.
\end{split}\end{equation}
As such, in models with singlet VLQs the current experimental data~\cite{ParticleDataGroup:2020ssz} translates these expressions into the following constraint,
\begin{equation}\label{eq:rtd-constraint}
    |F^u_{qt}|<(3.337\pm0.099)\times10^{-2}\,.
\end{equation}
Notice that the constraint on $|F^u_{qt}|$ will always be dominated by one of the decay channels in~\cref{eq:rtd-branchingratio}.
At the moment, $t\rightarrow Zq$ is the most restrictive.%
\footnote{
A sizeable branching ratio for \(t \to Zc\) can be generated in models with VLQs attempting to accommodate the $W$-mass anomaly, see e.g.~\cite{Crivellin:2022fdf}.
}
When only a single up-type VLQ is introduced, this bound becomes
\begin{equation}\label{eq:rtd-constraintU}
    |\Theta_q\Theta^*_t|=\theta_{q4}\theta_{34}<(3.337\pm0.099)\times10^{-2}\,.
\end{equation}
However, these observables provide no constraints to models with only down-type VLQs.

In the presence of VLQs, the flavour-changing radiative decays with a photon or a gluon in the final state, $t \to \gamma q$ and $t \to g q$ ($q=u,c$), can also be enhanced by orders of magnitude with respect to SM predictions, see e.g.~\cite{Aguilar-Saavedra:2002lwv,Balaji:2021lpr}. In general, these signals are out of reach of current experiments.

\subsubsection{\texorpdfstring{$B_q$}{Bq} decays}
In the SM, the one-loop generated $B_q\rightarrow\mu^+\mu^-$ ($q=d,s$) decays face additional suppression from the CKM matrix and helicity factors~\cite{Blake:2016olu}, thus being extremely rare.
After accounting for NLO electroweak~\cite{Bobeth:2013tba} and NNLO QCD~\cite{Hermann:2013kca} corrections, its predictions, $\br(B_d\rightarrow\mu^+\mu^-)=(1.12\pm0.12)\times10^{-10}$ and $\br(B_s\rightarrow\mu^+\mu^-)=(3.52\pm0.15)\times10^{-9}$~\cite{Blake:2016olu}, leave plenty of room for new physics.

In extensions of the SM with singlet VLQs, the contributions to these decays yield~\cite{Morozumi:2018cnc}
\begin{equation}\label{eq:bd-rBq}
    \br\big(B_q\rightarrow\mu^+\mu^-\big)=                                                                  \tau_{B_q}\frac{G^2_F}{16\pi}\bigg(\frac{\alpha_\text{em}}{\pi s^2_W}\bigg)^2f^2_{B_q}m_{B_q}m^2_\mu                \sqrt{1-\frac{4m^2_\mu}{m^2_{B_q}}}|\eta^2_Y|\big|\lambda^t_{qb}Y_0(x_t)+\Delta^{B_q}_{\mu\mu}\big|^2\,,
\end{equation}
where the decay constants $f_{B_d}=(190.0\pm1.3)\;\unit{MeV}$ and $f_{B_s}=(230.3\pm1.3)\;\unit{MeV}$ were determined from lattice computations with $N_f=2+1+1$~\cite{FlavourLatticeAveragingGroupFLAG:2021npn}, the factor $\eta_Y=1.0113$ accounts for QCD corrections in a scheme where NLO electroweak corrections can be neglected~\cite{Buras:2012ru}, $\tau_{B_q}$ denotes the meson lifetime, and $\Delta^{B_q}_{\mu\mu}$ encodes the NP effects.
After considering the current experimental constraints~\cite{ParticleDataGroup:2020ssz},~\cref{eq:bd-rBq} leads to
\begin{equation}\label{eq:bd-constraints}\begin{split}
    \big|\lambda^t_{db}Y_0(x_t)+\Delta^{B_d}_{\mu\mu}\big|^2&=(0.8\pm1.0)\times10^{-4}\,,\\
    \big|\lambda^t_{sb}Y_0(x_t)+\Delta^{B_s}_{\mu\mu}\big|^2&=(1.47\pm0.20)\times10^{-3}\,.
\end{split}\end{equation}
The error in the first constraint is caused by the large uncertainty associated with the experimental measure of $\br(B_d\rightarrow\mu^+\mu^-)$.

In models with only down-type VLQs, the factors $\Delta^{B_q}_{\mu\mu}$ are given by~\cite{Aguilar-Saavedra:2002phh}
\begin{equation}\label{eq:bd-deltaD}
    \Delta^{B_q}_{\mu\mu}=-\frac{\pi s^2_W}{\alpha_\text{em}}F^d_{qb}\,.
\end{equation}
After inserting this expression into~\cref{eq:bd-constraints}, we obtain the constraints
\begin{equation}\label{eq:bd-constraintsD}\begin{split}
    \big|F^d_{db}-(8.19+3.40i)\times10^{-5}\big|^2&=(0.8\pm1.0)\times10^{-8}\,,\\
    \big|F^d_{sb}+4.21\times10^{-4}\big|^2&=(1.48\pm0.20)\times10^{-7}\,.
\end{split}\end{equation}
The best-fit values for the CKM matrix in the SM\cite{ParticleDataGroup:2020ssz} were employed in the derivation of the bounds above.
In models with a single down-type VLQ, \cref{eq:bd-constraintsD} could be developed further with
\begin{equation}\label{eq:bd-FD}
    F^d_{db}=-\Theta_d\Theta^*_b=-\theta_{14}\theta_{34}e^{-i\delta_{14}}\,,\quad                            F^d_{sb}=-\Theta_s\Theta^*_b=-\theta_{24}\theta_{34}e^{-i\delta_{24}}\,,
\end{equation}
where we have expanded $\Theta_d$ as $(\Theta_d,\Theta_s,\Theta_b)^T$ (cf.~\cref{sec:ph-parameterizations}).

When exclusively up-type VLQs are present, the NP contributions look like\cite{Nardi:1995fq,Vysotsky:2006fx,Kopnin:2008ca,Picek:2008dd,Botella:2017caf}\footnote
{The flavour-violating electroweak Penguin diagrams which produce the contribution proportional to $N(x_i,x_j)$ are often neglected in the literature (for example, see\cite{Aguilar-Saavedra:2002phh}).
Here, they are included to restore decoupling for large VLQ masses.}
\begin{equation}\label{eq:bd-deltaU}
    \Delta^{B_q}_{\mu\mu}=                                                                            \sum^{n_u}_{k=1}\lambda^{U_k}_{qb}Y_0(x_{U_k})+\sum_{i,j}V^*_{iq}\big(F^u-1\big)_{ij}V_{jb}N(x_i,x_j)\,,
\end{equation}
where the sums over $i,j$ cover every up-type quark in the theory (light-$u$ and heavy-$U$).
By considering the limit 
$F^u\to \diag(1,1,1,0)$ of no deviations from unitarity,
we can combine the previous relation with \cref{eq:bd-constraints} to constrain models with a single up-type VLQ $T$,
\begin{equation}\label{eq:bd-constraintsU}\begin{split}
    \big|\lambda^T_{db}\big[Y_0(x_T)-N(x_T)\big]+(8.15+3.39i)\times10^{-3}\big|^2&=(0.8\pm1.0)\times10^{-4}\,,\\
    \big|\lambda^T_{sb}\big[Y_0(x_T)-N(x_T)\big]-4.19\times10^{-2}\big|^2&=(1.47\pm0.20)\times10^{-3}\,.
\end{split}\end{equation}
Like before, we employed the SM best-fit values for the CKM matrix~\cite{ParticleDataGroup:2020ssz} to obtain these expressions.
Of course,~\cref{eq:bd-constraintsU} may always be developed further with the parameterizations introduced in~\cref{sec:ph-parameterizations}, which establish
\begin{equation}\label{eq:bd-LU}
    \lambda^T_{db}=\Theta_u\Theta^*_t=\theta_{14}\theta_{34}e^{-i\delta_{14}}\,,\quad                           \lambda^T_{sb}=\Theta_c\Theta^*_t=\theta_{24}\theta_{34}e^{-i\delta_{24}}\,,
\end{equation}
with the vector $\Theta_u$ expanded as $(\Theta_u,\Theta_c,\Theta_t)^T$.

The analyses performed in this subsubsection can, of course, be extended to the decays $B_q\rightarrow e^+e^-$ and $B_q\rightarrow\tau^+\tau^-$ by replacing $m_\mu$ with $m_e,m_\tau$ in~\cref{eq:bd-rBq}.
However, the current experimental results $\br(B_d\rightarrow e^+e^-)<2.5\times10^{-9}$, $\br(B_d\rightarrow\tau^+\tau^-)<2.1\times10^{-3}$, $\br(B_s\rightarrow e^+e^-)<9.4\times10^{-9}$ and $\br(B_s\rightarrow\tau^+\tau^-)<6.8\times10^{-3}$\cite{ParticleDataGroup:2020ssz} lead to weak bounds which are covered by~\cref{eq:bd-constraints}.
As such, at the moment phenomenological analyses of VLQ models can ignore these decays.

\subsubsection{\texorpdfstring{$K$}{K} decays}
The rare kaon decays $K^+\rightarrow\pi^+\nu\bar{\nu}$, $K_L\rightarrow\mu^+\mu^-$ and $K_L\rightarrow\pi^0\bar{\nu}\nu$ provide significant constraints to the mixing in the $sd$ sector of any model with singlet VLQs.
Here, we will discuss them by presenting fractions of branching ratios where most theoretical uncertainties are cancelled.
With that in mind, one usually employs the isospin symmetry of QCD to describe the first decay in the following manner~\cite{Aguilar-Saavedra:2002phh},
\begin{equation}\label{eq:kd-rKp}
    \frac{\br(K^+\rightarrow\pi^+\nu\bar{\nu})}{\br(K^+\rightarrow\pi^0e^+\bar{\nu})}=                   \frac{\alpha^2_\text{em}r_{K^+}}{2\pi^2s^4_W|V_{us}|^2}                                                      \sum_{l=e,\mu,\tau}\big|\lambda^c_{sd}X^l_{NL}+\lambda^t_{sd}\eta^X_tX_0(x_t)+\Delta^{K^+}_{\pi\nu\nu}\big|^2\,,
\end{equation}
where $r_{K^+}=0.901$~\cite{Marciano:1996wy} captures all isospin breaking corrections.
At NLO, the charm contributions to this ratio are $X^{e,\mu}_{NL}=(10.6\pm1.5)\times10^{-4}$ and $X^\tau_{NL}=(7.1\pm1.4)\times10^{-4}$~\cite{Buchalla:1998ba}, while the QCD correction to the top contributions is given by $\eta^X_t=0.994$~\cite{Buchalla:1993bv}.

In models with only up (U) or down (D) singlet VLQs we have
\cite{Nardi:1995fq,Aguilar-Saavedra:2002phh,Vysotsky:2006fx,Kopnin:2008ca,Picek:2008dd,Botella:2017caf}
\begin{equation}\label{eq:delta-Kp}\begin{split}
    \Delta^{K^+}_{\pi\nu\nu}&=                                                                        \sum^{n_u}_{k=1}\eta^X_{U_k}\lambda^{U_k}_{sd}X_0\big(x_{U_k}\big)+                                             \sum_{i,j}\eta^X_iV^*_{is}(F^u-1)_{ij}V_{jd}N(x_i,x_j)\,,\qquad\text{(U)}\,,\\
    \Delta^{K^+}_{\pi\nu\nu}&=-\frac{\pi s^2_W}{\alpha_\text{em}}F^d_{sd}\,,\qquad\text{(D)}\,,\\
\end{split}\end{equation}
where the sums over $i,j$ cover every up-type quark (light-$u$ and heavy-$U$) in the theory.
For the heavy quarks, the slow-running nature of QCD is often used to consider $\eta^X_{U_k}\simeq\eta^X_t$.

In order to describe the second rare decay, one usually compares the short-distance contributions to $K_L\rightarrow\mu^+\mu^-$ with the branching ratio for $K^+\rightarrow\mu^+\nu$~\cite{Aguilar-Saavedra:2002phh},
\begin{equation}\label{eq:kd-rKL}
    \frac{\br(K_L\rightarrow\mu^+\mu^-)_{\rm SD}}{\br(K^+\rightarrow\mu^+\nu)}=                      \frac{\tau_{K_L}}{\tau_{K^+}}\frac{\alpha^2_\text{em}}{\pi^2s^4_W|V_{us}|^2}                                \big[Y_{NL}\,\mathrm{Re}\,\lambda^c_{sd}+\eta^Y_tY_0(x_t)\,\mathrm{Re}\,\lambda^t_{sd}+          \Delta^{K_L}_{\mu\mu}\big]^2\,.
\end{equation}
At NLO, the QCD corrections to the box diagrams which generate this expression are $Y_{NL}=(2.94\pm0.28)\times10^{-4}$~\cite{Buchalla:1998ba} and $\eta^Y_t=1.012$~\cite{Buchalla:1993bv}.
Meanwhile, we consider the conservative bound $\br(K_L\rightarrow \mu^+\mu^-)_{\rm SD}< 2.5\times10^{-9}$~\cite{Isidori:2003ts}.
The NP effects provided by models with up (U) or down (D) type VLQs look like
\cite{Nardi:1995fq,Aguilar-Saavedra:2002phh,Vysotsky:2006fx,Kopnin:2008ca,Picek:2008dd,Botella:2017caf}
\begin{equation}\label{eq:delta-KL}\begin{split}
    \Delta^{K_L}_{\mu\mu}&=                                                                           \sum^{n_u}_{k=1}\eta^Y_{U_k}\mathrm{Re}\big(\lambda^{U_k}_{sd}\big)Y_0\big(x_{U_k}\big)+                \sum_{i,j}\eta^Y_i\mathrm{Re}\Big[V^*_{is}(F^u-1)_{ij}V_{jd}\Big]N(x_i,x_j)\,,\qquad\text{(U)}\,,\\
    \Delta^{K_L}_{\mu\mu}&=-\frac{\pi s^2_W}{\alpha_\text{em}}\mathrm{Re}\,F^d_{sd}\,,\qquad\text{(D)}\,.
\end{split}\end{equation}
As before, we may consider $\eta^Y_{U_k}\simeq\eta^Y_t$ for heavy $U_k$.

Finally, the third rare decay, $K_L\rightarrow\pi^0\bar{\nu}\nu$, is often measured against the SM expectations~\cite{Botella:2021uxz},
\begin{equation}\label{eq:kd-rKl}
    \frac{\br(K_L\rightarrow\pi^0\bar{\nu}\nu)}{\br(K_L\rightarrow\pi^0\bar{\nu}\nu)_\mathrm{SM}}=              \bigg|\frac{\mathrm{Im}[\lambda^c_{sd}X_0(x_c)+\lambda^t_{sd}X_0(x_t)+\Delta^{K_L}_{\pi\nu\nu}]}          {\mathrm{Im}[\lambda^c_{sd}X_0(x_c)+\lambda^t_{sd}X_0(x_t)]}\bigg|^2\,,
\end{equation}
where $\br(K_L\rightarrow\pi^0\bar{\nu}\nu)_\mathrm{SM}=(3.0\pm0.6)\times10^{-11}$~\cite{Buras:2004uu}.
Since the $Z$-mediated FCNC in the down sector are heavily constrained by the mixing in neutral meson systems, this observable should be dominated by the contributions from the up-type VLQs.
Then, we can simply consider\cite{Nardi:1995fq,Vysotsky:2006fx,Kopnin:2008ca,Picek:2008dd,Botella:2017caf}
\begin{equation}\label{eq:delta-Kl}
    \Delta^{K_L}_{\pi\nu\nu}=                                                                         \sum^{n_u}_{k=1}\lambda^{U_k}_{sd}X_0\big(x_{U_k}\big)+\sum_{i,j}V^*_{is}(F^u-1)_{ij}V_{jd}N(x_i,x_j)
\end{equation}
for the U-models and $\Delta^{K_L}_{\pi\nu\nu}=0$ for the D-models.

At the moment~\cite{ParticleDataGroup:2020ssz,KOTO:2018dsc,KOTO:2020prk},~\cref{eq:kd-rKp,eq:kd-rKL,eq:kd-rKl} constrain models with VLQs in the following manner,
\begin{equation}\label{eq:kd-constraints}\begin{split}
    \sum_{l=e,\mu,\tau}\big|\lambda^c_{sd}X^l_{NL}+\lambda^t_{sd}\eta^X_tX_0(x_t)+\Delta^{K^+}_{\pi\nu\nu}\big|^2&=(3.7\pm2.4)\times10^{-6}\,,\\
    \big[Y_{NL}\,\mathrm{Re}\,\lambda^c_{sd}+\eta^Y_tY_0(x_t)\,\mathrm{Re}\,\lambda^t_{sd}+          \Delta^{K_L}_{\mu\mu}\big]^2&<(4.753\pm0.041)\times10^{-7}\,,\\
    \Big|\mathrm{Im}\big[\lambda^c_{sd}X_0(x_c)+\lambda^t_{sd}X_0(x_t)+\Delta^{K_L}_{\pi\nu\nu}\big]\Big|^2&<  (4.4\pm1.0)\times10^{-6}\,.
\end{split}\end{equation}
Notice that even though these observables probe the same couplings, none can be neglected.
Since the second only fixes their real part, the others are needed to constrain the imaginary component of the NP.
In models with extra down(up)-type VLQs, the first (third) becomes dominant.
Such an intricate scenario may, of course, change in the future as the experimental sensitivity to each decay improves.

In the D-models, where all VLQs are of the down-type, \cref{eq:kd-constraints} becomes
\begin{equation}\label{eq:kd-constraintsD}\begin{split}
    \sum_{l=e,\mu,\tau}\big|F^d_{sd}+(5.06-2.10i)\times10^{-6}+2.22\times10^{-3}X^l_{NL}\big|^2&=               (3.7\pm2.4)\times10^{-10},\\
    \big[\mathrm{Re}\,F^d_{sd}+4.01\times10^{-6}\big]^2&<(4.797\pm0.041)\times10^{-11}\,,
\end{split}\end{equation}
when the best-fit values for the SM CKM matrix are employed\cite{ParticleDataGroup:2020ssz}.
With the parameterizations introduced in~\cref{sec:ph-parameterizations}, for theories with a single down-type VLQ these bounds can be expanded upon through
\begin{equation}\label{eq:kd-FD}
    F^d_{sd}=-\Theta_s\Theta^*_d=-\theta_{14}\theta_{24}e^{i(\delta_{14}-\delta_{24})}\,,
\end{equation}
where again $\Theta_s=(\Theta_d)_s$.

In models with a single up-type VLQ $T$,~\cref{eq:kd-constraints} reads
\begin{equation}\label{eq:kd-constraintsU}\begin{split}
    \big\{\mathrm{Re}\,\lambda^T_{sd}\big[Y_0(x_T)-N(x_T)\big]-3.94\times10^{-4}\big\}^2&<(4.641\pm0.040)\times10^{-7}\,,\\
    \big|\mathrm{Im}\,\lambda^T_{sd}\big[X_0(x_T)-N(x_T)\big]+2.11\times10^{-4}\big|^2&<(4.4\pm1.0)\times10^{-6}\,,
\end{split}\end{equation}
in the limit $F^u\to \diag(1,1,1,0)$.
As before, we employed the SM best-fit values for the CKM matrix~\cite{ParticleDataGroup:2020ssz} in~\cref{eq:kd-constraintsU}.
Such an expression can be further developed with
\begin{equation}\label{eq:kd-LU}
    \lambda^T_{sd}=\Theta_c\Theta^*_u=\theta_{14}\theta_{24}e^{i(\delta_{14}-\delta_{24})}\,,
\end{equation}
by using the parameterizations in~\cref{sec:ph-parameterizations}.

\subsubsection{\texorpdfstring{$K^0$}{K0} system}\label{sec:kaon}
As shown in~\cref{sec:gaugeinteractions}, models with VLQs have tree-level $Z$ and Higgs-mediated FCNC which contribute to the $K^0$--$\overline{K^0}$ mixing.
However, the latter are suppressed by a factor of about $m^2_K/m^2_h\sim10^{-5}$.
As such, at LO the short-range NP contributions to this particular mixing look like~\cite{Aguilar-Saavedra:2002phh}
\begin{equation}\label{eq:km-m12}
    M^K_{12}=B_K\frac{G^2_Ff^2_Km^2_Wm_K}{12\pi^2}\Big[ \eta^K_{cc}S_0(x_c)\big(\lambda^c_{ds}\big)^2\!+ \eta^K_{tt}S_0(x_t)\big(\lambda^t_{ds}\big)^2\!+ 2\eta^K_{ct}S_0(x_c,x_t)\lambda^c_{ds}\lambda^t_{ds}+                     \Delta_{K^0}\Big]\,,
\end{equation}
where $\eta^K_{cc}=1.38\pm0.20$, $\eta^K_{ct}=0.47\pm0.04$ and $\eta^K_{tt}=0.57\pm0.01$ are QCD correction factors evaluated at NLO~\cite{Buchalla:1995vs}, while 
$B_K=0.717\pm0.018 \,(\text{stat.})\pm0.016\, (\text{syst.})$ and $f_K=(155.7\pm0.3)\;\unit{MeV}$ are, respectively, the kaon bag factor and decay constant determined from lattice computations with $N_f=2+1+1$~\cite{FlavourLatticeAveragingGroupFLAG:2021npn}.
Meanwhile, the NP effects are controlled by
\begin{equation}\label{eq:delta-K}\begin{split}
    \Delta_{K^0}&=                                                                                         \sum^{n_u}_{k=1}\eta^K_{U_kU_k}S_0(x_{U_k})\big(\lambda^{U_k}_{ds}\big)^2+                                2\sum_{q=c,t}\sum^{n_u}_{k=1}\eta^K_{qU_k}S_0(x_q,x_{U_k})\lambda^q_{ds}\lambda^{U_k}_{ds}\,,\qquad\text{(U)}\,,\\
    \Delta_{K^0}&=                                                                          -8F^d_{ds}\Big[\eta^K_ZY_0(x_c)\lambda^c_{ds}+\eta^K_{tt}Y_0(x_t)\lambda^t_{ds}\Big]+                       \frac{4\pi s^2_W}{\alpha_\text{em}}\eta^K_Z\big(F^d_{ds}\big)^2\,,\qquad\text{(D)}\,.
\end{split}\end{equation}
Like for the $K$ decays, we can also use the slow-running nature of QCD here to consider the approximations $\eta^K_{qU_k}\simeq\eta^K_{qt}$ and $\eta^K_{U_kU_k}\simeq\eta^K_{tt}$.
Meanwhile, $\eta^K_Z=0.60$ is a QCD correction factor estimated at NLO~\cite{Buchalla:1995vs}.

Since a complete evaluation of kaon physics requires a knowledge of long-distance physics that we do not possess, we cannot simply fit $M^K_{12}$ to the relevant data.
Nevertheless, we constrain NP by assuming that their short-distance contributions do not saturate the experimental bound, i.e.
\begin{equation}\label{eq:km-constraintM}
    \big(\Delta m_K\big)_\text{NP}=2\Big|\big(M^K_{12}\big)_\text{NP}\Big|<\big(\Delta m_K\big)_\text{exp}\,.
\end{equation}
Meanwhile, in~\cite{Brod:2019rzc} most uncertainties in the determination of $\epsilon_K$ were cancelled by using the unitarity of the CKM matrix.
Since the result, $|(\epsilon_K)_\text{SM}|=(2.16\pm0.18)\times10^{-3}$, is in good agreement with the current experimental measurement, $|\epsilon_K|=(2.257\pm0.018)\times10^{-3}$~\cite{ParticleDataGroup:2020ssz}, it is reasonable to demand
\begin{equation}\label{eq:km-constraintE}
    \Big|\big(\epsilon_K\big)_\text{NP}\Big|=                                                                 \frac{\kappa_\epsilon}{\sqrt{2}\Delta m_K}\Big|\text{Im}\big(M^K_{12}\big)_{NP}\Big|<              \Big|\big(\epsilon_K\big)_\text{exp}\Big|-\Big|\big(\epsilon_K\big)_\text{SM}\Big|\,,
\end{equation}
in which $\kappa_\epsilon=0.92\pm0.02$~\cite{Buras:2008nn}.
At the moment, with a theoretical uncertainty of about $10\%$, $(\epsilon_K)_\text{SM}$ is still compatible with $(\epsilon_K)_\text{exp}$.
For that reason, some authors prefer to limit the contributions to $(\epsilon_K)_\text{NP}$ with one tenth of the latter~\cite{Brod:2019rzc,Botella:2021uxz}.

With the current experimental data~\cite{ParticleDataGroup:2020ssz},~\cref{eq:km-constraintM,eq:km-constraintE} become
\begin{equation}\label{eq:km-constraint}\begin{split}
    \big|\Delta_{K^0}\big|&<(2.714\pm0.092)\times10^{-5}\,,\\
    \big|\text{Im}\,\Delta_{K^0}\big|&<(0.8\pm1.5)\times10^{-8}\,.
\end{split}\end{equation}
Clearly, of the two, $\epsilon_K$ provides the most significant constraint for these models.
Nevertheless, $\Delta m_K$ cannot be ignored since $\epsilon_K$ is not sensitive to real (i.e.~non-imaginary) NP contributions.

In models with only down-type VLQs, the constraints in~\cref{eq:km-constraint} become
\begin{equation}\label{eq:km-constraintD}\begin{split}
    \Big|F^d_{ds}\big[F^d_{ds}+(7.07+2.70i)\times10^{-6}\big]\Big|&<(1.136\pm0.039)\times10^{-7}\,,\\
    \Big|\mathrm{Im}\Big\{F^d_{ds}\big[F^d_{ds}+(7.07+2.70i)\times10^{-6}\big]\Big\}\Big|&<(3.3\pm6.3)\times10^{-11}\,,
\end{split}\end{equation}
provided that the SM best-fit values for the CKM matrix are employed~\cite{ParticleDataGroup:2020ssz}.
Meanwhile, when a single VLQ is present,
\begin{equation}\label{eq:km-FD}
    F^d_{ds}=-\Theta_d\Theta^*_s=-\theta_{14}\theta_{24}e^{i(\delta_{14}-\delta_{24})}
\end{equation}
can be written with the parameterizations in~\cref{sec:ph-parameterizations}.

In models where only one up-type VLQ singlet $T$ was introduced,~\cref{eq:km-constraint} is controlled by
\begin{equation}\label{eq:km-constraintU}
    \Delta_{K^0}= \lambda^T_{ds}\Big[0.57S_0(x_T)\lambda^T_{ds}-(3.77+1.56i)\times10^{-4}S_0(x_t,x_T)-0.206S_0(x_c,x_T)\Big]\,,
\end{equation}
where with the parameterizations introduced in~\cref{sec:ph-parameterizations} we have
\begin{equation}\label{eq:km-LU}
    \lambda^T_{ds}=\Theta_u\Theta^*_c=\theta_{14}\theta_{24}e^{-i(\delta_{14}-\delta_{24})}\,.
\end{equation}
Like before, the SM best-fit values for the CKM matrix were employed in these expressions~\cite{ParticleDataGroup:2020ssz}.

\subsubsection{\texorpdfstring{$D^0$}{D0} system}\label{sec:D0sys}
In the $D^0$ system, the SM contribution to neutral meson mixing is mediated by box diagrams with internal down-type quarks.
As such, the GIM mechanism renders it negligible~\cite{Datta:1984jx,Donoghue:1985hh}.
Then, in models with VLQs $M^D_{12}$ is entirely controlled by the NP.
After neglecting the Higgs-mediated FCNC, we estimate their effects through~\cite{Aguilar-Saavedra:2002phh}
\begin{equation}\label{eq:dm-m12}
    M^D_{12}=B_D\frac{G^2_Ff^2_Dm^2_Wm_D}{12\pi^2}\Delta_{D^0}\,,
\end{equation}
where $f_D=(212.0\pm0.7)\;\unit{MeV}$ is the $D^0$ decay constant~\cite{FlavourLatticeAveragingGroupFLAG:2021npn} and its bag factor is estimated to be $B_D=1.0\pm0.3$~\cite{Aguilar-Saavedra:2002phh}.
Here, and only here, $\lambda^q_{cu}=V_{cq}V^*_{uq}$.
Then, since small quark masses lead to $S_0(x_{d,s,b},x_{D_k})\ll1$ and the CKM matrix is such that $|\lambda^q_{cu}|\ll1$, at LO the new effects $\Delta_{D^0}$ can be approximated in the following manner,
\begin{equation}\label{eq:delta-D}\begin{split}
    \Delta_{D^0}&=\sum^{n_d}_{k=1}S_0(x_{D_k})\big(\lambda^{D_k}_{cu}\big)^2\,,\qquad\text{(D)}\,,\\
    \Delta_{D^0}&=\frac{4\pi s^2_W}{\alpha_\text{em}}\eta^D_Z\big(F^u_{cu}\big)^2\,,\qquad\text{(U)}\,.
\end{split}\end{equation}
At the moment, $\eta^D_Z=0.59$~\cite{Buchalla:1995vs} is the only known QCD correction to this process.

Even with NP, the mixing in $D^0$--$\overline{D^0}$ systems should be dominated by non-perturbative long-range effects~\cite{Wolfenstein:1985ft}.
As such, models with VLQs must satisfy
\begin{equation}\label{eq:dm-constraintM}
    \big(\Delta m_D\big)_\text{NP}=2\Big|\big(M^D_{12}\big)_\text{NP}\Big|<\big(\Delta m_D\big)_\text{exp}\,.
\end{equation}
After considering the current experimental results~\cite{ParticleDataGroup:2020ssz}, this bound can be translated into
\begin{equation}\label{eq:dm-constraint}
    \big|\Delta_{D^0}\big|<(5.0\pm2.8)\times10^{-6}\,.
\end{equation}
In~\cite{Aguilar-Saavedra:2002phh}, it was argued that this constraint can be ignored for $D_k$ with a mass below $1\;\unit{TeV}$.

In the U-models where no down-type VLQs are present,~\cref{eq:dm-constraint} is given by
\begin{equation}\label{eq:dm-constraintU}
    |F^u_{cu}|^2<(2.1\pm1.2)\times10^{-8}\,.
\end{equation}
In the simplest case of
$n_u=1$, we may also write
\begin{equation}\label{eq:dm-FU}
    F^u_{cu}=\Theta_c\Theta^*_u=\theta_{14}\theta_{24}e^{i(\delta_{14}-\delta_{24})}
\end{equation}
with the parameterizations shown in~\cref{sec:ph-parameterizations}.

In extensions of the SM with a single down-type VLQ $B$,~\cref{eq:dm-constraint} looks like
\begin{equation}\label{eq:dm-constraintD}
    S_0(x_B)\big|\lambda^B_{cu}\big|^2<(5.0\pm2.8)\times10^{-6}\,.
\end{equation}
After employing the parameterizations displayed in~\cref{sec:ph-parameterizations}, we may also write
\begin{equation}\label{eq:dm-LD}
    \lambda^B_{cu}=\Theta_d\Theta^*_s=\theta_{14}\theta_{24}e^{-i(\delta_{14}-\delta_{24})}\,.
\end{equation}

\subsubsection{\texorpdfstring{$B^0_q$}{B0q} systems}\label{sec:B0q}
In models with singlet VLQs, the contributions from the Higgs-mediated FCNC to $B^0_q$--$\overline{B^0_q}$ ($q=d,s$) mixing are suppressed by a factor of $m^2_{B_q}/m^2_h\sim10^{-3}$ versus those which arise from the $Z$ interactions.
As such, at LO we have~\cite{Aguilar-Saavedra:2002phh}
\begin{equation}\label{eq:bm-m12}
    M^{B_q}_{12}=
    B_{B_q}\frac{G^2_Ff^2_{B_q}m^2_Wm_{B_q}}{12\pi^2}
    \Big[\eta^{B_q}_{tt}S_0(x_t)\big(\lambda^t_{qb}\big)^2+\Delta_{B_q}\Big]\,,
\end{equation}
where the NLO QCD corrections are given by $\eta^{B_d}_{tt}=\eta^{B_s}_{tt}=0.55$~\cite{Buchalla:1993bv}.
Also, we employ the bag factors $B_{B_d}=1.222\pm0.061$ and $B_{B_s}=1.232\pm0.053$ alongside the decay constants $f_{B_d}=(190.0\pm1.3)\;\unit{MeV}$ and $f_{B_s}=(230.3\pm1.3)\;\unit{MeV}$ which were determined through lattice computations with $N_f=2+1+1$~\cite{FlavourLatticeAveragingGroupFLAG:2021npn}.
Finally, we have
\begin{equation}\label{eq:delta-Bq}\begin{split}
    \Delta_{B_q}&=                                                                                         \sum^{n_u}_{k=1}\eta^{B_q}_{U_kU_k}S_0(x_{U_k})\big(\lambda^{U_k}_{qb}\big)^2+                        2\sum^{n_u}_{k=1}\eta^{B_q}_{tU_k}S_0(x_t,x_{U_k})\lambda^t_{qb}\lambda^{U_k}_{qb}\,,\qquad\text{(U)}\,,\\
    \Delta_{B_q}&=                                                                            -8F^d_{qb}\eta^{B_q}_{tt}Y_0(x_t)\lambda^t_{qb}+\frac{4\pi s^2_W}{\alpha_\text{em}}\eta^{B_q}_Z\big(F^d_{qb}\big)^2\,,      \qquad\text{(D)}\,,
\end{split}\end{equation}
where $\eta^{B_d}_Z=\eta^{B_s}_Z=0.57$~\cite{Buchalla:1995vs}.
As always, for VLQs with masses above $m_t$ we remember the slow-running nature of QCD and assume $\eta^{B_q}_{U_kU_k}\simeq\eta^{B_q}_{tU_k}\simeq\eta^{B_q}_{tt}$.

By now, it is well-known that both $B^0_q$ systems can be described with the perturbative interactions of~\cref{eq:bm-m12}.
As such, they are the only system for which the amplitudes and phases of $M_{12}$ may be fitted to data.
In~\cite{Lenz:2012az}, this was done in a model-independent fashion.
After introducing
\begin{equation}\label{eq:bm-constraintD}
    \Delta_q=M^{B_q}_{12}/\big(M^{B_q}_{12}\big)_\text{SM}\equiv
    1+\big(M^{B_q}_{12}\big)_\text{NP}/\big(M^{B_q}_{12}\big)_\text{SM}\,,
\end{equation}
the authors of~\cite{Lenz:2012az} find
$\text{Re}(\Delta_d)=0.82\pm0.14$, $\text{Im}(\Delta_d)=-0.199\pm0.062$, $\text{Re}(\Delta_s)=0.97\pm0.13$ and $\text{Im}(\Delta_s)=0.00\pm0.10$.
Thus, at the moment
\begin{equation}\label{eq:bm-constraints}\begin{split}
    \text{Re}\Big[\Delta_{B_d}/\eta^{B_d}_{tt}S_0(x_t)\big(\lambda^t_{db}\big)^2\Big]&= -0.18\pm0.14\,,\\
    \text{Im}\Big[\Delta_{B_d}/\eta^{B_d}_{tt}S_0(x_t)\big(\lambda^t_{db}\big)^2\Big]&= -0.199\pm0.062\,,\\
    \text{Re}\Big[\Delta_{B_s}/\eta^{B_s}_{tt}S_0(x_t)\big(\lambda^t_{sb}\big)^2\Big]&= -0.03\pm0.13\,,\\
    \text{Im}\Big[\Delta_{B_s}/\eta^{B_s}_{tt}S_0(x_t)\big(\lambda^t_{sb}\big)^2\Big]&= 0.00\pm0.10
\end{split}\end{equation}
must be satisfied by every model with singlet VLQs.

In a model with singlet VLQs of the down-type, the constraints in~\cref{eq:bm-constraints} become
\begin{equation}\label{eq:bm-constraintsD}\begin{split}
    \mathrm{Re}\Big\{(0.707-0.707i)F^d_{db}\big[F^d_{db}-(1.58+0.66i)\times10^{-4}\big]\Big\}&=                (-8.1\pm6.3)\times10^{-8}\,,\\
    \mathrm{Im}\Big\{(0.707-0.707i)F^d_{db}\big[F^d_{db}-(1.58+0.66i)\times10^{-4}\big]\Big\}&=                (-8.9\pm2.8)\times10^{-8}\,,\\
    \mathrm{Re}\Big[F^d_{sb}\big(8.37\times10^{-4}+F^d_{sb}\big)\Big]&=(-0.3\pm1.4)\times10^{-6}\,,\\
    \mathrm{Im}\Big[F^d_{sb}\big(8.37\times10^{-4}+F^d_{sb}\big)\Big]&=(0.0\pm1.1)\times10^{-6}\,,
\end{split}\end{equation}
with the SM best-fit values for the CKM matrix~\cite{ParticleDataGroup:2020ssz}.
When $n_d=1$, we can also write
\begin{equation}\label{eq:bm-FD}
    F^d_{qb}=-\Theta_{d_q}\Theta^*_b=-\theta_{q4}\theta_{34}e^{-i\delta_{q4}}
\end{equation}
when the parameterizations introduced in~\cref{sec:ph-parameterizations} are employed.

When one VLQ $T$ of the up-type is introduced into the SM,~\cref{eq:bm-constraints} is controlled by
\begin{equation}\label{eq:bm-constraintsU}\begin{split}
    \Delta_{B_d}&=\lambda^T_{db}\Big[0.55S_0(x_T)\lambda^T_{db}+(8.70+3.60i)\times10^{-3}S_0(x_t,x_T)\Big]\,,\\
    \Delta_{B_s}&=\lambda^T_{sb}\Big[0.55S_0(x_T)\lambda^T_{sb}-4.60\times10^{-2}S_0(x_t,x_T)\Big]\,,
\end{split}\end{equation}
where we may replace the following expression when the parameterizations in~\cref{sec:ph-parameterizations} are employed,
\begin{equation}\label{eq:dm-LU}
    \lambda^T_{qb}=\Theta_{u_q}\Theta^*_t=\theta_{q4}\theta_{34}e^{-i\delta_{q4}}\,.
\end{equation}
As always, we used the SM best-fit values for the CKM matrix~\cite{ParticleDataGroup:2020ssz} in~\cref{eq:bm-constraintsU}.

\subsubsection{Heavy vector-like quarks}
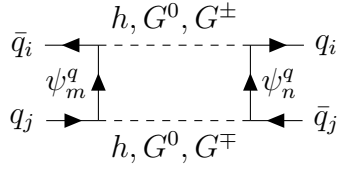
\begin{figure}[!t]\centering
    \begin{tikzpicture}\begin{feynman}
        \vertex                 (u1){$\bar{q}_i$};\vertex[right=1cm of u1](u2);
        \vertex[below=1cm of u1](d1){$q_j$};      \vertex[right=1cm of d1](d2);
        \vertex[right=4cm of u1](u4){$q_i$};      \vertex[right=3cm of u1](u3);
        \vertex[right=4cm of d1](d4){$\bar{q}_j$};\vertex[right=3cm of d1](d3);
        \diagram{
        (d1)--[fermion](d2)--[fermion,edge label={$\psi_m^q$}](u2)--[fermion](u1),
        (d4)--[fermion](d3)--[fermion,edge label'={$\psi_n^q$}](u3)--[fermion](u4),
        (u2)--[scalar,edge label={$h,G^0,G^\pm$}](u3),
        (d3)--[scalar,edge label={$h,G^0,G^\mp$}](d2)};
    \end{feynman}\end{tikzpicture}
    \caption{Scalar-box contributions to neutral meson mixing, where $\psi^q_k$ is the $k$-th entry of $\psi^q$ (see text).}\label{fig:hvlq-box}
\end{figure}

Throughout~\cref{sec:kaon,sec:D0sys,sec:B0q}, we have considered the gauge-mediated one-loop contributions to neutral meson mixing alongside the tree-level NP effects.
Now, we will focus our attention on the scalar-mediated box diagrams of~\Cref{fig:hvlq-box}, which become important for large VLQ masses~\cite{Ishiwata:2015cga,Bobeth:2016llm}.
With the notation introduced in~\cref{sec:notation}, the Lagrangian which controls these Feynman diagrams looks like
\begin{equation}\begin{split}\label{eq:hvlq-lagrangian}
    \mathscr{L}_Y\supset&                                                                                  -\frac{\sqrt{2}}{v}G^+\Big[\bar{\psi}^u\big(VD_d\gamma_R-D_uV\gamma_L\big)\psi^d\Big]                  +\frac{\sqrt{2}}{v}G^-\Big[\bar{\psi}^d\big(V^\dagger D_u\gamma_R-D_dV^\dagger\gamma_L\big)\psi^u\Big]\\
    &-i\epsilon_q\frac{G^0}{v}\Big[\bar{\psi}^q\big(F^qD_q\gamma_R-D_qF^q\gamma_L\big)\psi^q\Big]                 -\frac{h}{v}\Big[\bar{\psi}^q\big(F^qD_q\gamma_R+D_qF^q\gamma_L\big)\psi^q\Big]\,.
\end{split}\end{equation}
We have defined the column vectors $\psi^q$ ($q=u,d$), with $\psi^u = (u,c,t,U_1,\ldots,U_{n_u})^T$ and $\psi^d = (d,s,b,D_1,\ldots,D_{n_d})^T$.
In the second line of this equation, there are implicit sums over $q=d,u$ with $\epsilon_d=-\epsilon_u=1$.
By working in the $R_\xi$ gauge, where $m^2_{G^0}=\xi m^2_Z$ and $m^2_{G^\pm}=\xi m^2_W$, we may express the contributions from the neutral scalars to the effective Hamiltonian in the following manner,
\begin{equation}\label{eq:hvlq-Hneutral}
    \effHam=\frac{m^2_mm^2_n}{64\pi^2v^4}                                                                   \Big[\Upsilon_3\big(m^2_h,m^2_h\big)+\Upsilon_3\big(\xi m^2_Z,\xi m^2_Z\big)+2\Upsilon_3\big(m^2_h,\xi m^2_Z\big)\Big]\Lambda^m_{ij}\Lambda^n_{ij}\big(\bar{q}_i\gamma^\mu\gamma_Lq_j\big)^2\,.
\end{equation}
Once again, this expression includes implicit sums over all fermions $\psi^q_{m,n}$.
We have also introduced $\Lambda^m_{ij}=F^{*q}_{mi}F^q_{mj}$ alongside the function $\Upsilon_3(m^2_a,m^2_b)\equiv\Upsilon_3[m^2_a,m^2_b,m^2_m,m^2_n]$, where
\begin{equation}\begin{split}\label{eq:hvlq-Ups}
    \Upsilon_3&[m^2_a,m^2_b,m^2_c,m^2_d]\equiv\\
    &\frac{m^4_a\log m^2_a}{2(m^2_a-m^2_b)(m^2_a-m^2_c)(m^2_a-m^2_d)}                                     +\frac{m^4_b\log m^2_b}{2(m^2_b-m^2_a)(m^2_b-m^2_c)(m^2_b-m^2_d)}\\
    +&\frac{m^4_c\log m^2_c}{2(m^2_c-m^2_a)(m^2_c-m^2_b)(m^2_c-m^2_d)}                                    +\frac{m^4_d\log m^2_d}{2(m^2_d-m^2_a)(m^2_d-m^2_b)(m^2_d-m^2_c)}\,.
\end{split}\end{equation}
Meanwhile, the charged scalar contributions to neutral meson mixing in~\cref{fig:hvlq-box} equate to
\begin{equation}\label{eq:hvlq-Hcharged}
    \effHam=\frac{m^2_mm^2_n}{16\pi^2v^4}                                                                        \Upsilon_3\big(\xi m^2_W,\xi m^2_W\big)\lambda^m_{ij}\lambda^n_{ij}\big(\bar{q}_i\gamma^\mu\gamma_Lq_j\big)^2\,,
\end{equation}
with $\lambda^m_{ij}=V^*_{mi}V_{mj}$ ($\lambda^m_{ij}=V_{im}V^*_{jm}$) for down-type (up-type) mesons. 

In models with a heavy VLQ $Q$, where $m_Q\gg m_{W,Z}$,~\cref{eq:hvlq-Ups} is replaced with
\begin{equation}\label{eq:hvlq-UpsUps}
    \Upsilon_3[m^2_Q,m^2_Q,0,0]=\frac{1}{2m^2_Q}
\end{equation}
at leading order, implying that the previous Hamiltonians may be combined into
\begin{equation}\label{eq:hvlq-Heff}
    \effHam=\frac{m^2_Q}{32\pi^2v^4}                            \Big[
        \big(\lambda^Q_{ij}\big)^2
        +\big(\Lambda^Q_{ij}\big)^2        
    \Big]                         \big(\bar{q}_i\gamma^\mu\gamma_Lq_j\big)^2\,,
\end{equation}
which scales as $(\oY_d)^4/m_Q^2$ in terms of the VLQ Yukawas, thus decoupling for large $m_Q$.
Hence, in theories with a heavy VLQ,~\cref{eq:km-m12,eq:dm-m12,eq:bm-m12} should be replaced by
\begin{equation}\begin{split}\label{eq:hvlq-m12}
    M^K_{12}&=-\frac{f^2_Km^2_Qm_K}{96\pi^2v^4}                                    \Big[\big(\Lambda^Q_{ds}\big)^2+\big(\lambda^Q_{ds}\big)^2\Big]\,,\\
    M^D_{12}&=-\frac{f^2_Dm^2_Qm_D}{96\pi^2v^4}                                    \Big[\big(\Lambda^Q_{cu}\big)^2+\big(\lambda^Q_{cu}\big)^2\Big]\,,\\
    M^{B_q}_{12}&=-\frac{f^2_{B_q}m^2_Qm_{B_q}}{96\pi^2v^4}                        \Big[\big(\Lambda^Q_{qb}\big)^2+\big(\lambda^Q_{qb}\big)^2\Big]\,.
\end{split}\end{equation}
In~\cite{Ishiwata:2015cga}, these expressions were shown to be favoured over the ones displayed throughout~\cref{sec:kaon,sec:D0sys,sec:B0q} for $m_Q\gtrsim10\,\mathrm{TeV}$.
Note that through the equivalence theorem, the calculations of the previous sections involving the $W$ boson exchange already contain the corresponding Goldstone contribution of~\cref{fig:hvlq-box}; the contribution involving the $Z$ boson and the physical Higgs are missing.

\subsubsection{The ratio \texorpdfstring{$\epsilon'/\epsilon$}{ε'/ε}}
Experimentally, direct CP violation in the $K$ sector is measured by the parameter $\epsilon'$ (for a precise definition, see~\cite{Branco:1999fs}).
However, since its theoretical determination is plagued with many uncertainties, one usually favours the ratio~\cite{Aguilar-Saavedra:2002phh}
\begin{equation}\label{eq:eps-ratio}
    \frac{\epsilon'}{\epsilon}=F_{\epsilon'}(x_t)\mathrm{Im}\,\lambda^t_{sd}+\Delta_{\epsilon'}\,,
\end{equation}
where we have introduced the function~\cite{Buras:2001pn}
\begin{equation}\label{eq:eps-Fx}
    F_{\epsilon'}(x)=P_0+P_XX_0(x)+P_YY_0(x)+P_ZZ_0(x)+P_EE_0(x)\,.
\end{equation}
In the NDR scheme~\cite{Buras:2000qz}, the coefficients of the Inami-Lim functions above are given by
\begin{equation}\label{eq:eps-FxP}\begin{split}
    P_0&=-3.167+12.409B^{(1/2)}_6+1.262B^{(3/2)}_8\,,\\
    P_X&=0.540+0.023B^{(1/2)}_6\,,\\
    P_Y&=0.387+0.088B^{(1/2)}_6\,,\\
    P_Z&=0.474-0.017B^{(1/2)}_6-10.186B^{(3/2)}_8\,,\\
    P_E&=0.188-1.399B^{(1/2)}_6+0.459B^{(3/2)}_8\,,
\end{split}\end{equation}
while the bag factors $B^{(1/2)}_6=1.36\pm0.23$ and $B^{(3/2)}_8=0.79\pm0.05$ were found with lattice QCD~\cite{Aebischer:2020jto}.
Finally, the NP contributions from these models to~\cref{eq:eps-ratio} look like
\begin{equation}\label{eq:delta-Eps}\begin{aligned}
    \Delta_{\epsilon'}\! &=   \!                                 \sum^{n_u}_{k=1}\!F_{\epsilon'}(x_{U_k})\mathrm{Im}\,\lambda^{U_k}_{sd}\!+                                     \!(P_X\!+\!P_Y\!+\!P_Z)\!\sum_{i,j}\mathrm{Im}\big[V^*_{is}(F^u\!-\!1)_{ij}V_{jd}\big]N(x_i,x_j)\,,\,\,\,\,\text{(U)}\,,\\
    \Delta_{\epsilon'}\!&=-\frac{\pi s^2_W}{\alpha_\text{em}}\big(P_X+P_Y+P_Z\big)\mathrm{Im}\,F^d_{sd}\,,\quad\text{(D)}\,.
\end{aligned}\end{equation}
At the moment,
all models with VLQ singlets must satisfy
\begin{equation}\label{eq:eps-constraint}
    \Delta_{\epsilon'}=(2.9\pm4.4)\times10^{-4}
\end{equation}
due to the current experimental value for $\epsilon'/\epsilon=(1.66\pm0.23)\times10^{-3}$~\cite{ParticleDataGroup:2020ssz}.

When the SM is extended with just down-type VLQs,~\cref{eq:eps-constraint} may be written as
\begin{equation}\label{eq:eps-constraintD}
    \mathrm{Im}\,F^d_{sd}=(4.5\pm6.8)\times10^{-7}\,.
\end{equation}
Like before, when $n_d=1$ we can develop this expression with
\begin{equation}\label{eq:eps-FD}
    F^d_{sd}=-\Theta_s\Theta^*_d=-\theta_{14}\theta_{24}e^{i(\delta_{14}-\delta_{24})}
\end{equation}
by using the parameterizations introduced in~\cref{sec:ph-parameterizations}.

In models with a single up-type VLQ $T$, the constraint in~\cref{eq:eps-constraint} is translated into
\begin{equation}\label{eq:eps-constraintU}
    \mathrm{Im}\,\lambda^T_{sd}\Big[F_{\epsilon'}(x_T)-\big(P_X\!+\!P_Y\!+\!P_Z\big)N(x_T)\Big]=                (2.9\pm4.4)\times10^{-4}\,,
\end{equation}
in the limit $F^u\to\diag(1,1,1,0)$.
Here, the parameterizations of~\cref{sec:ph-parameterizations} produce
\begin{equation}\label{eq:eps-LU}
    \lambda^T_{sd}=\Theta_c\Theta^*_u=\theta_{14}\theta_{24}e^{i(\delta_{14}-\delta_{24})}\,.
\end{equation}

\subsubsection{\texorpdfstring{$Z\to bb$: $R_b$}{Z -> bb: Rb}}
Up to this point, all observables shown were impacted by the $Z$-mediated FCNC.
However, introducing VLQs into the SM has further effects.
In particular, it modifies the diagonal couplings of the down-type quarks to the $Z$ boson.
As always, such changes are strongly constrained by the ratio
\begin{equation}\label{eq:Rb-Rdef}
    R_b=\frac{\Gamma(Z\rightarrow b\bar{b})}{\Gamma(Z\rightarrow\text{hadrons})}\,,
\end{equation}
which currently lies in agreement with the SM prediction~\cite{ALEPH:2005ab,Haller:2018nnx}.
As such, one usually employs the approximate formula~\cite{Aguilar-Saavedra:2013qpa}
\begin{equation}\label{eq:Rb-Rconstraint}
    R_b=R^\text{SM}_b(1+0.5118\delta c_{L_d}+0.5118\delta c_{L_s}-1.8178\delta c_{L_b})\,,
\end{equation}
with $R^\text{SM}_b=0.21582\pm0.00011$~\cite{Haller:2018nnx} to constrain these models.
The parameters $\delta c_{L_q}$ which we have introduced in~\cref{eq:Rb-Rconstraint} are typically defined through the Lagrangian
\begin{equation}\label{eq:Rb-Ldef}
    \mathscr{L}_{Zqq}=-\frac{g}{2c_W}\bar{q}\gamma^\mu(c_{L_q}\gamma_L+c_{R_q}\gamma_R)bZ_\mu\,,
\end{equation}
where $c_{L_q}=c^\text{SM}_{L_q}+\delta c_{L_q}$.
While up-type VLQs provide one-loop contributions to these couplings via top corrections to the effective $Zb_Lb_L$ vertices~\cite{Bernabeu:1987me,Bamert:1996px}, the resulting constraints do not compete with those provided by the electroweak observables~\cite{Aguilar-Saavedra:2013qpa}.
Thus, here we will only consider the tree-level result
\begin{equation}\label{eq:Rb-Cdef}
    \delta c_{L_q}=1-F^d_{qq}\,.
\end{equation}
With the current experimental value, $R^\text{exp}_b = 0.21629 \pm 0.00066$~\cite{ALEPH:2005ab}, 
models with down-type VLQs are constrained in the following manner,
\begin{equation}\label{eq:Rb-constraint}
    F^d_{bb}-0.2815(F^d_{dd}+F^d_{ss})=0.4381\pm0.0017\,.
\end{equation}
In models with only one down-type VLQ, this constraint can be further developed with
\begin{equation}\label{eq:Rb-FD}
    F^d_{qq}=1-|\Theta_q|^2=1-\theta^2_{q4}\,,
\end{equation}
where we employed the parameterizations introduced in~\cref{sec:ph-parameterizations}.

\subsubsection{Oblique parameters}
All models which modify the gauge couplings of quarks must face the multiple constraints imposed by electroweak precision data~\cite{ALEPH:2010aa}.
Typically, one describes NP in that sector by evaluating the oblique parameters introduced in~\cite{Peskin:1990zt,Peskin:1991sw,Maksymyk:1993zm}.
The more important ones, $S$, $T$ and $U$, are given in models with singlet VLQs by\cite{Lavoura:1992np}
\begin{equation}\label{eq:oblique-parameters}\begin{split}
    S&=\frac{3}{2\pi}\Bigg[                                                       \sum_{\alpha,i}\big|V_{\alpha i}\big|^2\psi(y_\alpha,y_i)-                         \sum_{j<i}\big|F^d_{ij}\big|^2\chi(y_i,y_j)-                              \sum_{\beta<\alpha}\big|F^u_{\alpha\beta}\big|^2\chi(y_\alpha,y_\beta)\Bigg]\,,\\
    T&=\frac{3}{16\pi s^2_Wc^2_W}\Bigg[                                           \sum_{\alpha,i}\big|V_{\alpha i}\big|^2\theta(y_\alpha,y_i)-                       \sum_{j<i}\big|F^d_{ij}\big|^2\theta(y_i,y_j)-                            \sum_{\beta<\alpha}\big|F^u_{\alpha\beta}\big|^2\theta(y_\alpha,y_\beta)\Bigg]\,,\\
    U&=-\frac{3}{2\pi}\Bigg[                                                      \sum_{\alpha,i}\big|V_{\alpha i}\big|^2\chi(y_\alpha,y_i)-                         \sum_{j<i}\big|F^d_{ij}\big|^2\chi(y_i,y_j)-                              \sum_{\beta<\alpha}\big|F^u_{\alpha\beta}\big|^2\chi(y_\alpha,y_\beta)\Bigg]\,,
\end{split}\end{equation}
where sums over Greek (Latin) indices cover every up-type (down-type) quark in the theory.
We have also introduced the functions
\begin{equation}\label{eq:oblique-functions}\begin{split}
    \psi(y_\alpha,y_i)&\equiv\frac{1}{3}-\frac{1}{9}\log\frac{y_\alpha}{y_i}\,,\\
    \chi(y_1,y_2)&\equiv\frac{5(y^2_1+y^2_2)-22y_1y_2}{9(y_1-y_2)^2}+ \frac{3y_1y_2(y_1+y_2)-(y^3_1+y^3_2)}{3(y_1-y_2)^3}\log\frac{y_1}{y_2}\,,\\
    \theta(y_1,y_2)&\equiv(y_1+y_2)-\frac{2y_1y_2}{y_1-y_2}\log\frac{y_1}{y_2}\,,
\end{split}\end{equation}
with $y_i=m^2_i/m^2_Z$.

In models with a single down-type VLQ $B$ which only mixes with the third generation, one often parameterizes the mixing matrix through~\cite{Lavoura:1992qd}
\begin{equation}\label{eq:oblique-parameterizationD}
    V\supset\begin{pmatrix}
        V_{tb}&V_{tB}
    \end{pmatrix}=\begin{pmatrix}
        c&s
    \end{pmatrix}\,,
\end{equation}
in which we have introduced $c=\cos\theta$ and $s=\sin\theta$.
As a consequence,~\cref{eq:oblique-parameters} produces
\begin{equation}\label{eq:oblique-parametersD}\begin{split}
    \Delta S&=                                                                           \frac{3}{2\pi}s^2\Big[\psi(y_t,y_B)-c^2\chi(y_B,y_b)-\psi(y_t,y_b)\Big]\,,\\
    \Delta T&=                                                                           \frac{3}{16\pi s^2_Wc^2_W}s^2\Big[\theta(y_t,y_B)-c^2\theta(y_B,y_b)-\theta(y_t,y_b)\Big]\,,\\
    \Delta U&=                                                                          -\frac{3}{2\pi}s^2\Big[\chi(y_t,y_B)-c^2\chi(y_B,y_b)-\chi(y_t,y_b)\Big]\,,\\
\end{split}\end{equation}
where $(\Delta S,\Delta T,\Delta U)=(S-S_\text{SM},T-T_\text{SM},U-U_\text{SM})$ with $S_\text{SM}=3\psi(y_t,y_b)/2\pi$, $T_\text{SM}=3\theta(y_t,y_b)/16\pi s^2_Wc^2_W$ and $U_\text{SM}=-3\chi(y_t,y_b)/2\pi$.

In extensions of the SM where only one up-type VLQ $T$ which mixes exclusively with the third generation is introduced, the situation is analogous.
As such, after parameterizing the CKM matrix with
\begin{equation}\label{eq:oblique-parameterizationU}
    V\supset\begin{pmatrix}
        V_{tb}\\V_{Tb}
    \end{pmatrix}=\begin{pmatrix}
        c\\s
    \end{pmatrix}\,,
\end{equation}
\cref{eq:oblique-parameters} provides the following constraints,
\begin{equation}\label{eq:oblique-parametersU}\begin{split}
    \Delta S&=                                                                           \frac{3}{2\pi}s^2\Big[\psi(y_T,y_b)-c^2\chi(y_T,y_t)-\psi(y_t,y_b)\Big]\,,\\
    \Delta T&=                                                                           \frac{3}{16\pi s^2_Wc^2_W}s^2\Big[\theta(y_T,y_b)-c^2\theta(y_T,y_t)-\theta(y_t,y_b)\Big]\,,\\
    \Delta U&=                                                                          -\frac{3}{2\pi}s^2\Big[\chi(y_T,y_b)-c^2\chi(y_T,y_t)-\chi(y_t,y_b)\Big]\,.
\end{split}\end{equation}

In the literature, the contributions of these models to $\Delta U$ are often neglected.
In order to understand why, we evaluate~\cref{eq:oblique-parametersU} in the heavy VLQ limit, $y_T\gg1$,
\begin{equation}\label{eq:oblique-parametersH}\begin{split}
    \Delta S\simeq-\frac{s^2}{6\pi}\big(1-3c^2\big)\log y_T\,,\quad                      \Delta T\simeq\frac{3y_ts^2c^2}{8\pi s^2_Wc^2_W}\log y_T\,,\quad                     \Delta U\simeq\frac{s^4}{2\pi}\log y_T\,.\\
\end{split}\end{equation}
After remembering~\cref{eq:uparam-approxT} to establish $s^2\sim y^{-1}_T$, these relations predict a small $\Delta U$,
\begin{equation}\label{eq:oblique-parametersHy}
    \Delta S,\Delta T\sim y^{-1}_T\log y_T\,,\qquad\Delta U\sim y^{-2}_T\log y_T\,.
\end{equation}
For this reason, it is safe to consider only $\Delta S$ and $\Delta T$ in the phenomenological analysis of a model with singlet VLQs.%
\footnote{A similar argument can be applied to~\cref{eq:oblique-parametersD}.}

\subsection{Global fits}\label{sec:ph-fits}
\providecommand{\tY}{\tilde{Y}}

In this subsection we discuss some global fits that were performed considering the presence of one singlet VLQ of down or up-type and a set of the observables presented in the last subsection. For the down-type case we will mainly use the results of~\cite{Cherchiglia:2021vhe}, where we have defined the Yukawa matrix connecting the VLQ of down-type $B$ to the SM quarks as $\tY^{B}$.
This matrix corresponds to $\oYd$ in the WB that $Y_d$ is diagonal and $\oM_d=0$
(cf.~the second matrix in~\cref{eq:WB2}).
Using the approximation introduced in~\cref{sec:devunit}, which is valid for small deviations from unitarity, we can use~\cref{eq:uparam-approxT} and relate
\eq{
\tY^B_i=(\oYd)_i
\simeq \frac{\sqrt{2}M_B}{v}(\Theta_d)_i
\simeq \frac{\sqrt{2}M_B}{v}F^d_{4i}\,.
}
In~\cite{Bobeth:2016llm}, the authors performed a series of global fits, considering not only singlet VLQ of down-type but also doublets and triplets. In their notation, the product $\tY^{B}_{i}\tY^{B*}_{j}$ was denoted as $\Lambda_{ij}$.

We begin with the $(sd)$ sector, whose main constraints come from the ratio $\epsilon'/\epsilon$ and the short-distance contributions
$\br(K_L\to\mu\bar{\mu})_{\rm SD}$.
Regarding the latter, one can use~\cref{eq:kd-rKL} to estimate
\begin{equation}
    \br(K_L\rightarrow\mu^+\mu^-)_{\rm SD}\sim 5.93\times10^{-3}\left(\Delta^{K_L}_{\mu\mu} - 3.59\times10^{-4}\right)^{2}\,.
\end{equation}
For a down-type VLQ,~\cref{eq:delta-KL} provides us
\begin{equation}
    \Delta^{K_L}_{\mu\mu} \sim 92.94 \;\mathrm{Re} (\Theta_{s} \Theta_{d}^{*}) \sim \;92.94 \frac{v^{2}}{2 M_{B}^{2}} \mathrm{Re} (\tY^B_s \tY^{B*}_d)\,.
\end{equation}
Thus, given the current experimental constraint $\br(K_L\rightarrow \mu^+\mu^-)_{\rm SD}< 2.5\times10^{-9}$, one can estimate the following bound for a VLQ with a mass around $1.5\;\unit{TeV}$
\begin{equation}
   -2\times10^{-4}<\mathrm{Re} (\tY^B_s \tY^{B*}_d)<7\times10^{-4}\,.
\end{equation}

Regarding $\epsilon'/\epsilon$, one can use~\cref{eq:delta-Eps} to obtain
\begin{equation}
    \Delta_{\epsilon'}\sim  6.06\times10^{2} \; \mathrm{Im}\,(\Theta_{s} \Theta_{d}^{*}) \sim 6.06\times10^{2} \frac{v^{2}}{2 M_{B}^{2}} \mathrm{Im} (\tY^B_s \tY^{B*}_d)\,.
\end{equation}
Once again, for a VLQ with a mass of a few TeV, one obtains the bound 
\begin{equation}
    |\mathrm{Im} (\tY^B_s \tY^{B*}_d)|\lesssim 10^{-4}\,.
\end{equation}

\begin{figure}[!t]
\centering
\includegraphics[scale=0.4]{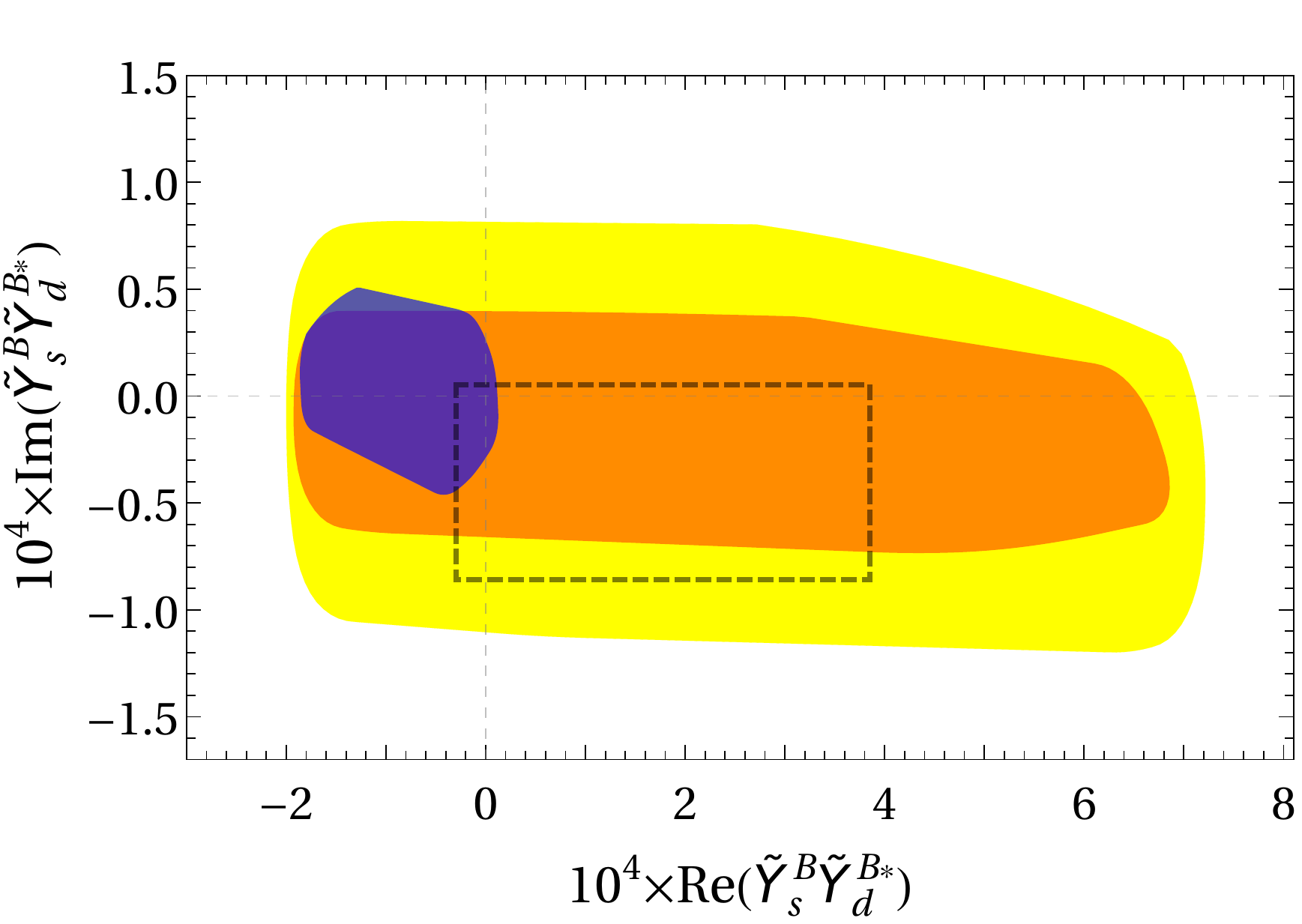}
\caption{\label{fig:Lambda.sd}%
Allowed regions for the product $\tY^{B}_{s}\tY^{B*}_{d}$ at 95\% (68\%) C.L.~are shown in yellow (orange) in a scenario with only one singlet VLQ of down-type. In blue we show the 95\% C.L.~allowed region for a specific model that can address the strong CP problem. In all cases, we adopt $M_B=1.4\;\unit{TeV}$. The dashed region was extracted from Ref.~\cite{Bobeth:2016llm}, where a global fit was performed considering $M_B=1\;\unit{TeV}$.
}
\end{figure}

In~\Cref{fig:Lambda.sd} we show the allowed regions for the real and imaginary parts of the product $\tY^{B}_{s}\tY^{B*}_{d}$. As anticipated, the region is vertically constrained by the ratio $\epsilon'/\epsilon$ while horizontally the observable $\br(K_L\to\mu\bar{\mu})_{\rm SD}$ provides the main constraint. In all subsequent plots, we show in yellow (orange) the 95$\%$ (68$\%$) C.L.~allowed region for a VLQ of down-type with mass $M_{B}=1.4\;\unit{TeV}$. The dashed contour is extracted from~\cite{Bobeth:2016llm} which performs the global fit choosing $M_{B}=1.0\;\unit{TeV}$. Finally, the blue region corresponds to the 95$\%$ C.L.~allowed region for a more specific type of VLQ that are connected to the Nelson-Barr mechanism to solve the strong CP problem. These type of VLQ, denoted VLQ of Nelson-Barr type, will be explained in detail in~\cref{sec:NB_VLQ}. The important point to be noticed is that the allowed parameter region for this kind of VLQ is a subset of the parameter region for the generic VLQ of down-type.

\begin{figure}[!t]
\centering
\includegraphics[scale=0.4]{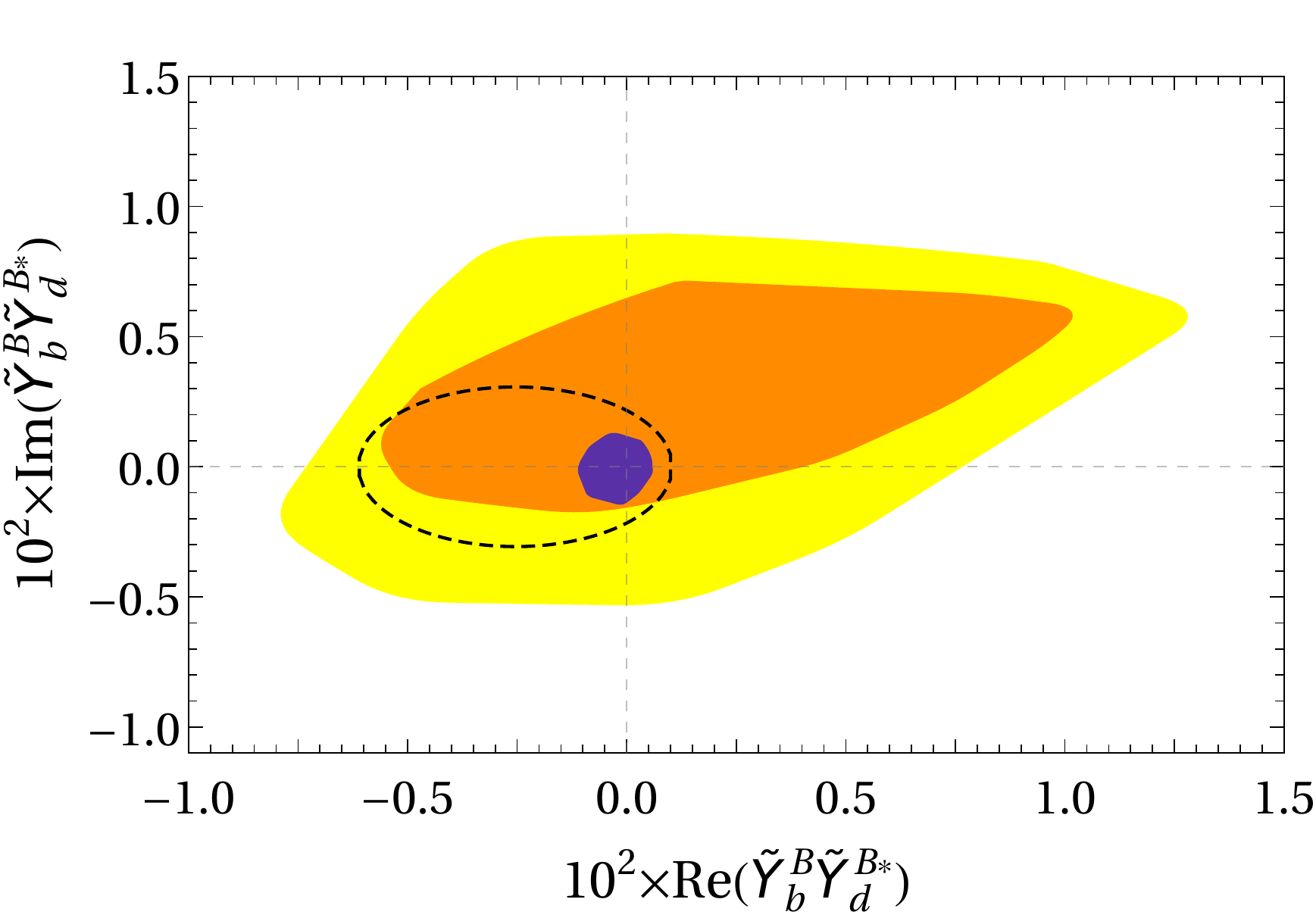}
\caption{\label{fig:Lambda.bd}
Allowed regions for the product $\tY^{B}_{b}\tY^{B*}_{d}$ at 95\% (68\%) C.L.~are shown in yellow (orange) in a scenario with only one singlet VLQ of down-type. In blue we show the 95\% C.L.~allowed region for a specific model that can address the strong CP problem. In all cases, we adopt $M_B=1.4\;\unit{TeV}$. The dashed region was extracted from Ref.~\cite{Bobeth:2016llm}, where a global fit was performed considering $M_B=1\;\unit{TeV}$.
}
\end{figure}

Regarding the ($bd$) sector, for the mass range of a few TeV, the most constraining observable is $\br(B^+\to \pi^+\mu\bar{\mu})$ which was not included in the global fit performed in Ref.~\cite{Cherchiglia:2021vhe}. This explains why the region depicted in~\Cref{fig:Lambda.bd} is much wider than the dashed contour extracted from~\cite{Bobeth:2016llm}. Nevertheless, the allowed region for VLQs of Nelson-Barr type is completely enclosed in the dashed contour.

\begin{figure}[t]
\centering
\includegraphics[scale=0.4]{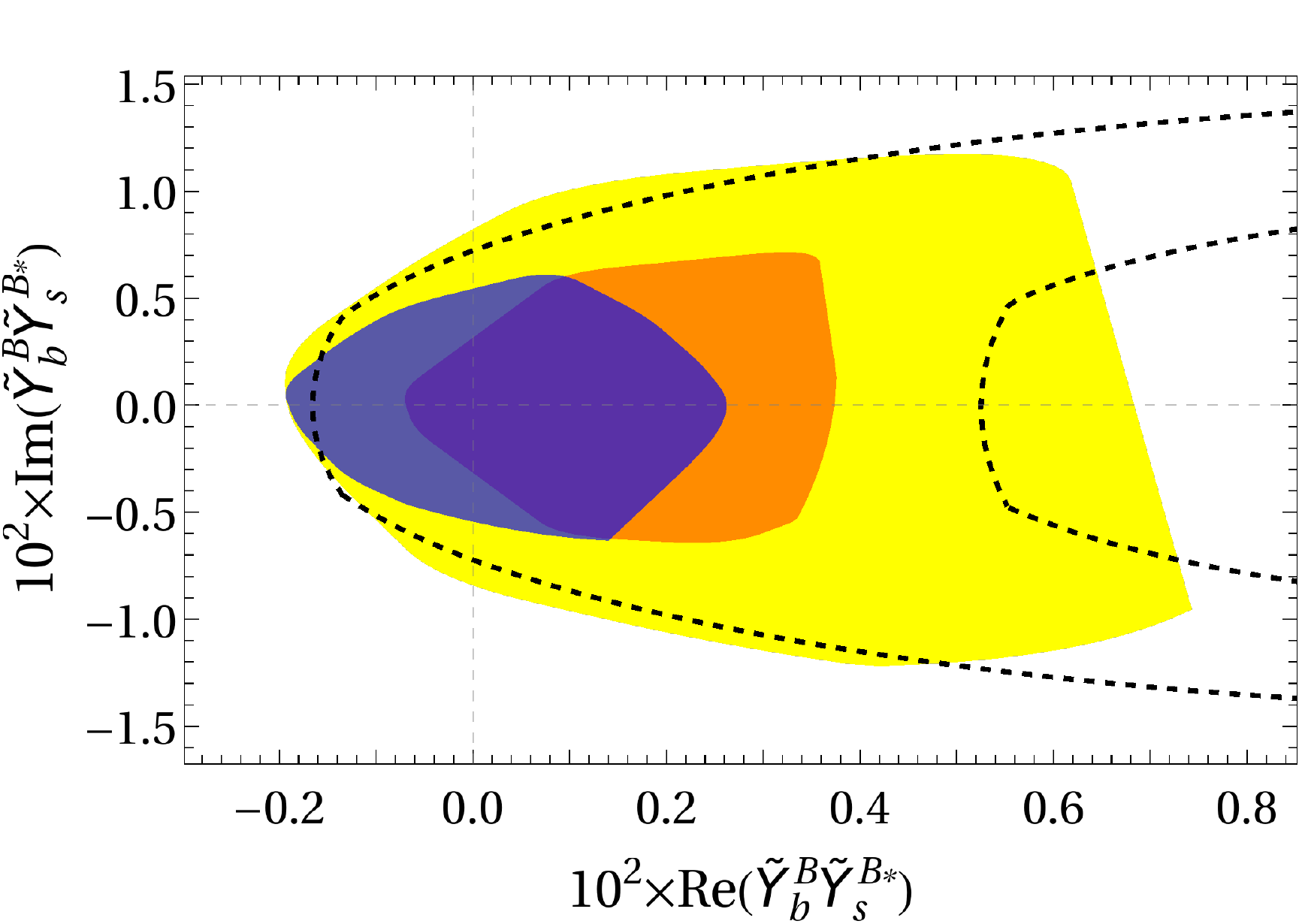}
\caption{\label{fig:Lambda.bs}
Allowed regions for the product $\tY^{B}_{b}\tY^{B*}_{s}$ at 95\% (68\%) C.L.~are shown in yellow (orange) in a scenario with only one singlet VLQ of down-type. In blue we show the 95\% C.L.~allowed region for a specific model that can address the strong CP problem. In all cases, we adopt $M_B=1.4\;\unit{TeV}$. The dashed region was extracted from Ref.~\cite{Bobeth:2016llm}, where a global fit was performed considering $M_B=1\;\unit{TeV}$.
}
\end{figure}

Finally, the ($bs$) sector is mainly constrained by $\br(B_s\to\mu\bar{\mu})$ explaining why the yellow and dashed regions are largely compatible in~\Cref{fig:Lambda.bs}. The possible enlargement is due to the higher mass range chosen in~\cite{Cherchiglia:2021vhe}. 
One can clearly see the dashed curves follow the ring shape of the shifted circular pattern of the second constraint in~\eqref{eq:bd-constraintsD}, which we reproduce below in terms of $\tY^{B}$:
\begin{equation}
\bigg|\frac{v^{2}}{2 M_{B}^{2}} (\tY^B_s \tY^{B*}_b)
-0.000442\bigg|^2=(1.48\pm0.20)\times10^{-7}\,.
\end{equation}
Thus, for a VLQ of mass around $1.5\;\unit{TeV}$, one obtains the bound
\begin{equation}
    0.023\lesssim\Big|\tY^B_s \tY^{B*}_b - 0.029\Big|  \lesssim 0.027\,.
\end{equation}

\clearpage

Regarding the scenario with VLQs of up-type, a global fit was performed in~\cite{Branco:2021vhs} aiming to provide an explanation for the non-unitarity of the first row of the CKM. As discussed in this reference, the presence of the VLQs can alleviate the present tension, being a motivation for the introduction of such particles to the SM. Using the parameterization introduced in~\cref{eq:SVLQ-FU}, one obtains allowed regions for the new mixing angles and phases as in~\Cref{fig:correlation}. In this figure, the lighter (darker) solid green areas represent allowed regions at $2\sigma$ ($3\sigma$) level after using the constraints discussed in the previous subsection. The mass of the VLQ is assumed to be higher than $1\;\unit{TeV}$. The dashed lines come from constraints on the experimental values of the CKM elements only, while the solid lines include further constraints such as meson mixing, and $|\epsilon_K|$. It is instructive to notice that, given the present tension with the first row of the CKM, the angle $\theta_{14}$ must be non-null. It comes only from considering the experimental determination of the elements of the CKM matrix, as the dashed lines show. We emphasize that this feature is usually not considered in present experimental searches where the VLQ couples only to the third family, which may allow one to circumvent present bounds on the VLQ mass. Finally, for the plots shown in~\mbox{\cref{fig:correlation}} the more conservative bound $\big|\left(\epsilon_K\right)_\text{NP}\big|< \big|\left(\epsilon_K\right)_\text{exp}\big|$ was considered. Another approach, as discussed at end of~\mbox{\cref{sec:kaon}}, is to consider $\big|\left(\epsilon_K\right)_\text{NP}\big|\lesssim 0.1\big|\left(\epsilon_K\right)_\text{exp}\big|$ 
as in~\cite{Botella:2021uxz}. The main result is that the region with $s_{24}=s_{34}=0$ is not allowed in this case.

\afterpage{
\begin{figure}[!ht]
\centering
\includegraphics{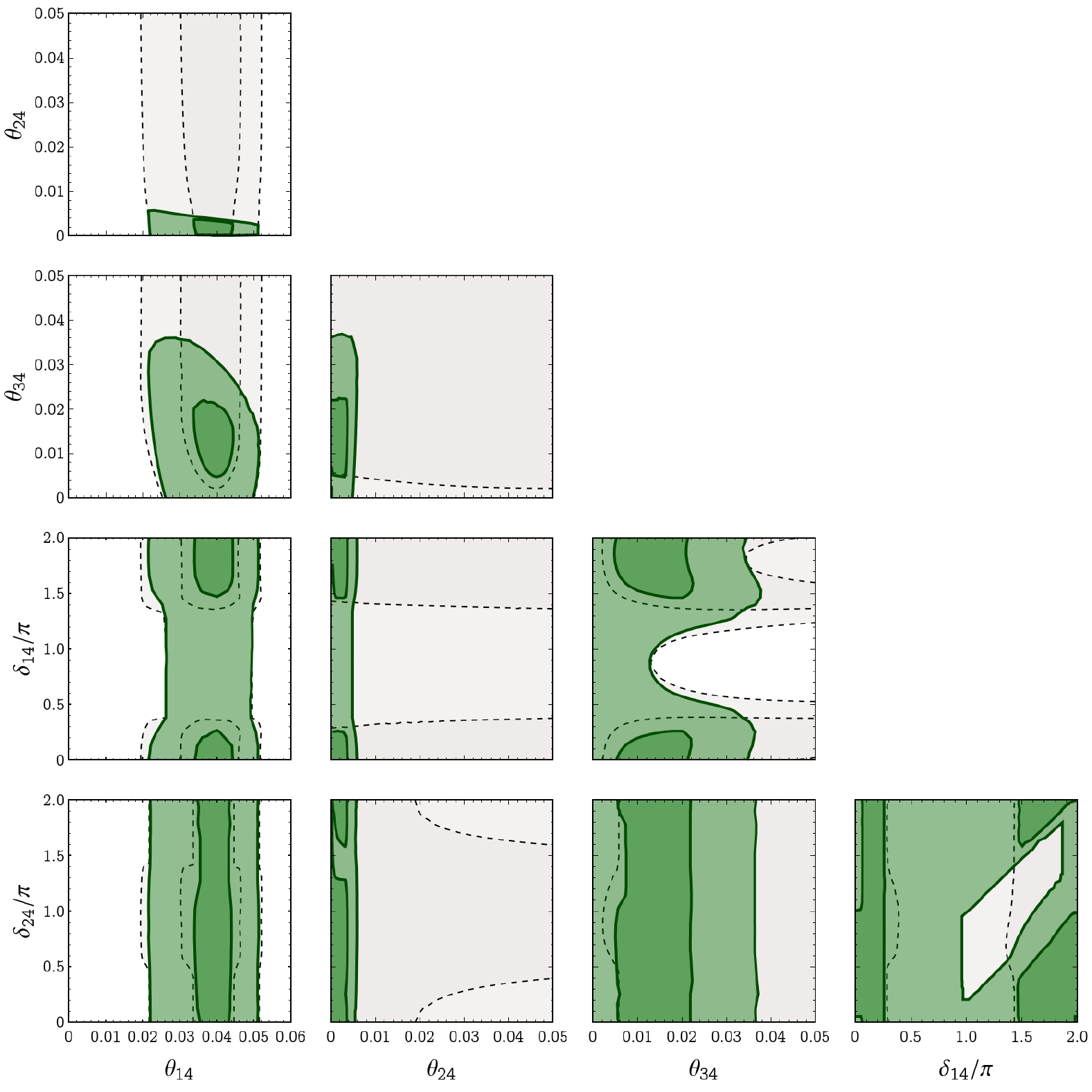}
\caption{\label{fig:correlation}
Allowed regions for the mixing angles and new phases due to the inclusion of VLQ of up-type. Lighter (darker) solid green areas represent allowed regions at $2\sigma$ ($3\sigma$). Taken from Ref.~\cite{Branco:2021vhs}.}
\end{figure}
\clearpage
}

\subsection{VLQs in the SMEFT}
\label{sec:SMEFT}

As VLQs are constrained to be above the TeV scale, an effective field theory description is often appropriate and sufficient at energies below the TeV scale.
As we integrate out the heavy VLQs to match onto the SM effective field theory (SMEFT)~\cite{Buchmuller:1985jz,Grzadkowski:2010es}, the information of the VLQs in the low energy theory will be imprinted in the Wilson coefficients of the dimension 6 effective operators and the correlations among them  (see~\cite{Criado:2019mvu} for the phenomenology of dimension 5 operators that can be built with VLQs and SM fields).
At present, it is even possible to automatize the process of matching at tree-level (and one-loop) of general models to the SMEFT~\cite{Carmona:2021xtq,Fuentes-Martin:2022jrf}.
One of the advantages of using an EFT description is that we can use the renormalization group equations (RGEs) of the SMEFT~\cite{Alonso:2013hga} to resum large logs between the matching scale and the low-energy scale of the processes of interest. 

Only four operators of dimension 6 in the SMEFT are generated at tree level, as will be seen in what follows.
In the SMEFT, we use the Warsaw basis~\cite{Grzadkowski:2010es}.
In the VLQ model, we use the basis where $\,\oM_u=0$ and $\,\oM_d=0$ in~\eqref{eq:baremasses} while all terms in~\eqref{eq:lagrangian} are present. The VLQ mass matrices $M_u$ and $M_d$ are diagonal.
Integrating out the heavy vector-like quarks $D_{rR},D_{rL}$ and $U_{rR},U_{rL}$ we obtain~\cite{delAguila:2000rc}
\eqali{
\label{dim6:1}
\frac{\lag_6}{\Lambda^2}
&=+\bar{Q}_{iL}\Phi (\oYd)_{ir}\frac{1}{M^2_{D_r}}({\oYd}^\dag)_{rj}i\slashed{D}(\Phi^\dag Q_{jL})
\cr&\quad
+\bar{Q}_{iL}\tilde{\Phi} (\oYu)_{ir}\frac{1}{M^2_{U_r}}({\oYu}^\dag)_{rj}i\slashed{D}(\tilde{\Phi}^\dag Q_{jL})
\,.
}
Let us define the combination of coefficients above:
\eq{
G^d_{ij}\equiv (\oYd)_{ir}\frac{1}{M^2_{D_r}}({\oYd}^\dag)_{rj}\,,\quad
G^u_{ij}\equiv (\oYu)_{ir}\frac{1}{M^2_{U_r}}({\oYu}^\dag)_{rj}\,.
}
The combinations $(\Phi^\dag Q_{jL})$ and $(\tilde{\Phi}^\dag Q_{jL})$ have the same quantum numbers as $D_{rL}$ and $U_{rL}$, respectively.
After applying the covariant Leibniz rule for the covariant derivative and Fierz rearrangement, we see that the terms with $D_\mu\Phi$ already give the operators in the Warsaw basis.
The terms involving $D_\mu q_L$ need to be simplified with the equations of motion
after ensuring the Hermiticity of the whole Lagrangian.
The result is the appearance of only four operators:
\eqali{
\label{dim6:smeft}
\frac{\lag_6}{\Lambda^2}
&=
\frac{C^{(1)}_{\phi q}}{\Lambda^2}\cO^{(1)}_{\phi q}
+\frac{C^{(3)}_{\phi q}}{\Lambda^2}\cO^{(3)}_{\phi q}
+\frac{C_{d\phi}}{\Lambda^2}\cO_{d\phi}
+\frac{C_{u\phi}}{\Lambda^2}\cO_{u\phi}\,.
}
We show the operators in~\cref{tab:VLQ:op} and the Wilson coefficients in~\cref{tab:VLQ:wilson}.
From the latter, it should be clear that only three operators arise if we only consider either up-type or down-type VLQs.
The definition of the operators is the same as in Ref.~\cite{Grzadkowski:2010es} with $Q_F=\cO_F$, where e.g.~$F=\phi q$ is the label.
In flavour space, the usual convention is to take $Y_d$ diagonal and $Y_u=V^\dag \hat{Y}_u$ with $\hat{Y}_u$ being diagonal and $V$ is the CKM. 
Note that Ref.~\cite{delAguila:2000rc} uses for $\cO^{(1)}_{\phi q}$ and $\cO^{(3)}_{\phi q}$
the convention of Ref.~\cite{Buchmuller:1985jz} which differs slightly of the convention in~\cite{Grzadkowski:2010es}, used here.
The results here match the extensive work of Ref.~\cite{deBlas:2017xtg} which uses the latter convention.
Note that our convention for the covariant derivative is
\eq{
D_\mu=\partial_\mu+ig\frac{\tau_a}{2}W^a_\mu+ig'YB_\mu\,.
}
%
\begin{table}[t!]
  \centering
  \begin{tabular}{lc}
    \toprule    
    $\cO^{(1)}_{\phi q}$ & $(\Phi^\dag i\te{D}_\mu\Phi)(\bar{Q}^i_L\gamma^\mu Q^j_L)$
\\[2mm]
$\cO^{(3)}_{\phi q}$ & $(\Phi^\dag i\te{D}^I_\mu\Phi)(\bar{Q}^i_L\tau^I\gamma^\mu Q^j_L)$
\\[2mm] 
$\cO_{d\phi}$ & $(\Phi^\dag\Phi)(\bar{Q}^i_L\Phi d^j_R)$
\\[2mm]
$\cO_{u\phi}$ & $(\Phi^\dag\Phi)(\bar{Q}^i_L\tilde{\Phi} u^j_R)$
\\
\bottomrule
  \end{tabular}
  \caption{Dimension-6 operators generated by VLQs.}
  \label{tab:VLQ:op}
\end{table}
%
\begin{table}[t!]
  \centering
  \begin{tabular}{lcccc}
    \toprule   
    Operator &  $\cO^{(1)}_{\phi q}$ & $\cO^{(3)}_{\phi q}$ & $\cO_{d\phi}$ & $\cO_{u\phi}$
\\
\midrule
\rule{0cm}{1.5em}%
$D\colon C^F_{ij}/\Lambda^2$
    & $-\frac{1}{4}G^d_{ij}$ & $-\frac{1}{4}G^d_{ij}$ & $+\frac{1}{2}G^d_{ik}(Y_d)_{kj}$ & --- 
\\[4mm]
$U\colon C^F_{ij}/\Lambda^2$
    & $+\frac{1}{4}G^u_{ij}$ & $-\frac{1}{4}G^u_{ij}$ & --- & $+\frac{1}{2}G^u_{ik}(Y_u)_{kj}$
\\[2mm] \bottomrule
  \end{tabular}
  \caption{Wilson coefficients of generated dimension-6 operators.}
  \label{tab:VLQ:wilson}
\end{table}
%

After the EWSB the new dimension six operators in~\eqref{dim6:1} modifies the quark interactions to $W,Z$ and $h$ as~\cite{delAguila:2000rc}
\eqali{
\label{smeft:deviation:WZh}
-\lag^W&=\frac{g}{\sqrt{2}}\bar{u}^i_L V_{ij}\gamma^\mu d^j_LW_\mu^+ + \text{h.c.}\,,
\\
-\lag^Z&=\frac{g}{2c_w}\left(
    \bar{u}^i_L F^{u}_{ij}\gamma^\mu u^j_L-\bar{d}^i_L F^{d}_{ij}\gamma^\mu d^j_L
    -2s^2_w J^\mu_\text{em}
    \right)Z_\mu\,,
\\    
-\lag^h&=\bar{u}^i_L g^{uh}_{ij}u^j_R\frac{h}{\sqrt{2}}+\bar{d}^i_L g^{dh}_{ij}d^j_R\frac{h}{\sqrt{2}}+ \text{h.c.}\,,
}
where
\eqali{
\label{smeft:deviation:WZh:C}
V_{ij}&=V^{sm}_{ik}\left[\delta_{ij}+\frac{v^2}{\Lambda^2}C^{(3)}_{\phi q} \right]_{kj}\,,
\\
F^{u}_{ij}&=\delta_{ij}-\frac{v^2}{\Lambda^2}\left(
    C^{(1)}_{\phi q}-C^{(3)}_{\phi q} \right)_{ij}\,,
\\
F^{d}_{ij}&=\delta_{ij}+\frac{v^2}{\Lambda^2}\left(
    C^{(1)}_{\phi q}+C^{(3)}_{\phi q} \right)_{ij}\,,
\\
g^{uh}_{ij}&=\left[\frac{\sqrt{2}}{v}M_u-\frac{v^2}{\Lambda^2}C^{u\phi}\right]_{ij}\,,
\\
g^{dh}_{ij}&=\left[\frac{\sqrt{2}}{v}M_d-\frac{v^2}{\Lambda^2}C^{d\phi}\right]_{ij}\,.
}
The CKM matrix $V^{sm}$ is given by the mismatch in diagonalization of the effective mass matrices $M_u$ and $M_d$.
We can choose a basis where $M_d=\hat{M}_d$ while $M_u={V^{sm}}^\dag \hat{M}_u$, with hatted matrices being diagonal.
Note that these mass matrices already take into account the effects of the new operators~\cite{Buchmuller:1985jz}:
\eqali{
M_u&=\frac{v}{\sqrt{2}}\left[Y_u-\frac{v^2}{2\Lambda^2}C^{u\phi}\right]\,,
\cr
M_d&=\frac{v}{\sqrt{2}}\left[Y_d-\frac{v^2}{2\Lambda^2}C^{d\phi}\right]\,,
}
where the Yukawa couplings $Y_u$ and $Y_d$ are the ones appearing in the original dimension four terms 
\eqref{eq:lagrangian} without VLQs. The extraction of the CKM parameters in the general SMEFT is discussed in Ref.~\cite{Descotes-Genon:2018foz}.

The deviations in~\eqref{smeft:deviation:WZh:C}, with Wilson coefficients given in~\cref{tab:VLQ:wilson},
match the direct diagonalization of~\cref{sec:gaugeinteractions} in leading order in $1/M$ with $M$ being the heavy VLQ masses.

Beyond the simple picture above, there are two potentially important effects: (i) operators that are generated at one-loop and (ii) operators that are induced by operator mixing by the running down from the VLQ scale to the electroweak scale.
For (i), some operators in the SMEFT indeed can be only generated at one-loop or higher at any BSM scenario~\cite{Arzt:1994gp}. 

Let us follow Ref.~\cite{Bobeth:2016llm} to see what happens with down-type VLQs $D_r$. In this case, we expect contributions to $\Delta F=2$ processes implying that four-fermion operators ($\psi^4$) will play a major role. 
Concerning four-fermion operators, singlet VLQs induce left-handed currents only, generating the operators
\eqali{
\label{QQQQ}
\cO^{(1)}_{QQ}&=(\bar{Q}_L\gamma_\mu Q_L)(\bar{Q}_L\gamma^\mu Q_L)\,,
\cr
\cO^{(3)}_{QQ}&=(\bar{Q}_L\gamma_\mu \tau^I Q_L)(\bar{Q}_L\gamma^\mu \tau^I Q_L)\,.
}
These operators are induced at one-loop by box diagrams with VLQs and Higgses exchanged, and induce $\Delta F=2$ processes at low energy. 
The Wilson coefficients are proportional to $(\oYd)_{ir}^2(\oYd)_{jr}^{*2}/M^2$. For singlet VLQ of up-type we have the same contribution exchanging $\oYd$ by $\oYu$~\cite{Ishiwata:2015cga}.
These operators are also induced by tree-level $Z$ exchange of the operators $\cO^{(1)}_{\phi q}$ and $\cO^{(3)}_{\phi q}$. The latter effect is relevant for VLQ of TeV masses but it becomes subdominant for a $10\;\unit{TeV}$ VLQ.

Concerning the renormalization group evolution from the VLQ scale down to the electroweak scale, the tree-level generated operators in~\cref{tab:VLQ:op} will induce the following classes of operators that are not generated at tree-level matching:
\eq{
\Phi^6\,,\quad
\Phi^4D^2\,,\quad
\psi^4\,.
}
As usual, the most relevant interaction for operator mixing is the top Yukawa~\cite{Jenkins:2013wua}.
The latter will generate in leading log approximation an additional contribution to the operators in~\eqref{QQQQ} with coefficient proportional to $V_{3i}^*V_{3j}(\oYd)_{ir}(\oYd)_{jr}^*/M_r^2$.
For $D$, these effects are cancelled for $\Delta F=2$ processes~\cite{Bobeth:2016llm}. 
Another relevant flavour conserving effect comes from QCD running, which can be directly applied to the one-loop box contribution.

As an example of operators that are only generated at one-loop, we have the dipole (penguin) operators such as~\cite{Arnan:2016cpy,Morozumi:2018cnc}
\eq{
(\bar{Q}_L\sigma^{\mu\nu}d_R)\Phi B_{\mu\nu}
\,,\quad
(\bar{Q}_L\tau^I\sigma^{\mu\nu}d_R)\Phi W^I_{\mu\nu}
\,,\quad
(\bar{Q}_L\sigma^{\mu\nu}\lambda_a d_R)\Phi G^a_{\mu\nu}
\,.
}
These operators are relevant for the transition $b\to s\gamma$ and some Wilson coefficients of the low energy effective Lagrangian can be found in Ref.~\cite{Arnan:2016cpy}.
Analogous operators with $u_R$ exist as well as semileptonic four-fermion operators.

\vfill
\clearpage

\section{Spontaneous CPV and the Strong CP Problem }
\label{sec:strongCP}

\subsection{Generating a complex CKM from a vacuum phase}
\label{sec:complexCKM}
At present, there is experimental evidence that the CKM matrix is complex even 
if one allows for the presence of new physics~\cite{Botella:2005fc}. However, 
the origin of CP violation is not known. The breaking of CP may arise from the 
introduction of complex Yukawa couplings leading to CP violation at the Lagrangian 
level. This is the situation one encounters in the SM with only one scalar doublet.
Alternatively, one may impose CP invariance at the Lagrangian level but have 
CP spontaneously broken by the vacuum of an extended scalar sector. In this case, 
in order to be in agreement with experiment, the vacuum phase(s) should be able to generate a complex CKM matrix.

Models with vector-like quarks provide some of the most plausible scenarios 
with CP violation generated spontaneously. In order to understand the above 
statement, let us recall some of the various scenarios which have been proposed for 
having CP spontaneously broken.

The idea of spontaneous CP violation was first introduced by T.~D.~Lee~\cite{Lee:1973iz} who considered an extension of the SM with two Higgs doublets. For three families of quarks in the original Lee model, the vacuum phase is able to generate a complex CKM matrix. However Lee's model has dangerous scalar mediated FCNC at tree level which render the model unrealistic. If one tries to eliminate these tree-level scalar FCNC by introducing, for example, an extra unflavoured symmetry of the Lagrangian, it turns out that one eliminates the possibility of having a CP violating vacuum in the context of two Higgs doublet models, while this is still possible in the context of three Higgs doublets~\cite{Branco:1980sz}.
Recently, it was pointed out~\cite{Nebot:2018nqn} that one can 
have a viable model with spontaneous CP violation in the context of two Higgs doublets. 
The model is built in the framework of a generalized Branco-Grimus-Lavoura 
(BGL) model~\cite{Branco:1996bq} with a flavoured $\ZZ_2$ symmetry. 
In the sequel it will be shown that in the context of VLQ models, one may have a 
viable scenario for spontaneous CP violation in a much simpler (plausible) model. 
In its simplest implementation only a very minimal extension of the Standard Model 
is required with the addition of a vector-like down isosinglet quark and a 
complex singlet scalar~\cite{Bento:1990wv}. The complex singlet scalar acquires a complex VEV breaking CP spontaneously at a high energy scale. It is through the coupling of the vector-like quark to this singlet and the mixing of this quark with the SM-like quarks that this phase generates a complex $V_{CKM}$ matrix. None of the main features of this scenario depend on the type of vector-like quark being up or down or on the number of vector-like quarks introduced.

The field content of the model first introduced in Ref.~\cite{Bento:1990wv} is
\begin{equation}
\left( 
\begin{array}{c}
u^0 \\ 
 d^0
\end{array}
\right)_{iL}\,, \ \ u^0_{iR}\,,  \ \ d^0_{\alpha R}\,, \ \ D^0_L \ \ (i= 1,2,3, 
\ \   \alpha = 1, ..., 4)\,, \ \ \Phi\,, \ \  S\,,
\end{equation}
together with the gauge bosons and the leptons, where $\Phi$ is a Higgs doublet and $S$ is a complex singlet scalar. In this subsection, we will denote the right-handed VLQ as $d^0_{4 R}$, allowing us to simplify some of the equations that follow. In terms of the notation introduced in~\cref{sec:notation}, we have $D^0_{R}=d^0_{4 R}$.

The first illustration of the generation of a complex CKM from a vacuum phase in this framework only had one large mass scale~\cite{Bento:1990wv}. This was achieved by forbidding bare mass terms of the type $ D^0_L d^0_{\alpha R} $  through a discrete $\ZZ_2$ symmetry under which $D^0_L,$ and $S$ are odd and all other fields are even, as a result the two different mass scales are the VEVs of $ \phi$ and $S$:
\begin{equation}
\langle {\Phi}^0 \rangle = \frac{v}{\sqrt 2}\,, \ \   \ \ \   
\langle S \rangle = \frac{V \exp (i \beta )}{\sqrt 2}\,.
\label{eq:vevphiS}
\end{equation}
CP invariance is imposed at the Lagrangian level, therefore all coefficients are real. The scalar potential leading to the above VEVs is the most general $SU(2) \times U(1) \times \ZZ_2$-invariant potential:
\begin{equation}
V = V_0 \ (\Phi, S) + (\mu^2 + \lambda_1 \ S^\ast S + \lambda_2 \ \Phi^\dagger \Phi) (S^2 + S^{\ast 2}) + \lambda_3 \ (S^4 + S^{\ast 4} )\,,
\end{equation}
where $V_0$ contains all terms that are phase independent and includes the SM Higgs potential. It was shown in Ref.~\cite{Bento:1990wv} that in general the VEVs of~\cref{eq:vevphiS} violate CP spontaneously. This violation occurs at a high energy scale since it is associated to the VEV of a singlet of the gauge group. This can be understood by checking that, after spontaneous symmetry breaking, the phase-dependent part of $V$ can be expressed in the form:
\begin{equation}
V(v, V, \beta) =  a \cos (2 \beta) + b \cos (4 \beta)\,.
\end{equation}
In the region of parameters where $b$ is positive and ($|a| < 4 b$) the absolute minimum of the potential is at
\begin{equation}
\beta = \frac{1}{2} \arccos (a/4b)
\end{equation}
and CP is spontaneously broken.

The Yukawa interactions of the quarks are given by:
\begin{equation} 
 {\cal L}_Y = - \sqrt{2} (\overline{u^0} \ \overline{d^0})_L^i (g_{i\alpha } 
\Phi \, d^0_{\alpha R}+ 
h_{ij}  \tilde{\Phi}\, u^0_{jR}) - \sqrt{2} (f_\alpha \, S + f^\prime_\alpha \, S^* ) \, \overline{D^0_L} d^0_{\alpha R} + \text{h.c.}\,,
\end{equation}
where all coefficients are real. We work in the weak basis where the up quark mass matrix is diagonal. The down quark mass matrix is of the general form
\begin{equation}
{ \cal M}_d =  \begin{pmatrix}
m_d & \overline{m}_d \\[2mm]
\overline{M}_d & M_d
\end{pmatrix} \,,
\end{equation}
where $(\overline{M}_{d})_j = f_j \, V e^{i\beta} + 
f^\prime_j \, V e^{-i\beta}$, and $M_d = f_4 \, V e^{i\beta} + 
f^\prime_4 \, V e^{-i\beta}$. All terms in the fourth row of ${ \cal M}_d $ are potentially complex. There is a parameter redundancy in the mass matrix ${ \cal M}_d$. As already pointed out in~\cref{sec:basis_VLQ}, one can choose to go to a weak basis where either $ \overline{m}_d$ or $\overline{M}_d$ are zero blocks. However, imposing certain symmetries may automatically lead to different zero sectors in this matrix, in the basis where the symmetry is imposed. A choice of weak basis does not change the physics whereas symmetries have physical implications.

The matrix ${\cal M}_d  { \cal M}^\dagger_d $ is diagonalized by a unitary matrix $ U_L \equiv \mathcal{V}_L $, and the diagonalization condition can be expressed in the form:
\begin{equation} \label{eq:KRST}
{\cal M}_d  { \cal M}^\dagger_d \ U_L = U_L \begin{pmatrix}
d^2_d & \\[2mm]
 & D^2_d \end{pmatrix}\,, \qquad  \mbox{with } U_L = 
 \begin{pmatrix}
 K & R \\[2mm]
S & T   \end{pmatrix}\,,
\end{equation}
leading to
\begin{subequations}\begin{align}
(m_d\ m_d^\dagger + \overline{m}_d \ \overline{m}_d^\dagger)\ K +
(m_d \ \overline{M}_d^\dagger   + \overline{m}_d\  M_d^\dagger) \ S &= K \ d_d^2 \,,
\label{eq:a} \\
(m_d\ m_d^\dagger + \overline{m}_d \ \overline{m}_d^\dagger)\ R +
(m_d \ \overline{M}_d^\dagger   + \overline{m}_d\  M_d^\dagger) \ T &= R \  D_d^2 \,,
\label{eq:b} \\
(\overline{M}_d\ m_d^\dagger  + M_d \ \overline{m}_d^\dagger )\ K + 
(\overline{M}_d\ \overline{M}_d^\dagger  + M_d \  {M}_d^\dagger) \ S &= 
S \ d_d^2 \,,
\label{eq:c}  \\
(\overline{M}_d\ m_d^\dagger  + M_d \ \overline{m}_d^\dagger )\ R + 
(\overline{M}_d\ \overline{M}_d^\dagger  + M_d \  {M}_d^\dagger) \ T &= 
T \  D_d^2 \,.
\label{eq:d} 
\end{align}\end{subequations}
The above expressions are valid in general for any model with the addition of a single vector-like down quark. Notice that in this case the term ($\overline{M}_d\ \overline{M}_d^\dagger  + M_d \  {M}_d^\dagger)$ is just a number and, as shown in what follows, it is very approximately equal to the square of the mass of the heavy quark $D$. 

Assuming $V$ to be much larger than $v$ we obtain from~\cref{eq:c}, to a very good approximation:
\begin{equation}
S \simeq - 
(\overline{M}_d\ \overline{M}_d^\dagger  + M_d \  {M}_d^\dagger)^{-1}
(\overline{M}_d\ m_d^\dagger  + M_d \ \overline{m}_d^\dagger )\ K  \,.
\label{pri}
\end{equation}
This equation implies that the entries of $S$, in this model, are suppressed by the ratio $v/V$.
From~\cref{eq:d} we find:
\begin{equation}
D_d^2 \simeq (\overline{M}_d\ \overline{M}_d^\dagger  + M_d \  {M}_d^\dagger)\,,
\label{seg}
\end{equation}
implying that the mass of the heavy down quark is of order $V$. From~\cref{eq:b} one can readily conclude that the entries of $R$ are also suppressed by the ratio $v/V$. As $V$ goes to infinity the mass of the heavy quark also goes to infinity while deviations from unitarity of the block $K$ disappear due to the growing suppression of $S$ and $R$. Finally from~\cref{eq:a}, replacing $S$ with the help of~\cref{pri,seg} one obtains:
\begin{equation}
K^{-1} \, \mathfrak{h}_\text{eff}\, K = d_d^2\,, 
\end{equation}
with
\begin{equation}
\mathfrak{h}_\text{eff} = m_d\ m_d^\dagger + \overline{m}_d \ \overline{m}_d^\dagger - 
 (m_d \ \overline{M}_d^\dagger   + \overline{m}_d\  M_d^\dagger)
 D_d^{-2}
(\overline{M}_d\ m_d^\dagger  + M_d \ \overline{m}_d^\dagger )\,.
\label{14}
\end{equation}
It is the block $K$ of the matrix $U_L$ that plays the role of the $V_\text{CKM}$ of the Standard Model by mixing the SM-like quarks, whenever $V \gg v$. This becomes clear when replacing $S$ given by~\cref{pri} into~\cref{eq:a}. This is also clear from the fact that the matrix $R$ in~\cref{eq:KRST} is suppressed by the ratio $v/V$ and therefore the SM-like quarks have very suppressed mixing with the vector-like quarks.

All terms in~\cref{14} are of order $v^2$ which means that all of them may contribute to $\mathfrak{h}_\text{eff}$ on an equal footing. The matrices $m_d$ and $\overline{m}_d$ are both real while $ M_d$ and $ \overline{M}_d$ are complex in the case of spontaneous CP violation. The masses of the SM-like quarks are obtained, to a good approximation, from the diagonalization of $\mathfrak{h}_\text{eff}$. There is no suppression of the last term of~\cref{14} when $V$ goes to infinity. This term is complex and may generate a complex mass matrix at low energies. This matrix remains complex with an unsuppressed phase irrespective of how large the scale $V$ is.

\subsection{Solution of the strong CP problem without axions}
\label{sec:strongCPnoax}

The Standard Model is a remarkable theory, capable of accounting for a variety of processes observed in particle accelerators such as the LHC. Nevertheless, some aspects
cannot be explained by this framework which are commonly accounted as problems:  
the quadratic sensitivity of the Higgs boson mass to high new physics scales (weak scale naturalness problem); 
the hierarchy between the masses of the different fermions (mass hierarchy problem);  
the absence of dark matter candidates (dark matter problem); 
the difference in proportion to the observed matter and antimatter in the Universe (the matter-antimatter asymmetry problem), 
to cite a few. Another issue is related to a more intricate subject, connected to the fact that Nature \emph{does} distinguish between right and left as well as between particle and antiparticle, at least in the energies we have access to. 
This astonishing fact was first observed in 1964, in experiments involving kaon mesons~\cite{Christenson:1964fg}. 
However, only in 1973 an elegant and economical description of the phenomena could be put forward, known as the Kobayashi-Maskawa (KM) mechanism~\cite{Kobayashi:1973fv}. 

In practical terms, it amounts to a complex phase parameter in the mixing matrix that connects the interaction of the $W$ boson, up-type quarks and down-type quarks when the fermion fields are written in the 
mass eigenstate basis.
Rephrased in an equivalent form, the mass matrix of the quarks must be intrinsically complex to comply with the observed CP violation in Nature. Although models that seek to explain the origin of the complexity of these matrices can be written (for instance the one presented on the last section), it does not pose a problem per se. 
The real problem occurs when one realizes the existence of another CP (and P) violating parameter arising from the non-trivial vacuum structure of QCD:
\eq{
\label{L:theta.qcd}
\lag \;\supset \;\theta_{\text{QCD}} \frac{g^{2}_{s}}{64 \pi^{2}}\epsilon^{\mu\nu\alpha\beta}G_{\mu\nu}^{a}G_{\alpha\beta}^{a}\,,
}
with $\epsilon^{0123}=1$.
The parameter $\theta_{\rm QCD}$ itself is not physical because the rephasing of each quark field $q_{jL}\to e^{-i\alpha^L_j}q_{jL}$ and $q_{jR}\to e^{-i\alpha^R_j}q_{jR}$ modifies the path integral measure~\cite{Fujikawa:1979ay} which exactly induces  
\eq{
\lag\to \lag+\left(\sum_{j}\alpha^L_j-\sum_j\alpha^R_j\right)\frac{g^{2}_{s}}{64 \pi^{2}}\epsilon^{\mu\nu\alpha\beta}G_{\mu\nu}^{a}G_{\alpha\beta}^{a}\,.
}
Thus the physical parameter is the reparameterization-invariant combination
\eq{
\bar{\theta} = \theta_{\text{QCD}}- \theta_{\text{weak}}\,,
}
which in the SM involves the quark Yukawa couplings in
\eq{
\theta_{\text{weak}} = 
\arg\det Y_{u} +\arg \det Y_{d}\,. 
}
We use the basis where $Y_d$ accompanies $\bar{q}_{iL}d_{jR}$ and similarly for $Y_u$.
In the presence of VLQs, the previous relation generalizes to
\eq{
\label{theta.weak:gen}
\theta_{\text{weak}}= 
\arg\det \mathcal{M}_u + \arg \det \mathcal{M}_d\,,
}
where $\cM_d,\cM_u$ are the full mass matrices~\eqref{eq:genmass}.

It is the fact that $\bar{\theta}$ is tiny ($\bar{\theta}\lesssim 10^{-10}$~\cite{Pospelov:2005pr,Baker:2006ts,Afach:2015sja,Graner:2016ses}) that constitutes the Strong CP problem, to be added to the set of issues that the Standard Model cannot account for. More interestingly, the Strong CP problem is quite unique in two ways: (1) it cannot be dismissed based on the anthropic principle~\cite{Ubaldi:2008nf,Dine:2018glh} (no appreciable difference would occur in nuclear physics if the present value of $\bar{\theta}$ were changed by several orders of magnitude); (2) no additional symmetry at Lagrangian level is restored when $\bar{\theta}=0$, and yet its value is small.
More detailed discussion of the problem, focusing mainly on the axion solution, as well as original references can be found in various review articles~\cite{Kim:1986ax,Cheng:1987gp,Dine:2000cj,Kim:2008hd,Hook:2018dlk}.

From the point of view of model building, the Strong CP problem can be addressed from three different perspectives: (i) the promotion of $\bar{\theta}$ to a dynamical field (the axion), which couples to the QCD gluon potential being dynamically driven to zero at potential minima~\cite{Peccei:1977hh,Peccei:1977ur}; (ii) the up quark is considered massless, allowing $\bar{\theta}$ to be removed by a global axial symmetry (strongly disfavoured by lattice calculations~\cite{Aoki:2013ldr,Alexandrou:2020bkd}); (iii) CP (or P) is a conserved symmetry at fundamental level, whose observed violation is due to a spontaneous symmetry breaking.
In the latter case, a calculable $\thetabar$ is induced by radiative corrections and its tiny value may be potentially justified~\cite{Georgi:1978xz,Beg:1978mt,Mohapatra:1978fy,Segre:1979dx,Barr:1979as,Nelson:1983zb,Barr:1984qx}. See also~\cite{Barr:1996wx,Hiller:2001qg,Cheung:2007bu,
Schwichtenberg:2018aqc,Egana-Ugrinovic:2018znw, Cherchiglia:2019gll,Mimura:2019yfi,Choi:2019omm,Evans:2020vil,Perez:2020dbw, Valenti:2021rdu,Valenti:2021xjp, Fujikura:2022sot,Girmohanta:2022giy,Bai:2022nat, Perez:2023zin,Camara:2023nyf} for recent proposals.
Following the last approach, the Nelson-Barr mechanism is one of the simplest ways to guarantee $\bar{\theta}=0$ at tree level upon explicit CP conservation~\cite{Nelson:1983zb,Barr:1984qx}. Having in mind the connection to VLQs, we will mainly focus on the last approach in this section.

\subsection{The general Nelson-Barr proposal}

The Nelson-Barr~\cite{Nelson:1983zb,Barr:1984qx} proposal to solve the strong CP problem is based on spontaneously broken CP 
with $\thetabar$  naturally vanishing
at tree level even after the breaking.
The challenge is to keep the radiative corrections small enough~\cite{Dine:2015jga,Perez:2020dbw}.
The transmission of spontaneously broken CP from a CP breaking sector to the SM is accomplished by heavy vector-like quarks that mix with ordinary quarks.
The first model was proposed in Ref.~\cite{Nelson:1983zb} and was based on an $SU(5)$ gauge group.
And soon after it was realized by Barr~\cite{Barr:1984qx,Barr:1984fh} that to guarantee $\thetabar=0$ at tree level it was sufficient that these heavy vector-like quarks obey a set of conditions on their couplings with the SM.

Barr considers the SM with additional heavy VLQs (real representation of SM gauge group) and his conditions~\cite{Barr:1984qx} are the following:
\begin{enumerate}
\item VEVs that break the SM gauge group cannot break CP and they only connect the usual quark fields.
\item VEVs that break CP spontaneously cannot break the SM gauge group and they can only connect SM quark fields with the additional VLQs.
\end{enumerate}
VEVs may be replaced by bare mass terms in condition 2.

There is an implicit assumption that these VLQs need to mix with the SM quarks.
Condition 2 tell us that CP should be broken by scalars that are SM singlets.
Therefore, to couple with these scalars and with the usual SM quark fields $q_L\sim (\bs{2},1/6)$, $d_R\sim (\bs{1},-1/3)$ and $u_R\sim (\bs{1},2/3)$, the possible VLQs that can mix with SM quarks can only have quantum numbers identical to the SM ones as $Q_L,Q_R\sim (\bs{2},1/6)$, $D_L,D_R\sim (\bs{1},-1/3)$ and $U_L,U_R\sim (\bs{1},2/3)$.
So we only have three out of the seven irreps available in~\cref{tab:VLQ:irreps} of~\cref{sec:notation}.

The only scalar representation that couples to ordinary quarks and obeys condition 1 is the usual Higgs doublet $\Phi\sim (\bs{2},1/2)$.%
\footnote{With more than one Higgs doublet, one needs to ensure that their VEVs do not break CP.} 
Condition 1 allows for the usual quark Yukawa couplings
\eq{
\label{lag:NB:sm}
-\lag\supset\bar{q}_L\Phi d_R+\bar{q}_L\tilde{\Phi}u_R+ \text{h.c.} 
\,,
}
but forbids the light-heavy and heavy-heavy couplings
\eq{
\label{NB:mixing:FR_RR}
\bar{q}_L\Phi D_R\,,\,\,
\bar{q}_L\tilde{\Phi}U_R\,,\,\,
\bar{Q}_L\Phi d_R\,,\,\,
\bar{Q}_L\tilde{\Phi}u_R\,,\,\, 
\bar{Q}_L\Phi D_R\,,\,\,
\bar{Q}_L\tilde{\Phi}U_R\,.
}
We write only the operator forms without the couplings.
Denoting by $S$ the scalars in condition 2, we can add to~\eqref{lag:NB:sm} the light-heavy CP breaking mixing terms
\eq{
\label{lag:NB:FR}
\bar{q}_LSQ_R+\bar{D}_L S d_R+\bar{U}_L S u_R +  \text{h.c.} 
\,,
}
and the CP conserving terms
\eq{
\label{lag:NB:FR2}
\bar{Q}_LQ_R+\bar{D}_L D_R+\bar{U}_L U_R +  \text{h.c.} \,.
}
CP breaking heavy-heavy and light-light mixing terms are forbidden by condition 2.

Analysing the charge $-1/3$ mass matrix, conditions 1 and 2 enforce the zeros in
\begin{align}
\label{Barr:M:d}
\begin{blockarray}{cccc}
 & d_R & Q_R & D_R \\ [2.5mm] 
\begin{block}{c(ccc)}
  \bar{q}_L & (\bs{2},1/2) & (\bs{1},0)e^{i\alpha} & 0 \topstrutN  \\[2.5mm] 
  \bar{Q}_L & 0 & (\bs{1},0) & 0   \\[2.5mm] 
  \bar{D}_L & (\bs{1},0)e^{i\alpha} & 0 & (\bs{1},0)  \botstrutN \\[2.5mm] 
\end{block}
\end{blockarray}\,\,.
 \end{align}
The conditions also ensure that CP breaking phases only appear in the positions including the factor $e^{i\alpha}$ coming from the VEV of $S$.

It is clear that the  mass matrix of the quarks with charge $-1/3$ in~\eqref{Barr:M:d} has real determinant, even with complex entries. These complex entries should and indeed can account for the CP violation in the SM.
As a similar consideration applies for the charge $2/3$ sector, $\bar{\theta}=0$ at tree-level.

The position of the zeros, of the real and of the complex entries in the matrix above are crucial to guarantee $\bar{\theta}=0$ at tree-level and still generate CP violation.
The position of the zeros can be enforced by gauge or global symmetries and the conditions above can be embedded in Grand Unified groups~\cite{Nelson:1983zb,Barr:1984qx}.
In the latter case, vector-like representations may arise from real representations of the larger gauge group.
Within the SM gauge group, the simplest choice is the addition of a global $\ZZ_2$ symmetry under which the heavy quarks and the CP breaking scalars are odd. This symmetry can always be imposed, but given the last terms in 
\eqref{NB:mixing:FR_RR} it is only sufficient if singlet and doublet VLQs are not simultaneously present.
Adding only one down-type singlet VLQ leads exactly to the minimal model proposed by Bento, Branco, Parada~\cite{Bento:1991ez} which we will describe in detail in the next section.     
This $\ZZ_2$ was also used recently in~\cite{Cherchiglia:2020kut} to define VLQ of Nelson-Barr type that are electroweak singlets.
Larger symmetries such $\ZZ_n$ or $U(1)$ can be used~\cite{Dine:2015jga} as well.

\subsection{A minimal model: the Bento-Branco-Parada model}
\label{sec:BBP}

The \gls*{BBP} model was introduced in Ref.~\cite{Bento:1991ez}. 
The Lagrangian is chosen to be CP invariant. The model resembles the one described in~\cref{sec:complexCKM}, the field content is the same, but a different $\ZZ_2$ symmetry is imposed in order to naturally suppress strong CP a la Barr and Nelson. Under the new $\ZZ_2$ symmetry all fields of the SM transform trivially and all new fields ($D^0_L$, $D^0_R$, and $S$) are odd. The scalar potential is the same as in the first subsection, leading to the possibility of having spontaneous CP violation.
The new $\ZZ_2$ symmetry allows for a bare mass term for the vector-like quark. Therefore, there are two new mass scales in this scenario, other than the scale of electroweak symmetry breaking, the scale of the VEV of the new scalar singlet and the scale of the bare mass term of the quark singlet. The Yukawa interactions of the quarks are now given by:
\begin{equation}
\label{BBP:yukawa} 
 {\cal L}_Y = - \sqrt{2} (\overline{u^0} \ \overline{d^0})_L^i (g_{ij} \phi \, d^0_{jR} + 
h_{ij}  \tilde{\phi}\, u^0_{jR}) - M_d \overline{D^0_L}\, D^0_R - \sqrt{2} (f_i \, S + f^\prime_i \, S^* ) \, \overline{D^0_L} d^0_{iR} + \text{h.c.}\,,
\end{equation}
with ($i,j = 1, 2, 3$) and the down quark mass matrix is now of the form
\begin{equation}
{ \cal M}_d = \begin{pmatrix}
m_d & 0 \\[2mm]
\overline{M}_d & M_d
\end{pmatrix}
\end{equation}
in the weak basis where the symmetry is imposed. The only contribution to $M_d$ is the bare mass term, which is real, since the $\ZZ_2$ symmetry forbids the coupling $ \overline{D_L} D_R S$ as well as $ \overline{D_L} D_R S^*$. 

\vskip 2mm

From~\cref{eq:a,eq:c,eq:d} we now obtain:
\begin{equation}
S \simeq - \frac{1}{D_d^2}(\overline{M}_d\ m_d^\dagger)\ K \,,
\label{newS}
\end{equation}
with 
\begin{equation}
D_d^2 \simeq (\overline{M}_d\ \overline{M}_d^\dagger  + M_d^2)\,,
\label{newD}
\end{equation}
where here $(\overline{M}_{d})_j = f_j \, V e^{i\beta} + 
f^\prime_j \,V e^{-i\beta}$
and 
\begin{equation}
\mathfrak{h}_\text{eff} =  m_d\ m_d^\dagger - 
\frac{1}{D_d^2} (m_d \ \overline{M}_d^\dagger )
(\overline{M}_d\ m_d^\dagger)\,.
\label{new14}
\end{equation}
Here, $\mathfrak{h}_\text{eff}$ stands for a particular case of $h_{\rm eff}$ defined in~\cref{eq:heff,eq:explicitheff}, with $\overline{m}=0$.
These expressions show that the square of the mass of the heavy quark (given by $D_d^2$) is the sum of a term proportional to $V^2$, where $V$ is the scale of the VEV of the scalar singlet, and another term proportional to the square of the bare mass term. As a result, this mass grows with both scales. From the matrix $S$ given by~\cref{newS}, and analogously from $R$, one sees that
the suppression of deviations from unitarity of $V_\text{CKM}$ occurs irrespective of which one of these scales dominates. However, as for the generation of a complex phase in $V_\text{CKM}$ from spontaneous CP violation, there is no such suppression, provided that the scale of the bare mass term of the quark singlet does not dominate over the scale of the VEV of the scalar singlet. Since both these scales contribute to the mass of the heavy quark its mass may go to infinity, while at the same time CP violation at low energies is not suppressed. \\

The parameter $\overline{\theta}$ associated to the strong CP violation is, as explained in~\cref{sec:strongCPnoax}, the sum of two components $\overline{\theta} = \theta_\text{QCD} - \theta_{\text{weak}} $ where $\theta_{\text{weak}}= \arg(\det { \cal M}_d \ \times \det {m_u}) $ in this case (see~\cref{theta.weak:gen} for the general expression). In the present framework, since CP is a symmetry imposed on the Lagrangian, on the one hand we have $\theta_\text{QCD}=0$. On the other hand, the zeros in the $3 \times 1$ block of ${ \cal M}_d $, which are a consequence of the $\ZZ_2$ symmetry, guarantee that the determinant of ${ \cal M}_d $ is real. The determinant of $m_u$ is also real as a result of imposing CP conservation at the Lagrangian level. 
In Ref.~\cite{Bento:1991ez} the higher order corrections to $\overline{\theta}$ are evaluated. Since $\theta_{\text{weak}}$ is zero at tree level, as a result of a symmetry, higher order corrections to $\overline{\theta}$ are finite and calculable~\cite{Georgi:1974yw}. In Ref.~\cite{Bento:1991ez} it was shown that it is possible to comply with the limit on the electrical dipole moment without the need to impose very small Yukawa couplings. This is an elegant feature of this framework even though having very small Yukawa couplings would be natural in the sense of 't Hooft~\cite{tHooft:1979rat}, since it would increase the symmetry of the Lagrangian. At one loop level the contributions to $\overline{\theta}$ are suppressed both by Yukawa couplings and the ratio $v/V$ which suggests a connection between the strong CP problem and the hierarchy problem. We are thus led to a solution of the strong CP problem~\cite{tHooft:1976rip,tHooft:1976snw} of the type proposed by Nelson~\cite{Nelson:1983zb} and Barr~\cite{Barr:1984qx}. \\

Obviously this scenario may also work with simple variations, in particular by introducing additional down vector-like quarks or by replacing the vector-like down isosinglet quark by a vector-like up isosinglet quark.

\subsection{Implications of vector-like quarks of Nelson-Barr type}
\label{sec:NB_VLQ}
\providecommand{\cM}{\mathscr{M}}
\providecommand{\cY}{\mathscr{Y}}

We have seen in~\cref{sec:BBP} the minimal implementation of the NB scheme within the SM: the BBP model.
This model only required the addition of one down-type VLQ $D_L,D_R$ and one complex singlet scalar $S$ to the SM field content.
The complex VEV of $S$, breaking CP spontaneously, then induced the complex mass mixing connecting the heavy $D$ quark with the SM down-type quarks, which in turn was the only source of CKM CP violation.
This feature tells us that, once the CP breaking scale is fixed, the VLQ mass cannot be arbitrarily large, otherwise the necessary CP violation cannot be transmitted to the SM.

The BBP model illustrates a generic feature of the NB scheme which generically involves the presence of two scales: the scale 
$\Lambda_{\rm CP}$ of spontaneous CP breaking ($V$ in the BBP model) and the scale $M_{\rm VLQ}$ of VLQs transmitting the CP breaking to the SM.
Even though the model enforces $\thetabar$ to vanish at tree-level, one-loop corrections are generically generated and these must be small.
To achieve this purpose, both the Yukawa coupling of the CP breaking scalar connecting the SM quarks with the VLQs as well as the mixing between the SM Higgs with the CP breaking scalars must be suppressed. The latter will require a high scale $\Lambda_{\rm CP}$, which implies that the CP breaking scalars are 
expected to be out of experimental reach but, in contrast, the VLQs may be at TeV scale.

The last consideration leads us to consider scenarios where only the VLQs in the NB scheme have low enough mass to be probed.
This motivates us to define VLQs of \emph{Nelson-Barr type}~\cite{Cherchiglia:2020kut}, NB-VLQs for short, as VLQs that are responsible for transmitting the CP breaking to the SM in the NB scheme.
More concretely, focusing on the case of down-type VLQs, we define singlet NB-VLQ of down-type as 
the ones where the Yukawa couplings in~\eqref{eq:lagrangian} and the bare mass terms in~\eqref{eq:baremasses}
comes from a structure of the form%
\footnote{In Ref.~\cite{Cherchiglia:2020kut}, the VLQ is denoted by $B_L,B_R$ instead of $D_L,D_R$.}
\eqali{
\label{yuk:NB}
-\lag&=\bar{Q}^0_{Li}(\cY_d)_{ij} \Phi d^0_{Rj}+\bar{Q}^0_{Li}(\cY_u)_{ij} \tilde{\Phi}u^0_{Rj}
\cr
&\quad +
\bar{D}^0_{Lr}(\cM_{Dd})_{rj} d^0_{Rj}+\bar{D}^0_{Lr}(\cM_D)_{rs} D^0_{Rs}+ \text{h.c.}\,,
}
with the additional requirement that $\cY^u,\cY^d$ are \textit{real} $3\times 3$ matrices, $\cM_D$ is a real mass matrix and only $\cM_{Dd}$ is complex. In the BBP model, the latter is induced by the CP breaking VEV. 
Note that in the notation of~\cref{eq:lagrangian,eq:baremasses} we have
\eqali{
\label{notation:NB}
Y_d&=\cY_d\,,\quad
Y_u=\cY_u\,,\quad
\oYd=0\,,\quad
\oYu=0\,,
\cr
\oM_d&=\cM_{Dd}\,,\quad
M_d=\cM_D\,,\quad
\oM_u=0\,,\quad
M_u=0\,.
}
A similar definition applies to up-type NB-VLQs.

Similarly to the BBP model, the structure in~\eqref{yuk:NB} follows from CP conservation and a $\ZZ_2$ symmetry%
\footnote{A larger $\ZZ_n$ or $U(1)$~\cite{Dine:2015jga}, a non-Abelian global or gauge symmetry~\cite{Nelson:1983zb,Barr:1984qx} can be also used.
Note that the definition based on $\ZZ_2$ is not sufficient when singlet and doublet VLQs are simultaneously present.}
under which only $D_{L,R}$ are odd, and only $\cM_{Dd}$ breaks CP and $\ZZ_2$ softly (spontaneously) satisfying the Barr conditions that guarantees $\bar{\theta}=0$ at tree-level.
In this way, there is no need to detail the CP breaking sector and all the effects concerning CP breaking will be encoded in the soft breaking mass term $\cM_{Dd}$.

In the following, we will identify few more generic features of these NB-VLQs responsible for transmitting CP violation to the SM in the NB scheme.
The specific structure of real and complex couplings in~\eqref{yuk:NB} contrasts with the case of \emph{generic} VLQs, unrelated to the strong CP problem and the origin of CP violation, where all couplings may be present and may be complex.
Following the recent works~\cite{Cherchiglia:2020kut,Cherchiglia:2021vhe}, we will see that the NB case in~\eqref{yuk:NB} depends on \emph{one less parameter} than in the generic case and that, for the case of a single down-type (or up-type) NB-VLQ,  
the mixing of the heavy VLQ with the SM up-type quarks will roughly follow the same hierarchy as the CKM $b$-quark mixing:
\eq{
\label{ViB:3f}
|V_{uB}|:|V_{cB}|:|V_{tB}|\,\sim\, |V_{ub}|:|V_{cb}|:|V_{tb}|\,\simeq\, 0.004:0.04:1\,.
}
This hierarchy largely renders the model flavour safe as the most restrictive flavour constraints of flavour changing among the first and second families are naturally suppressed.

\subsubsection{One parameter less}

The Lagrangian~\eqref{yuk:NB} contains in $\cY^u,\cY^d,\cM_{D},\cM_{Dd}$ the number of parameters
\eq{
\label{params:NB:gen}
N_{\rm general}=2\times 9 + n_d^2 + 6n_d\,.
}
We initially consider only the case $n_d>0$ with $n_u=0$. We add up-type singlets in the end.
Due to the (softly broken) CP symmetry, the reparameterization transformations in 
\eq{
SO(3)_{q_L}\otimes SO(3)_{d_R}\otimes SO(3)_{u_R}\otimes SO(n_d)_{D_L}\otimes SO(n_d)_{D_R}\,,
}
reduces the number of parameters by
\eq{
3\times 3+ n_d(n_d-1)\,,
\label{params:NB:SO}
}
still leaving the couplings as well as the original CP symmetry with the same structure.
At last, in the basis where $\cM_D$ is diagonal, one can still use a common rephasing in $D_{Lr},D_{Rr}$ to remove $n_d$ phases in $\cM_{Dd}$ and we arrive at~\cite{Cherchiglia:2020kut}
\eq{
N_{\rm phys}=9+6n_d\,.
}
If we add up-type singlet NB-VLQs, we arrive similarly at
\eq{
\label{params:nb-vlq}
N_{\rm phys}=9+6n_d+6n_u\,.
}

We conclude that singlet NB-VLQs are described by \emph{one parameter less} if compared to the case of generic VLQs, see~\cref{sec:par_counting}. For $n_d=n_u=0$ we obtain the 9 parameters of unrealistic CP conserving SM compared to the actual 10. For $n_u=0$ and $n_d=1$, we obtain 15 parameters compared to the case of a generic down-type VLQ. 
We emphasize this information in~\cref{tab:param:VLQ=1}.
For each additional NB-VLQ we gain 6 more parameters.
\begin{table}[t!]
  \centering
  \begin{tabular}{lcc}
    \toprule   
  & \# of parameters & \# of CP odd
\\
\midrule
SM          & 10                &1\\[1mm]
One generic VLQ  & 16                 &3\\[1mm]
One NB-VLQ       & 15                 &1\\
\bottomrule
  \end{tabular}
  \caption{Number of parameters in the flavour sector of the SM and with the addition of one VLQ.}
  \label{tab:param:VLQ=1}
\end{table}
%

We can also specify how many of the parameters in~\eqref{params:nb-vlq} are CP odd.
The comparison to the CP conserving limit is non-trivial~\cite{Cherchiglia:2020kut}.
In that limit, the number of parameters should match the generic case in~\eqref{eq:fsp-counting} disregarding the imaginary parts, i.e.~$9+4(n_d+n_u)$ parameters.%
\footnote{We obtain the same number of parameters by considering real $\cM_{Dd}$ in~\eqref{yuk:NB} and, analogously, real couplings for up-type NB-VLQs.}
Naively, we would think that the remaining $2(n_d+n_u)$ parameters would correspond to the number of CP odd parameters.
That is not the case as one among of these $2n_d$ parameters (and one among $2n_u$ parameters) is \emph{CP even} and becomes \emph{unphysical} in the CP conserving limit.
We can explain this property in an explicit basis where the first row of $\cM_{Dd}$ is $(\cM_{Dd})_{1i}\sim (x_1+iy_1,x_2,0)$ after using real orthogonal transformations on $d_{iR}$. From rephasing of $D_{1L}$ we would expect two imaginary entries but here we have only one. Moreover, in the limit $y_1\to 0$, we can rotate $x_2$ away and a remaining $SO(2)$ freedom in the $(d_{2R},d_{3R})$ space eliminates one CP even parameter.
So for only one NB-VLQ, e.g.~$n_u=0$ and $n_d=1$, we have \emph{only one} CP phase controlling all CP violation.
In contrast, the generic VLQ case depends on three CP violating parameters~\cite{Branco:1986my}.
The counting for this case is given in~\cref{tab:param:VLQ=1}.

As one parameter less is needed to describe NB-VLQs as compared to generic VLQs, the various parameters cannot be independent and correlations are expected to appear~\cite{Cherchiglia:2020kut}.

\subsubsection{Flavour alignment}

Let us first discuss how the requirement for reproducing the CKM mixing in the SM, including its CP violation phase, imposes non-trivial constraints on the couplings of the VLQs with the SM. 
The requirement comes from the effective down-type $3\times 3$ mass matrix squared in~\eqref{new14}, which in the notation of~\eqref{notation:NB}, becomes
\eq{
\label{Heff:NB}
\mathfrak{h}_{\rm eff}=\frac{v^2}{2}\cY_d\left[\id_3-\cM_{Dd}^\dag H_D^{-1}\cM_{Dd}\right]\cY_d^T\,,
}
where
\eq{
\label{HD}
H_D\equiv\cM_{Dd}{\cM_{Dd}}^\dag+\cM_D\cM_D^T\,,
}
is the VLQ mass squared ($D_d^2$) or mass matrix for more families of down-type VLQs.
A non-trivial requirement appears because the effective mass matrix~\eqref{Heff:NB} should reproduce the SM structure
\eq{
\label{Heff:NB=SM}
\cY_d\left[\id_3-\cM_{Dd}^\dag H_D^{-1}\cM_{Dd}\right]\cY_d^T=V_\text{CKM}^\dag\diag(y_d^2,y_s^2,y_b^2)V_\text{CKM}\,,
}
where $y_d,y_s,y_b$ are the SM down-type Yukawa couplings.
We are assuming the basis where the up-type Yukawa is diagonal, hence the appearance of the CKM matrix $V_\text{CKM}$.%
\footnote{Two more physical phases might be present in front of $V_\text{CKM}$; see Ref.~\cite{Cherchiglia:2020kut} for details.}

It is clear that the bare mass term $\cM_D$ in~\eqref{HD} cannot be arbitrarily larger than the CP violating contribution $\cM_{Dd}$ because the CP violating contribution would vanish in~\eqref{Heff:NB=SM} and the CKM phase would not be reproduced.
In contrast, in the opposite limit of $\cM_{Dd}$ arbitrarily larger than $\cM_D$, the effective mass matrix squared~\eqref{Heff:NB=SM} would gain one or more zero eigenvalues (depending on $n_d$) and the SM Yukawa is not reproduced as well.

To get more quantitative, it is helpful to compare two bases for the size $(3+n_d)$ mass matrix after EWSB:
\eq{
\label{mass.matrix}
\text{NB:}\quad\cM^{d+D}=\mtrx{\frac{v}{\sqrt{2}}\cY_d & 0\cr \cM_{Dd}  & \cM_D}
\,,\quad
\text{generic:}\quad M^{d+D}=\mtrx{\frac{v}{\sqrt{2}}Y_d & \frac{v}{\sqrt{2}}Y_D\cr 0 & M_D}\,.
}
The first form is obtained from the NB structure while the second is generic and any theory with down-type singlet VLQs can be cast in this form, cf.~\cref{sec:wbzero}.
Moreover, the form $M^{d+D}$ has a simple interpretation in the leading seesaw approximation:%
\footnote{In this approximation, the block-diagonalized form of $M^{d+D}$ is simply given by neglecting the $Y_D$ term off the diagonal. At this order, $Y_D$ will only contribute to the VLQ mixing with the SM but not to the masses.}
$M_D$ is the VLQ mass while $Y_d$ is the SM down-type Yukawa matrix which contains the CKM CP violation.

We can connect the two bases in~\eqref{mass.matrix} through a unitary transformation in the space $(d_R,D_R)$. An explicit analytical formula can be written if necessary.
For our purposes, we only need the exact relations
\subeqali{
\label{MB}
M_DM_D^\dag&=H_D=\cM_{Dd}\cM_{Dd}^\dag+\cM_D\cM_D^T\,,
\\
\label{YB:NB}
Y_D &=\cY_d w\,,
\\
\label{Yd:NB}
Y_dY_d^\dag &=\cY_d\left(\id_3-ww^\dag\right)\cY_d^T\,.
}
where we have defined
\eq{
\label{def:w}
w\equiv \cM_{Dd}^\dag {M_D}^{\dag -1}\,.
}

Considering $Y_d$ as known, we need to invert~\cref{Yd:NB} and solve for $\cY_d$.
This inversion process is non-trivial and the explicit formula was found in Ref.~\cite{Cherchiglia:2020kut} for $n_d=1$.
It reads
\eq{
\label{cYd:sol}
\cY_d=\left(\re(Y_dY_d^\dag)\right)^{1/2}\mathcal{O}
\mtrx{1&&\cr &\frac{1}{\sqrt{1-b^2}}&\cr &&\frac{1}{\sqrt{1-a^2}}}
\,,
}
where $a,b$ comes from
\eq{
w=(0,ib,a)^T\,,
}
after a basis choice. The matrix $\mathcal{O}$ is a real orthogonal matrix in the (23) block.

We note immediately that the first term in~\eqref{cYd:sol} is completely determined by $Y_d$ of the SM and it is thus \emph{hierarchical}, although the overall scale is not fixed.
This hierarchy is then inherited by the Yukawa coupling $Y_D$, cf.~\eqref{YB:NB}, connecting the VLQ with the SM quarks and the Higgs.
The hierarchy~\eqref{ViB:3f} in the mixing is a reflection of the hierarchy of $Y_D$.
Then, considering the VLQs are constrained by direct searches to be above one TeV~\cite{ATLAS:2018ziw,CMS:2020ttz} (see also~\cref{sec:searches}), the presence of one NB-VLQ of down type is largely flavour safe.%
\footnote{We should stress that the hierarchy~\eqref{ViB:3f} following from~\eqref{cYd:sol} arises for typical points. 
As shown in Ref.~\cite{Cherchiglia:2021syq}, by appropriately choosing certain parameters, one can decouple the heavy VLQ from one or even two families of SM quarks, thus leading to a very different pattern.}
The constraints on $(\tY_D\tY_D^\dag)_{ij}$, where $\tY_D=(V^\dag Y_D)$, were shown in~\cref{sec:ph-fits}.
A brief discussion on the case of up-type NB-VLQs can be found in Ref.~\cite{Cherchiglia:2021vhe}. The parameterization presented can also be extended to the case of two NB-VLQs of up- or down-type~\cite{Alves:2023cmw}.

\subsection{Solving the strong CP problem with non-conventional CP}

In~\cref{sec:BBP}, a minimal model (BBP) that solves the strong CP problem with VLQs was presented. The SM particle content was enlarged only by one singlet VLQ with $Q=-1/3$ and one complex singlet scalar, while only a $\ZZ_2$ symmetry was added on top of the symmetries of the SM. 
By construction, $\btheta$ is null at tree-level as the determinant of the tree level Yukawa $Y_d$ is real.
However, if present, higher order corrections $\delta Y_d$ contribute as
\eq{
\delta\btheta=-\im\tr[Y_d^{-1}\delta Y_d]\,.
}

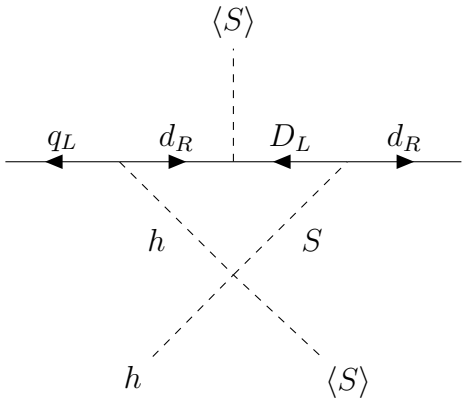
\begin{figure}[!t]
\begin{center}
\begin{tikzpicture}
\begin{feynman}
\vertex (a);
\vertex [right=of a] (b);
\vertex [right=of b] (c);
\vertex [right=of c] (d);
\vertex [right=of d] (e);
\vertex [above=of c] (f) {$\langle S\rangle$};
\vertex [below=of c] (g);
\vertex [below right=of g] (i) {$\langle S\rangle$};
\vertex [below left=of g] (h) {$h$};
\diagram* {
(a) -- [anti fermion, edge label=$q_{L}$ ] (b) -- [fermion, edge label=$d_{R}$] (c) -- [anti fermion, edge label=$D_{L}$] (d) -- [fermion, edge label=$d_{R}$] (e),
(c) -- [scalar] (f),
(b) -- [scalar, edge label'=$h$] (g) -- [scalar, edge label'=$S$] (d),
(h) -- [scalar] (g),
(i) -- [scalar] (g),
};
\end{feynman}
\end{tikzpicture}
\caption{One-loop threshold correction to $Y_d$ in the BBP model.}
\label{fig:dine}
\end{center}
\end{figure}

In the BBP model, a threshold correction $\delta Y_d$ arises already at one-loop from the diagram depicted in~\Cref{fig:dine}, giving
\eqali{
\label{BBP:theta}
\delta\btheta \sim \frac{1}{16\pi^2\Lambda^2_{CP}}\aver{S_a}\im\tr[F^{(a)}{}^\dag F^{(b)}]\lambda^\Phi_{bc}\aver{S_c}\,,
}
where $\Lambda_{CP}$ is the CP breaking scale and there is an implicit summation over indices.
We see the contribution depends on the Yukawa interactions $F^{(a)}_j$ of the VLQs with the SM and the Higgs portal couplings $\lambda^\Phi_{bc}$.
We are separating $\sqrt{2}S=S_1+iS_2$ and writing the relevant interactions as
\eq{
\label{BBP:portals}
-\lag\subset \bar{D}_L(F^{(a)}S_a)d_R + \ums{2}\lambda^\Phi_{ab}S_aS_b|\Phi|^2\,.
}
For example, compared to~\eqref{BBP:yukawa}, $F^{(1)}_j=f_j+f_j'$ is real and $F^{(2)}_j=i(f_j-f_j')$ is purely imaginary.
The $\ZZ_2$ and CP symmetries act on $S_a$ as two separate $\ZZ_2$ symmetries.
Therefore, the only non-vanishing Higgs portal couplings are $\lambda^\Phi_{11}$ and $\lambda^\Phi_{22}$.
As only $a\neq b$ contributes,~\cref{BBP:theta} will be proportional to the difference $\lambda^\Phi_{22}-\lambda^\Phi_{11}$.
A similar contribution also appears proportional to the mass mixing between $S_1$ and $S_2$.
As can be seen from the expression~\eqref{BBP:theta}, to have a value of $\btheta$ compatible with experimental constraints of $\mathcal{O}(10^{-10})$, very suppressed portal couplings ($F$ or $\lambda^\Phi$) are required.

In order to avoid one-loop threshold corrections to $\btheta$, one possibility is to modify the symmetry content of the BBP model.
In Ref.~\cite{Cherchiglia:2019gll}, the choice was to assume an order four CP symmetry known as CP4 symmetry~\cite{Ivanov:2015mwl} under which only the BSM fields transform non-conventionally. To realize this choice, two singlet VLQs $D_{1}, D_{2}$ are required while one can still consider one complex scalar singlet $S$. 
Like in the three-Higgs-doublet model (3HDM) context where it was discovered~\cite{Ivanov:2015mwl}, the CP4 symmetry allows complex parameters without signalling CP violation. Here the Yukawa couplings are complex and we can read them from
\eqref{BBP:portals} as $F^{(a)}_{rj}$ where $r=1,2$ denote $B_{1L},B_{2L}$; both $F^{(1)}$ and $F^{(2)}$ are complex.
However, as the CP4 symmetry acts on the CP breaking scalar as 
\eq{
S\to -iS\,,
}
the Higgs portal interaction goes with $|S|^2$ implying the portal couplings obey $\lambda^\Phi_{22}=\lambda^\Phi_{11}$.
The symmetry also guarantees $\aver{S_2}=0$. The result is that the one-loop correction to $\btheta$ given by~\eqref{BBP:theta} vanishes. The contribution involving the $S_1$-$S_2$ mixing also vanishes because this mixing is absent at tree-level. It only appears at one-loop and then the first contribution to $\btheta$ arises only at 2-loops. So the necessary suppression on the portal couplings to suppress $\btheta$ is less severe than in the original BBP model. The full calculation showing the vanishing of the one-loop contribution can be seen in Ref.~\cite{Cherchiglia:2019gll}.
 
It is important to comment that the one-loop contributions we discussed here refer to \emph{reducible} contributions to $\btheta$ that can be made arbitrarily small as we suppress the VLQ Yukawa couplings.
In contrast, there are \emph{irreducible} contributions that cannot be made small as they depend on the couplings that transmit the CP breaking to the SM. These contributions arise first at 3-loops when the NB-VLQs are weak singlets~\cite{Valenti:2021rdu} and at 2-loops when they are weak doublets~\cite{Vecchi:2014hpa}.%
\footnote{We are considering only non-decoupling contributions.}
The latter already tends to give too large contributions to $\btheta$.
There are also naturalness issues regarding the quality of the CP conserving couplings against higher order CP breaking effects~\cite{Dine:2015jga}. It is possible to build models addressing that, also tying the scale of the CP conserving bare mass of the VLQs with the CP breaking mixing term~\cite{Valenti:2021xjp}.

\subsection{A common origin for all CP violations}

In this section we describe a framework~\cite{Branco:2003rt} in which all manifestations of CP violation would have a common origin, starting from a CP conserving Lagrangian and having CP violation generated spontaneously. In order to generate CP spontaneously the scalar sector of the Standard Model must be extended. The framework described here is an extension to the leptonic sector of the Bento-Branco-Parada model~\cite{Bento:1991ez}. This model deals with two of four aspects of CP violation: CP violation in quark sector and a possible solution to the strong CP problem. The other two additional aspects considered in this extension are: accounting for the possibility of having CP violation at low energies in the leptonic sector and also leptonic CP violation at high energies thus opening the door for the possibility of leptogenesis~\cite{Fukugita:1986hr} as the mechanism generating the observed baryon asymmetry of the Universe (BAU).
CP violation in the leptonic sector has not yet been observed, however the fact that it has already been observed in other sectors of the Lagrangian makes is very plausible and natural for CP to be violated in this sector.

In the leptonic sector in the context of seesaw~\cite{Minkowski:1977sc,Yanagida:1979as,GellMann:1980vs,Glashow:1979nm,Mohapatra:1979ia}, with the introduction of three right handed neutrinos, which are singlets of all gauge interactions, one obtains the following mass terms after spontaneous symmetry breakdown:
\begin{equation}
\begin{aligned}
{\cal L}_m  &= - \overline{{\nu}_{L}^0} m \nu_{R}^0 +
\frac{1}{2} \nu_{R}^{0T} C^\dagger M \nu_{R}^0 -
\overline{l_L^0} m_l l_R^0  + 
{\rm h.c.} \\
&= \frac{1}{2} n_{L}^{T} C^\dagger {\cal M}^* n_L -
\overline{l_L^0} m_l l_R^0  + {\rm h.c.}\,.
\label{eq:lm}
\end{aligned}
\end{equation}
The extension of the Bento-Branco-Parada model to the leptonic sector requires that the transformation of the leptonic fields under the additional $\ZZ_2$ symmetry be defined. The
fields that are not invariant under $\ZZ_2$ transform as:
\begin{equation}
\begin{aligned}
D^0 &\rightarrow -D^0\,, \quad S \rightarrow -S\,,  \\
{\psi_L^0} &\rightarrow i \psi_L^0\,, \quad
e_R^0 \rightarrow i e_R^0\,,\quad   
\nu_{R}^0 \rightarrow i \nu_{R}^0\,,
\label{eq:zds}
\end{aligned}
\end{equation}
where $\psi_L^0$ denotes the left handed lepton doublets, $e_R^0$ and $\nu_{R}^0$ are 
the right handed charged lepton and neutrino singlets. The transformations in the first 
row of this equation are the ones given in~\cref{sec:BBP}. The initial $\ZZ_2$ symmetry  
is thus promoted to a $\ZZ_4$ symmetry with the extension of this model to the leptonic sector.
In the quark sector of the model all SM fields are invariant under the initial $\ZZ_2$ symmetry
and this remains for the new $\ZZ_4$ symmetry, whereas in the leptonic sector the SM fields are
not invariant under the symmetry. The Yukawa terms for the quark sector are those given in~\cref{BBP:yukawa}. For the leptonic sector we have:
\begin{equation}
{\cal L}_l = -{\overline {\psi_L^0}} G_l \phi \ e_R^0 -
{\overline {\psi_L^0}} G_{\nu}  \tilde{\phi } \ \nu _R^0 +
\frac{1}{2} {\nu} _R^{0T} C^\dagger ({f_\nu} S + 
+ {f_\nu}^{\prime} S^\ast )\nu _R^0 +  \text{h.c.}\,.   
\label{eq:ll}
\end{equation} 
All new coefficients $G_l$ and $G_\nu$ are taken to be real. The $\ZZ_4$ symmetry 
prevents the existence of bare Majorana mass terms for the neutrinos, which otherwise 
would imply the appearance of an additional mass scale, however terms of this form are 
generated after spontaneous symmetry breakdown through the coupling to the scalar 
singlet $S$. In the quark sector a bare mass term of the form 
$ M_d {\overline {D_L^0}} D_R^0$ is allowed by 
the symmetry. The VEVs of $\phi$ and $S$ are given by~\cref{eq:vevphiS}. 
The leptonic mass matrices of~\cref{eq:lm} are given by:
\begin{equation}
\begin{aligned}
{\cal M} &= \begin{pmatrix}
0 & m \\[2mm]
m^T & M
\end{pmatrix}\,, \ \ m_l = \frac{v}{\sqrt 2} G_l\,, 
\ \  m =  \frac{v}{\sqrt 2} G_{\nu} \,, \\[2mm]
M &=  \frac{V}{\sqrt 2}( f_{\nu}^+ \cos (\alpha) +
i f_{\nu}^- \sin (\alpha) ) \,,
\label{eq:mmm}
\end{aligned}
\end{equation}
with $f_{\pm}^{\nu} \equiv f_{\nu}  \pm  
{f_{\nu} }^{\prime}$. 
In the weak basis where $m_l$ is to chosen to be real and diagonal the light neutrino masses 
$d_{\nu}$ and the low energy leptonic mixing, 
$U_\text{PMNS}$, are obtained to an excellent approximation by:
\begin{equation}
-K^\dagger m \frac{1}{M} m^T K^* =d_{\nu}\,, 
\end{equation}
where $U_\text{PMNS}$ can be identified to $K$. The matrix $m$ is real while $M$ is a generic complex matrix therefore in general $K$ will contain three CP violating phases, after eliminating three factorizable phases, one of Dirac type and two Majorana phases. In the seesaw framework the heavy neutrino masses are very approximately given by the eigenvalues of the matrix $M$. It is always possible to choose a weak basis in which both $m_l$ and $M$ are chosen to be real and diagonal at the same time. In this weak basis the decay of the heavy Majorana neutrino $N^j$ into charged leptons $l_i^\pm$ ($i$ = e, $\mu$, $\tau$) generates a lepton-number asymmetry given by~\cite{Flanz:1994yx,Plumacher:1996kc,Covi:1996wh,Buchmuller:1997yu}:
\begin{equation}
A^j = \frac{g^2}{{M_W}^2} \sum_{k \ne j} 
{\rm Im} \left((m^\dagger m)_{jk} (m^\dagger m)_{jk} \right)
\frac{1}{16 \pi} 
\left( I(x_k)+ \frac{\sqrt{x_k}}{1-x_k} \right)
\frac{1}{(m^\dagger m)_{jj}}\,,
\label{eq:rmy}
\end{equation}
where $x_k$ is defined as $x_k=\frac{{M_k}^2}{{M_j}^2}$, $M_k$ are the heavy neutrino masses, and 
$ I(x_k)=\sqrt{x_k} \left(1+(1+x_k) \log(\frac{x_k}{1+x_k}) \right)$. These expressions apply to the case of unflavoured leptogenesis. In this framework the lepton-number asymmetry is only sensitive to the CP-violating phases appearing in $m^\dagger m$. Notice that although the matrix $m$ in~\cref{eq:mmm} is real, once we change to the weak basis where the matrix $M$ is diagonal real and positive the phases appearing in $M$ are shifted to the matrix $m$. In general one can thus generate the CP violation required by leptogenesis~\cite{Branco:2003rt}.

\vfill
\clearpage

\section{Production and Decays of VLQs}
\label{sec:searches}

Heavy VLQs may be produced in pairs (pair production), via QCD interactions, or solitarily (single production), via electroweak interactions. Their specific production mechanism and subsequent decay modes at high energy are hallmark features of the different VLQ models.

Most experimental analyses assume that VLQs decay only or mainly to third-generation quarks. Under this assumption, detailed studies have been carried out in earlier works, with a specific focus on the pair production \cite{Aguilar-Saavedra:2009xmz} and single production \cite{Aguilar-Saavedra:2013qpa} of VLQs. Both of these works highlighted the relation between the decay widths, electroweak observables, and the production modes of VLQs, taking into considering different types of VLQs,
such as singlets, doublets, and triplets of $SU(2)_L$.

The decay widths of isosinglet VLQs, the focus of this review, via the $Z$, the $h$ and the $W$ boson respectively, may be expressed, at tree-level, in the following manner,%
\footnote{
Radiative decays of VLQ singlets with a photon or a gluon in the final state are discussed, e.g., in Refs.~\cite{Cacciapaglia:2010vn,Balaji:2021lpr}.
}
\begin{align}\label{eq:hd-fullwidths}
    \Gamma\big(Q_i\rightarrow Z\psi^q_j\big)&=  \frac{G_F M^3_i}{16\pi\sqrt{2}}\lambda[r_Z,r_j] \Big[\big(1-r^2_j\big)^2+r^2_Z\big(1+r^2_j\big)-2r^4_Z\Big]\big|F^q_{ji}\big|^2\,,\\
    \Gamma\big(Q_i\rightarrow h\psi^q_j\big)&=  \frac{G_F M^3_i}{16\pi\sqrt{2}}\lambda[r_h,r_j] \Big[\big(1+r^2_j\big)^2+4r^2_j-r^2_h\big(1+r^2_j\big)\Big]\big|F^q_{ji}\big|^2\,,\\
    \Gamma\big(U_i\rightarrow W^+\psi^d_j\big)&= \frac{G_F M^3_i}{8\pi\sqrt{2}}\lambda[r_W,r_j] \Big[\big(1-r^2_j\big)^2+r^2_W\big(1+r^2_j\big)-2r^4_W\Big]\big|V_{ij}\big|^2\,,\\
    \Gamma\big(D_i\rightarrow W^-\psi^u_j\big)&= \frac{G_F M^3_i}{8\pi\sqrt{2}}\lambda[r_W,r_j] \Big[\big(1-r^2_j\big)^2+r^2_W\big(1+r^2_j\big)-2r^4_W\Big]\big|V_{ji}\big|^2\,,
\end{align}
where the vector $\psi^q$ contains the SM and heavy quarks, as defined after~\cref{eq:hvlq-lagrangian}, with $\psi^q_j$ being its $j$-th entry.
In the ratio $r_j=m_j/M_i$, the $m_j$ is the mass of the final-state quark and $M_i$ the mass of the VLQ --- the decaying particle. The function $\lambda[r,r']$ was given in~\cref{eq:lambda-function}, and for the Fermi constant $G_F$ one has, at tree level, $G_F=\frac{\sqrt{2}g^2}{8m_W^2}$. 
The quark mixing matrix $V$ was defined in~\cref{eq:W-coupling}, while the matrices $F^u = V V^\dagger$ and $F^d = V^\dagger V$, controlling the FCNC in the up sector and down sector, respectively, were defined in~\cref{eq:Fud}.

For sufficiently heavy quarks decaying into SM quarks, with all $r \ll 1$, the expressions above reduce to
\begin{align}\label{eq:hd-widths}
    \Gamma\big(Q_i\rightarrow Z q_j\big)\simeq \frac{G_F M^3_i}{16\pi\sqrt{2}}\big|F^q_{ji}\big|^2&\,,\quad            \Gamma\big(U_i\rightarrow W^+ d_j\big)\simeq \frac{G_F M^3_i}{8\pi\sqrt{2}}\big|V_{ij}\big|^2\,,\\
    \Gamma\big(Q_i\rightarrow h q_j\big)\simeq \frac{G_FM^3_i}{16\pi\sqrt{2}}\big|F^q_{ji}\big|^2&\,,\quad            \Gamma\big(D_i\rightarrow W^-u_j\big) \simeq \frac{G_FM^3_i}{8\pi\sqrt{2}}\big|V_{ji}\big|^2\,,
\end{align}
which can be used to derive the following useful relations,
\begin{align}\label{eq:BR-relation:U}
    \frac{\Gamma(U_i\rightarrow Zu_j)}{|F^u_{ji}|^2}&\simeq          \frac{\Gamma(U_i\rightarrow h  u_j)}{|F^u_{ji}|^2}\simeq          \frac{\Gamma(U_i\rightarrow W^+ d_j)}{2|V_{ij}|^2}\,,\\
\label{eq:BR-relation:D}
    \frac{\Gamma(D_i\rightarrow Z d_j)}{|F^d_{ji}|^2}&\simeq          \frac{\Gamma(D_i\rightarrow h d_j)}{|F^d_{ji}|^2}\simeq
    \frac{\Gamma(D_i\rightarrow W^- u_j)}{2|V_{ji}|^2}\,.
\end{align}

It is important to note the relation between the decay and single production of VLQs.
In a single production event, the vertex involving the heavy quark is of the same form as that of one of its decays.
Evidently, single production also depends on the flavour couplings between VLQ, SM quarks and $Z$, $W$ and $h$ bosons. These couplings consist of the elements of the matrices $F^q$ and $V$, previously defined. 

\subsection{Historic searches}
\label{sec:defsu2brs}
In the past decades, various searches for heavy quarks have been undertaken at the Tevatron and at the LHC. Although no evidence for the existence of heavy quarks has yet been obtained, experimental bounds are becoming stricter.

Heavy quarks can be chiral (``fourth generation'') or vector-like. Furthermore, vector-like quarks may be $SU(2)$ singlets, doublets or triplets. All these heavy quarks will lead to different experimental signals, since, in general, they have different couplings, decay channels, and different BRs for specific channels.
Nonetheless, the analysis of heavy quark searches may be interpreted in such a way as to provide mass bounds either for heavy chiral quarks or for VLQs, as it was done in~\cite{ATLAS:2012qe}.
This can be achieved by employing a model-independent framework~\cite{Buchkremer:2013bha,Barducci:2014ila}, or resorting to specific models.

Most model-independent bounds are obtained from pair production searches, since the production mechanism of the heavy quarks is due to QCD. The final result of a pair production search is usually an exclusion, ruling out, at a certain confidence level (e.g.~95\% C.L.), VLQ masses below a given lower bound.

Single production of VLQs is very model-dependent, since the production mechanism of the VLQ is electroweak. Hence, the results of the searches for VLQs are usually expressed in terms of bounds on cross sections and couplings of the heavy quarks to SM quarks, for different masses of the heavy quarks.
Within the context of a specific model, it is then possible to set bounds on VLQ masses at certain confidence levels. 
Note that if the coupling between light quarks and VLQs is large, single production of VLQs is expected to occur with a higher cross section than pair production~\cite{Atre:2008iu,Atre:2011ae} in proton colliders like the LHC.
More recently, a semi-analytical approach for reinterpreting, in an almost model-independent fashion, experimental searches for VLQ single production was put forward~\cite{Roy:2020fqf}. 
Also, analyses beyond the narrow-width approximation have been performed for both single~\cite{Carvalho:2018jkq,Deandrea:2021vje,Roy:2022wjj} and pair~\cite{Moretti:2016gkr} production.

Following the outline of Ref.~\cite{Okada:2012gy}, here we give an overview of all searches of heavy quarks at Tevatron and at the LHC. We focus in this review on $SU(2)$ singlet VLQs, and the searches involving these particles will be presented, highlighting the underlying assumptions which lead to specific mass bounds. 
In particular, mass bounds will only be included when the assumptions of the search roughly correspond to an $SU(2)$ singlet VLQ with realistic BRs. For instance, we will not list mass bounds obtained under the assumption $\br(T \rightarrow Wb) = 1$, since $T$ would also be expected to decay into $h$ and $Z$ with comparable strengths, as can be seen from~\cref{eq:BR-relation:U}.

As pointed out, most searches are performed under the assumption that VLQs decay only or mainly to third-generation quarks.
Such an assumption may not be warranted, for instance, in scenarios where VLQs play a role in solving the CKM unitarity problem (see~\cref{sec:CKMUP}).
Nonetheless, this heavier-generation
decay hypothesis --- evident in~\cref{eq:3rdgen} below --- has become part of a certain ``conventional wisdom'' in the literature.

\begin{figure}[!t]
\centering
\includegraphics[scale=0.42]{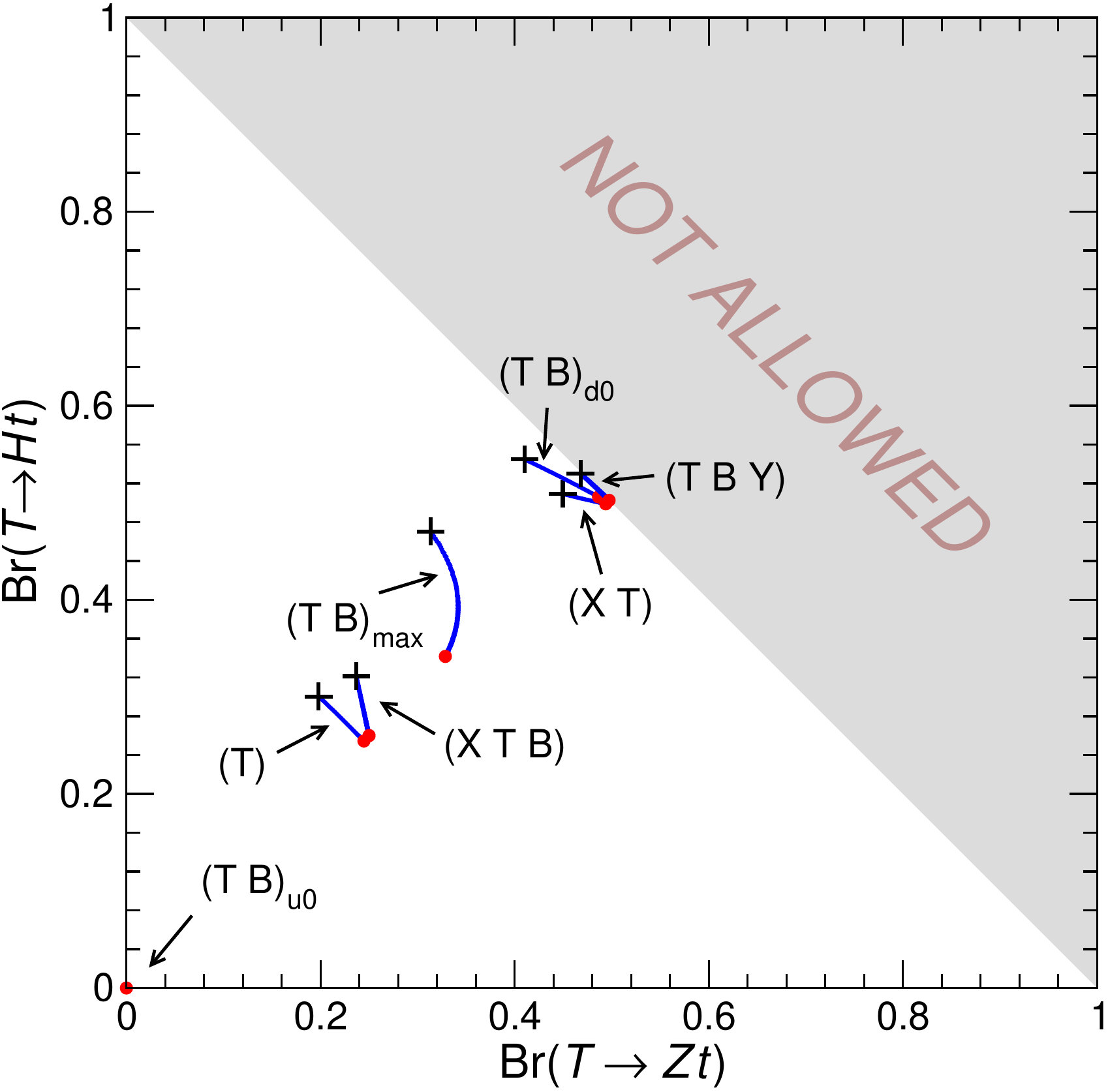}
\includegraphics[scale=0.42]{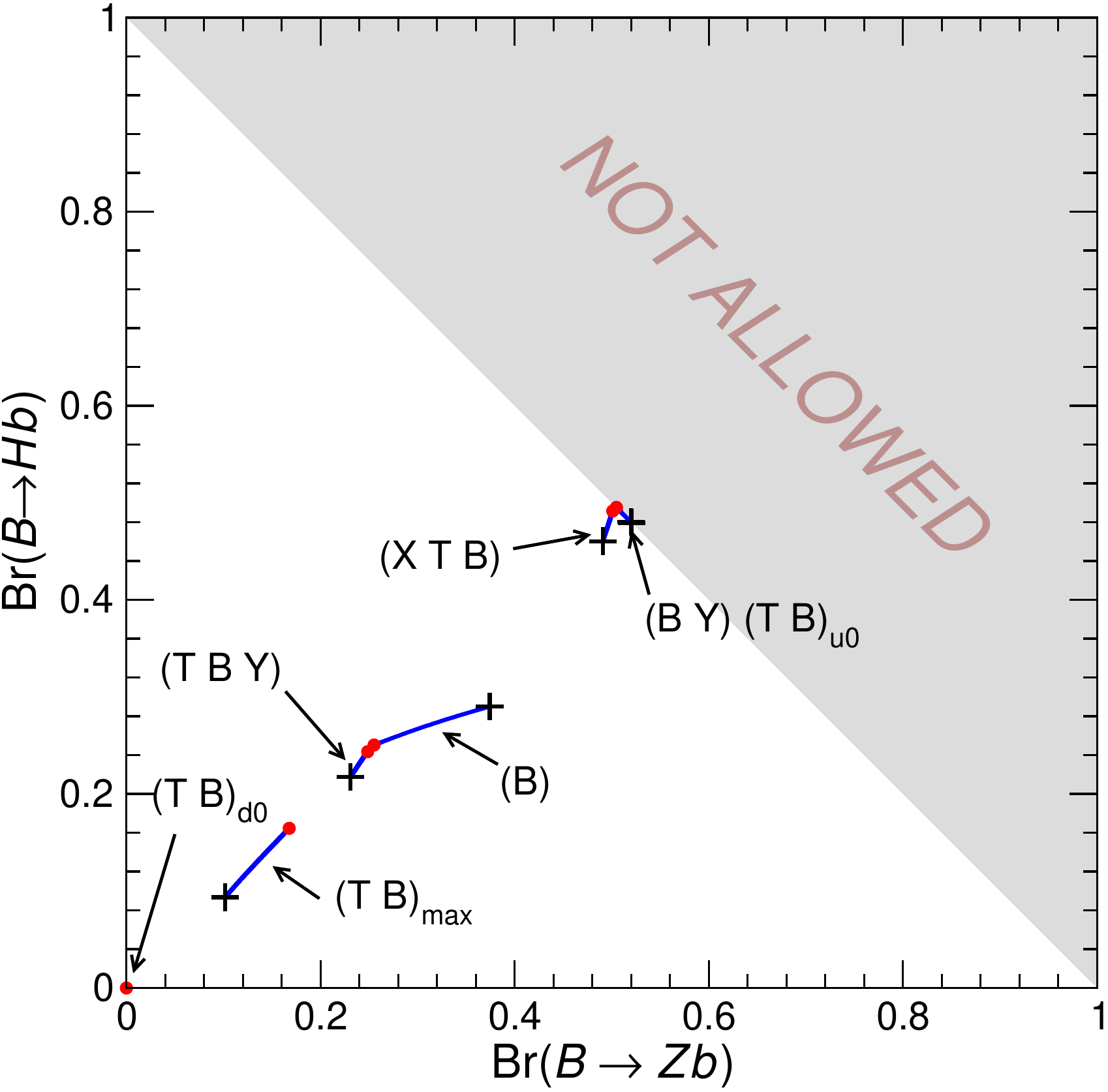}
\caption{\label{fig:branching-ratio}%
Allowed branching ratios for singlet VLQs $T$ (left) and $B$ (right) assuming they only couple with the third family quarks. VLQs in non-trivial $SU(2)_L$ multiplets are also shown. Red dots correspond to $M_{T,B}=2\;\unit{TeV}$ and are 
representative of the asymptotic high-mass value.
The crosses indicate $M_T=640\;\unit{GeV}$ or $M_B=590\;\unit{GeV}$.
The blue lines refer to intermediate masses.
Taken from Ref.~\cite{Aguilar-Saavedra:2013qpa}.
}
\end{figure}

\Cref{fig:branching-ratio} illustrates the possible BRs of VLQs, depending on their gauge group representation, 
when they are only allowed to decay to third-generation quarks. Although these results
depend on their masses, it is very clear that $SU(2)$ singlet VLQs that are only allowed to decay to third-generation quarks have a well-defined region of allowed BRs. These are very close to the points where
\begin{equation}
\begin{aligned} \label[pluralequation]{eq:SU2BRs}
     \br(T \rightarrow W b) &= 0.5\,,\,\, \br(T \rightarrow Zt)= 0.25\,,\,\, \br(T \rightarrow ht) = 0.25\,, \text{ and}\\[2mm]
     \br(B \rightarrow W t) &= 0.5\,,\,\, \br(B \rightarrow Zb)= 0.25\,,\,\, \br(B \rightarrow hb) = 0.25 \,.
\end{aligned}
\end{equation}
These limiting values can be seen in the relations of~\cref{eq:BR-relation:U,eq:BR-relation:D}, taking $|F_{ji}|\simeq |V_{ji}|$ or $|F_{ji}|\simeq |V_{ij}|$, valid when the index $i$ refers to the heavy quark with high mass and $j$ refers to the third-generation quark, neglecting CKM mixing, cf.~\cref{eq:uparam-approxV,eq:duFCNC-approx}.

For consistency, we call these BRs ``$SU(2)$ singlet (VLQ) BRs'', as this is also the nomenclature used in most of the literature. However, one should keep in mind that these BRs only correspond to the allowed BRs of $SU(2)$ singlet VLQs when these can decay exclusively to third-generation quarks.
Under such a strong assumption, but allowing for deviations from the values of~\cref{eq:SU2BRs}, one has instead the weaker constraints
\begin{equation}
\begin{aligned}\label[pluralequation]{eq:3rdgen}
    \br(T \rightarrow W b) + \br(T \rightarrow Zt) + \br(T \rightarrow ht) &= 1\,, \text{ and}\\[2mm]
    \br(B \rightarrow W t) + \br(B \rightarrow Zb) + \br(B \rightarrow hb) &= 1 \,.
\end{aligned}
\end{equation}
It is important to keep in mind that these
are assumptions that may be violated by the appearance of other relevant decay channels in extended scenarios, see e.g.~\cite{Aguilar-Saavedra:2017giu,Chala:2017xgc,Kim:2018mks,Alhazmi:2018whk,Kim:2019oyh,Criado:2019mvu,Wang:2020ips,Banerjee:2022xmu,Bhardwaj:2022wfz,Verma:2022nyd,Bardhan:2022sif}.
For example, for a VLQ $T$ with suppressed mixing with SM quarks, decays to $tg$ and $t\gamma$ may dominate, requiring a different search strategy~\cite{Kim:2018mks,Alhazmi:2018whk,Criado:2019mvu}.
Another example is the aforementioned possibility of decay to SM quarks other than the third generation, e.g., 
$T \rightarrow W s$.
This motivates searches for VLQs which decay predominantly to light quarks $u,d,s$~\cite{ATLAS:2015lpr,CMS:2017asf}.
However, even with additional decay channels, it is possible to recast mass limits from pair production searches~\cite{Aguilar-Saavedra:2017giu,Criado:2019mvu}.

Let us finally comment on notation. Many sources use the variables $c_{L}^{Wb}$ ($c_{L}^{Wt}$) or $\sin \theta_L^{(b)}$ ($\sin \theta_L^{(t)}$) to quantify the strength of the mixing between a up (down) VLQ and its third-generation down (up) counterpart. They are related by
\begin{equation}
    c_{L}^{Wq} \simeq \sqrt{2} \sin \theta_L^{(q)}\,, \quad (q=b,t)\,,
\end{equation}
with $\sin \theta_L$ corresponding, in general, to the element $ V_{Tb}$ ($V_{tB}$) of the quark mixing matrix.

\vskip 2mm
We compile, in~\cref{tab:expup,tab:expdown} of the following section, the results of the most stringent isosinglet VLQ searches done to date, as a summary of the discussion that follows in the text.
The searches are divided into three sets, corresponding to the Tevatron (\cref{sec:Tevatron}), LHC Run 1 (\cref{sec:LHC1}) and LHC Run 2 (\cref{sec:LHC2}).
\Cref{sec:Tevatron} 
includes the pre-LHC searches done at the Tevatron particle accelerator, in the USA. These were performed by the CDF and D0 experiments, which gathered data from 1983 to 2011. The LHC Run 1 and Run 2 segments include the LHC searches to date. These were performed by the ATLAS and CMS experiments, which gathered data from 2009 to 2013, and from 2015 to 2018, for the LHC Runs 1 and 2, respectively.
Future VLQ search prospects are discussed in~\cref{sec:future}.

\vfill
\clearpage


\fancypagestyle{lscapedplain}{%
  \fancyhf{}
  \renewcommand{\headrulewidth}{0pt}
  \fancyfoot{%
    \begin{tikzpicture}[remember picture,overlay]
      \node[outer sep=2cm,above,rotate=90] at (current page.east) {\normalsize\thepage};
      \end{tikzpicture}}
}

\newgeometry{left=2.5cm,right=3.5cm,bottom=2.5cm,top=2.5cm}

\footnotesize
\begin{landscape}
\pagestyle{lscapedplain}

\subsection{Summary tables}
\renewcommand{\arraystretch}{1.2}
\begin{longtable}{cccc c m{4.8cm} >{\centering \arraybackslash}m{3.4cm} c}
  \caption{
  A compilation of the most stringent $SU(2)_L$ singlet \textbf{up-type VLQ} searches done to date. Here, ``$SU(2)$ singlet BRs'' corresponds to $\br(T \rightarrow ht)=\br(T \rightarrow Zt)=0.25$ and $\br(T \rightarrow Wb)=0.5$ (see text), $p_T$ is transverse momentum, $E^\text{miss}_T$ is the magnitude of missing transverse momentum and $H_T$ denotes the scalar sum of the $p_T$ of all selected jets and leptons. For VLQ searches assuming decays to SM bosons and light quarks, one has $q=d,s$. Single production bounds depend on both the VLQ mass and the couplings to SM particles.
  }
  \label{tab:expup} \\
    \toprule   
    Experiments (int. &  $\sqrt{s}$ &   Type  of   &  Decay  & \multirow{2}{*}{Assumptions} & \centering \multirow{2}{*}{Selected events}& 95\% C.L.~bound & \multirow{2}{*}{Refs.} \\
    luminosity in $\unit{fb}^{-1}$) &  (TeV) & production & channels & & & (masses in GeV)
    \\
\midrule \endfirsthead
\caption{(cont.)}\\
    \toprule   
    Experiments (int. &  $\sqrt{s}$ &   Type  of   &  Decay  & \multirow{2}{*}{Assumptions} & \centering \multirow{2}{*}{Selected events}& 95\% C.L.~bound & \multirow{2}{*}{Refs.} \\
    luminosity in $\unit{fb}^{-1}$) &  (TeV) & production & channels & & & (masses in GeV) 
    \\
\midrule \endhead
\bottomrule \endfoot
\makecell{CMS (1 and 5),\\ATLAS (1)} & 7 &  Pair  & \makecell{$T \rightarrow ht$\\ $T \rightarrow Wb$\\ $T \rightarrow Zt$} & \makecell[l]{$\hphantom{+}\br(T \rightarrow ht)$\\$+\br(T \rightarrow Wb)$\\$+\br(T \rightarrow Zt) = 1$} & Reinterpretation of chiral quark searches. See reference and references therein & \makecell{$m_T >419$, for \\ ``$SU(2)$ singlet BRs''} & \mbox{\cite{Rao:2012sx,Rao:2012gf}}
\\[7mm]
CMS (19.7) & 8 & Pair  &
\makecell{$T \rightarrow ht$\\ $T \rightarrow Wb$\\ $T \rightarrow Zt$} &
\makecell[l]{$\hphantom{+}\br(T \rightarrow ht)$\\$+\br(T \rightarrow Wb)$\\$+\br(T \rightarrow Zt) = 1$} & 
Single lepton, multilepton (same sign and opposite sign), and trilepton states
& 
\makecell{$m_T >696$, for \\ ``$SU(2)$ singlet BRs''} &
\mbox{\cite{Bhattacharya:2013poa,Majumder:2013jqa}}
\\[7mm]
CMS (19.7) & 8 & Pair  &
\makecell{$T \rightarrow ht$\\ $T \rightarrow Wb$\\ $T \rightarrow Zt$} &
\makecell[l]{$\hphantom{+}\br(T \rightarrow ht)$\\$+\br(T \rightarrow Wb)$\\$+\br(T \rightarrow Zt) = 1$} & 
Single lepton, multilepton, 2 all-hadronic channels optimized for $Wb$ and $ht$, and a $h\rightarrow \gamma \gamma$ state
& 
\makecell{$m_T >740$, for \\ ``$SU(2)$ singlet BRs''} &
\mbox{\cite{CMS:2015lzl}}
\\[7mm]
CMS (19.7) & 8 & Pair  &
\makecell{$T \rightarrow hu$\\ $T \rightarrow Wq$\\ $T \rightarrow Zu$} &
\makecell[l]{$\hphantom{+}\br(T \rightarrow hu)$\\$+\br(T \rightarrow Wq)$\\$+\br(T \rightarrow Zu) = 1$} & 
Single lepton, dilepton and multilepton events
& 
\makecell{$m_T >430-845$, \\ depending on the BRs} & \mbox{\cite{CMS:2017asf}}
\\[7mm]
ATLAS (20.3) & 8 & Pair  &
$T \rightarrow Zt$ &
\makecell[l]{$\hphantom{+}\br(T \rightarrow ht)$\\$+\br(T \rightarrow Wb)$\\$+\br(T \rightarrow Zt) = 1$} & 
High $p_T$ $Z$ boson candidate reconstructed from a pair of oppositely-charged same-flavour leptons (electrons or muons)
& 
\makecell{$m_T >655$, for \\ ``$SU(2)$ singlet BRs''} & \mbox{\cite{ATLAS:2014vpn}}
\\[7mm]
ATLAS (20.3) & 8 & Pair  &
\makecell{$T \rightarrow ht$\\ $T \rightarrow Wb$\\ $T \rightarrow Zt$} &
\makecell[l]{$\hphantom{+}\br(T \rightarrow ht)$\\$+\br(T \rightarrow Wb)$\\$+\br(T \rightarrow Zt) = 1$} & 
Same-sign leptons (pairs of same charge high $p_T$ leptons + b-jets + $E^\text{miss}_T$) & 
\makecell{$m_T >590$, for \\ ``$SU(2)$ singlet BRs''} & \mbox{\cite{ATLAS:2015uaw}}
\\[7mm]
ATLAS (20.3) & 8 & Pair  &
\makecell{$T \rightarrow hu$\\ $T \rightarrow Wq$\\ $T \rightarrow Zu$} &
\makecell[l]{$\hphantom{+}\br(T \rightarrow hu)$\\$+\br(T \rightarrow Wq)$\\$+\br(T \rightarrow Zu) = 1$} & 
One charged lepton, large $E_T^\text{miss}$ and 4 non-$b$-tagged jets
& 
\makecell{$m_T >340-690$, \\ depending on the BRs} & \mbox{\cite{ATLAS:2015lpr}}
\\[7mm]
ATLAS (20.3) & 8 &  \makecell{Single \\ ($Wb \rightarrow T$)}  &
$T \rightarrow Wb$ & \makecell[c]{$\br(T \rightarrow Wb)$\\$=0.5$} & 
1 isolated electron or muon, $\geq 2$ small-$R$ jets, $\geq 1$ large-$R$ jet, 1 $b$-tagged jet and $E^\text{miss}_T$ &
\makecell{$|V_{Tb}|<0.42\, (0.85)$, for\\ $m_T < 700\,(m_T = 1200)$}
& \mbox{\cite{ATLAS:2016scx}}
\\[7mm]
CMS (2.3) & 13 &  \makecell{Single \\ ($Wb/Zt \rightarrow T$)}   &
$T \rightarrow ht$ &
\makecell[l]{$\hphantom{+}\br(T \rightarrow ht)$\\$+\br(T \rightarrow Wb)$\\$+\br(T \rightarrow Zt) = 1$} & 
$t$ decay includes an electron or a muon, while $h$ decays into a pair of $b$ quarks
& 
\makecell{$m_T \notin [700,1800]$,\\ for $V_{Tb}=0.5$ and 
\\ ``$SU(2)$ singlet BRs'' } &
\mbox{\cite{CMS:2016edj}}
\\[7mm]
CMS (2.3) & 13 &  \makecell{Single \\ ($Wb/Zt \rightarrow T$)}   &
$T \rightarrow ht$ &
\makecell[l]{$\hphantom{+}\br(T \rightarrow ht)$\\$+\br(T \rightarrow Wb)$\\$+\br(T \rightarrow Zt) = 1$} & 
$t$ and $h$ highly Lorentz-boosted, appearing as single
hadronic jets + boosted $h$ and $t$ tagging techniques
& 
\makecell{$m_T \notin [1000,1800]$,\\ for $V_{Tb}=0.5$ and 
\\ ``$SU(2)$ singlet BRs'' } &
\mbox{\cite{CMS:2016jce}}
\\[7mm]
CMS (2.3) & 13 &  \makecell{Single \\ ($Wb \rightarrow T$)}   &
$T \rightarrow Zt$ &
\makecell[l]{$\hphantom{+}\br(T \rightarrow ht)$\\$+\br(T \rightarrow Wb)$\\$+\br(T \rightarrow Zt) = 1$} & 
$Z$ boson decaying leptonically, accompanied by a $t$ decaying hadronically
& 
\makecell{$m_T \notin [700,1700]$,\\ for $V_{Tb}=0.5$ and 
\\ ``$SU(2)$ singlet BRs'' } &
\mbox{\cite{CMS:2017gsh}}
\\[7mm]
CMS (2.6) & 13 &  Pair   &
\makecell{$T \rightarrow ht$\\ $T \rightarrow Wb$\\ $T \rightarrow Zt$} &
\makecell[l]{$\hphantom{+}\br(T \rightarrow ht)$\\$+\br(T \rightarrow Wb)$\\$+\br(T \rightarrow Zt) = 1$} & 
$\geq 1$ lepton and several jets, $W$ or $h$ bosons decaying hadronically with large $p_T$
& \makecell{$m_T >860$, for \\ ``$SU(2)$ singlet BRs''} & 
\mbox{\cite{CMS:2017ked}}
\\[7mm]
CMS (35.9) & 13 & Pair  &
\makecell{$T \rightarrow ht$\\ $T \rightarrow Wb$\\ $T \rightarrow Zt$} &
\makecell[l]{$\hphantom{+}\br(T \rightarrow ht)$\\$+\br(T \rightarrow Wb)$\\$+\br(T \rightarrow Zt) = 1$} & 
Single lepton, 2 same-charge leptons, or $\geq 3$ leptons
& \makecell{$m_T >1200$, for \\ ``$SU(2)$ singlet BRs''} & \mbox{\cite{CMS:2018zkf}}
\\[7mm]
CMS (35.9) & 13 & Pair  &
\makecell{$T \rightarrow ht$\\ $T \rightarrow Wb$\\ $T \rightarrow Zt$} &
\makecell[l]{$\hphantom{+}\br(T \rightarrow ht)$\\$+\br(T \rightarrow Wb)$\\$+\br(T \rightarrow Zt) = 1$} & 
Two oppositely-charged electrons or muons, coming from
the decay of a $Z$ boson, and jets
& \makecell{$m_T >1095$, for \\ ``$SU(2)$ singlet BRs''} & \mbox{\cite{CMS:2018wpl}}
\\[7mm]
CMS (35.9) & 13 & Pair  &
\makecell{$T \rightarrow ht$\\ $T \rightarrow Wb$\\ $T \rightarrow Zt$} &
\makecell[l]{$\hphantom{+}\br(T \rightarrow ht)$\\$+\br(T \rightarrow Wb)$\\$+\br(T \rightarrow Zt) = 1$} & 
Fully hadronic final states, categorized according to jet multiplicities and the scalar sum of jet momenta 
& \makecell{$m_T >960-980$, for \\ ``$SU(2)$ singlet BRs''} & \mbox{\cite{CMS:2019eqb}}
\\[7mm]
CMS (35.9) & 13 & \makecell{Single \\ ($Wb/Zt \rightarrow T$)}  &
\makecell{$T \rightarrow ht$\\ $T \rightarrow Zt$} &
\makecell[l]{$\hphantom{+}\br(T \rightarrow ht)$\\$+\br(T \rightarrow Wb)$\\$+\br(T \rightarrow Zt) = 1$} & 
Fully hadronic final states. Hadronic $t$ decay and primarily the $\overline{b}b$ decay of $h$ and $Z$ bosons
& \makecell{$m_T \notin [700,1000]$, for \\ ``$SU(2)$ singlet BRs'',\\ width-dependent } 
& \mbox{\cite{CMS:2019afi}}
\\[7mm]
ATLAS (36.1) & 13 &  Pair   &
$T \rightarrow Zt$ &
\makecell[l]{$\hphantom{+}\br(T \rightarrow ht)$\\$+\br(T \rightarrow Wb)$\\$+\br(T \rightarrow Zt) = 1$} & 
1 lepton, jets and sizeable $E^\text{miss}_T$
& 
\makecell{$m_T >870$, for \\ ``$SU(2)$ singlet BRs''} &
\mbox{\cite{ATLAS:2017vdo}}
\\[7mm]
ATLAS (36.1) & 13 & Pair  &
$T \rightarrow Wb$ &
\makecell[l]{$\hphantom{+}\br(T \rightarrow ht)$\\$+\br(T \rightarrow Wb)$\\$+\br(T \rightarrow Zt) = 1$} & 
Lepton-plus-jets, including $\geq 1$ $b$-tagged jet and a large-$R$ jet
& \makecell{$m_T >1170$, for \\ ``$SU(2)$ singlet BRs''} & \mbox{\cite{ATLAS:2017nap}}
\\[7mm]
ATLAS (36.1) & 13 & Pair  &
\makecell{$T \rightarrow ht$\\ $T \rightarrow Wb$\\ $T \rightarrow Zt$} &
\makecell[l]{$\hphantom{+}\br(T \rightarrow ht)$\\$+\br(T \rightarrow Wb)$\\$+\br(T \rightarrow Zt) = 1$} & 
Isolated electron or muon with high $p_T$, large $E_T^\text{miss}$ and multiple jets 
& \makecell{$m_T >1190$, for \\ ``$SU(2)$ singlet BRs''} & \mbox{\cite{ATLAS:2018cye}}
\\[7mm]
ATLAS (36.1) & 13 &  Pair   &
$T \rightarrow Zt$ &
\makecell[l]{$\hphantom{+}\br(T \rightarrow ht)$\\$+\br(T \rightarrow Wb)$\\$+\br(T \rightarrow Zt) = 1$} & 
Dilepton with $\leq 1$ large-$R$ jet,
dilepton with $\geq 2$ large-$R$ jets, 
trilepton
& 
\makecell{$m_T >1030$, for \\ ``$SU(2)$ singlet BRs''} &
\mbox{\cite{ATLAS:2018tnt,Vale:2018bpf}}
\\[7mm]
ATLAS (36.1) & 13 & Pair  &
\makecell{$T \rightarrow ht$\\ $T \rightarrow Wb$\\ $T \rightarrow Zt$} &
\makecell[l]{$\hphantom{+}\br(T \rightarrow ht)$\\$+\br(T \rightarrow Wb)$\\$+\br(T \rightarrow Zt) = 1$} & 
$\geq 2$ leptons, with a same-charge pair, $\geq 1$ $b$-tagged jet, sizeable $E_T^\text{miss}$, and large $H_T$
& \makecell{$m_T >980$, for \\ ``$SU(2)$ singlet BRs''} & \mbox{\cite{ATLAS:2018alq}}
\\[7mm]
ATLAS (36.1) & 13 & Pair  &
\makecell{$T \rightarrow ht$\\ $T \rightarrow Wb$\\ $T \rightarrow Zt$} &
\makecell[l]{$\hphantom{+}\br(T \rightarrow ht)$\\$+\br(T \rightarrow Wb)$\\$+\br(T \rightarrow Zt) = 1$} & 
Multiple jets, $b$-tagged jets, $t$-tagged jets,
$h$-tagged jets, and $E_T^\text{miss}$.
1 lepton, $\geq 3$ jets and $E_T^\text{miss}$
& \makecell{$m_T >1170$, for \\ ``$SU(2)$ singlet BRs''} & \mbox{\cite{Nikiforou:2018geu}}
\\[7mm]
ATLAS (36.1) & 13 &  \makecell{Single \\ ($Wb \rightarrow T$)}   &
$T \rightarrow Wb$ &
\makecell[l]{$\hphantom{+}\br(T \rightarrow ht)$\\$+\br(T \rightarrow Wb)$\\$+\br(T \rightarrow Zt) = 1$} & 
1 isolated electron or muon, a high-$p_T$ $b$-tagged jet, $E^\text{miss}_T$ and $\geq 1$ forward jet
& 
\makecell{$|V_{Tb}|<0.18\, (0.35)$,\\ for $m_T = 800\,(1200)$}
&
\mbox{\cite{ATLAS:2018dyh}}
\\[7mm]
ATLAS (36.1) & 13 &  \makecell{Single \\ ($Wb \rightarrow T$)}   &
$T \rightarrow Zt$ &
\makecell[l]{$\hphantom{+}\br(T \rightarrow ht)$\\$+\br(T \rightarrow Wb)$\\$+\br(T \rightarrow Zt) = 1$} & 
$Z$ boson decaying into neutrinos, resulting in a mono-top signature
& 
\makecell{$|V_{Tb}|<0.7$,\\ for $m_T < 1400$  } &
\mbox{\cite{ATLAS:2018cjd}}
\\[7mm]
CMS (138) & 13 &  Pair   &
\makecell{$T \rightarrow ht$\\ $T \rightarrow Wb$\\ $T \rightarrow Zt$} &
\makecell[l]{$\hphantom{+}\br(T \rightarrow ht)$\\$+\br(T \rightarrow Wb)$\\$+\br(T \rightarrow Zt) = 1$} & 
Single lepton, 2 same-charge leptons, or $\geq 3$ leptons
& 
\makecell{$m_T >1480$, for \\ ``$SU(2)$ singlet BRs''} &
\mbox{\cite{CMS:2022fck}}
\\[7mm]
CMS (138) & 13 &  \makecell{Single \\ ($Wb \rightarrow T$)}    &
$T \rightarrow ht$ &
\makecell[l]{$\hphantom{+}\br(T \rightarrow ht)$\\$+\br(T \rightarrow Wb)$\\$+\br(T \rightarrow Zt) = 1$} & 
Diphoton decay of the Higgs, hadronic and leptonic decay modes of the top
& 
\makecell{$m_T > 960$,\\ for $V_{Tb}=0.18$ and 
\\ ``$SU(2)$ singlet BRs'' } &
\mbox{\cite{CMS:2023agg}}
\\[7mm]
ATLAS (139) & 13 & Pair  &
$T \rightarrow Zt$ &
\makecell[l]{$\hphantom{+}\br(T \rightarrow ht)$\\$+\br(T \rightarrow Wb)$\\$+\br(T \rightarrow Zt) = 1$} & 
2 or 3 leptons plus jets, with a signal region built thanks to a deep neural
network tagger for jet classification
& \makecell{$m_T >1270$, for \\ ``$SU(2)$ singlet BRs''} 
& \mbox{\cite{Vale:thesis}}
\\[7mm]
ATLAS (139) & 13 &  \makecell{Single \\ ($Wb/Zt \rightarrow T$)}    &
$T \rightarrow ht$ &
\makecell[c]{$\br(T \rightarrow ht)$\\$=0.25$} &
Fully hadronic final states, large-$R$ jets in the reconstruction
& 
\makecell{$|V_{Tb}|<0.25$,\\ for $m_T < 1400$} &
\mbox{\cite{ATLAS:2022ozf}} 
\\[7mm]
ATLAS (139) & 13 &  Pair   &
$T \rightarrow Zt$ &
\makecell[l]{$\hphantom{+}\br(T \rightarrow ht)$\\$+\br(T \rightarrow Wb)$\\$+\br(T \rightarrow Zt) = 1$} & 
Pair of same-flavour leptons with opposite charges,
$b$-tagged jets,
and high-$p_T$ large-$R$ jets
& 
\makecell{$m_T >1270$, for \\ ``$SU(2)$ singlet BRs''} &
\mbox{\cite{ATLAS:2022hnn}}
\\[7mm]
ATLAS (139) & 13 & Pair  &
\makecell{$T \rightarrow ht$\\ $T \rightarrow Wb$\\ $T \rightarrow Zt$} &
\makecell[l]{$\hphantom{+}\br(T \rightarrow ht)$\\$+\br(T \rightarrow Wb)$\\$+\br(T \rightarrow Zt) = 1$} & 
One charged lepton, large $E_T^\text{miss}$ and $\geq 4$ jets, including $\geq 1$ $b$-tagged jet
& \makecell{$m_T >1260$, for \\ ``$SU(2)$ singlet BRs''} 
& \mbox{\cite{ATLAS:2022tla}}
\\[7mm]
ATLAS (139) & 13 &  \makecell{Single \\ ($Wb/Zt \rightarrow T$)}    &
$T \rightarrow Zt$ &
\makecell[l]{$\hphantom{+}\br(T \rightarrow ht)$\\$+\br(T \rightarrow Wb)$\\$+\br(T \rightarrow Zt) = 1$} &
Two oppositely-charged electrons or muons, coming from the decay of a $Z$ boson, and jets
& 
\makecell{$|V_{Tb}|<0.45$,\\ for $m_T < 1975$ and 
\\ ``$SU(2)$ singlet BRs'' } &
\mbox{\cite{ATLAS:2023uah}} 
\\[7mm]
ATLAS (139) & 13 &  \makecell{Single \\ ($Wb/Zt \rightarrow T$)}    &
\makecell{$T \rightarrow ht$\\  $T \rightarrow Zt$} &
\makecell[l]{$\hphantom{+}\br(T \rightarrow ht)$\\$+\br(T \rightarrow Wb)$\\$+\br(T \rightarrow Zt) = 1$} &
Single lepton plus jets, including multiple jets and $\geq 1$ $b$-tagged jet
& 
\makecell{$|V_{Tb}|<0.4$,\\ for $m_T < 2300$ and 
\\ ``$SU(2)$ singlet BRs'' } &
\mbox{\cite{ATLAS:2023pja}} 
\\[7mm]
\end{longtable}

\clearpage

\renewcommand{\arraystretch}{1.2}
\begin{longtable}{cccc c m{4.8cm} >{\centering \arraybackslash}m{3.4cm} c}
  \caption{
    A compilation of the most stringent $SU(2)_L$ singlet \textbf{down-type VLQ} searches done to date. Here, ``$SU(2)$ singlet BRs'' corresponds to $\br(B \rightarrow hb)=\br(B \rightarrow Zb)=0.25$ and $\br(B \rightarrow Wt)=0.5$ (see text), $p_T$ is transverse momentum, $E^\text{miss}_T$ is the magnitude of missing transverse momentum and $H_T$ denotes the scalar sum of the $p_T$ of all selected jets and leptons. For VLQ searches assuming decays to SM bosons and light quarks, one has $q=d,s$. Single production bounds depend on both the VLQ mass and the couplings to SM particles.
  }
  \label{tab:expdown} \\
    \toprule   
    Experiments (int. &  $\sqrt{s}$ &   Type  of   &  Decay  & \multirow{2}{*}{Assumptions} & \centering \multirow{2}{*}{Selected events}& 95\% C.L.~bound & \multirow{2}{*}{Refs.} \\
    luminosity in $\unit{fb}^{-1}$) &  (TeV) & production & channels & & & (masses in GeV) 
    \\
\midrule \endfirsthead
\caption{(cont.)}\\
    \toprule   
    Experiments (int. &  $\sqrt{s}$ &   Type  of   &  Decay  & \multirow{2}{*}{Assumptions} & \centering \multirow{2}{*}{Selected events}& 95\% C.L.~bound & \multirow{2}{*}{Refs.} \\
    luminosity in $\unit{fb}^{-1}$) &  (TeV) & production & channels & & & (masses in GeV) 
    \\
\midrule \endhead
\bottomrule \endfoot
CMS (19.7) & 8 & Pair &
\makecell{$B \rightarrow hb$\\ $B \rightarrow Wt$\\ $B \rightarrow Zb$} &
\makecell[l]{$\hphantom{+}\br(B \rightarrow hb)$\\$+\br(B \rightarrow Wt)$\\$+\br(B \rightarrow Zb) = 1$} &
Single lepton, multilepton (same sign and opposite sign), $\geq 3$ leptons, all-hadronic final state targeting $h \rightarrow b \overline{b}$
&
\makecell{$m_B >810$, for \\ ``$SU(2)$ singlet BRs''} & \mbox{\cite{CMS:2015hyy}}
\\[7mm]
CMS (19.7) & 8 & Pair  &
\makecell{$B \rightarrow hq$\\ $B \rightarrow Wu$\\ $B \rightarrow Zq$} &
\makecell[l]{$\hphantom{+}\br(B \rightarrow hq)$\\$+\br(B \rightarrow Wu)$\\$+\br(B \rightarrow Zq) = 1$} &
Single lepton, dilepton, and multilepton events
& 
\makecell{$m_B >430-845$, \\ depending on the BRs} &
\mbox{\cite{CMS:2017asf}}
\\[7mm]
ATLAS (20.3) & 8 & Pair &
$B \rightarrow Zb$ &
\makecell[l]{$\hphantom{+}\br(B \rightarrow hb)$\\$+\br(B \rightarrow Wt)$\\$+\br(B \rightarrow Zb) = 1$} &
High-$p_T$ $Z$ boson reconstructed from pair of oppositely-charged same-flavour leptons (electrons or muons) 
&
\makecell{$m_B >685$, for \\ ``$SU(2)$ singlet BRs''} & \mbox{\cite{ATLAS:2014vpn}}
\\[7mm]
ATLAS (20.3) & 8 & Pair &
\makecell{$B \rightarrow hb$\\ $B \rightarrow Wt$\\ $B \rightarrow Zb$} &
\makecell[l]{$\hphantom{+}\br(B \rightarrow hb)$\\$+\br(B \rightarrow Wt)$\\$+\br(B \rightarrow Zb) = 1$} &
Lepton-plus-jets (electron or muon + jets + high $E^\text{miss}_T$)
&
\makecell{$m_B >640$, for \\ ``$SU(2)$ singlet BRs''} & \mbox{\cite{ATLAS:2015vzd}}
\\[7mm]
ATLAS (20.3) & 8 & Pair &
\makecell{$B \rightarrow hb$\\ $B \rightarrow Wt$\\ $B \rightarrow Zb$} &
\makecell[l]{$\hphantom{+}\br(B \rightarrow hb)$\\$+\br(B \rightarrow Wt)$\\$+\br(B \rightarrow Zb) = 1$} &
Same-sign-leptons (pairs of same-charge high-$p_T$ leptons + $b$-jets + $E^\text{miss}_T$) 
&
\makecell{$m_B >620$, for \\ ``$SU(2)$ singlet BRs''} & \mbox{\cite{ATLAS:2015uaw}}
\\[7mm]
ATLAS (20.3) & 8 & Pair  &
\makecell{$B \rightarrow hq$\\ $B \rightarrow Wu$\\ $B \rightarrow Zq$} &
\makecell[l]{$\hphantom{+}\br(B \rightarrow hq)$\\$+\br(B \rightarrow Wu)$\\$+\br(B \rightarrow Zq) = 1$} &
One charged lepton, large $E_T^\text{miss}$ and 4 non-$b$-tagged jets
& 
\makecell{$m_B >340-690$, \\ depending on the BRs} &
\mbox{\cite{ATLAS:2015lpr}}
\\[7mm]
CMS (2.3) & 13 & \makecell{Single\\($Wt \rightarrow B$)} &
$B \rightarrow Zb$ &
\makecell[l]{$\hphantom{+}\br(B \rightarrow hb)$\\$+\br(B \rightarrow Wt)$\\$+\br(B \rightarrow Zb) = 1$} &
$Z$ boson decaying leptonically, accompanied by a $b$ decaying hadronically
& 
\makecell{$m_B \notin [700,1700]$,\\ for $V_{tB}=0.5$ and 
\\ ``$SU(2)$ singlet BRs'' } &
\mbox{\cite{CMS:2017gsh}}
\\[7mm]
CMS (2.6) & 13 & Pair &
\makecell{$B \rightarrow hb$\\ $B \rightarrow Wt$\\ $B \rightarrow Zb$} &
\makecell[l]{$\hphantom{+}\br(B \rightarrow hb)$\\$+\br(B \rightarrow Wt)$\\$+\br(B \rightarrow Zb) = 1$} &
$\geq 1$ lepton and several jets, $W$ or $h$ bosons decaying hadronically with large $p_T$
& 
\makecell{$m_B >730$, for \\ ``$SU(2)$ singlet BRs''} & 
\mbox{\cite{CMS:2017ked}}
\\[7mm]
CMS (35.9) & 13 & Pair &
\makecell{$B \rightarrow hb$\\ $B \rightarrow Wt$\\ $B \rightarrow Zb$} &
\makecell[l]{$\hphantom{+}\br(B \rightarrow hb)$\\$+\br(B \rightarrow Wt)$\\$+\br(B \rightarrow Zb) = 1$} &
Single lepton, 2 same-charge leptons, or $\geq 3$ leptons
& 
\makecell{$m_B >1170$, for \\ ``$SU(2)$ singlet BRs''} & 
\mbox{\cite{CMS:2018zkf}}
\\[7mm]
CMS (35.9) & 13 & Pair &
\makecell{$B \rightarrow hb$\\ $B \rightarrow Wt$\\ $B \rightarrow Zb$} &
\makecell[l]{$\hphantom{+}\br(B \rightarrow hb)$\\$+\br(B \rightarrow Wt)$\\$+\br(B \rightarrow Zb) = 1$} &
Two oppositely-charged electrons or muons, coming from the decay of a $Z$ boson, and jets
& 
\makecell{$m_B >955$, for \\ ``$SU(2)$ singlet BRs''} & 
\mbox{\cite{CMS:2018wpl}}
\\[7mm]
CMS (35.9) & 13 & Pair &
\makecell{$B \rightarrow hb$\\ $B \rightarrow Wt$\\ $B \rightarrow Zb$} &
\makecell[l]{$\hphantom{+}\br(B \rightarrow hb)$\\$+\br(B \rightarrow Wt)$\\$+\br(B \rightarrow Zb) = 1$} &
Fully hadronic final states, categorized according to jet multiplicities and the scalar sum of jet momenta
& 
\makecell{$m_B >890-910$, for \\ ``$SU(2)$ singlet BRs''} & 
\mbox{\cite{CMS:2019eqb}}
\\[7mm]
ATLAS (36.1) & 13 & Pair &
$B \rightarrow Wt$ &
\makecell[l]{$\hphantom{+}\br(B \rightarrow hb)$\\$+\br(B \rightarrow Wt)$\\$+\br(B \rightarrow Zb) = 1$} &
Lepton-plus-jets, including $\geq 1$ $b$-tagged jet and a large-$R$ jet
& 
\makecell{$m_B >1080$, for \\ ``$SU(2)$ singlet BRs''} & 
\mbox{\cite{ATLAS:2017nap}}
\\[7mm]
ATLAS (36.1) & 13 & Pair &
$B \rightarrow Wt$ &
\makecell[l]{$\hphantom{+}\br(B \rightarrow hb)$\\$+\br(B \rightarrow Wt)$\\$+\br(B \rightarrow Zb) = 1$} &
Lepton-plus-jets, with a high-$p_T$ isolated electron or muon, large $E_T^\text{miss}$, and multiple jets (at least one $b$-tagged)
& 
\makecell{$m_B >1170$, for \\ ``$SU(2)$ singlet BRs''} & 
\mbox{\cite{ATLAS:2018mpo}}
\\[7mm]
ATLAS (36.1) & 13 & Pair &
$B \rightarrow Zb$ &
\makecell[l]{$\hphantom{+}\br(B \rightarrow hb)$\\$+\br(B \rightarrow Wt)$\\$+\br(B \rightarrow Zb) = 1$} &
Dilepton with $\leq 1$ large-$R$ jet, dilepton with $\geq 2$ large-$R$ jets, trilepton
&
\makecell{$m_B >1010$, for \\ ``$SU(2)$ singlet BRs''} & \mbox{\cite{ATLAS:2018tnt,Vale:2018bpf}}
\\[7mm]
ATLAS (36.1) & 13 & Pair &
\makecell{$B \rightarrow hb$\\ $B \rightarrow Wt$\\ $B \rightarrow Zb$} &
\makecell[l]{$\hphantom{+}\br(B \rightarrow hb)$\\$+\br(B \rightarrow Wt)$\\$+\br(B \rightarrow Zb) = 1$} &
$\geq 2$ leptons, with a same-charge pair, $\geq 1$ $b$-tagged jet, sizeable $E_T^\text{miss}$, and large $H_T$
& 
\makecell{$m_B >1000$, for \\ ``$SU(2)$ singlet BRs''} & 
\mbox{\cite{ATLAS:2018alq}}
\\[7mm]
ATLAS (36.1) & 13 & Pair &
\makecell{$B \rightarrow hb$\\ $B \rightarrow Wt$\\ $B \rightarrow Zb$} &
\makecell[l]{$\hphantom{+}\br(B \rightarrow hb)$\\$+\br(B \rightarrow Wt)$\\$+\br(B \rightarrow Zb) = 1$} &
Multiple jets, $b$-, $t$- and
$h$-tagged jets, and $E_T^\text{miss}$. 1 lepton, $\geq 3$ jets and $E_T^\text{miss}$
& 
\makecell{$m_B >1080$, for \\ ``$SU(2)$ singlet BRs''} & 
\mbox{\cite{Nikiforou:2018geu}}
\\[7mm]
CMS (138) & 13 &  Pair   &
\makecell{$B \rightarrow hb$\\ $B \rightarrow Wt$\\ $B \rightarrow Zb$} &
\makecell[l]{$\hphantom{+}\br(B \rightarrow hb)$\\$+\br(B \rightarrow Wt)$\\$+\br(B \rightarrow Zb) = 1$} & 
Single lepton, 2 same-charge leptons, or $\geq 3$ leptons
& 
\makecell{$m_B >1470$, for \\ ``$SU(2)$ singlet BRs''} &
\mbox{\cite{CMS:2022fck}}
\\[7mm]
ATLAS (139) & 13 & Pair &
$B \rightarrow Zb$ &
\makecell[l]{$\hphantom{+}\br(B \rightarrow hb)$\\$+\br(B \rightarrow Wt)$\\$+\br(B \rightarrow Zb) = 1$} &
2 or 3 leptons plus jets, with a signal region built thanks to a deep neural
network tagger for jet classification
& 
\makecell{$m_B >1200$, for \\ ``$SU(2)$ singlet BRs''} & 
\mbox{\cite{Vale:thesis}}
\\[7mm]
ATLAS (139) & 13 & Pair &
$B \rightarrow Zb$ &
\makecell[l]{$\hphantom{+}\br(B \rightarrow hb)$\\$+\br(B \rightarrow Wt)$\\$+\br(B \rightarrow Zb) = 1$} &
Pair of same-flavour leptons with opposite charges, $b$-tagged jets, and high-$p_T$ large-$R$ jets
&
\makecell{$m_B >1200$, for \\ ``$SU(2)$ singlet BRs''} & \mbox{\cite{ATLAS:2022hnn}}
\\[7mm]
ATLAS (139) & 13 & Pair  &
\makecell{$B \rightarrow hb$\\ $B \rightarrow Wt$\\ $B \rightarrow Zb$} &
\makecell[l]{$\hphantom{+}\br(B \rightarrow hb)$\\$+\br(B \rightarrow Wt)$\\$+\br(B \rightarrow Zb) = 1$} & 
One charged lepton, large $E_T^\text{miss}$ and $\geq 4$ jets, including $\geq 1$ $b$-tagged jet
& \makecell{$m_T >1330$, for \\ ``$SU(2)$ singlet BRs''} 
& \mbox{\cite{ATLAS:2022tla}}
\\[7mm]
\end{longtable}

\end{landscape}

\restoregeometry
\normalsize

\subsubsection{Tevatron}
\label{sec:Tevatron}
The CDF and D0 experiments collected data from the Tevatron at $\sqrt{s}=1.96\;\unit{TeV}$. 
Most of the heavy quark searches done at CDF and D0 considered the pair production of chiral fourth generation quarks~\cite{CDF:2007qrj,CDF:2009gat,CDF:2011cqh,D0:2011fzt,CDF:2011lng}. 
Other chiral-quark searches involving dark matter particles were also performed~\cite{CDF:2011bsz,CDF:2011sxg}.
A specific search for the single production of a heavy chiral quark that decays predominantly to lighter generations was performed by CDF~\cite{cdfpublicnote}.

Only one explicit search for VLQs was done, at D0~\cite{D0:2010ckq}. However, motivated by a model which was popular at the time~\cite{Atre:2008iu}, this search was performed considering two degenerate doublet VLQs. Since it is a single production search, its bounds cannot be extrapolated to an $SU(2)$ singlet model.

It is out of the scope of this work to convert the mass bounds obtained in these searches to mass bounds on isosinglet VLQs.
As this cannot be done in a straightforward manner, these bounds will not be listed.

\subsubsection{LHC Run 1}
\label{sec:LHC1}
In Run 1, the ATLAS and CMS experiments collected data from the LHC at $\sqrt{s} = 7\;\unit{TeV}$ and $\sqrt{s} = 8\;\unit{TeV}$.

Some of the first heavy-quark searches were motivated by supersymmetry~\cite{ATLAS:2011yhy,ATLAS:2012ppg}. These focused on a pair-produced exotic top partner that decayed 100\% of the time to a top quark and a long-lived undetected neutral particle. 
Other exotic top/bottom partner searches include~\cite{ATLAS:2012tkh}, considering a pair-produced exotic top partner that decays 100\% of the time to a bottom quark and a $W$ boson, and~\cite{ATLAS:2012isv}, searching for a pair-produced exotic top (bottom) partner that decays 100\% of the time to $d,s,b$ quarks ($u,c$ quarks) and a $W$ boson. Searches for fourth-generation chiral bottom~\cite{ATLAS:2012iws,ATLAS:2012aw,CMS:2012jki,ATLAS:2012hpa} (top~\cite{CMS:2012ab,CMS:2012mir,ATLAS:2012qe}) quarks were performed in both ATLAS and CMS, assuming that the heavy quarks decay 100\% of the time to a top (bottom) quark and a $W$ boson.

In Refs.~\cite{Rao:2012sx,Rao:2012gf}, several searches for a chiral fourth-generation quark were reinterpreted as searches for a VLQ top quark. In these analyses, decays to $hq$ and $Zq$ were also included, as well as decays to $Wq$, without assuming that the branching ratio is $1$ for the third-generation quarks. 
A mass bound was obtained in~\cite{Rao:2012sx}, for the case approximating an $SU(2)$ singlet up quark, with
\begin{equation*}
    m_T > 419\;\unit{GeV}\quad {(95\%\text{ C.L.})}\,.
\end{equation*}

Searches similar to the one done in D0~\cite{D0:2010ckq} were performed with data from ATLAS Run 1~\cite{ATLAS:2011tvb,ATLAS:2012apa}. These searches looked for singly-produced degenerate doublet VLQs that decay predominantly to light quarks (first and second generations), which enhances single production without spoiling electroweak precision or meson mixing observables. This is possible due to some cancellations that stem from the fact that these VLQs are in two degenerate $SU(2)$ doublets, resembling the VLQs in the model of Ref.~\cite{Atre:2008iu}. 
These bounds cannot be extrapolated to the case of an $SU(2)$ singlet.

In Ref.~\cite{ATLAS:2012wxk} a search for a VLQ $SU(2)$ singlet bottom quark that decays 100\% of the time to a $Z$ boson and a bottom quark was performed.
Ref.~\cite{CMS:2011lcm} (Ref.~\cite{CMS:2012jwa}) contains a search for a pair-produced top (bottom) VLQ that decays 100\% of the time to the top (bottom) quark and a $Z$ boson. 

A singly-produced $SU(2)$ singlet top VLQ search was performed in Ref.~\cite{ATLAS:2016scx}. With the assumption that $\br(T \rightarrow W b)= 0.5$, which is more realistic and consistent with many VLQ singlet models, the excluded region was, at 95\% C.L.,
\begin{equation*}
|V_{Tb}|>0.42\,,~\textrm{for}~m_T < 700\;\unit{GeV} \,,
\end{equation*}
and
\begin{equation*}
    |V_{Tb}|>0.85\,,~\textrm{for}~m_T=1.2\;\unit{TeV}\,.
\end{equation*}

In Ref.~\cite{ATLAS:2015uaw} a search for pair-produced top (bottom) VLQ singlets with ``$SU(2)$ singlet BRs'' --- as defined in~\cref{eq:SU2BRs} --- was performed. Although these are much more realistic than the BRs used in previous studies, we repeat that they are not consistent with 
models where VLQs decay significantly
into the lighter generations. These decays are constrained by meson mixing and electroweak data but may still be sizeable~\cite{Atre:2008iu,Branco:2021vhs}. The 95\% C.L.~observed lower limits on the top (bottom) VLQ mass were 
\begin{equation*}
    m_T > 590\;\unit{GeV}\,,~ m_B > 620\;\unit{GeV}\,.
\end{equation*}
A similar analysis was presented in Ref.~\cite{ATLAS:2015vzd} for an $SU(2)$ singlet bottom-like VLQ, obtaining, at the same C.L.,~the mass bound
\begin{equation*}
    m_B > 640\;\unit{GeV}\,.
\end{equation*}

Searches for pair-produced top (bottom) VLQ singlets under the 
weaker assumption of~\cref{eq:3rdgen}
were also performed~\cite{ATLAS:2014vpn,ATLAS:2015ktd}. In Ref.~\cite{ATLAS:2015ktd}, the 95\% C.L.~observed lower limits on the top (bottom) VLQ mass are in a range between 715 and $950\;\unit{GeV}$ (575 and $813\;\unit{GeV}$) for all possible values of the branching ratios into the three decay modes. In Ref.~\cite{ATLAS:2014vpn}, the results are used to set lower mass limits of 
\begin{equation*}
     m_T > 655\;\unit{GeV} \,,~m_B > 685 \;\unit{GeV}\,,
\end{equation*}
at 95\% C.L., on vector-like $T$ ($B$) quarks, when assuming  ``$SU(2)$ singlet VLQ BRs''. 
Searches~\cite{ATLAS:2015lpr,CMS:2017asf} consider top or bottom VLQs, $Q=T,B$, that decay to $W,Z,h$ and the {light quarks} $q=u,d,s$. The results are given as a function of the branching ratios $\br(Q \rightarrow Wq)$ versus $\br(Q \rightarrow hq)$, with the branching ratio to $Zq$ fixed by the requirement $\br(Q\rightarrow Zq)= 1 - \br(Q \rightarrow Wq) - \br(Q \rightarrow hq)$. 
Note that this analysis considers VLQs that may decay to the lighter generations, which may be the case in general. Depending on the BRs, the exclusion region in~\cite{ATLAS:2015lpr} ranges from
\begin{equation*}
    m_{T,B} >340\;\unit{GeV}\quad \textrm{to}\quad m_{T,B} >690\;\unit{GeV} \quad {(95\%\text{ C.L.})}\,.
\end{equation*}
As for~\cite{CMS:2017asf}, the exclusion region ranges from
\begin{equation*}
  m_{T,B} >430\;\unit{GeV}\quad \textrm{to}\quad m_{T,B} >845\;\unit{GeV}\quad {(95\%\text{ C.L.})}\,.
\end{equation*}
A search for a pair-produced VLQ top (down) quark was performed in Refs.~\cite{Bhattacharya:2013poa,Majumder:2013jqa,CMS:2015lzl} (Ref.~\cite{CMS:2015hyy}), considering three decay modes: $h$, $Z$ and $W$. The expected 95\% C.L.~lower mass bounds are between 790 and $890\;\unit{GeV}$ (740 and $900\;\unit{GeV}$) depending on the branching fraction of the $T$ ($B$) quark. 
In Ref.~\cite{CMS:2015lzl}, the obtained mass limit for a up-type VLQ with, roughly, ``$SU(2)$ singlet BRs'' was
\begin{equation*}
    m_T > 740\;\unit{GeV} \quad {(95\%\text{ C.L.})}\,,
\end{equation*}
while for~\cite{Bhattacharya:2013poa,Majumder:2013jqa}, with ``$SU(2)$ singlet BRs'', a weaker bound was obtained,
\begin{equation*}
    m_T > 696\;\unit{GeV}\quad {(95\%\text{ C.L.})}\,.
\end{equation*}
In Ref.~\cite{CMS:2015hyy}, the obtained bound for a down-type VLQ with, roughly, ``$SU(2)$ singlet BRs'' was
\begin{equation*}
    m_B > 810\;\unit{GeV}\quad {(95\%\text{ C.L.})}\,.
\end{equation*}

Finally, a search for a pair-produced top VLQ was considered in Ref.~\cite{CMS:2015jwh}, focusing on the decay mode to Higgs and the top quark. If the $T$ quark decays exclusively to $ht$, the observed lower limit on the mass of the $T$ quark is $745\;\unit{GeV}$ (at $95\%$ C.L.). This exclusion is relevant in cases where the decay to Higgs and top dominates over the others.

\subsubsection{LHC Run 2}
\label{sec:LHC2}
In Run 2, the ATLAS and CMS experiments collected data from the LHC at $\sqrt{s} = 13\;\unit{TeV}$.

Searches for singly-produced top VLQ, $T$, were performed in~\cite{ATLAS:2018dyh,ATLAS:2018cjd} under the assumption of ``$SU(2)$ singlet VLQ BRs''. As it is a single production search, the value of $V_{Tb}$, the mixing matrix element that connects the bottom quark to the top VLQ, is of relevance, since the production is electroweak and proportional to the square of this parameter.
In Ref.~\cite{ATLAS:2018cjd}, the excluded region is 
\begin{equation*}
  m_T  < 1400\;\unit{GeV}\,,~\text{for}~ |V_{Tb}|>0.7\quad {(95\%\text{ C.L.})}\,.
\end{equation*}
In~\cite{ATLAS:2018dyh}, the allowed region {(at 95\% C.L.)} is between
\begin{equation*}
   m_T = 800\;\unit{GeV}\,,~\text{for}~ |V_{Tb}|<0.18\,,
\end{equation*}
and
\begin{equation*}
 m_T =1200\;\unit{GeV}\,,~\text{for}~ |V_{Tb}|<0.35 \,.
\end{equation*}

A search for a pair-produced top or bottom VLQ decaying to a $Z$ boson and a third-generation quark was performed in~\cite{ATLAS:2017vdo,ATLAS:2018tnt,Vale:2018bpf,ATLAS:2022hnn}, assuming ``$SU(2)$ singlet VLQ BRs''. In~\cite{ATLAS:2018tnt,Vale:2018bpf}, the 95\% C.L.~obtained mass bounds were
\begin{equation*}
     m_T > 1030\;\unit{GeV}\,,~ m_B >1010\;\unit{GeV}\,,
\end{equation*}
while in~\cite{ATLAS:2017vdo}, for a top VLQ, it was
\begin{equation*}
     m_T > 870\;\unit{GeV}\quad {(95\%\text{ C.L.})}\,,
\end{equation*}
and in~\cite{ATLAS:2022hnn}, for also a bottom VLQ, they were
\begin{equation*}
m_T > 1270\;\unit{GeV}\,,~ m_B >1200\;\unit{GeV}\quad {(95\%\text{ C.L.})}\,.
\end{equation*}

A search~\cite{CMS:2017gsh} for a singly-produced (via $W$ boson interactions) top (bottom) VLQ decaying to a $Z$ boson and third-generation quarks was performed, assuming ``$SU(2)$ singlet VLQ BRs''.
The excluded region at 95\% C.L.~is
\begin{equation*}
    m_{T,B} \in [700,1700]\;\unit{GeV}\,,~\text{for}~|V_{Tb}|=|V_{tB}|=0.5\,.
\end{equation*}

A search~\cite{CMS:2016edj,CMS:2016jce} for a singly-produced top VLQ, that decays to a $h$ boson and third-generation quarks was carried out, assuming ``$SU(2)$ singlet VLQ BRs''. In~\cite{CMS:2016jce}, the excluded region at 95\% C.L.~is
\begin{equation*}
    m_T \in [1000,1800]\;\unit{GeV}\,,~\text{for}~|V_{Tb}|=0.5\,,
\end{equation*}
 where the production is via $W$ boson interactions.
In~\cite{CMS:2016edj}, the {excluded} region at 95\% C.L.~is
\begin{equation*}
    m_T \in [700,1800]\;\unit{GeV}\,,~\text{for}~V_{Tb}=F_{tT}=0.5\,,
\end{equation*}
and the production is via a $W$ or $Z$ boson interaction. 

Searches for a pair-produced top or bottom VLQ decaying to a $W$, $Z$ or $h$ boson and third-generation quarks were performed in~\cite{CMS:2017ked,ATLAS:2018cye,CMS:2018zkf,ATLAS:2018alq,Nikiforou:2018geu,CMS:2018wpl,CMS:2019eqb}, for ``$SU(2)$ singlet VLQ BRs''. 
In~\cite{Nikiforou:2018geu}, the 95\% C.L.~lower bounds
\begin{equation*}
m_T > 1170\;\unit{GeV}
     \,,~ m_B >1080\;\unit{GeV}
\end{equation*}
were obtained, while in~\cite{ATLAS:2018alq} the bounds were
\begin{equation*}
     m_T > 980\;\unit{GeV}\,,~ 
     m_B >1000\;\unit{GeV}
     \quad {(95\%\text{ C.L.})}\,,
\end{equation*}
both improving on the previous bound of Ref.~\cite{CMS:2017ked},
\begin{equation*}
    m_T > 860\;\unit{GeV} \quad {(95\%\text{ C.L.})}\,,
\end{equation*}
for a top VLQ.
Considering other possible branching fraction combinations for $T$ quarks, but still assuming~\cref{eq:3rdgen}, lower bounds on $m_T$ can be set in a range from $790$ to $940$ GeV, for combinations with $\br(T \rightarrow th) + \br(T \rightarrow Wb) \geq 0.4$. 
Limits are also set on the pair production of a bottom VLQ, which can be excluded for masses $m_B$ up to $730\;\unit{GeV}$.
In~\cite{ATLAS:2018cye}, the 95\% C.L.~lower bound
\begin{equation*}
    m_T > 1190\;\unit{GeV}
\end{equation*}
 was obtained, while in~\cite{CMS:2018wpl} the 95\% C.L.~lower mass bounds were  
\begin{equation*}
m_T > 1095\;\unit{GeV}\,,~
     m_B >955\;\unit{GeV} \,,
\end{equation*}
and in~\cite{CMS:2019eqb} the 95\% C.L.~bounds were  
\begin{equation*}
m_T > 960-980\;\unit{GeV}\,,~
     m_B >890-910\;\unit{GeV}\,.
\end{equation*}
Further, in~\cite{CMS:2018zkf}, the 95\% C.L.~lower bounds were
\begin{equation*}
m_T > 1200\;\unit{GeV} \,, ~
     m_B >1170\;\unit{GeV} \,.
\end{equation*}

A search for a top (bottom) VLQ decaying to a $W$ boson and a third-generation quark was performed in Ref.~\cite{ATLAS:2017nap} (Refs.~\cite{ATLAS:2017nap,ATLAS:2018mpo}), assuming ``$SU(2)$ singlet VLQ BRs''. In~\cite{ATLAS:2017nap}, the 95\% C.L.~lower mass bound obtained were
\begin{equation*}
    m_T > 1170\;\unit{GeV}\,,~ m_B > 1080\;\unit{GeV}\,,
\end{equation*}
while in~\cite{ATLAS:2018mpo}, for the down-type VLQ, one finds
\begin{equation*}
    m_B > 1170\;\unit{GeV}\,. 
\end{equation*}

A search for a singly-produced up VLQ that decays to $Z$ or $h$ bosons and third-generation quarks was performed in Ref.~\cite{CMS:2019afi}, assuming ``$SU(2)$ singlet VLQ BRs''. The excluded region was
\begin{equation*}
    m_T \in [700,1000]\;\unit{GeV}\,.
\end{equation*}

In Ref.~\cite{Vale:thesis} a pair production search for a top (bottom) VLQ that decays into $Zt$ ($Zb$) was performed, assuming ``$SU(2)$ singlet VLQ BRs'' (decays only to third-generation quarks). This is one of the first searches that uses the complete LHC Run 2 dataset, amounting to $139\;\unit{fb}^{-1}$. The 95\% C.L.~lower mass bounds were
\begin{equation*}
     m_T > 1270\;\unit{GeV}\,, ~
     m_B >1200\;\unit{GeV}
     \,.
\end{equation*}

A search for a singly-produced down-type VLQ that decays to $W$ bosons and a top quark was performed in Ref.~\cite{CMS:2018dcw}. Depending on the VLQ type, coupling, and decay width to $Wt$,
mass bounds up to $1660\;\unit{GeV}$ were obtained. 

A search for a pair-produced down-type VLQ decaying to a $h$ boson and a bottom quark was performed in Ref.~\cite{CMS:2018kcw}. The results of this search were upper
limits on the product of the cross section and the branching fraction of the $B$ quark, considering masses extending from $700$ to $1800\;\unit{GeV}$. This search is performed under the hypothesis of a singlet or doublet $B$ quark of
narrow width decaying to $hb$ with a branching fraction of approximately 25\%.

More recent analyses include searches for pair-produced singlet VLQs~\cite{CMS:2022fck,ATLAS:2022tla}, for ``$SU(2)$ singlet BRs'', resulting in the lower bounds~\cite{CMS:2022fck}
\begin{equation*}
m_T > 1480\;\unit{GeV}\,,~ m_B >1470\;\unit{GeV}\quad {(95\%\text{ C.L.})}\,,
\end{equation*}
and~\cite{ATLAS:2022tla}
\begin{equation*}
m_T > 1260\;\unit{GeV}\,,~ m_B >1330\;\unit{GeV}\quad {(95\%\text{ C.L.})}\,,
\end{equation*}
as well as searches for a singly-produced up-type VLQ~\cite{ATLAS:2022ozf,CMS:2023agg,ATLAS:2023uah,ATLAS:2023pja}. 
The latter exclude
\begin{equation*}
|V_{Tb}|>0.25\,,~\textrm{for}~m_T < 1400\;\unit{GeV}\quad {(95\%\text{ C.L.})}\,,
\end{equation*}
under the sole assumption that $\br(T\to ht) = 0.25$~\cite{ATLAS:2022ozf}, while for ``$SU(2)$ singlet BRs'' one excludes masses~\cite{CMS:2023agg}
\begin{equation*}
m_T < 960\;\unit{GeV}\,,~\textrm{for}~V_{Tb}= 0.18 \quad {(95\%\text{ C.L.})}\,,
\end{equation*}
and couplings~\cite{ATLAS:2023uah}
\begin{equation*}
|V_{Tb}|>0.45\,,~\textrm{for}~m_T < 1975\;\unit{GeV}\quad {(95\%\text{ C.L.})}\,,
\end{equation*}
and~\cite{ATLAS:2023pja}
\begin{equation*}
|V_{Tb}|>0.4\,,~\textrm{for}~m_T < 2300\;\unit{GeV}\quad {(95\%\text{ C.L.})}\,.
\end{equation*}

\subsection{Future search prospects}
\label{sec:future}

Most of the recent searches provide results for a range of BRs
without the restrictions of~\cref{eq:SU2BRs,eq:3rdgen}, i.e.~of assuming ``$SU(2)$ singlet VLQ BRs'' or decays exclusively to third-generation quarks. 
Nevertheless, if a VLQ decays predominantly to lighter generations, it may not be seen in a search focused on the phenomenology of a VLQ that decays to third-generation quarks, as its BRs may be very small in the channels used in this search. Searches for VLQs that decay predominantly to lighter generations (as in~\cite{ATLAS:2015lpr,CMS:2017asf}) are thus complementary to those where VLQs are assumed to decay predominantly to third-generation quarks.

According to~\cite{Liu:2016jho}, in the High-Lumi LHC, a $5\sigma$ discovery reach for the single production of an up-VLQ with mass $m_T\sim 1\;\unit{TeV}$ is attainable for couplings to SM quarks $q$ as low as $|V_{Tq}| \sim 0.2$.
For a pair-produced down-type VLQ, the $5\sigma$ discovery reach is possible for masses around $730\;\unit{GeV}$~\cite{Paul:2020mul}. Scenarios in which an up-type VLQ couples only to first-generation quarks may also be probed~\cite{Cui:2022hjg}.

For the FCC, it is argued~\cite{Paul:2020mul} that a $5\sigma$ discovery reach for a pair-produced down-type VLQ can be increased up to $2980\;\unit{GeV}$, considering only the $4l + 2j$ decay channel. It is also shown that the FCC-hh generically requires about two orders of magnitude less integrated luminosity than High-Lumi LHC for the discovery of a down-type VLQ, at a given mass. For $V_{Tb}\sim 0.3$, up-type VLQs with masses up to $2.1\;\unit{TeV}$ can be probed at the $5\sigma$ level, at the FCC-hh, through single production followed by a decay to $ht$, while 95\% C.L.~exclusion limits can reach $2.6\;\unit{TeV}$~\cite{Tian:2021nmj}.

The prospects for CLIC are also very positive, providing an opportunity to search for VLQs via electron-positron interactions. 
In Ref.~\cite{Han:2021lpg}, one can find a study for the $3\;\unit{TeV}$ CLIC with an integrated luminosity of $5\,\unit{ab}^{-1}$, for a singly-produced down-type VLQ that decays exclusively to third-generation quarks. The VLQ is produced in the process $e e^+ \rightarrow \overline{B}b$ and decays via $B \rightarrow Zb$.
For couplings to the third-generation $|V_{tB}| \sim 0.3$, the discovery region is found to be $m_B \in [1200,1900]\;\unit{GeV}$.

Finally, in~\cite{AlAli:2021let} it is argued
how a future high-energy muon collider may bring about several advantages compared to other projects being considered.
Specifically, as pointed out in appendix A.3 therein, the production cross section of a TeV-mass VLQ can be significantly enhanced when the VLQ is produced via $\mu^-\mu^+$ annihilation, for $\sqrt{s}<100\;\unit{TeV}$.

\vfill
\clearpage

\section{Summary}

We have presented a systematic review of the main features of vector-like quarks (VLQs), i.e.~quarks with the right-handed and left-handed components having the same transformation properties under the SM gauge group. VLQs are one of the simplest additions to the SM, with automatic anomaly cancellation. They can bring about interesting and in principle testable new-physics effects that may be detected in the next round of experiments at LHC and High-Lumi LHC. VLQs may also populate the eventual desert between the electroweak scale and the Planck scale, without worsening the hierarchy problem. 

We focused on the simplest case, which is that of isosinglet VLQs. We considered a general and exact parameterization for the physical matrices in VLQ models, from which the most common approximations found in the literature can be derived, allowing also for the study of other limits. We pointed out the analogy between VLQs and sterile neutrinos in the seesaw context. We studied weak basis transformations involving VLQs and discussed the new CP-odd invariants which are relevant in their presence. We presented a careful treatment of the deviations of unitarity which arise in the presence of VLQs, making use of the above parameterization. VLQs may provide a solution to the unitarity anomaly in the first row of the CKM matrix, which may be seen as a possible hint for their existence. Generic consequences on flavour and electroweak observables were reviewed and detailed expressions were collected. Particular attention was given to the main phenomenological features for the case of a single additional VLQ (either up or down). Current bounds coming from collider searches were reviewed, together with the assumptions involved.

We emphasized through specific examples that VLQs may be easily used as messengers of CP violation to the SM through their mixing with SM quarks, thereby being important ingredients to models of spontaneous CP violation. This kind of models provides a simple solution to the strong CP problem, based on spontaneous CP violation, which does not require axions. In its simplest incarnation, only one complex singlet scalar needs to be added besides a VLQ. With the addition of right-handed neutrinos, it is even possible to allow for a common origin of all CP violations, namely CP violation in the quark and lepton sectors as well as the CP violation needed to generate the baryon asymmetry of the universe through leptogenesis. 
VLQs may leave possible imprints on low-energy observables. For the spontaneous CP violation models discussed in this review, VLQs act as the messengers of CP violation, transmitting it to the SM. 
Nevertheless, VLQ models generically lead to flavour-changing neutral currents and deviations from unitarity which are naturally suppressed by the ratio of SM-quark and VLQ masses. In the framework of effective field theories, these suppressed effects arise as dimension-6 effective operators that are induced by the introduction of VLQs. We reviewed the tree-level matching and other effects such as operator mixing were also briefly discussed.

All these features of VLQs motivate an intense search programme. Their main feature --- their mixing with SM quarks being suppressed by their heavy masses --- makes them a simple and well-motivated ingredient that can be added to the SM, potentially solving other problems. This also implies they may easily conceal themselves, making the search difficult. This search is nevertheless worth the effort.

\section*{Acknowledgements}

We would like to thank M.~Pérez-Victoria for reading the manuscript and for valuable comments and suggestions.
We thank F.J.~Botella for numerous and interesting conversations and we are indebted to him and to J.F.~Bastos for helping improve~\cref{sec:pheno}.
We also thank J.A.~Aguilar-Saavedra, R.~Benbrik, S.~Heinemeyer and M.~Pérez-Victoria
for allowing us to reproduce a figure from their work~\cite{Aguilar-Saavedra:2013qpa}. We thank A.~Onofre, N.~Castro and N.~Leonardo for useful discussions.
A.C.~acknowledges support from National Council for Scientific and Technological Development – CNPq through projects 166523\slash2020-8 and 201013\slash2022-3. 
C.C.N.~acknowledges partial support by Brazilian Fapesp, grant 2014/19164-6, and
CNPq, grant 312866/2022-4. 
This work was partially supported by Fundação para a Ciência e a Tecnologia (FCT, Portugal) through the projects 
No.~PTDC/FISPAR/29436/2017,
No.~CERN/FIS-PAR/0004/2019,
No.~CERN/FIS-PAR/0008/2019,
No.~CERN/FIS-PAR/
\allowbreak 0002/2021,
No.~CERN/FIS-PAR/0019/2021,
and CFTP-FCT Unit No.~UIDB/00777/2020 and No.~UIDP/00777/2020, 
which are partially funded through Programa Operacional Ciência Tecnologia Inovação (POCTI) (Fundo Europeu de Desenvolvimento Regional (FEDER)), Programa Operacional Competitividade e Internacionalização (COMPETE), Quadro de Referência Estratégica Nacional (QREN) and European Union (EU). 
J.M.A.~acknowledges support from FCT through the PhD grant SFRH/BD/139937/2018.
P.M.F.P.
acknowledges support from FCT through the PhD grant SFRH/BD/145399/2019.

\vfill
\clearpage

\appendix

\section{VLQs from UV completions}
\label{app:UV}
\subsection{VLQs and GUTs}
While a detailed analysis of \gls*{UV} complete models is beyond our scope, here we comment on some potential UV motivations for the addition of VLQs. As stated in the introduction, VLQs may arise in the context of \glspl*{GUT}, as is the case of models based on the exceptional simple Lie group $E_6$ (see e.g.~\cite{Gursey:1975ki,Gursey:1981kf,Hewett:1988xc}).
In the $E_6$ GUT, each SM family of left-handed fermions is embedded in the $\mathbf{27}$ irreducible  representation, which under the SM gauge group decomposes as (see also~\cite{Branco:1999fs})
\begin{equation}
\begin{aligned}
    \mathbf{27} \,\leadsto\,\, &(\mathbf{3},\mathbf{2})_{1/6} \,\oplus\,  (\mathbf{\bar{3}},\mathbf{1})_{-2/3} \,\oplus\,  (\mathbf{\bar{3}},\mathbf{1})_{1/3}
    \,\oplus\, 
    (\mathbf{1},\mathbf{2})_{-1/2} \,\oplus\, (\mathbf{1},\mathbf{1})_1 \\
\,\oplus\,\,\, &(\mathbf{3}, \mathbf{1})_{-1/3} \,\oplus\, (\mathbf{\bar{3}}, \mathbf{1})_{1/3} \,\oplus\, (\mathbf{1},\mathbf{2})_{1/2} \,\oplus\, (\mathbf{1},\mathbf{2})_{-1/2} \,\oplus\, (\mathbf{1},\mathbf{1})_0 \,\oplus\, (\mathbf{1},\mathbf{1})_0\,.
\end{aligned}
\end{equation}
Apart from the usual 15 SM chiral fields (first line), one sees that a VLQ isosinglet down-type quark, a vector-like isodoublet of leptons and two isosinglet neutrinos are also present, for each generation of fermions.

While the masses of GUT-derived VLQs are not necessarily close to the TeV scale, scenarios can be devised where they are light and populate the desert between the electroweak and unification scales~\cite{Bowick:1981zx,Ramond:1981jx}.
For instance, in the $SU(6)$ GUT model of Ref.~\cite{Dutta:2016ach} ($SU(6) \subset E_6$), VLQs along with SM fermions are part of the anomaly-free representation $\mathbf{15} \,\oplus\, \mathbf{\bar{6}} \,\oplus\, \mathbf{\bar{6}'}$.  
In this context, $SU(6)$ breaks down to the group $SU(3)_c \times SU(2)_L \times U(1)_Y \times U(1)_X$ at the GUT scale, while masses for the vector-like fermions are generated at the TeV scale, following the breaking of the $U(1)_X$ factor. 
A similar separation of scales may be achieved for one or two generations of down-type VLQs in
the framework of an anomaly-free supersymmetric chiral $E_6 \times SU(2)_F \times U(1)_F$ GUT containing the SM~\cite{Morais:2020odg,Morais:2020ypd}.
Finally, vector-like matter at intermediate scales may assist gauge coupling unification~\cite{Giudice:2012zp} (see also~\cite{Gogoladze:2010in}). In Ref.~\cite{Lee:2016wiy}, a model based on $SU(5) \times SU(5)$ relies on the properties of isosinglet VLQs --- at the TeV and at an intermediate scale --- to suppress proton decay and achieve unification. The $SO(10)$-based models of Ref.~\cite{Fukuyama:2019zun} consider instead TeV-scale isosinglet and doublet VLQs as relevant for the successful unification of gauge couplings.

\subsection{VLQs and extra dimensions}

Vector-like fermions are an integral part of of some of the most prominent models of extra dimensions, such as the famous \gls*{RS} warped geometry models~\cite{Randall:1999ee}. We focus on an RS-based model as developed in Ref.~\cite{Moreau:2006np} (see references therein for further detail and other ED
models; see also~\cite{delAguila:2000kb}). In these ED models with just an extra fifth
dimension --- described by the coordinate $y$ --- it is impossible to assign
different group representations to left- and right-handed chiralities.
Keeping e.g.~the same electroweak symmetry group $SU(2)_L\times U(1)_Y$ as in the SM, we now
have a full quark doublet $q(x^{\mu },y)$ (both chiralities), instead of only the usual SM
left-handed doublet $q_{L}(x^{\mu })$. The same applies
to the right-handed quarks: full up-quark singlets $u(x^{\mu
},y)$ and down-quark singlets $d(x^{\mu },y)$ must be present, instead of the SM right-handed $u_{R}(x^{\mu })$ and $d_{R}(x^{\mu })$.

Due to the compactification of the fifth spacial dimension, each field
becomes associated with a tower of \gls*{KK}
modes. Crucially, when integrating out the fifth
dimension to obtain an effective 4-dimensional action, only the right-handed part of the zero mode of $u(x^{\mu },y)$ will survive within each family, remaining
massless (before EWSB). The same applies to the down-type
quarks. Thus, one effectively obtains a right-handed up-quark field $u_{R}^{(u_{0})}(x^{\mu })$, which plays the role of the usual SM singlet.
The other (higher) modes, $u^{(u_{1})}(x^{\mu })$, $u^{(u_{2})}(x^{\mu })$, $u^{(u_{3})}(x^{\mu }), \ldots$ acquire large vector-like masses, which can be of $\mathcal{O}(\text{TeV})$, and correspond to vector-like quarks. In other words, we will have the higher-order singlets $u_{R}^{(u_{1})}(x^{\mu })$ and $u_{L}^{(u_{1})}(x^{\mu })$ which pair up (and so on, for all higher orders).
Following a similar KK reduction for the $SU(2)$ doublet quark $q(x^{\mu },y)$, and after EWSB triggered by the VEV of a Higgs scalar, we obtain a left-handed up-quark zero mode $u_{L}^{(q_{0})}(x^{\mu })$ and a left-handed down-quark $d_{L}^{(q_{0})}(x^{\mu })$,
corresponding to the usual left-handed SM quarks. We also obtain an infinite tower of
left- and right-handed KK massive excited states, $u_{L}^{(q_{n})}(x^{\mu })$
and $u_{R}^{(q_{n})}(x^{\mu })$ in the up sector.
Schematically, the mass matrix for the up-type quarks is then of the form
\begin{equation} \label{eq:mED}
\begin{pmatrix}
    \,\overline{ u_{L}^{(q_{0})} } & \overline{ u_{L}^{(q_{1})} } & \overline{ u_{L}^{(u_{1})} } & \ldots\,{ }
\end{pmatrix}
\begin{pmatrix}
m^{(0,0)} & 0 & m^{(0,1)} & \ldots\,{ } \\[2mm]
m^{(1,0)} & M^{q_{1}} & m^{(1,1)} & \ldots\,{ } \\[2mm]
0 & 0 & M^{u_{1}} & \ldots\,{ } \\[2mm]
\vdots & \vdots & \vdots & \ddots\,{ }
\end{pmatrix}
\begin{pmatrix}
u_{R}^{(u_{0})} \\[2mm]
u_{R}^{(q_{1})} \\[2mm]
u_{R}^{(u_{1})} \\[2mm]
\vdots
\end{pmatrix}
\,,
\end{equation}
with a similar structure for the down quark sector. Note that the $m^{(i,j)}$ are proportional to the Higgs VEV and, like the vector-like masses $M^\psi$, may have a non-trivial structure in flavour space. Other structures are possible, depending on the details of the ED model under consideration.

\section{Improved Seesaw Expansion}
\label{app:improved_seesaw}

\providecommand{\tT}{\tilde{\Theta}}
\providecommand{\heff}{h_{\rm eff}}
\providecommand{\theff}{\tilde{h}_{\rm eff}}
\providecommand{\Heff}{H_{\rm eff}}
\providecommand{\tHeff}{\tilde{H}_{\rm eff}}

Here we connect the notation described in~\cref{sec:notation} and the exact parameterization presented in~\cref{sec:exact_param_X} with the quark seesaw expansion which is also frequently used. We develop an improved block-diagonalization procedure modifying Ref.~\cite{Grimus:2000vj}. The resulting series will be simpler and the expressions more compact.%
\footnote{The idea for this improved expansion can be already found in Ref.~\cite{Korner:1992zk} for the case of seesaw neutrinos.}

Let us repeat the mass matrix~\eqref{eq:genmass} in the left-right basis $\bar{\psi}_L\psi_R$ as
\begin{align}
\renewcommand{\arraystretch}{1}
\setlength{\extrarowheight}{6pt}
\cM \,=\,
\left(\begin{array}{c;{2pt/2pt}c}
m & \,\om
\\[1.5mm] \hdashline[2pt/2pt]
\oM & \,M\\[1mm]
\end{array}\right) 
=\mtrx{
\cM_m \cr \cM_M
}
\,,
\label{eq:genmass:ap}
\end{align}
where $m\sim 3\times 3$ and $M\sim n\times n$.
We suppress the labels $d,u$ for simplicity.
For singlet VLQs, one can always choose 
\eq{
\label{eq:Mbar0}
\oM=0 \text{~~and~~}
\cM_M=\mtrx{0 & M}
}
by redefining $\psi_R$.%
\footnote{For doublet VLQ, one can always choose $\om=0$ with a different hierarchy.}
In this basis $M$ is non-singular. We assume $\cM_M\cM_M^\dag\gg \cM_m\cM_m^\dag$ so that a seesaw expansion in $\eps\sim\cM_m/\cM_M$ will be possible. Therefore,
\eq{
\cM\cM^\dag=
\mtrx{\cM_m\cM_m^\dag & \cM_m\cM_M^\dag
\cr \cM_M\cM_m^\dag & \cM_M\cM_M^\dag}
\sim \mtrx{\eps^2 & \eps \cr \eps & 1}
\,.
}

\subsubsection*{Block diagonalization}
The block diagonalization of $\cM\cM^\dag$ can be performed as
\eq{
\label{block.diag:ap}
\cV_1^\dag \cM\cM^\dag \cV_1=\mtrx{\heff & 0 \cr 0& \Heff}\,,
}
where, differently from~\cref{eq:diagonalization}, the matrices $\heff$ and $\Heff$ are not diagonal in general. One interesting parameterization for the unitary matrix $\cV_1$ is~\cite{Grimus:2000vj,Hettmansperger:2011bt}:
\eq{
\label{block.V:grimus}
\cV_1=\mtrx{\sqrt{\id_3-\tT \tT^\dag} & \tT \cr -\tT^\dag & \sqrt{\id_n-\tT^\dag \tT}}\,,
}
where $\id_n$ is the identity of size $n$.
The imposition of zeros in~\cref{block.diag:ap} allows the systematic calculation of $\tT$ in powers of $\eps$~\cite{Grimus:2000vj}: $\tT=\tT_1+\tT_3+\dots$, where the label denotes the order in $\eps$ and only odd powers are necessary.

However, the outlined block-diagonalization procedure is cumbersome beyond the order of $\tT_3$ because of the square root function in~\eqref{block.V:grimus}. We can avoid this by using the parameterization
\eq{
\label{block.V:improved}
\cV_2=\mtrx{\id_3 & \Theta\cr -\Theta^\dag & \id_n}
\mtrx{\big[\id_3+\Theta\Theta^\dag\big]^{-1/2} & 0 \cr 0 & \big[\id_n+ \Theta^\dag\Theta\big]^{-1/2}}\,,
}
which is very similar to~\cref{eq:uparam-exact} except for the order of the matrices. One can check that $\cV_2^\dag\cV_2=\id_{3+n}$ and this parameterization is equally exact independently of the possibility of series expansion. By comparing~\eqref{block.V:improved} with~\eqref{block.V:grimus} one can relate $\Theta$ and $\tT$ if necessary.
Note that throughout the text, e.g.~in~\cref{sec:devunit}, $\Theta$ refers only to the leading order expansion, here denoted as $\Theta_1$,%
\footnote{To be precise, $\Theta$ in the text might also differ from $\Theta_1$ here by additional unitary rotations from right or left (cf.~\cref{eq:XvsR}).} 
while $\Theta$ here will be reserved to the exact matrix as defined above.

Let us denote the first matrix in~\eqref{block.V:improved} as $\cV_0$. Beware that it is not a unitary matrix. Then, as far as block diagonalization is concerned, only $\cV_0$ is necessary; the second piece in~\eqref{block.V:improved} only modifies the diagonalized blocks, restoring unitarity in $\cV_2$.

Imposing the zeros in~\eqref{block.diag:ap}, with $\cV_1$ replaced by $\cV_0$, we obtain the condition
\eq{
\cM_m\cM_m^\dag \Theta+\cM_m\cM_M^\dag-\Theta \cM_M\cM_m^\dag \Theta - \Theta\cM_M\cM_M^\dag=0\,,
}
which defines $\Theta$ exactly in terms of $\cM$. 
Rewriting it as 
\eq{
\label{F:iterate}
\Theta=\cM_m\cM_M^\dag S+(\cM_m-\Theta\cM_M)\cM_m^\dag \Theta S\,,
}
where
\eq{
\label{def:S}
S=(\cM_M\cM_M^\dag)^{-1}=(MM^\dag)^{-1}\,,
}
we can see that the first term is order $\eps$ while the second term is at least order $\eps^3$.
Note that $S=S^\dagger$.
In the last equality of~\eqref{def:S}, we write the result assuming the weak basis~\eqref{eq:Mbar0}.
The convention of specifying the result in this WB in the second equality will be used in what follows.

\Cref{F:iterate}, which has no square roots, can be used iteratively to determine $\Theta =\Theta_1+\Theta_3+\dots$ as 
\eqali{
\label{Fi:improved}
\Theta_1&=\cM_m\cM_M^\dag S=\om M^{-1}\,,
\cr
\Theta_3&=(\cM_m-\Theta_1\cM_M)\cM_m^\dag \Theta_1S
\cr
&=mm^\dag\om M^{-1}S\,,
\cr
\Theta_5&=(\cM_m-\Theta_1\cM_M)\cM_m^\dag \Theta_3S
\cr &\quad - \Theta_3\cM_M\cM_m^\dag \Theta_1S
\cr
&=mm^\dag[mm^\dag \Theta_1S-\Theta_1\Theta_1^\dag \Theta_1]S\,.
}
For example, $\Theta_1$ is obtained from~\eqref{F:iterate} by taking $\Theta = 0$ in the right-hand side,
$\Theta_3$ is obtained by taking $\Theta=\Theta_1$ and isolating to order $\eps^3$ and so on.
Numerically, one can apply~\eqref{F:iterate} iteratively at any desired order.

Block diagonalization is then achieved as~\eqref{block.diag:ap}, with $\cV_2$ in place of $\cV_1$.
In the block matrices, it is useful to separate the square root contribution in $\cV_2$ as
\eqali{
\label{D-Dtilde}
\heff&=[\id_3+\Theta\Theta^\dag]^{-1/2}\theff[\id_3+\Theta\Theta^\dag]^{-1/2}\,,\cr
\Heff&=[\id_n+\Theta^\dag\Theta]^{-1/2}\tHeff[\id_n+\Theta^\dag\Theta]^{-1/2}\,.
}
The central pieces $\theff,\tHeff$ are the non-zero blocks in the block-diagonal product $\cV_0^\dag \cM\cM^\dag \cV_0$. They do not depend on square roots and can be determined from
\eqali{
\label{eq:tDmM}
\theff&= \cM_m\cM_m^\dag+\Theta\cM_M\cM_M^\dag \Theta^\dag-(\Theta\cM_M\cM_m^\dag+ \text{h.c.})\,,
\\
\tHeff&= \cM_M\cM_M^\dag + (\cM_M\cM_m^\dag \Theta + \text{h.c.})+\Theta ^\dag \cM_m\cM_m^\dag \Theta \,.
}
The first equation has all terms at order $\eps^2$ in leading order while the second equation has terms of varying order starting with order $\eps^0$.
If we expand $\theff=\theff^{(0)}+\theff^{(2)}+\theff^{(4)}+\cdots$, with labels denoting the order of \emph{correction} in $\eps$, we can systematically calculate each term collecting the terms in $\Theta$. 
We use a similar notation for $\tHeff$.
The first terms read 
\eqali{
\label{ap:tD.exp}
\theff^{(0)}&=\cM_m[\id_{3+n}-\cM_M^\dag S\cM_M]\cM_m^\dag
\cr
&=mm^\dag\,,
\cr 
\theff^{(2)} 
&=(\Theta_1\cM_M-\cM_m)\cM_M^\dag \Theta_3^\dag+ \text{h.c.}
\cr
&=0\,,
\cr 
\tHeff^{(0)}&=\cM_M\cM_M^\dag=MM^\dag\,,
\cr
\tHeff^{(2)}&=\cM_M\cM_m^\dag \Theta_1+ \text{h.c.}
\cr
&=M\om^\dag\om M^{-1}+ \text{h.c.}\,.
}
Note that only corrections in even powers of $\eps$ arise and the subleading correction to $\theff$ vanishes in the weak basis~\eqref{eq:Mbar0}.

Now, we can easily write the subleading corrections for the block-diagonalized pieces \eqref{D-Dtilde}~\cite{Grimus:2000vj}:
\eqali{
\label{ap:D.exp}
\heff^{(2)} 
&=-\ums{2}\heff^{(0)}\Theta_1\Theta_1^\dag+ \text{h.c.}
\cr
&=-\ums{2}mm^\dag\om S\om^\dag + \text{h.c.}
\cr
\Heff^{(2)}&=+\ums{2}\Heff^{(0)}\Theta_1^\dag \Theta_1 + \text{h.c.}\,,
\cr
&=+\ums{2}M\om^\dag\om M^{-1}+ \text{h.c.}\,.
}
The leading terms are unchanged: $\heff^{(0)}=\theff^{(0)}$, $\Heff^{(0)}=\tHeff^{(0)}$.
Higher order corrections can be easily obtained if needed.
Note that $\heff$ in~\cref{sec:light-eff} denotes only the leading order $\heff^{(0)}=\theff^{(0)}$.

\subsubsection*{Nelson-Barr VLQs}
Instead of adhering to the WB choice of~\cref{eq:Mbar0}, one may consider the specific case, relevant for VLQs of Nelson-Barr type, where
\eq{
\om=0 \text{~~and~~} \cM_m=\mtrx{m&0}\,,
}
while $\oM$ is non-zero.
In this case,~\cref{def:S,Fi:improved,ap:D.exp} can be fully adapted with 
\eqali{
\Theta_1&\,\to\, m \oM\mkern2mu^\dag S\,,
\cr
S&\,\to\, (MM^\dag +\oM\:\oM\mkern2mu^\dag)^{-1}\,,
\cr
\heff^{(0)}&\,\to\, m[\id_3-\oM\mkern2mu^\dag S \oM]m^\dag\,,
\cr
\Heff^{(0)}&\,\to\,  MM^\dag+\oM\:\oM\mkern2mu^\dag\,.
}

\subsubsection*{Full diagonalization of the Hermitian matrix}

To complete the full diagonalization as in~\eqref{eq:diagonalization}, and not just the block diagonalization of~\eqref{block.diag:ap} (with $\cV_1 \to \cV_2$),
we need to include additional unitary matrices as
\eq{
\label{R.oR}
\cV=\cV_2\mtrx{R &\cr &\oR}\,,
}
where $R,\,\oR$ diagonalize $\heff,\Heff$, respectively.
Comparing to the exact parameterization in~\cref{sec:exact_param_X}, we obtain the exact relation
\eq{
\label{eq:XvsR}
X=R^\dag\,\Theta\, \oR\,,
}
where we can identify $R=V_K,\,\oR=V_{\oK}$ in~\eqref{eq:uparam-exactK}.
So these two exact parameterizations --- the one in~\eqref{eq:uparam-exact} and the other in~\eqref{R.oR} --- are equivalent. The difference is in the position of the block unitary matrix which makes it easier to consider the exact diagonalized masses for the former while the latter is more suited to consider the original Lagrangian parameters and write the expansion if needed.

\subsubsection*{Right-handed rotation and block diagonalization}

For completeness, we also list the necessary expansion for $\cV_R$: 
\eqali{
\label{F:VR}
\Theta_2&=m^\dag\om(M^\dag M)^{-1}\,,
\cr
\Theta_4&=m^\dag mm^\dag\om(M^\dag M)^{-2}-m^\dag\om(M^\dag M)^{-1}\om^\dag\om(M^\dag M)^{-1}\,.
}
These can be obtained by applying the previous formulae to $\cM^\dag\cM$.
We use the same notation as before but we just list the results in the WB~\eqref{eq:Mbar0} since otherwise the seesaw expansion would not be valid --- all entries of $\cM^\dag\cM$ would be of the same order. Because of the different hierarchy, the powers in $\eps$ change and we have $\Theta=\Theta_2+\Theta_4+\dots$.
The blocks along the diagonal of the new $\cV_2^\dag \cM^\dag\cM \cV_2$ can be expanded according to
\eqali{
\heff^{(0)}&=m^\dag m\,,
\cr
\heff^{(2)}&=-m^\dag \om(M^\dag M)^{-1}\om^\dag m\,,
\cr
\heff^{(4)}&=
\Theta_2\om^\dag \om\Theta_2^\dag-\left(\ums{2}m^\dag m \Theta_2\Theta_2^\dag+ \text{h.c.}\right)\,,
\cr
\Heff^{(0)}&=M^\dag M\,,
\cr
\Heff^{(2)}&=\om^\dag\om\,,
\cr
\Heff^{(4)}&=+\ums{2}\Theta_2^\dag m^\dag \om+ \text{h.c.}\,.
}
These simple equations can be used to check the eigenvalues of the diagonalized blocks.

Considering~\cref{eq:diagonalization} and~\cref{F:VR,Fi:improved} to order $\eps^3$, we can now write a simple block-diagonalization formula at subleading order, in the vanishing $\,\oM$ WB:
\eqali{
\mtrx{\id_3-\ums{2}\Theta_1\Theta_1^\dag & -\Theta_1-\Theta_3+\ums{2}\Theta_1\Theta_1^\dag\Theta_1\cr \Theta_1^\dag+ \Theta_3^\dag -\ums{2}\Theta_1^\dag\Theta_1\Theta_1^\dag& \id_n-\ums{2}\Theta_1^\dag\Theta_1}
\mtrx{m & \om \cr0 & M}
\mtrx{\id_3 & \Theta_2\cr -\Theta_2^\dag & \id_n}&
\cr
\hspace{3em}=\mtrx{m-\ums{2}\Theta_1\Theta_1^\dag m & 0
\cr\rule{0ex}{1.5em}
0 & M+\ums{2}\Theta_1^\dag \Theta_1 M
}\,.
}
The next order corrections at block-entries (11), (12), (21) and (22) are $M$ times $\eps^5,\eps^5,\eps^4$ and $\eps^4$,  respectively.
We can see that the corrections decrease $m$ but increase $M$.

\vfill
\pagebreak

\providecommand{\href}[2]{#2}\begingroup\raggedright\endgroup

\end{document}